\documentclass[aoas]{imsart}
\RequirePackage[authoryear]{natbib}
\RequirePackage{amsthm,amsmath,amsfonts,amssymb}
\RequirePackage{graphicx}
\RequirePackage{verbatim,enumitem,lipsum,float,url,leftidx}
\RequirePackage{color,mathabx,subcaption,tikz,xr}
\usetikzlibrary{positioning}
\startlocaldefs
\newcommand{\indep}{\;\, \rule[0em]{.03em}{.67em} \hspace{-.25em}
	\rule[0em]{.65em}{.03em}
	\hspace{-.25em}\rule[0em]{.03em}{.67em}\;\,}

\newtheorem{Th}{\underline{\bf Theorem}}

\newtheorem{Pro}{Proposition}
\newtheorem{Lem}{\underline{\bf Lemma}}

\def\bse{\begin{eqnarray*}}
	\def\ese{\end{eqnarray*}}
\def\be{\begin{eqnarray}}
	\def\ee{\end{eqnarray}}
\def\bsq{\begin{equation*}}
	\def\esq{\end{equation*}}
\def\bq{\begin{equation}}
	\def\eq{\end{equation}}

\def\var{\hbox{var}}

\def\wh{\widehat}
\def\wt{\widetilde}
\def\eff{_{\rm eff}}

\def\n{\nonumber}

\def\vecl{\mbox{vecl}}

\def\sumi{\sum_{i=1}^n}
\def\sumj{\sum_{j=1}^n}
\def\trans{^{\rm T}}

\def\bb{\boldsymbol\beta}

\def\bg{\boldsymbol\gamma}

\def\0{{\bf 0}}
\def\A{{\bf A}}
\def\a{{\bf a}}
\def\B{{\bf B}}

\def\c{{\bf c}}

\def\g{{\bf g}}

\def\h{{\bf h}}

\def\b{{\bf b}}
\def\I{{\bf I}}

\def\K{{\bf K}}

\def\T{{\bf T}}

\def\bS{{\bf S}}

\def\m{{\bf m}}
\def\v{{\bf v}}

\def\T{{\bf T}}

\def\X{{\bf X}}
\def\x{{\bf x}}
\def\I{{\bf I}}

\def\z{{\bf z}}

\def\bLam{{\boldsymbol \Lambda}}
\def\blam{{\boldsymbol \lambda}}
\def\bkappa{\boldsymbol\kappa}

\def\brho{\boldsymbol\rho}

\def\bq{\begin{equation}}
	\def\eq{\end{equation}}
\def\pr{\hbox{pr}}
\def\wh{\widehat}
\def\wt{\widetilde}
\def\trans{^{\rm T}}

\def\log{\hbox{log}}

\def\squarebox#1{\hbox to #1{\hfill\vbox to #1{\vfill}}}

\def\var{\hbox{var}}

\def\bse{\begin{eqnarray*}}
	\def\ese{\end{eqnarray*}}
\def\be{\begin{eqnarray}}
	\def\ee{\end{eqnarray}}
\def\bsq{\begin{equation*}}
	\def\esq{\end{equation*}}
\def\bq{\begin{equation}}
	\def\eq{\end{equation}}
\def\pr{\hbox{pr}}
\def\wh{\widehat}
\def\wt{\widetilde}

\def\log{\hbox{log}}

\def\trans{^{\rm T}}

\def\boxit#1{\vbox{\hrule\hbox{\vrule\kern6pt\vbox{\kern6pt#1\kern6pt}\kern6pt\vrule}\hrule}}

\endlocaldefs

\begin{document}

\begin{frontmatter}
\title{Evaluation of transplant benefits with the U.S. Scientific Registry of Transplant Recipients by semiparametric regression of mean residual life}
\runtitle{Evaluation of transplant by mean residual life}

\begin{aug}
\author[A]{\fnms{Ge}~\snm{Zhao}\ead[label=e1]{gzhao@pdx.edu}},
\author[B]{\fnms{Yanyuan}~\snm{Ma}\ead[label=e2]{yzm63@psu.edu}}
\author[C]{\fnms{Huazhen}~\snm{Lin}\ead[label=e3]{linhz@swufe.edu.cn}}
\and
\author[D]{\fnms{Yi}~\snm{Li}\ead[label=e4]{yili@med.umich.edu}}

\address[A]{Department of Mathematics and Statistics, Portland State University\printead[presep={,\ }]{e1}}

\address[B]{Department of Statistics, Pennsylvania State University\printead[presep={,\ }]{e2}}
\address[C]{Center of Statistical Research,  School of Statistics, Southwestern University of
	Finance and Economics, China\printead[presep={,\ }]{e3}}
\address[D]{Department of Biostatistics, University of Michigan,
	Ann Arbor\printead[presep={,\ }]{e4}}
\end{aug}

\begin{abstract}
Kidney transplantation is the most effective renal replacement therapy for end stage renal disease patients. With the severe shortage of kidney supplies and for the clinical effectiveness of transplantation, patient's life expectancy post transplantation is
used to prioritize patients for transplantation; however, severe comorbidity conditions and old age are the most dominant factors
that negatively impact post-transplantation life expectancy,  effectively
precluding sick or old patients from receiving transplants.  
It would be crucial to design objective measures to quantify the transplantation
benefit by comparing the mean residual life with and without a transplant, after adjusting for comorbidity and demographic conditions. To address this urgent need, we propose a new class of semiparametric  covariate-dependent mean residual life models. Our method estimates covariate effects semiparametrically efficiently and the mean
residual life function nonparametrically, enabling us to  predict the residual life increment potential for any given patient. Our method potentially leads to a more fair system that  prioritizes patients who would have the largest residual life gains. Our
analysis of the kidney transplant data from the U.S. Scientific
Registry of Transplant Recipients also suggests that a single index of
covariates  summarize well the impacts of multiple covariates, which may facilitate interpretations of each covariate's effect. Our subgroup analysis further disclosed inequalities in survival gains across groups defined by
race, gender and insurance type (reflecting socioeconomic status).
\end{abstract}

\begin{keyword}
\kwd{Mean residual life}
\kwd{kidney transplant}
\kwd{semiparametric method}
\end{keyword}

\end{frontmatter}


\section{Introduction}

 About 15\% of American adults have chronic kidney disease \citep{saran2016us}, suffering worsened kidney functions with less fluid filtrated by the glomerular, and losing kidney functions gradually but permanently
over the cause of months or years.  According to the glomerular
filtration rate (GFR), chronic kidney disease 
is  classified into five stages, where
stage four (GFR between 15 and 29 ml/min/1.73$m^2$)
and stage five (GFR less than 15 ml/min/1.73$m^2$) kidney
diseases are considered to be  end stage renal disease (ESRD),  one of the most lethal diseases globally \citep{ferri2017,feng2019efficacy}. In the U.S.,
more than 600,000 individuals are living with ESRD, about 100,000 new ESRD cases are diagnosed and 50,000 deaths occur each year \citep{salerno2021covid}.

The most common treatment for ESRD is renal replacement
therapy, including dialysis and kidney transplant.
As dialysis  only provides partial kidney functions,  dialysis patients tend to have shorter
survival than those receiving kidney transplants, which often lead to a longer and a better quality of life  \citep{evans1985quality,wolfe1999comparison,liem2007quality}. Due to severe shortages in kidney supplies, however, there are far more ESRD patients who
need kidney
transplants than donors available in the U.S. \citep{tonelli2011systematic}. For example, the U.S. Scientific
Registry
of Transplant Recipients (SRTR)
reports that among 247,123 patients awaiting kidney transplants during  2011-2018, only 139,270 patients actually received one, leaving the remaining 107,853 still waiting \citep{hart2021optn}.

Currently, decisions on patients' priority of
receiving  kidney transplants are based on 
the Estimated Post-Transplant Survival (EPTS) score, which predicts  a patient's life expectancy  post transplantation by using a Cox model with
age, diabetes status, prior solid organ transplant and time on
dialysis as predictors \citep{time2012guide}.
Pre-existing conditions such as 
diabetes, prior solid organ transplants and long dialysis vintage are associated with
shorter survival \citep{cosio1998patient,meier2000effect,kasiske2001evaluation}; thus patients with these conditions tend to have a lower priority for  transplantation \citep{cosio1998patient,molnar2011association}. On the other hand, younger age  is found to be associated with better outcomes  and 
younger patients are likely to
have a higher  priority for  transplantation.  Thus, age and severe comorbidity conditions have effectively become the most dominant factors when deciding on who to receive transplants, which may preclude older and sicker patients from benefiting from transplantation \citep{jassal2005baseline,gore2009disparities,weng2010barriers}. A more comprehensive system, however,  should  give a higher priority to those who would
 benefit more from transplantation among patients with similar conditions, and in the meantime triage candidates who may gain little or even suffer a loss in life expectancy. We propose to quantify the transplant
benefit by comparing the improvement of the patient's
expected residual life with and without transplantation.
The expected residual
life characterizes the mean of the remaining survival time
given that a patient has survived up to a certain time \citep{hall1981mean}. Compared to
overall survival,  the residual life expectancy provides a real time 
assessment of transplant benefits at any given time when a
kidney becomes available  \citep{lin2016semiparametrically}.
As demographic and clinical  conditions may be confounders  affecting  survival and should be 
adjusted for when assessing transplant benefits \citep{cosio1998patient,carrero2018sex},
we  aim at modeling and evaluating a patient's potential
residual life expectancy with or without transplant,
based on the patient's covariate profile.

Much work on mean residual life models has been sparked by  \cite{oakesdasu1990}. For example,
\cite{maguluri1994estimation}  proposed a univariate
proportional mean residual life model;
\cite{oakes2003inference} established the theoretical properties
of the methods in \cite{oakesdasu1990};
\cite{chencheng2005} estimated the coefficients of covariates  in a   proportional mean
residual life model by a partial-score approach, analogous to the partial likelihood
approach; 
\cite{chenetal2005} employed the inverse probability weighting
approach for inference; 
\cite{muller2005time} extended the
mean residual life model to incorporate time-varying covariates; \cite{chen2006linear}
proposed an extended Buckley-James estimator to estimate a linear residual life model and 
\cite{chen2007} further proposed an additive mean residual life
model.
These works inspired median and  quantile residual life
models; see for example, \cite{jeong2008nonparametric,
	jung2009regression, ma2010semiparametric} and
\cite{ma2012analysis}.
However, all
these works imposed parametric dependency of residual life on  covariates as well as how long the patient has lived up to transplantation (or ``alive time'' hereafter). 
Violations of the model assumptions will lead
to biased estimates and incorrect inferences \citep{chen2007, chenetal2005}. Our preliminary analysis of the kidney transplant data from SRTR indicates that the mean residual life depends on alive time and  patients' other covariates, such as treatment history, commorbidty conditions and demographics, through a complicated form which is challenging to model
parametrically. 

We  propose a new class of nonparametric mean residual life models, with the goal of detecting the effects of patients' covariates
on the residual life  and  identifying the patients
who may benefit  most from transplantation. Our model does not impose any parametric assumptions on
the mean residual life function and, thus,  the hazard function, and 
extends the model in \cite{ma2012} and \cite{ma2013efficient} to
accommodate censoring in response. Moreover, given multiple covariates,  we also propose a flexible dimension reduction method to achieve a parsimonious model for efficiency and interpretability.
To derive the estimators,  we employ a semiparametric method, in combination with a martingale
treatment as in \cite{zhao2021efficient}, to derive a semiparametrically efficient  estimator  \citep{bickeletal1994} for
the effects of  covariates and an asymptotically normally distributed nonparametric
estimator of the mean residual life function.
We apply  the proposed method to analyze the SRTR kidney transplant data and quantify
transplantation gains by using the residual life expectancy. 
 Our
analysis  suggested that a single index of
covariates  summarize well the impacts of multiple covariates, which may facilitate interpretations of each covariate's effect. Our subgroup analysis further disclosed inequalities in survival gains across groups defined by
race, gender and insurance type (reflecting socioeconomic status). The results may inform the priority rules for kidney transplantation.

This paper is organized as follows. Section
\ref{sec:model} proposes the mean residual life model, and  Section \ref{sec:estimation} derives the  estimators for
the proposed model and discusses their properties.
We assess the finite
sample properties
of the methods by simulation studies
in Section \ref{sec:sim} and apply it to analyze the kidney
transplant data in
Section \ref{sec:app}. We conclude the paper with some
discussions in Section \ref{sec:conclusion}. We defer the regularity conditions and technical properties to the Supplementary Materials.

\section{Semiparametric regression of mean residual life}\label{sec:model}
Denote by $T$ the potential time lag from being waitlisted for transplantation (i.e., became eligible) to death in the absence of censoring and by $\X\in{\cal R}^p$ the baseline covariates, such as age, diabetes status, and prior solid organ transplant,  measured at the waitlisting time. Denote by $W$ the time lag from waitlisting to hypothetical transplant time that would have occurred in the absence of censoring. Our focus is to model the difference of the mean residual life with and without  transplant  at any time point $t$, given $\X$ and $W$.  

Let the indicator function $I(W\le t)$ 
describe the time-dependent transplant status, with $I(W\le t)=0$ and $1$
corresponding to ``Non-transplant'' and ``Transplant'' at time $t$, respectively.
Following the missing data literature, we use $WI(W \le t)$ to indicate the value of
$W$ {\em only} when the transplant occurs before $t$.
Given the history of transplantation status up to time t, i.e., $\{I(W \le t), WI(W \le t)\}$, we specify that the conditional hazard at $t$ depends only on the transplantation information at $t$, that is,
\be\label{eq:modelhazard}
&&\lim_{h \rightarrow 0}  h^{-1}P\{ t \le T \le t+h\mid   T\ge t, \X, I(W\le t),WI(W\le t)\}\n\\
&=&\lambda\{t,\X,I(W\le t),WI(W\le t)\}\n\\
&=&
\lambda_T(t-W,\X,W)I(W\le t)+\lambda_N(t,\X)\{1-I(W\le t)\},
\ee
where the subscripts ``$_T$'' and ``$_N$'' respectively stand for ``Transplant'' and  ``Non-transplant.''  Within the non-transplant group by time $t$, i.e. $t<W$, the hazard function depends on the time and covariates only; at and after transplantation, i.e. $t\ge W$, the hazard function is to be reset and is a function of $t-W$ (the time lag since transplantation)  because of immediate surgical risks \citep{humar2005surgical,hernandez2006retrospective} and long term benefits of receiving functional organs \citep{lin2016semiparametrically}. Additionally, $W$ is considered as an influential factor in $\lambda_T$ because, for example,  there is a clear  survival advantage in favor of preemptive kidney transplantation \citep{liem2009early}.

A naive mean residual life  \citep{maguluri1994estimation} could have been computed as
\be  
&& E(T-t\mid T\ge t, \X, I(W\le t),WI(W\le t)) \n \\
&=&e^{\Lambda_T(t-W, \X,W)}\int_t^{\infty}e^{-\Lambda_T(s-W, \X,W)}ds I(W\le t)\n\\
&&+\left\{e^{\Lambda_N(t, \X)}\int_t^{W}e^{-\Lambda_N(s, \X)}ds+\int_W^{\infty}e^{-\Lambda_T(s-W, \X)}ds \right\} I(W> t),
\nonumber
\ee
where $\Lambda_N(t,\X) = \int_0^t\lambda_N(s,\X)ds$ and $\Lambda_T(t,\X,W) = \int_0^t\lambda_T(s,\X,W)ds$ are the two cumulative hazard functions. 
 However, the conditioning part of this expectation looks beyond $t$  for  a prospective $W>t$, which is problematic as a patient would be guaranteed to survive at least up to $W$ when  $W>t$, coinciding with the notion that one cannot directly use
  time dependent treatment or, more broadly, ``internal'' time dependent covariates to predict survival \citep{kalbfleisch2011statistical}.   
 Instead, at each time $t$, we compute the mean residual life by strictly conditioning on the information available by then, that is, 
\be
&& E(T-t\mid T\ge t, \X, I(W\le t),  WI(W\le t)) \n \\
&=&e^{\Lambda_T(t-W,\X,W)}\int_t^\infty
e^{-\Lambda_T(s-W,\X,W)}ds I(W\le t)+e^{\Lambda_N(t,\X)} \int_t^ \infty e^{-\Lambda_N(s,\X)}ds
I(W> t),\n\\
\label{eq:meanResidualLife}
\ee
and will draw inference based on this valid model.

For ease of  notation, we rewrite as (\ref{eq:meanResidualLife})
as $m\{t,\X,I(W\le t),WI(W\le t)\} =m_T(t-W,\X,W)I(W\le t)+m_N(t,\X)\{1-I(W\le t)\}$,
where
\bse
m_T(t-W,\X,W) =e^{\Lambda_T(t-W,\X,W)}\int_t^\infty
e^{-\Lambda_T(s-W,\X,W)}ds
\ese
and 
\bse
m_N(t,\X)
=e^{\Lambda_N(t,\X)}\int_t^\infty e^{-\Lambda_N(s,\X)}ds,
\ese
which may facilitate evaluation of the benefits of transplant at any given time. 
Particularly, 
$m_T(t-W,\X,W) - m_N(t,\X)$ quantifies the gain (or loss) of life expectancy of  patients at $t$ with a transplant given at $W<t$ compared with those
 who would never receive a transplant; candidates with close to zero or a negative value of $m_T(t-W,\X,W) - m_N(t,\X)$ would benefit little from organ transplantation 
 and would have lower priorities in the waiting list \citep{chadban2020kdigo}.

To ensure estimability, we make a complete follow-up assumption
  \citep{tsiatis1990estimating,chenetal2005,chencheng2005,sun2009class}, that is,
 the failure time $T$ is supported on a finite range 
   $(0, \tau)$ with  $\tau <\infty$, where in practice $\tau$ is the maximum follow-up time;  
we relax this assumption in Supplement \ref{sec:relax}. We further assume the covariates $\X$ affect $T$ via index $\bb$, where $\bb\in{\cal R}^{p\times
d}$ is the coefficient matrix with $d \le p$. Then (\ref{eq:modelhazard}) and (\ref{eq:meanResidualLife}) can be re-expressed as
\be
&&\lambda\{t,\X,I(W\le t),WI(W\le t)\}\n\\
&=&\lambda_T(t-W,\bb\trans\X,W)I(W\le t)+\lambda_N(t,\bb\trans\X)\{1-I(W\le t)\},\label{eq:hazardDR}\\
&&m\{t,\X,I(W\le t),WI(W\le t)\}\n\\
&=&
m_T(t-W,\bb\trans\X,W)I(W\le t)+m_N(t,\bb\trans\X)\{1-I(W\le t)\},\label{eq:meanResidualLifeDR}
\ee
where $\lambda_T, \lambda_N, m_T$ and $m_N$ are unspecified positive functions, which 
need to be estimated. 
 The model stipulates that the conditional mean of $T-t$ depends on $\X$ via its $d$ indices, formed by projecting $\X$ to the columns of $\bb$, and the waiting time $W$.
When $d=1$, the model reduces to a single index model in terms of $\X$; when $
1<d<p$, it corresponds to a dimension reduction structure; when $d=p$, the model is completely
	nonparametric. Our analysis first focuses on a fixed $d$,
	followed by  selecting $d$ in a data driven fashion as
	discussed in Section~\ref{sec:app}.
Model (\ref{eq:meanResidualLifeDR}) is general: it includes the proportional mean
residual life model, i.e.,   $m\{t,\bb\trans\X,I(W\le t),WI(W\le t)\}=m_0(t)\exp(\bb\trans\X)$ \citep{oakesdasu1990}
as a special case by specifying
$m_N\{t,\bb\trans\X\}=m_0(t)e^{\bb\trans\X}$, $m_T\{t-W,\bb\trans\X,W\}=m_0(t)e^{\bb\trans\X+\alpha W}$ with $\alpha=0$, 
 and $d=1$;
it  reduces to the additive model $m_0(t)+\bb\trans\X$ \citep{chen2007} by specifying
$m_N\{t,\bb\trans\X\}=m_0(t)+\bb\trans\X$, $m_T\{t-W,\bb\trans\X,W\}=m_0(t)+\bb\trans\X+\alpha W$ with $\alpha=0$, and setting $d=1$.
By allowing $d$ to be larger than 1,  model
(\ref{eq:meanResidualLifeDR}) extends these classical models by allowing  more flexible forms such as
$m\{t,\bb\trans\X,I(W\le t),WI(W\le t)\}=m_0(t)\{\sum_{k=1}^d\exp(\bb_{\cdot,k}\trans\X)\}$
and
$m\{t,\bb\trans\X,I(W\le t),WI(W\le t)\}=m_0(t)+\sum_{k=1}^d\bb_{\cdot,k}\trans\X$,
where $\bb_{\cdot,k}$ is the $k$th column of $\bb$. These special cases implicitly
assume that transplant or the timing of transplantation does not impact survival.

We further assume that $T$ is subject to random
right censoring so that $C\indep T\mid
W,\X$, where $C$ is the censoring time and  we observe  $Z=\min(T,C)$ and $\Delta=I(T\le
C)$. Let the observed $\{\X_i,
Z_i,\Delta_i,  W_i \}$, $i=1, \dots, n$ be independently and identically distributed  realizations of $\{\X,
Z,\Delta, W \}$. To make  (\ref{eq:meanResidualLifeDR})  identifiable and estimable, we fix
the
upper $d\times d$ block of
$\bb$
to be $\I_d$, and
estimate the lower $(p-d)\times d$ block of
$\bb$.  Corresponding to the upper and lower parts of $\bb$, we write $\X=(\X_u\trans,
\X_l\trans)\trans$, where $\X_u\in{\cal R}^d$ and $\X_l\in{\cal R}^{p-d}$.

\section{A semiparametrically efficient estimator}\label{sec:estimation}

Denote the conditional survival function, cumulative hazard
function, hazard function and  probability density function
(pdf) of the censoring time $C$ by
$S_c(z,\X)=\pr (C\ge z\mid \X)$,
$\Lambda_c(z,\X)=-\log S_c(z,\X)$,
$\lambda_c(z,\X)=\partial\Lambda_c(z,\X)/\partial z$ and
$f_c(z,\X)=-\partial S_c(z,\X)/\partial z$ with
$z<\tau$, where
 $0<\tau <\infty$ is the upper bound of the follow-up time.
Let
	$p(\X)\equiv\pr(C\leq\tau\mid\X)$,
	$S_c(\tau,\X)=f_c(\tau,\X)=p(\X)$, and	
$\lambda_c(\tau, \X)=1$.
	Here,
	$\lambda_c(z,\X)$ and $f_c(z,\X)$ are absolutely
	continuous on $(0, \tau)$, but with
	a discontinuity point at $\tau$.

To estimate $ m_T(t-W,\bb\trans\X,W) - m_N(t,\bb\trans\X)$, which quantifies the gain (or loss)  of mean residual life after $t$ with transplant given at $W<t$, we need to estimate
$\bb$ and the functionals of $m_T$ and $m_N$, for which we consider a likelihood-based approach.

Under independent censoring, the pdf of $\{\X,Z,\Delta,W I(W \le Z)\}$ is
\be\label{eq:pdf}
&&f_{\X,Z,\Delta, W}\{\x,z,\delta,w\}\n\\
&=& \{\lambda_T(z-w,{\bb}\trans\x,w) \}^\delta
e^{-\int_0^w\lambda_N(s,{\bb}\trans\x)ds - \int_w^z\lambda_T(s-w,{\bb}\trans\x,w)ds   }
\lambda_c(z,\x)^{1-\delta}
e^{-\int_0^z\lambda_c(s,\x)ds}\n\\
&&\quad\times f_{\X,W}(\x,w) I(w \le  z) \n\\
& & +  \{\lambda_N(z,{\bb}\trans\x) \}^\delta
e^{-\int_0^z\lambda_N(s,{\bb}\trans\x)ds }
\lambda_c(z,\x)^{1-\delta}
e^{-\int_0^z\lambda_c(s,\x)ds}f_{\X}(\x)
I(w>z). \ee
We do not model $W$ because the surgery schedule mainly depends on the
organ availability  other than $\X$.  
We view the pdf in (\ref{eq:pdf}) as a semiparametric model
where
all unknown components, except for $\bb$, are infinite dimensional
nuisance parameters.  The parameters $\bb$ are
parameters of interest with a finite
dimension. We will  estimate
$\bb$ by using
a geometric approach, which
avoids decomposing
$\lambda(\cdot)$ to be $\lambda_*(z) e^{\bb\trans\X}$ as in a proportional hazards model.
This entails more flexibility for the model. 

 Let $Y(t) = I(Z\ge t)$ and $N(t) = I(Z\le t)\Delta$  be the at-risk and counting process, respectively. Define the filtration ${\cal F}_t$ to be $\sigma\{
N(u), Y(u), \X, I(W\le u),WI(W\le u), 0\le u < t\}$,  and let $M(t)= N(t)-\int_0^t
Y(s)\lambda\{s,{\bb}\trans\X,I(W\le s),WI(W\le s)\}ds$ be
the martingale with respect to ${\cal F}_t$. In ${\cal F}_t$, only  $W$ that is less than $t$ can be observed in order to maintain a valid filtration; otherwise, if $W>t$ were to be included, it would introduce unpredictable future information and violate the definition of ${\cal F}_t$ as a filtration. This principle also clarifies the design choice in model (\ref{eq:modelhazard}), where values of $W>t$ are treated as missing and not incorporated into the analysis. 
  
Define the score function
$
\bS_{\bb}\{\Delta,Z,\X, I(W\le Z),WI(W\le Z)\}
=  \\
\int_0^\infty
\left[\frac{\m_{12}\{s,{\bb}\trans\X,I(W\le s),WI(W\le s)\}}{m_1\{s,{\bb}\trans\X,I(W\le s),WI(W\le s)\}+1} \\
-\frac{\m_2\{s,{\bb}\trans\X,I(W\le s),WI(W\le s)\}
}{m\{s,{\bb}\trans\X,I(W\le s),WI(W\le s)\}}\right]\otimes\X_l
dM\{s,{\bb}\trans\X,I(W\le s),WI(W\le s)\},
$
where  $m_{1}(s,\v, \cdot,\cdot)\equiv
\partial m(s,\v,\cdot,\cdot)/\partial s =
  \partial m_T(s-W,\v,W)/\partial s I(W\le s)+\partial m_N(s,\v)/\partial s \{1-I(W\le s)\}$, $\m_{2}(s,\v,\cdot,\cdot)\equiv
\partial m(s,\v,\cdot,\cdot)/\partial \v=
\partial m_T(s-W,\v,W)/\partial \v \allowbreak I(W\le s)+\partial m_N(s,\v)/\partial \v \{1-I(W\le s)\}$, $\m_{12}(s,\v,\cdot,\cdot)\equiv
\partial\m_2\{s,\v,\cdot,\cdot\}/\partial s = 
\partial^2 m_T(s-W,\v,W)\allowbreak /\partial s\partial \v I(W\le s)+\partial ^2 m_N(s,\v)/\partial s\partial \v \{1-I(W\le s)\}$ almost surely, and  $\X_l$
is the lower $p-d$ components in $\X$.
 
 As derived in
Supplementary \ref{sec:score}, an efficient score is
\be
&&\bS\eff\{\Delta,Z,{\bb_0}\trans\X,I(W\le Z),WI(W\le Z)\}\n\\
&=&\int_0^\infty
\left\{\frac{\m_{12}\{s,{\bb_0}\trans\X,I(W\le s),WI(W\le s)\}}{m_1\{s,\bb_0\trans\X,I(W\le s),WI(W\le s)\}+1}
-\frac{\m_2\{s,{\bb_0}\trans\X,I(W\le s),WI(W\le s)\} }{m(s,{\bb_0}\trans\X,I(W\le s),WI(W\le s))}\right\}\n\\
&&\otimes
\left[\X_l-
\frac{E\left\{\X_l
S_c(s,\X)\mid{\bb_0}\trans\X I(W\le s),WI(W\le s)\right\}}{E\left\{S_c(s,\X)\mid{\bb_0}\trans\X,I(W\le s),WI(W\le s)\right\}}\right]\n\\
&&\quad dM\{s,{\bb_0}\trans\X,I(W\le s),WI(W\le s)\},\n
\ee
based on which, we construct a semiparametrically efficient estimator of $\bb$.

First, a consistent estimating equation can be obtained
from
$E[\bS\eff\{\Delta,Z,\X,I(W\le Z),WI(W\le Z)\}\mid\X]=\0$ as
the integrand in the above integral is predictable and $M\{s,\bb_0\trans\X,I(W\le s),WI(W\le s)\}$ is a martingale. Hence, to preserve the mean zero
property and to simplify the computation, we can replace
$
\frac{\m_{12}\{s,{\bb_0}\trans\X,I(W\le s),WI(W\le s)\}}{m_1\{s,{\bb_0}\trans\X,I(W\le s),WI(W\le s)\}+1}
-\frac{\m_2\{s,{\bb_0}\trans\X,I(W\le s),WI(W\le s)\}}{m\{s,{\bb_0}\trans\X,I(W\le s),WI(W\le s)\}}
$
by an arbitrary function of $s$ and $\bb_0\trans\X$, say $\g\{s,{\bb_0}\trans\X,I(W\le s),WI(W\le s)\}$, and
still obtain
\bse
&&E\left(\int_0^\infty\g\{s,{\bb_0}\trans\X,I(W\le s),WI(W\le s)\}\right.\\
&&\left.\otimes\left[\X_l-
\frac{E\left\{\X_l
	S_c(s,\X)\mid{\bb_0}\trans\X\right\}}
{E\left\{S_c(s,\X)\mid{\bb_0}\trans\X\right\}}\right]dM\{s,{\bb_0}\trans\X,I(W\le s),WI(W\le s)\}\right)=\0.
\ese
This provides a richer class of estimators than the estimator based on
$\bS\eff$ alone.

Second, we can obtain
\be\label{eq:useful}
\frac{E\left\{ \X_lY(t)\mid{\bb_0}\trans\X,I(W \le t),WI(W\le t)\right\}}
{E\left\{Y(t)\mid{\bb_0}\trans\X,I(W \le t),WI(W\le t)\right\}}
=\frac{E\left\{\X_l
	S_c(t,\X)\mid{\bb_0}\trans\X\right\}}
{E\left\{S_c(t,\X)\mid{\bb_0}\trans\X\right\}},
\ee
where we define (\ref{eq:useful})
to be
$E\left\{\X_l
	p(\X)\mid{\bb_0}\trans\X\right\}/E\left\{p(\X)\mid{\bb_0}\trans\X\right\}$ when $t>\tau$.
We then verify that
\be\label{eq:expecteff}
E\left(\int_0^\infty\g\{s,{\bb_0}\trans\X,I(W\le s),WI(W\le s)\}
\otimes\left[\X_l-
\frac{E\left\{\X_l
	S_c(s,\X)\mid{\bb_0}\trans\X\right\}}
{E\left\{S_c(s,\X)\mid{\bb_0}\trans\X\right\}}\right]dN(s)\right)=\0.\n\\
\ee
The proof of (\ref{eq:useful}) and
(\ref{eq:expecteff}) is given
in Supplement \ref{app:eff}.
This implies that we can construct estimating equations for different transplant status of the form
\be\label{eq:general}
&&\sumi \Delta_i\g\{Z_i,{\bb_0}\trans\X_i,I(W_i\le Z_i),W_i I(W_i\le Z_i)\}\n\\
&&\otimes\left[\X_{li}-\frac{\wh E\left\{\X_{li}
	Y_i(Z_i)\mid\bb\trans\X_i,I(W_i\le Z_i),W_iI(W_i\le Z_i)\right\}}
{\wh E\left\{Y_i(Z_i)\mid\bb\trans\X_i, I(W_i\le Z_i),W_iI(W_i\le Z_i)\right\}}\right]=\0\n\\
\ee
for any $\g(\cdot,\cdot)$ with
\be
&&\wh E\left\{Y_i(Z_i)\mid\bb\trans\X_i,I(W_i\le Z_i),W_iI(W_i\le Z_i)\right\}\n\\
&=& \frac{\sumj K_h(\bb\trans\X_j-\bb\trans\X_i,W_j-W_i)I(Z_j\ge Z_i)I(W_j\le Z_j)}
{\sumj K_h(\bb\trans\X_j-\bb\trans\X_i,W_j-W_i)I(W_j\le Z_j)}I(W_i\le Z_i)\n\\
&&+\frac{\sumj K_h(\bb\trans\X_j-\bb\trans\X_i)I(Z_j\ge Z_i)\{1-I(W_j\le Z_j)\}}
{\sumj K_h(\bb\trans\X_j-\bb\trans\X_i)\{1-I(W_j\le Z_j)\}}\{1-I(W_i\le Z_i)\},\label{eq:expectY}\\
&&\wh E\left\{\X_{li}Y_i(Z_i)\mid\bb\trans\X_i,I(W_i\le Z_i),W_iI(W_i\le Z_i)\right\}\n\\
&=& \frac{\sumj K_h(\bb\trans\X_j-\bb\trans\X_i,W_j-W_i)\X_{lj}I(Z_j\ge
	Z_i)I(W_j\le Z_j)}
{\sumj K_h(\bb\trans\X_j-\bb\trans\X_i,W_j-W_i)I(W_j\le Z_j)}I(W_i\le Z_i)\n\\
&&+\frac{\sumj K_h(\bb\trans\X_j-\bb\trans\X_i)\X_{lj}I(Z_j\ge
	Z_i)\{1-I(W_j\le Z_j)\}}
{\sumj K_h(\bb\trans\X_j-\bb\trans\X_i)\{1-I(W_j\le Z_j)\}}\{1-I(W_i\le Z_i)\},\label{eq:expectXY}
\ee
with $K_h(u)=K(u/h)/h$ for univariate $u$ and $K_h(u_1,u_2,...,u_q)=\prod_{i=1}^q K(u_i/h_i)/h_i$ for multivariate $(u_1,u_2,...,u_q)$ with a bandwidth vector
$h=(h_1, \ldots, h_q)$, where $K(\cdot)$ is the standard kernel function
so that $K(u) \ge 0$ and  $\int_{-\infty}^\infty K(u)du = 1$ 
\citep{wand1994fast}.
Here, $\wh E\{Y_i(Z_i)\mid \bb\trans\X_i,I(W_i\le Z_i),W_iI(W_i\le Z_i)\}\equiv
\wh E\{Y_i(t)\mid\bb\trans\X_i,I(W_i\le Z_i),W_iI(W_i\le Z_i)\}|_{t=Z_i}$
and similarly for other terms.

Third, 
we  obtain the
efficient estimator of $\bb$ by solving
\be\label{eq:eff}
&&\sumi \Delta_i
\left[\frac{\wh\m_{12}\{Z_i,{\bb}\trans\X_i,I(W_i\le Z_i),W_i I(W_i\le Z_i)\}}{\wh m_1\{Z_i,{\bb}\trans\X_i,I(W_i\le Z_i),W_i I(W_i\le Z_i)\}+1}\right.\n\\
&&\quad\left.-\frac{\wh\m_2\{Z_i,{\bb}\trans\X_i,I(W_i\le Z_i),W_i I(W_i\le Z_i)\}}{\wh m\{Z_i,{\bb}\trans\X_i,I(W_i\le Z_i),W_i I(W_i\le Z_i)\}}
\right]\n\\
&&\otimes\left[\X_{li}-
\frac{\wh E\left\{\X_{li}
	Y_i(Z_i)\mid\bb\trans\X_i,I(W_i\le Z_i),W_i I(W_i\le Z_i)\right\}}
{\wh E\left\{Y_i(Z_i)\mid\bb\trans\X_i,I(W_i\le Z_i),W_i I(W_i\le Z_i)\right\}}\right]=\0,\n\\
\ee
where $\wh m_1(t,\v,\cdot,\cdot)$, $\wh\m_2(t,\v,\cdot,\cdot)$,   $\wh\m_{12}(t,\v,\cdot,\cdot)$ are estimators for the derivatives of  $m(t, \v,\cdot,\cdot)$ with respect to the first two elements given in Supplement \ref{sec:nonpara}.

Finally, we estimate $m\{t,{\bb}\trans\X,I(W\le t),W I(W\le t)\}$ nonparametrically via $\wh\Lambda_T\{t-W,{\bb}\trans\X,W\}I(W\le t)+\Lambda_N\{t,{\bb}\trans\X\}\{1-I(W\le t)\}$ based on a kernel
smoothed version of the Nelson-Aalen estimator \citep{ramlau1983choice, andersenborgan}.  For any $t$, $W$ with $W<t$, and $\bb\trans\X$,  the estimators, $\wh\Lambda_T\{t-W,{\bb}\trans\X,W\}$ and $\wh\Lambda_N\{t,{\bb}\trans\X\}$, have the forms of 
\bse
\wh\Lambda_T(t,\bb\trans\X,W)&=&\sum_{i=1}^n\int_{0}^{t}
\frac{I(W_i\le s)K_h(\bb\trans\X_i-\bb\trans\X,W_i-W)}{\sumj
Y_j(s) I(W_j\le s) K_h(\bb\trans\X_j-\bb\trans\X,W_j-W)}
dN_i(s),\\
\wh\Lambda_N(t,\bb\trans\X)&=&\sum_{i=1}^n\int_{0}^{t}
\frac{ I(W_i> s) K_h(\bb\trans\X_i-\bb\trans\X)}{\sumj
Y_j(s) I(W_j> s) K_h(\bb\trans\X_j-\bb\trans\X)}
dN_i(s).
\ese

Following \cite{maguluri1994estimation},  we obtain
\be\label{eq:m}
& & \wh m_T(t-W,{\bb}\trans\X,W)\n\\
& = & 
e^{\wh\Lambda_T(t-W,{\bb}\trans\X,W)}\int_{t-W}^\infty
e^{-\wh\Lambda_T(s,{\bb}\trans\X,W)}ds;   {\rm when} \, W<t; \n\\
& & \wh m_N(t,{\bb}\trans\X)\n\\
& = & 
e^{\wh\Lambda_N(t,{\bb}\trans\X)}\int_t^\infty e^{-\wh\Lambda_N(s,{\bb}\trans\X)}ds;  {\rm when}  \, W>t.
\n\\
\ee

We show in Supplement \ref{sec:asymp} that the estimators of $\bb$ are root-$n$ consistent,  asymptotically normally distributed and semiparametrically
efficient, and the nonparametric estimators,  $\wh m_T( \cdot,\cdot,\cdot)$
and $\wh m_N(\cdot, \cdot)$,  are asymptotically normally distributed. 

\section{Simulation}\label{sec:sim}

The section features four simulation studies for evaluating the
finite sample performance of our
method  in estimating  $\bb$ and
$m(t,\bb\trans\X)$.   For comparisons, we additionally  
	implement a semiparametric proportional mean
	residual life model, denoted as ``PM''  \citep{chencheng2005}.
This competing method implicitly assumes $d=1$.

\noindent
{\bf Study 1:} 
We generate event times with  hazard functions of
$
\lambda_N(t,\bb\trans\X) = te^{\bb\trans\X}
$ and 
$
\lambda_T(t,\bb\trans\X,W) = \frac{10e^{\bb\trans\X+W}+1}{t+1}
$
so that the true mean residual life is
\bse
m_N(t,\bb\trans\X)&=&e^{\frac{t^2}{2e^{\bb\trans\X}}}\Phi\left(-\frac{t}{\sqrt{e^{\bb\trans\X}}}\right)\sqrt{2\pi},\\
m_T(t,\bb\trans\X,W)&=&\frac{t+1}{10e^{\bb\trans\X+W}},
\ese
where $\Phi$ is the probability function of standard normal distribution.
Each component in $\X$ is generated independently from the
standard normal
distribution. The waiting time $W$ is generated independently from a uniform distribution over $[0,10]$. We consider $d=1, p=9$ and
set the true parameters
to be $\bb=(1,-0.6 ,0.0,-0.3,-0.1,0.0,0.1,0.3,-0.5)\trans$. The sample size is $n=300$ and we randomly assign around one third of samples to take the transplant. 

\noindent
{\bf Study 2:}  
We generate event times with  hazard functions of 
$
\lambda_N(t,\bb\trans\X) = \frac{2t}{e^{\bb\trans\X}+t^2}
$ and 
$
\lambda_T(t,\bb\trans\X,W) = \phi\{\ln(t)-3-W/100+0.1(1-\sqrt{2}\bb\trans\X)^2\}/t[\Phi\{-\ln(t)+3+W/100-0.1(1-\sqrt{2}\bb\trans\X)^2\}]
$
so that the true mean residual life is
\bse
m_N(t,\bb\trans\X)&=&\left(1+\frac{t^2}{e^{\bb\trans\X}}\right)\left\{\frac{\pi}{2}-\tan^{-1}\left(\frac{t}{e^{\bb\trans\X}}\right)\right\},\\
m_T(t,\bb\trans\X,W)&=&\Phi\left\{-\ln(t)+3+W/100-0.1(1-\sqrt{2}\bb\trans\X)^2\right\}\\
&&\times\int_t^{\infty}\frac{1}{\Phi\{3+W/100-\ln(s)-0.1(1-\sqrt{2}\bb\trans\X)^2\}}ds,
\ese
where $\phi$ is the probability density function of standard normal distribution.
Each component in $\X$ is generated independently from the
standard normal
distribution and $W$ is generated independently from uniform distribution over $[0,200]$. We consider $d=1, p=9$ and
set the true parameters
to be $\bb = (1,-0.6,0,-0.3,-0.1,0,0.1,0.3,-0.5)\trans$. The sample size is $n=1000$ and we randomly assign around one third of samples to take the transplant.

\noindent
{\bf Study 3:} The hazard functions are generated from $
\lambda_N(t,\bb\trans\X) = t^{2/5}\sum_{i=1}^d e^{{\bb_i}\trans\X}
$ and 
$
\lambda_T(t,\bb\trans\X,W) = t^{7/5}W\sum_{i=1}^de^{{\bb_i}\trans\X}.
$
The corresponding mean residual lives are
\bse
m_N(t,\bb\trans\X)&=&e^{\frac{5}{7} t^{7/5} \sum_{i=1}^d e^{{\bb_i}\trans\X}}\int_t^{\infty}e^{-\frac{5}{7}s^{7/5}\sum_{i=1}^d e^{{\bb_i}\trans\X}}ds,\\
m_T(t,\bb\trans\X,W)&=&e^{\frac{5}{12} t^{12/5} \sum_{i=1}^d e^{{\bb_i}\trans\X}} W \int_t^{\infty}e^{-\frac{5}{12}s^{12/5}\sum_{i=1}^d e^{{\bb_i}\trans\X}}ds.
\ese
Each component in $\X$ is generated independently from the
standard normal
distribution. The waiting time $W$ is generated independently from a uniform distribution over $(0,1)$. We consider $d=2, p=6$ and
set the true parameters
to be
$\bb = (\bb_{\cdot 1},\bb_{\cdot 2}) = ((1,0,-0.65,\allowbreak -0.5,-0.25,0.25)\trans,(0,1,-0.5,0.5,-0.4,0.25)\trans)\trans$. The sample size is $n=2000$ and we randomly assign around one third of samples to take the transplant.

\noindent
{\bf Study 4:} This study mimics the real data application. The hazards are set to be  
\bse
\lambda_N(t,\bb\trans\X) = \frac{1}{200}e^{t/200+\arctan(\bb\trans\X)+\pi/2} -\frac{1}{200},
\ese
and 
\bse
\lambda_T(t,\bb\trans\X,W) = \frac{1}{300}e^{t/300+\arctan(\bb\trans\X-W/5+10)+\pi/2} -\frac{1}{300}.
\ese
The corresponding mean residual lives are
\bse
m_N(t,\bb\trans\X)&=&200e^{-t/200-\arctan(\bb\trans\X)-\pi/2},\\
m_T(t,\bb\trans\X,W)&=&300e^{-t/300-\arctan(\bb\trans\X-W/5+10)-\pi/2}.
\ese
We consider $d=1, p=9$ and
set the true parameters
to be
$\bb = (0.4,1,-0.4,-1.50,\allowbreak-1.1,1.4,-0.1,-0.7)\trans$. The transplantation time $W$ is generated from the uniform distribution on $(0,\max(T_N))$. The sample size is $n=2,000$ with a censoring rate of 26\%, and about half of the samples receive the transplantation during followup. Study 4 mimics the features of real data that the mean residual life functions are decreasing gradually as $t$ increases. The transplant group accounts for $W$: the improvement $m_T(t-W,\X,W) - m_N(t,\X)$ is negative when $W$ is close to 0, and  approaches 0 positively as $W$ increases.

\begin{table}[H]
	\caption{Results of  study 1, based on 1000 simulations with sample size 300.
		``Prop.'' is the semiparametric method, ``PM'' is the proportional mean residual life
		method. ``emp sd'' is the sample standard deviation of the corresponding estimators; ``est sd" is the estimated standard deviation; ``CP'' is the estimated  coverage probability of confidence intervals.}
	\label{tab:simu1}
	\begin{tabular}{cc|cccccccc}\hline
		\vspace{.2cm}
		& & $\beta_2$ & $\beta_3$ & $\beta_4$ & $\beta_5$ & $\beta_6$ & $\beta_7$ & $\beta_8$ & $\beta_9$ \\ 
		& truth  & -0.6 & 0.0 & -0.3 & -0.1 & 0.0 & 0.1 & 0.3 & -0.5\\\hline
		&  &\multicolumn{8}{c}{No censoring}\\	
Prop. & point estimate & -0.597 & 0.002 & -0.290 & -0.096 & 0.000 & 0.073 & 0.302 & -0.504\\
& emp sd & 0.229 & 0.442 & 0.180 & 0.438 & 0.437 & 0.437 & 0.171 & 0.206\\
& est sd & 0.164 & 0.474 & 0.166 & 0.470 & 0.474 & 0.469 & 0.166 & 0.165\\
& CP(\%) & 88.8 & 96.2 & 94.4 & 96.0 & 95.8 & 96.1 & 94.9 & 91.2\\
PM & point estimate & 0.442 & -0.978 & 5.697 & -20.47 & 0.986 & -13.60 & 5.893 & -3.780\\
& emp sd & 57.8 & 95.11 & 155.9 & 583.0 & 36.7 & 352.5 & 212.5& 156.5\\\hline
& &\multicolumn{8}{c}{20\% censoring}\\		
            Prop. & point estimate & -0.590 & -0.003 & -0.289 & -0.086 & 0.007 & 0.101 & 0.289 & -0.486\\
            & emp sd & 0.178 & 0.379 & 0.153 & 0.373 & 0.379 & 0.368 & 0.148 & 0.165\\
            & est sd & 0.156 & 0.406 & 0.146 & 0.404 & 0.407 & 0.406 & 0.145 & 0.152\\
            & CP(\%) & 92.8 & 96.7 & 94.1 & 96.7 & 97.1 & 96.9 & 95.6 & 94.8\\
            PM & point estimate & -0.774 & 0.067 & -0.236 & -0.027 & 0.150 & 0.176 & 0.339 & -0.577 \\
            & emp sd & 2.235 & 3.794 & 1.380 & 3.693 & 5.911 & 3.563 & 1.570 & 3.981\\\hline
            &  &\multicolumn{8}{c}{40\% censoring}\\
            Prop. & point estimate & -0.518 & 0.020 & -0.260 & -0.079 & 0.020 & 0.088 & 0.266 & -0.434\\
            & emp sd & 0.168 & 0.368 & 0.136 & 0.368 & 0.387 & 0.362 & 0.142 & 0.159\\
            & est sd & 0.149 & 0.392 & 0.140 & 0.396 & 0.391 & 0.389 & 0.140 & 0.144\\
            & CP(\%) & 89.0 & 96.9 & 94.7 & 97.4 & 96.0 & 96.2 & 94.9 & 89.9\\
            PM & point estimate & 7.493 & -7.316 & -1.349 & 9.838 & -21.588 & -19.11 & 2.057 & 5.720 \\
            & emp sd & 262.6 & 160.8 & 116.7 & 367.1 & 570.9 & 762.8 & 63.4 & 185.9 \\\hline
            \end{tabular}
\end{table}

\begin{figure}[H]
	\centering
	\includegraphics[width=14cm]{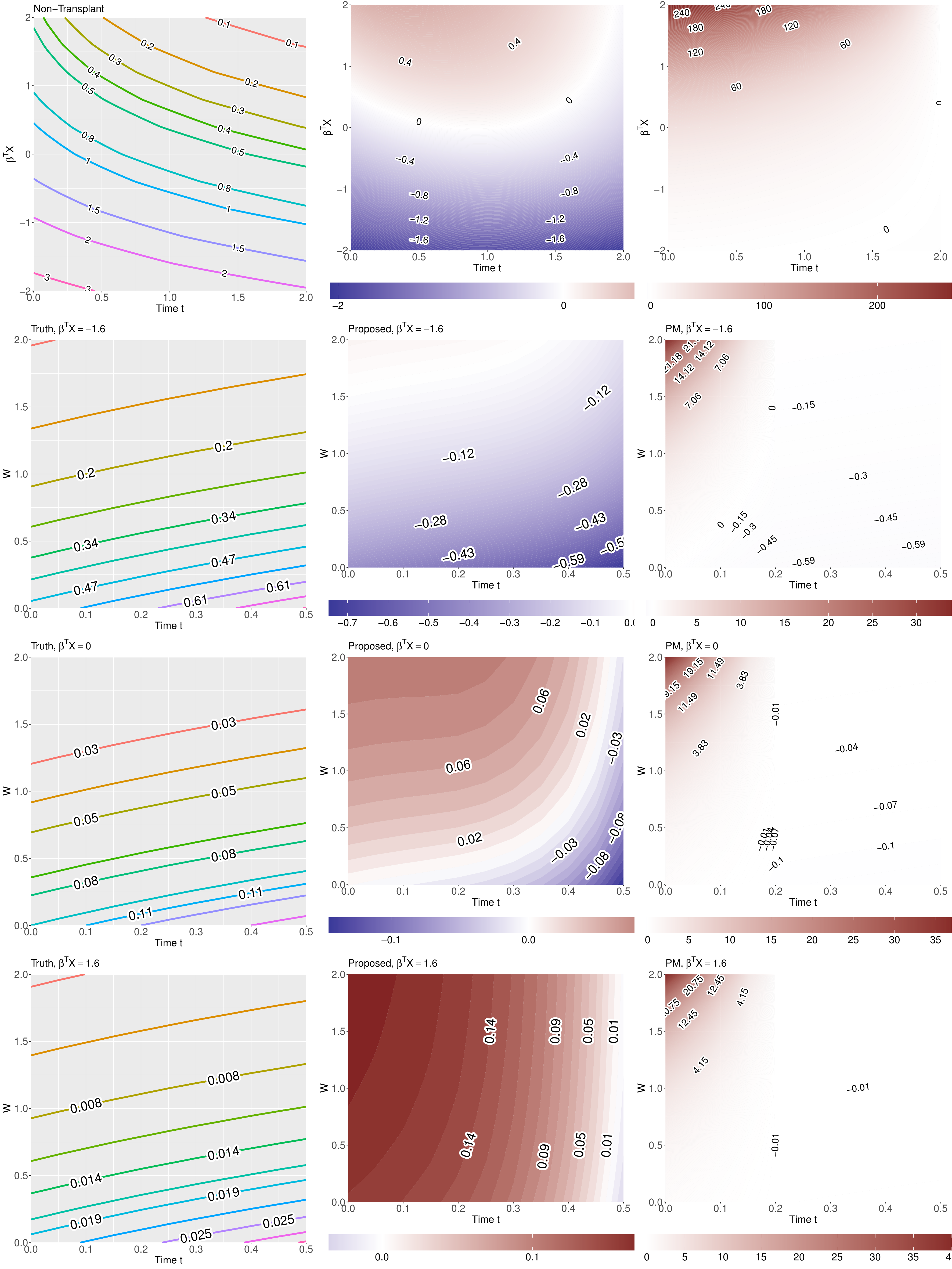}
	\caption{Performance of the semiparametric method on mean
		residual life
		function of study 1 without censoring.
        First row: $m_N(t,\bb\trans\X)$ and its estimations.
        Second row: $m_T(t,\bb\trans\X, W)$ and its estimations at $\bb\trans\X=-1.6$;
        Third row: $m_T(t,\bb\trans\X, W)$ and its estimations at $\bb\trans\X=0$;
        Fourth row: $m_T(t,\bb\trans\X, W)$ and its estimations at $\bb\trans\X=1.6$.
		First column: contour plot of the true $m_N(t,\bb\trans\X)$ and $\wh m_N(t,\bb\trans\X)$.
		Second column: contour plot of the residual of the proposed method;
        Third column: contour plot of the residual of the PM method;
	}
	\label{fig:simu1contour1}
\end{figure}
\begin{figure}[H]
	\centering
	\includegraphics[width=14cm]{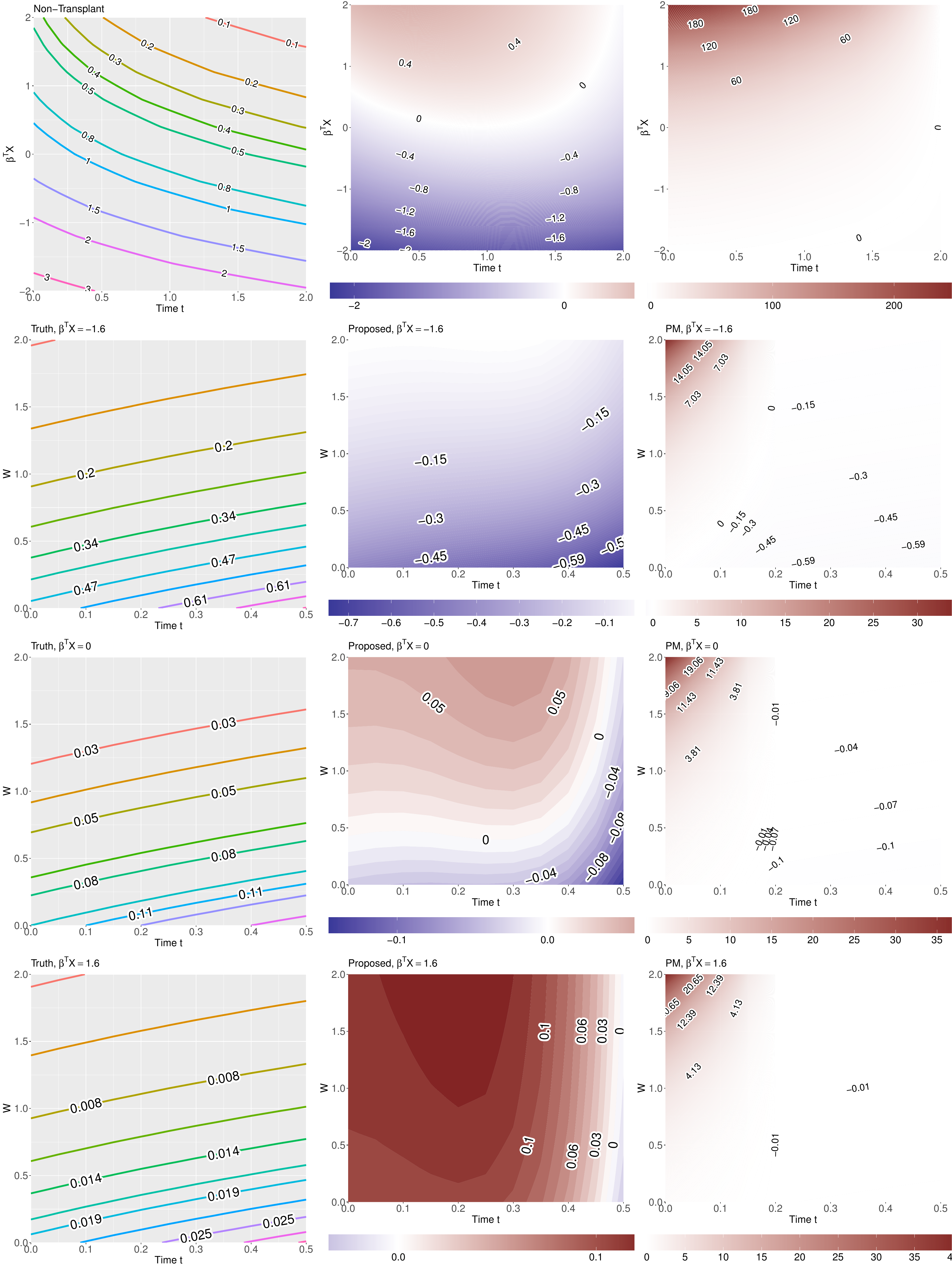}
	\caption{Performance of the semiparametric method on mean
		residual life
		function of study 1 at censoring rate 20\%.
        First row: $m_N(t,\bb\trans\X)$ and its estimations.
        Second row: $m_T(t,\bb\trans\X, W)$ and its estimations at $\bb\trans\X=-1.6$;
        Third row: $m_T(t,\bb\trans\X, W)$ and its estimations at $\bb\trans\X=0$;
        Fourth row: $m_T(t,\bb\trans\X, W)$ and its estimations at $\bb\trans\X=1.6$.
		First column: contour plot of the true $m_N(t,\bb\trans\X)$ and $\wh m_N(t,\bb\trans\X)$.
		Second column: contour plot of the residual of the proposed method;
        Third column: contour plot of the residual of the PM method;
	}
	\label{fig:simu1contour2}
\end{figure}
\begin{figure}[H]
	\centering
	\includegraphics[width=14cm]{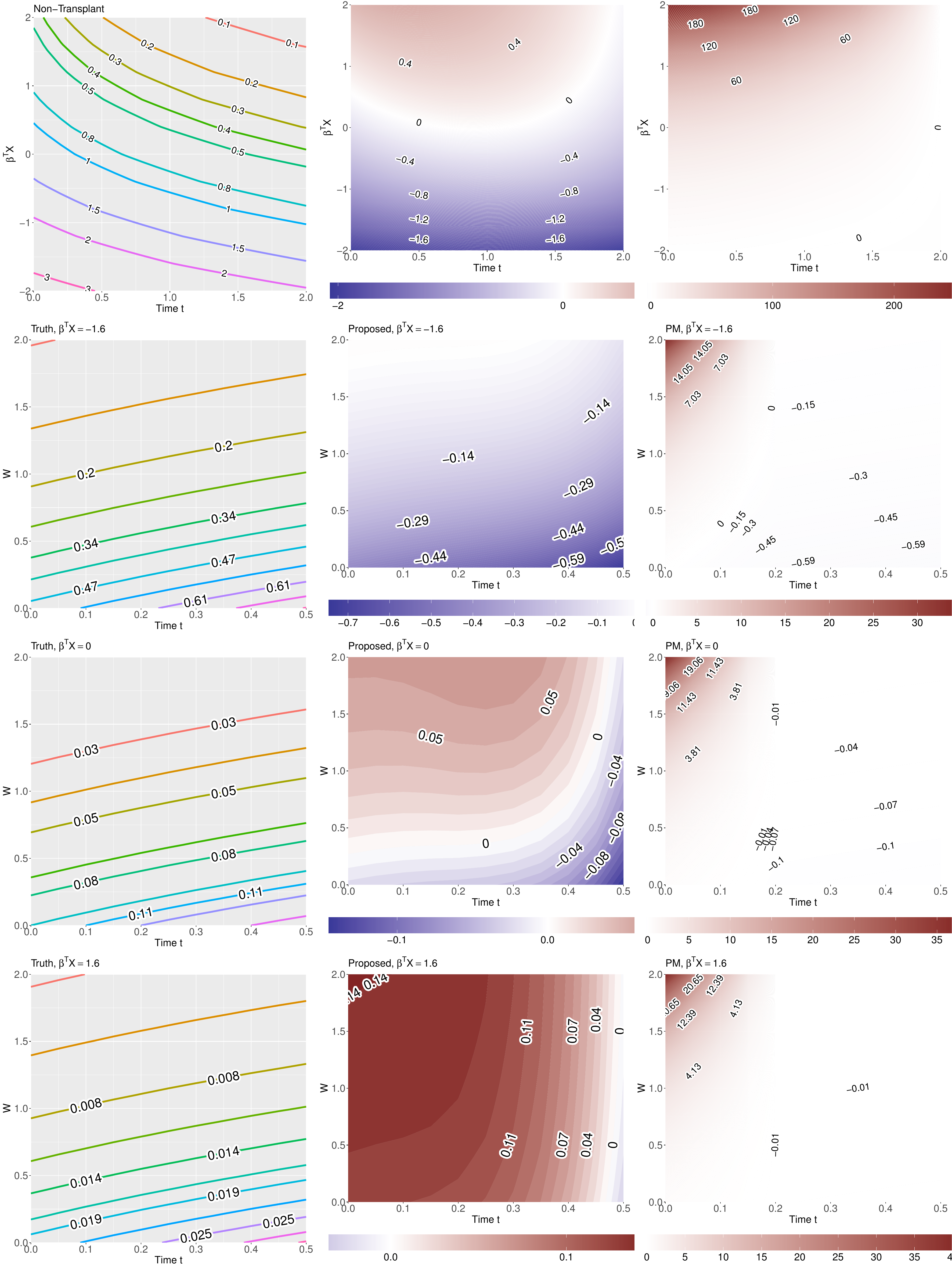}
	\caption{Performance of the semiparametric method on mean
		residual life
		function of study 1 at censoring rate 40\%.
        First row: $m_N(t,\bb\trans\X)$ and its estimations.
        Second row: $m_T(t,\bb\trans\X, W)$ and its estimations at $\bb\trans\X=-1.6$;
        Third row: $m_T(t,\bb\trans\X, W)$ and its estimations at $\bb\trans\X=0$;
        Fourth row: $m_T(t,\bb\trans\X, W)$ and its estimations at $\bb\trans\X=1.6$.
		First column: contour plot of the true $m_N(t,\bb\trans\X)$ and $\wh m_N(t,\bb\trans\X)$.
		Second column: contour plot of the residual of the proposed method;
        Third column: contour plot of the residual of the PM method;
	}
	\label{fig:simu1contour3}
\end{figure}

The results for the estimation of
$\bb$ under Study 1 are given in Table
\ref{tab:simu1}, with three
censoring rates, 0\%, 20\% and 40\%.
The proposed
method has much smaller biases and standard deviations, whereas
``PM" is
biased with larger standard deviations.
The performances of all of the estimators deteriorate when the censoring rate increases, though our method still outperforms the others.
We also
demonstrate the true and error plots
in Figure \ref{fig:simu1contour1}--\ref{fig:simu1contour3}, demonstrate
that our method fare well for
estimating $m(t,\bb\trans\x)$ when $t$ and $\bb\trans\x$ are not too extreme.
The contour plots reveal that bias
increases as censoring rate increases and the estimation
deteriorates when $t$ is large.
These results show an overall satisfactory  performance of our semiparametric method.
Figure \ref{fig:simu1contour1}--\ref{fig:simu1contour3} reveals that the performance of our method is better when $t$ is in the
interior of the range because more observations are
available for
the local estimation, as opposed to a larger $t$ with fewer observations available.
In contrast,
 regardless of the magnitude of $t$, the mean   residual
 life function estimated by ``PM'' is severely biased, as shown
in the last two rows from Figure \ref{fig:simu1contour1}--\ref{fig:simu1contour3}. This is because
this model assume a pre-determined functional
form of the mean residual life, which in this case is misspecified.

\begin{table}
	\caption{Results of  study 2, based on 1000
		simulations with sample size 1000. 
		``Prop.'' is the
		semiparametric method, ``PM'' is the
		proportional mean residual life
		method. 
		``emp sd'' is the sample standard deviation
		of the corresponding estimators; ``est sd" is the estimated standard deviation; ``CP'' is the estimated  coverage probability of confidence intervals.  }
	\label{tab:simu2}
	\begin{tabular}{cc|cccccccc}\hline
		\vspace{.2cm}
		& & $\beta_2$ & $\beta_3$ & $\beta_4$ & $\beta_5$ & $\beta_6$ & $\beta_7$ & $\beta_8$ & $\beta_9$ \\
		& truth & -0.60 & 0.00 & -0.30 & -0.10 & 0.00 & 0.10 & 0.30 & -0.50\\
		\hline
		& &\multicolumn{8}{c}{No censoring}\\
Prop. & point estimate & -0.611 & 0.002 & -0.308 & -0.101 & -0.005 & 0.101 & 0.310 & -0.505\\
& emp sd & 0.140 & 0.106 & 0.114 & 0.104 & 0.106 & 0.104 & 0.114 & 0.127\\
& est sd & 0.135 & 0.118 & 0.122 & 0.118 & 0.118 & 0.118 & 0.122 & 0.129\\
& CP(\%) & 95.0 & 97.4 & 97.5 & 98.1 & 97.4 & 97.6 & 97.3 & 95.8\\
PM & point estimate & -0.601 & 0.003 & -0.301 & -0.099 & 0.006 & 0.096 & 0.300 & -0.506\\
& emp sd & 0.069 & 0.083 & 0.073 & 0.074 & 0.083 & 0.088 & 0.097 & 0.084\\\hline
& &\multicolumn{8}{c}{20\% censoring}\\
Prop. & point estimate & -0.596 & 0.003 & -0.306 & -0.100 & 0.001 & 0.094 & 0.303 & -0.499\\
& emp sd & 0.140 & 0.116 & 0.126 & 0.113 & 0.109 & 0.110 & 0.123 & 0.137\\
& est sd & 0.135 & 0.119 & 0.123 & 0.119 & 0.119 & 0.119 & 0.123 & 0.130\\
& CP(\%) & 94.7 & 95.8 & 94.4 & 96.5 & 97.0 & 96.1 & 96.0 & 94.5\\
PM & point estimate & -0.604 & 0.000 & -0.305 & -0.114 & -0.023 & 0.092 & 0.299 & -0.501\\
& emp sd & 0.146 & 0.380 & 0.121 & 0.355 & 0.373 & 0.384 & 0.132 & 0.133\\\hline
		&  &\multicolumn{8}{c}{40\% censoring}\\
Prop. & point estimate & -0.596 & 0.001 & -0.300 & -0.098 & -0.004 & 0.098 & 0.295 & -0.498\\
& emp sd & 0.156 & 0.129 & 0.136 & 0.131 & 0.125 & 0.127 & 0.140 & 0.148\\
& est sd & 0.147 & 0.129 & 0.134 & 0.130 & 0.129 & 0.130 & 0.134 & 0.142\\
& CP(\%) & 95.1 & 95.0 & 95.4 & 94.8 & 95.9 & 95.6 & 93.5 & 93.9\\
PM & point estimate & -0.553 & 0.495 & -0.242 & 0.220 & 0.114 & -0.216 & 0.379 & -0.565 \\
& emp sd & 1.805 & 15.895 & 2.105 & 10.373 & 4.706 & 10.316 & 2.535 & 1.833 \\\hline
\end{tabular}
\end{table}

\begin{figure}[H]
	\centering
	\includegraphics[width=14cm]{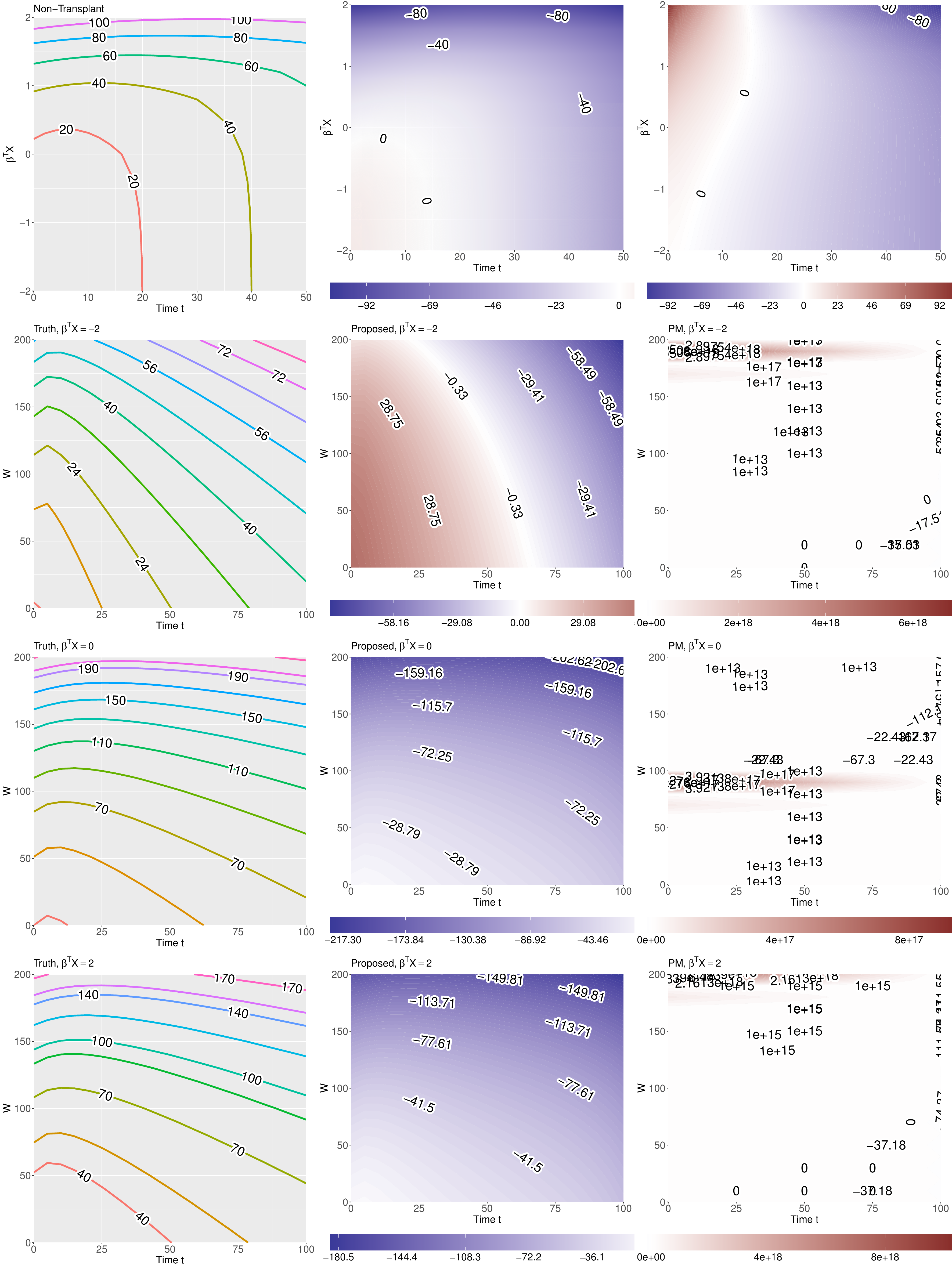}
	\caption{Performance of the semiparametric method on mean
		residual life
		function of study 2 without censoring.
        First row: $m_N(t,\bb\trans\X)$ and its estimations.
        Second row: $m_T(t,\bb\trans\X, W)$ and its estimations at $\bb\trans\X=-2$;
        Third row: $m_T(t,\bb\trans\X, W)$ and its estimations at $\bb\trans\X=0$;
        Fourth row: $m_T(t,\bb\trans\X, W)$ and its estimations at $\bb\trans\X=2$.
		First column: contour plot of the true $m_N(t,\bb\trans\X)$ and $\wh m_N(t,\bb\trans\X)$.
		Second column: contour plot of the residual of the proposed method;
        Third column: contour plot of the residual of the PM method;
	}
	\label{fig:simu2contour1}
\end{figure}
\begin{figure}[H]
	\centering
	\includegraphics[width=14cm]{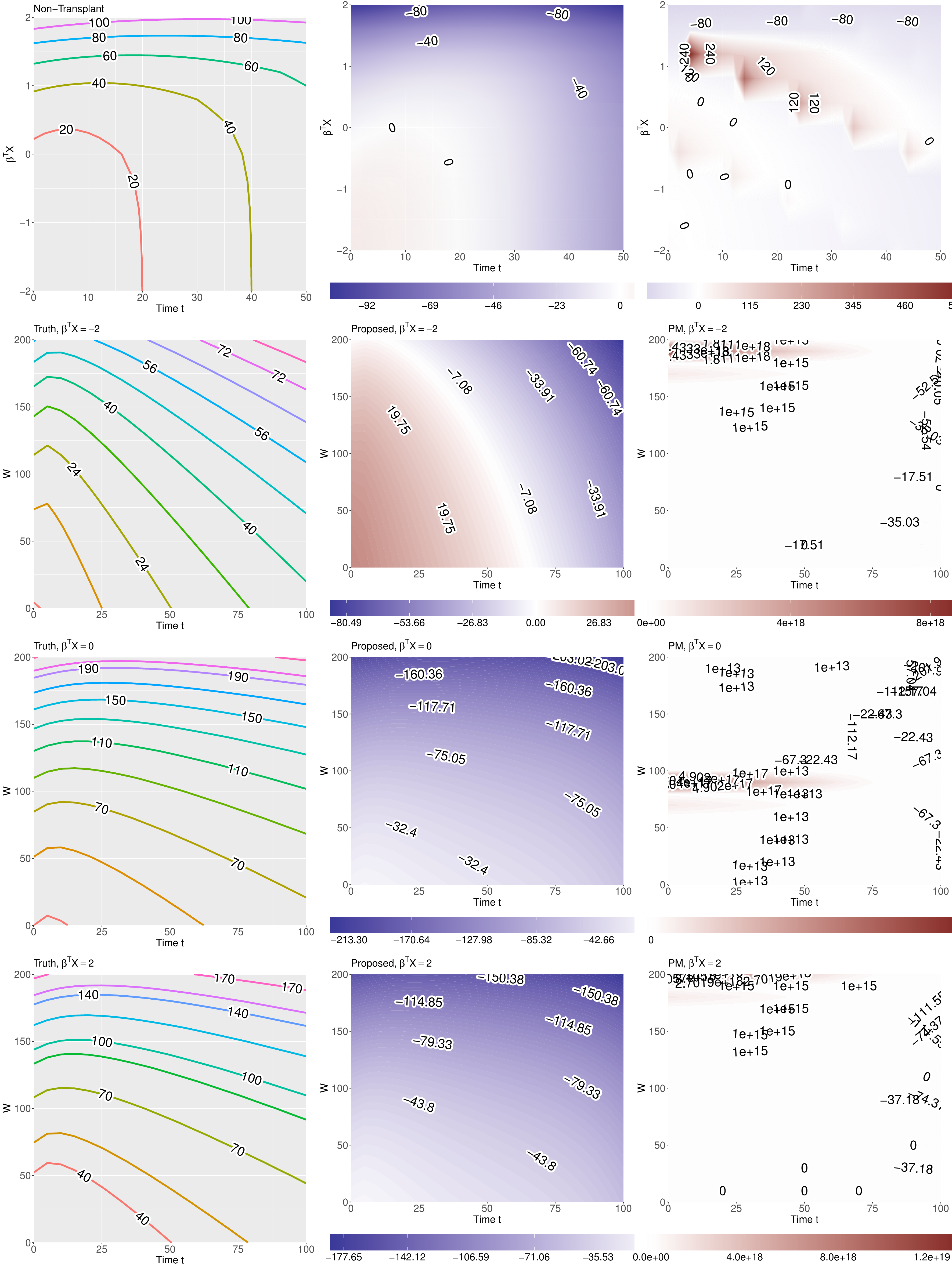}
	\caption{Performance of the semiparametric method on mean
		residual life
		function of study 1 at censoring rate 20\%.
        First row: $m_N(t,\bb\trans\X)$ and its estimations.
        Second row: $m_T(t,\bb\trans\X, W)$ and its estimations at $\bb\trans\X=-2$;
        Third row: $m_T(t,\bb\trans\X, W)$ and its estimations at $\bb\trans\X=0$;
        Fourth row: $m_T(t,\bb\trans\X, W)$ and its estimations at $\bb\trans\X=2$.
		First column: contour plot of the true $m_N(t,\bb\trans\X)$ and $\wh m_N(t,\bb\trans\X)$.
		Second column: contour plot of the residual of the proposed method;
        Third column: contour plot of the residual of the PM method;
	}
	\label{fig:simu2contour2}
\end{figure}
\begin{figure}[H]
	\centering
	\includegraphics[width=14cm]{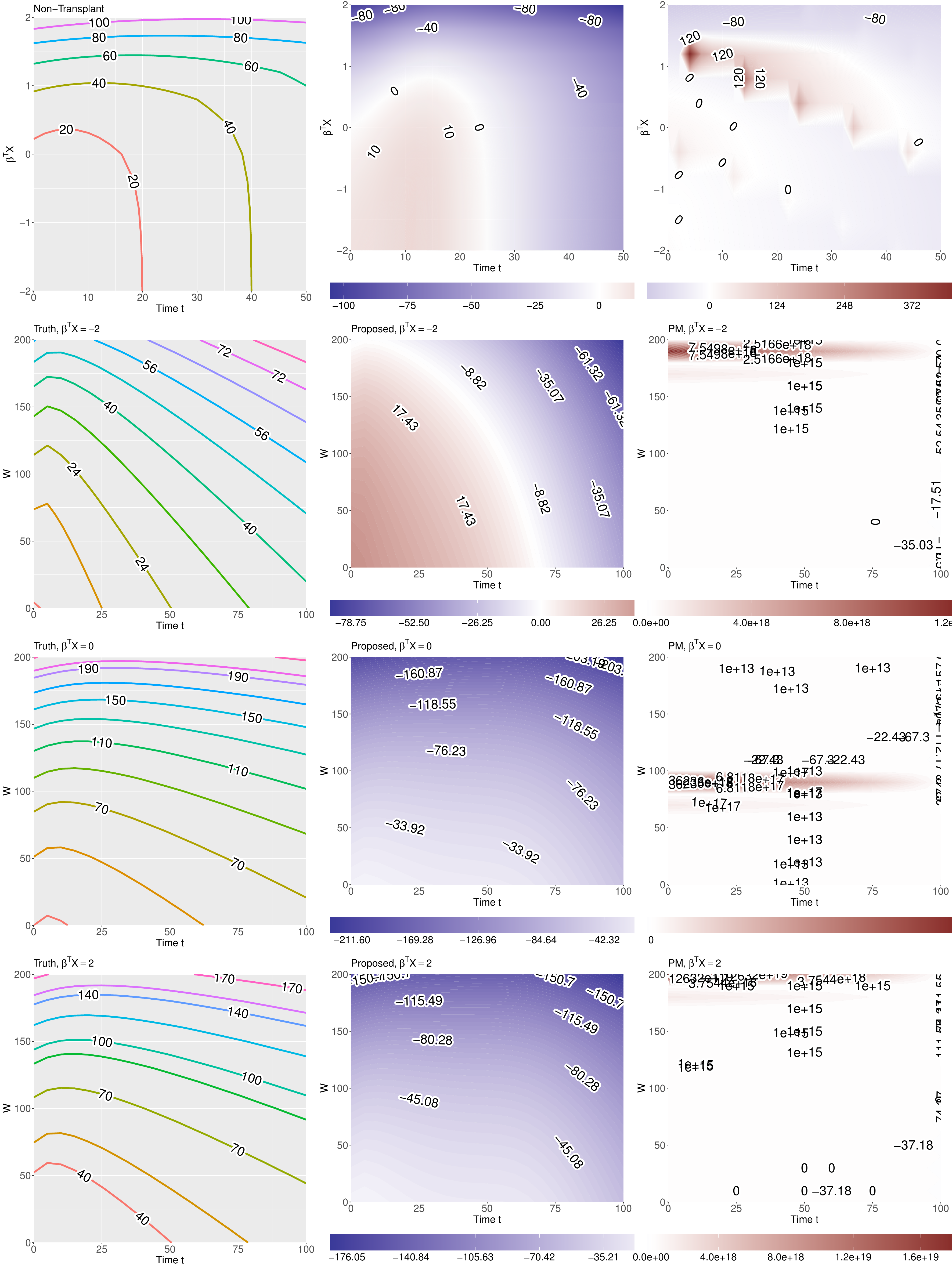}
	\caption{Performance of the semiparametric method on mean
		residual life
		function of study 1 at censoring rate 40\%.
        First row: $m_N(t,\bb\trans\X)$ and its estimations.
        Second row: $m_T(t,\bb\trans\X, W)$ and its estimations at $\bb\trans\X=-2$;
        Third row: $m_T(t,\bb\trans\X, W)$ and its estimations at $\bb\trans\X=0$;
        Fourth row: $m_T(t,\bb\trans\X, W)$ and its estimations at $\bb\trans\X=2$.
		First column: contour plot of the true $m_N(t,\bb\trans\X)$ and $\wh m_N(t,\bb\trans\X)$.
		Second column: contour plot of the residual of the proposed method;
        Third column: contour plot of the residual of the PM method;
	}
	\label{fig:simu2contour3}
\end{figure}

Tables \ref{tab:simu2}
and \ref{tab:simu3} report the results of Studies 2 and 3   related to $\wh\bb$,
respectively.
We also provide the error plots of $\wh m(t,\bb\trans\x)-m(t,\bb\trans\x)$ in Study 2 using a contour plot
in Figure \ref{fig:simu2contour1}--\ref{fig:simu2contour3}. The proposed method performs better than the competitor. For Study 3, we provide the error plots of $\wh m(t,{\bb_1}\trans\x,{\bb_2}\trans\x)-m(t,{\bb_1}\trans\x,{\bb_2}\trans\x)$ fixed at ${\bb_1}\trans\x = 0$ and ${\bb_2}\trans\x = 0$ in Figure \ref{fig:simu3contour}.
Similar to the conclusion in the first simulation
study, the performance of estimating $\bb$ by our proposed estimator
is satisfactory. The performance of the mean residual
life estimation is
better when $t$ is smaller,  deteriorates when
$t$ and $\bb\trans\x$ becomes extreme, and is better for smaller censoring rates.

	\begin{table}
	\caption{Results of study 3, based on 1000
		simulations with sample size 2000. 
		``emp sd'' is the sample standard deviation
		of the corresponding estimators; ``est sd" is the estimated standard deviation;
  ``CP'' is the estimated  coverage probability of confidence intervals.}
	\label{tab:simu3}
	\begin{tabular}{c|cccccccc}\hline
		\vspace{.2cm}
		& $\beta_{31}$ & $\beta_{41}$ & $\beta_{51}$ & $\beta_{61}$ & $\beta_{32}$ & $\beta_{42}$ & $\beta_{52}$ & $\beta_{62}$ \\ 
		truth & -0.65 & -0.50 & -0.25 & 0.25 & -0.50 & 0.40 & -0.40 & 0.25\\
		\hline
		&\multicolumn{8}{c}{No censoring}\\
point estimate & -0.662 & -0.551 & -0.237 & 0.253 & -0.492 & 0.465 & -0.407 & 0.251\\
emp sd & 0.152 & 0.117 & 0.136 & 0.129 & 0.169 & 0.121 & 0.137 & 0.129\\
est sd & 0.139 & 0.125 & 0.145 & 0.146 & 0.147 & 0.131 & 0.152 & 0.154\\
CP(\%) & 91.3 & 93.4 & 95.5 & 96.1 & 89.9 & 93.5 & 96.6 & 97.8\\
		&\multicolumn{8}{c}{20\% censoring}\\		
point estimate & -0.608 & -0.401 & -0.252 & 0.238 & -0.480 & 0.301 & -0.363 & 0.236\\
emp sd & 0.105 & 0.097 & 0.097 & 0.091 & 0.110 & 0.098 & 0.098 & 0.090\\
est sd & 0.103 & 0.094 & 0.107 & 0.107 & 0.108 & 0.098 & 0.111 & 0.112\\
CP(\%) & 93.4 & 81.2 & 94.9 & 96.1 & 93.1 & 81.8 & 95.9 & 98.0\\\hline
&\multicolumn{8}{c}{40\% censoring}\\		
point estimate & -0.587 & -0.420 & -0.234 & 0.227 & -0.456 & 0.316 & -0.357 & 0.228\\
emp sd & 0.084 & 0.071 & 0.080 & 0.073 & 0.089 & 0.082 & 0.078 & 0.074\\
est sd & 0.092 & 0.083 & 0.092 & 0.093 & 0.098 & 0.087 & 0.097 & 0.098\\
CP(\%) & 93.3 & 87.3 & 97.5 & 97.7 & 95.2 & 85.4 & 97.3 & 98.1\\\hline
\end{tabular}
\end{table}

\begin{figure}
	\centering
	\includegraphics[width=14cm]{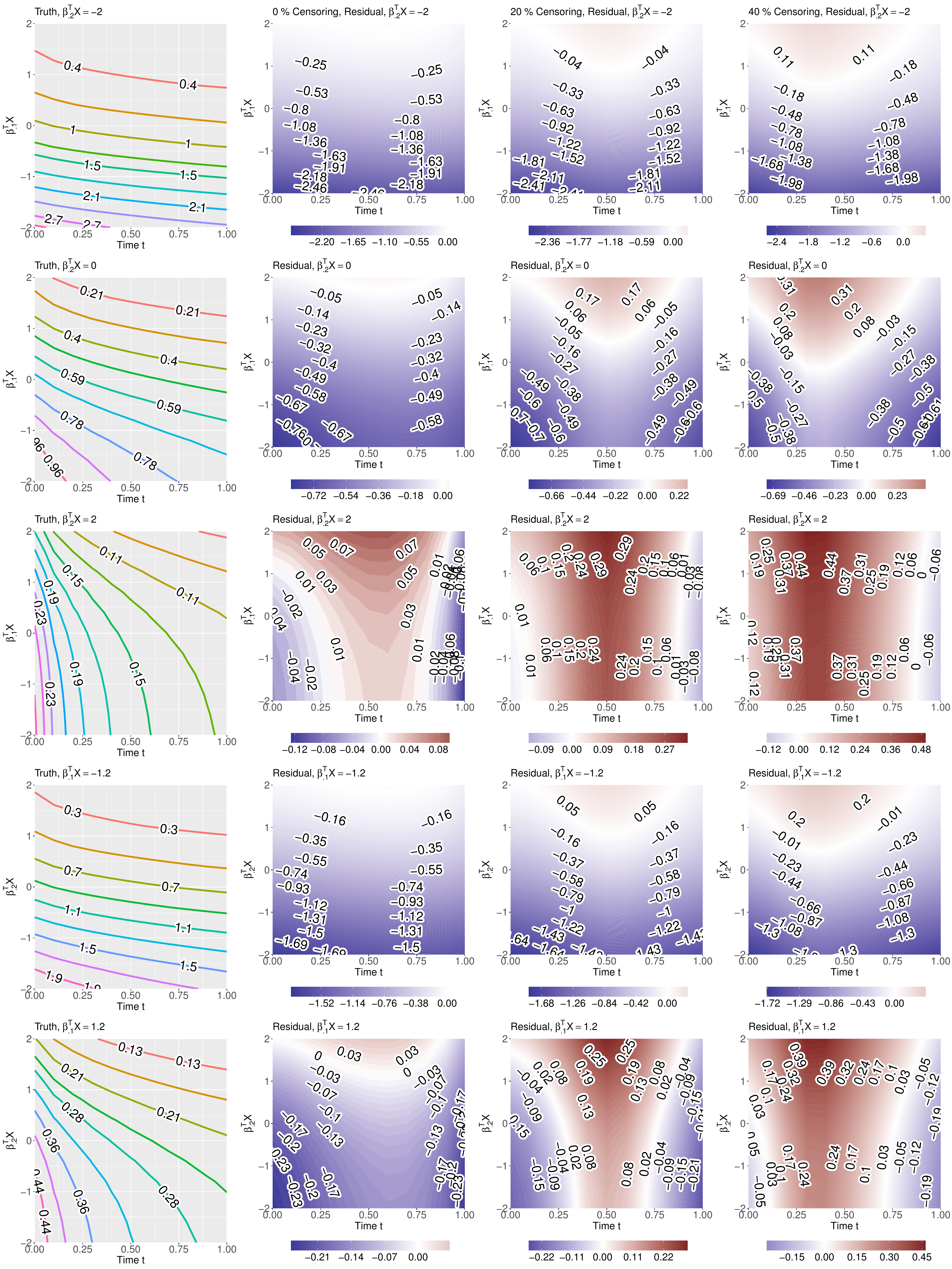}
	\caption{Performance of the proposed method on mean
		residual life
		function of the nontransplant group of study 3 at different $\bb\trans\X$.
		Left column: contour plots of true $m_N(t,0,{\bb}\trans\X)$;
        Column 2 to 4: contour plots of residual of $\wh m_N(t,{\bb_2}\trans\X,0)$ at censoring rate 0\%, 20\%, and 40\%.}	\label{fig:simu3contourN}
\end{figure}

\begin{figure}
	\centering
	\includegraphics[width=14cm]{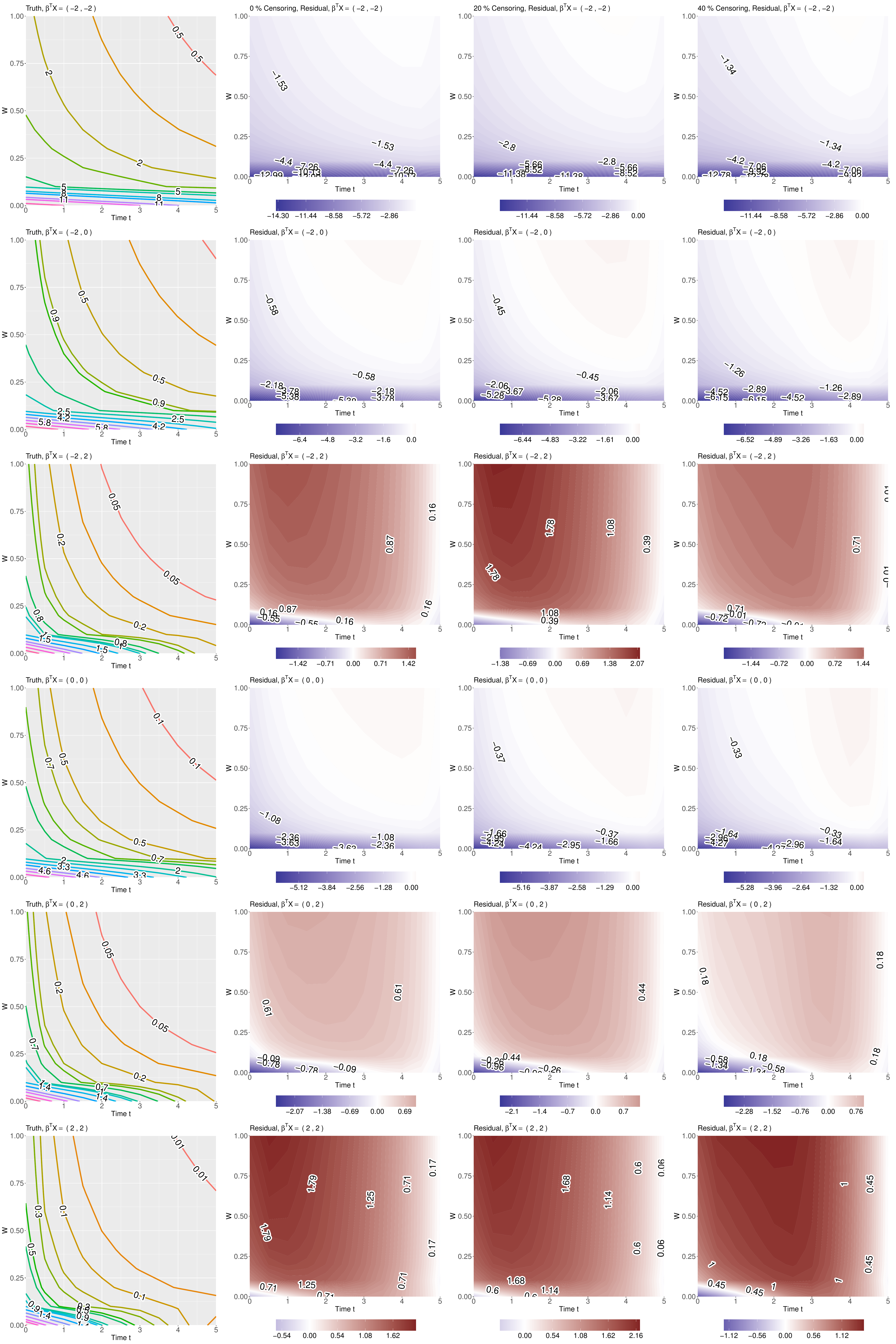}
	\caption{Performance of the proposed method on mean
		residual life
		function of the transplant group of study 3 at different $\bb\trans\X$.
		Left column: contour plots of true $m_T(t,0,{\bb}\trans\X)$;
        Column 2 to 4: contour plots of residual of $\wh m_T(t,{\bb_2}\trans\X,0)$ at censoring rate 0\%, 20\%, and 40\%.}	\label{fig:simu3contourT}
\end{figure}

\begin{figure}
	\centering
	\includegraphics[width=14cm]{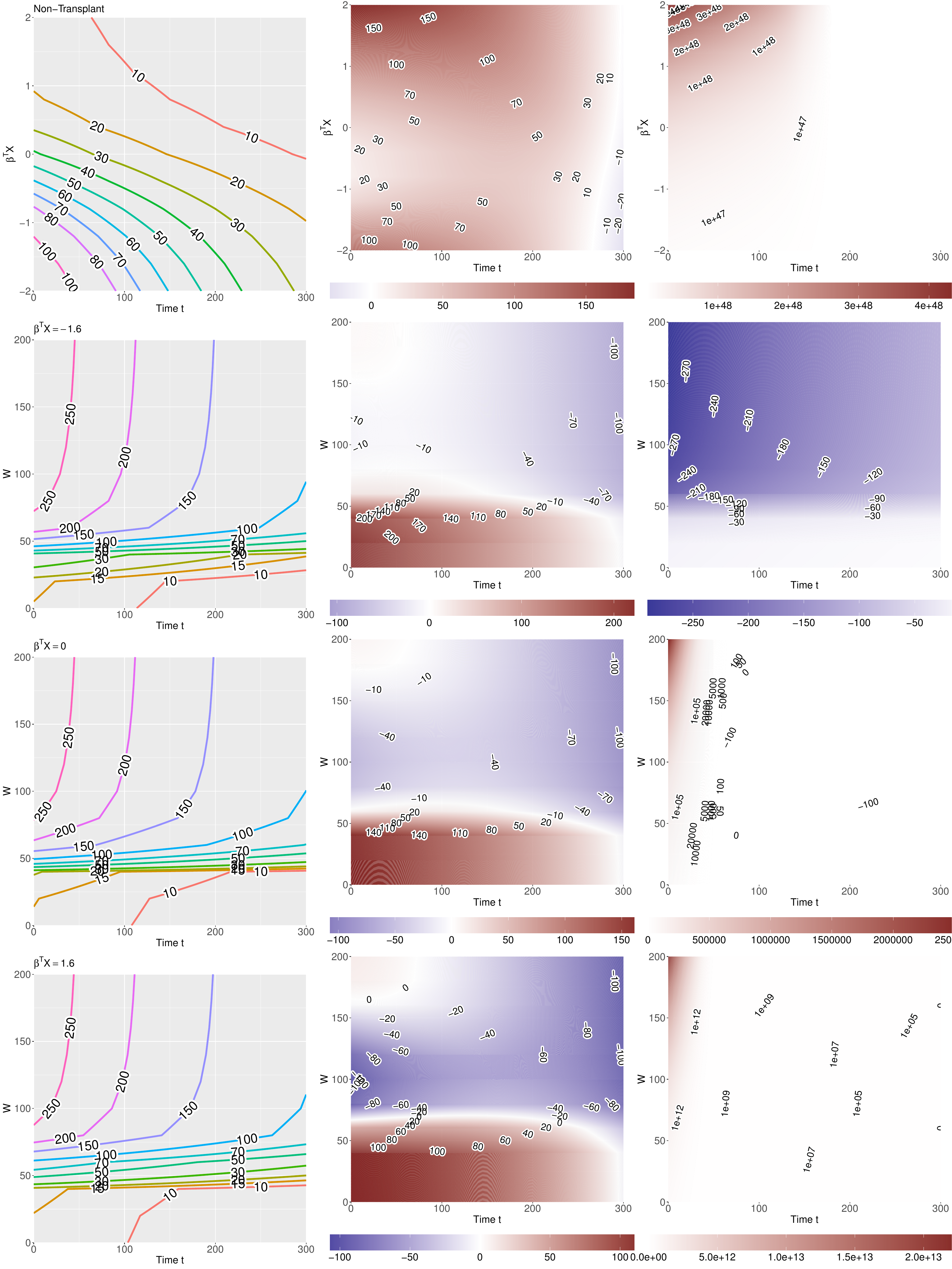}
	\caption{Performance of the mean
		residual life
		function estimation in study 4.
		Row 1: Performance of $m_N(t,{\bb_1}\trans\X)$;
  Row 2 to 4: $\wh m_T(t,{\bb}\trans\X,W)$ at ${\bb}\trans\X=-1.6,0,1.6$.
		Left column: Contour plot of true mean residual function. Middle column: Residual from the proposed method. Right column: Residual from the PM method.}
	\label{fig:simuMimic}
\end{figure}

The results for Study 4 are presented in Table \ref{tab:simuMimic} which displays the estimated vector $\wh\bb$ for both methods. Notably, the PM method exhibits a significantly larger bias compared to our method.  We  also assess the error of $\wh m(t-W,\bb\trans\x,W)-\wh m(t,\bb\trans\x)$ using a contour plot in Figure \ref{fig:simuMimic}. Our proposed method outperforms the competitor across various scenarios. This difference in performance is particularly evident when $\bb\trans\X>0$, where the PM method struggles to accurately estimate the mean residual life function of the transplanted objects. This discrepancy can be attributed to the PM method's limited ability to handle the nonlinear structure inherent in the mean residual life function.

	\begin{table}[H]
	\caption{Results of study 4, based on 1000
		simulations with sample size 2000. 
        ``emp sd'' is the sample standard deviation
		of the corresponding estimators; ``est sd" is the estimated standard deviation; ``CP'' is the estimated  coverage probability of confidence intervals.}
	\label{tab:simuMimic}
	\begin{tabular}{c|ccccccccc}\hline
		\vspace{.2cm}
&& $\beta_2$ & $\beta_3$ & $\beta_4$ & $\beta_5$ & $\beta_6$ & $\beta_7$ & $\beta_8$ & $\beta_9$ \\
& & 0.4 & 1 & -0.4 & -1.50 & -1.1 & 1.4 & -0.1 & -0.7\\ \hline
Prop. &point estimate & 0.409 & 0.912 & -0.396 & -1.263 & -0.945 & 1.344 & -0.109 & -0.606\\
&emp sd & 0.300 & 0.638 & 0.427 & 1.191 & 0.647 & 0.647 & 0.312 & 0.478\\
&est sd & 0.279 & 0.776 & 0.473 & 1.225 & 0.769 & 0.763 & 0.467 & 0.519\\
&CP(\%) & 90.3 & 96.5 & 97.5 & 94.8 & 95.7 & 97.4 & 97.6 & 92.0\\\hline
PM & point estimate & 0.646 & 0.950 & 0.205 & -0.776 & -0.303 & 0.790 & 1.557 & -0.090 \\
& emp sd & 22.847 & 6.882 & 2.545 & 3.545 & 1.879 & 4.259 & 8.725 & 1.556 \\\hline
	\end{tabular}
\end{table}

\section{Analysis of the Kidney Transplant Data}\label{sec:app}

We apply the proposed method to analyze a kidney transplant data
set from the
U.S. Scientific Registry of Transplant Recipients (SRTR) mentioned in the introduction. Briefly, the registry is
maintained by the United Network for Organ Sharing and Organ Procurement and
Transplantation Network (UNOS/OPTN) and includes all waitlisted
kidney transplant candidates and transplant recipients in the U. S. (\url{https://unos.org/}). For assessing possible benefits of  transplantation, we use the residual life to estimate how much longer a patient can
survive if she or he receives a transplant than otherwise.  

To avoid confounding cohort effects and also to have a sufficiently long followup,  we focus on the patients who were waitlisted in the same year of 2011. There were 43,140 patients in this cohort with an average followup  of 907 days after waitlisting. During the followup, a  total of  22,183
patients received kidney transplants. The
response variable is the survival time in days
($T_i$) starting from waitlisting.
Among patients who got a transplantation, 5.86\% of the observations were
censored, and the censoring rate was
26.43\% among those without a transplantation. The covariates $\X$ included in our analysis were
gender ($X_{1}$), race ($X_{2}$), max cold ischemia time
($X_{3}$),
insurance coverage ($X_{4}$), body mass index ($X_{5}$),
diagnosis
type ($X_{6}$), peak PRA/CPRA ($X_{7}$), previous malignancy
status
($X_{8}$) and diabetes indicator ($X_{9}$), all of which were used for computing the EPTS score \citep{time2012guide}. The waiting time $W$ is also considered in our model as proposed in (\ref{eq:modelhazard}) and (\ref{eq:meanResidualLife}).
Our analytical goal was to use model (\ref{eq:meanResidualLife}) to
quantify the potential residual life increment if a patient
receives a kidney transplant given the covariate profile. 
The model  mimics a real waitlisting to transplantation process by stipulating that all of the patients started by belonging in the non-transplant group,  while those who got a transplantation were viewed as censored at transplantation; once transplanted,  a patient would switch his or her membership to join the transplant group. 

To proceed, we first determine the number of indices $d$
using the Validated Information Criterion (VIC) \citep{ma2015validated},
where the smallest VIC value corresponds to the selected $d$ value.
In our analysis,  $d=1$ is chosen with the smallest $\text{VIC}=143.66$, indicating a single index is sufficiently informative; see  Table \ref{tab:app1}.
Subsequently, we normalize the index vector by fixing the first component (gender) at 1, and report  8 coefficient estimates.
All of the covariates, except for the max cold ischemia time ($X_{3}$) and the
	body mass index ($X_{5}$), have significant effects
	on the mean residual life, which agrees with the
	previous studies
	\citep{friedman2003demographics,webster2017chronic}. 
\begin{table}[b]
	\caption{Parameter estimation of the kidney transplant data. ``est." is the estimation of parameter, ``s.d."
		is the estimated standard deviation of $\wh\bb$.}
	\label{tab:app1}
	\centering
	\begin{tabular}{c|cccccccc}\hline
		&&&&&&&&\\[-1em]
		& $\wh\bb_{2}$& $\wh\bb_{3}$&
		$\wh\bb_{4}$&$\wh\bb_{5}$& $\wh\bb_{6}$& $\wh\bb_{7}$&
		$\wh\bb_{8}$& $\wh\bb_{9}$ \\
		&&&&&&&&\\[-1em]\hline
		&&&&&&&&\\[-1em]
        est. & -0.097 & -0.003 & -0.174 & -0.029 & -0.119 & 0.030 & -0.162 & 0.004\\
        s.d. &  0.011 & 0.007 & 0.010 & 0.007 &  0.009 & 0.005 & 0.016 & 0.011\\
        p-value & 0.000 & 0.866 &0.000 & 0.073 & 0.000 & 0.000 & 0.000 & 0.008 \\\hline
        \end{tabular}
\end{table}

The max cold ischemia time
($X_{3}$) that refers to the tolerable amount of time from when a kidney is removed from the donor to the time it is transplanted into the recipient. Although the max cold ischemia time reflects the patient's physiological conditions indirectly, it is not as significant as the real cold ischemia time in determining the post-operative risk \citep{iida2008minimal,kayler2011impact}.  BMI ($X_{5}$) is commonly suggested as a ``paradox'' risk factor in the literature \citep{kalantar2005survival,ahmadi2016association}. A popular
        explanation is that the BMI cannot differentiate between fat
        and muscle, thus high BMI patients may gain a survival
        advantage \citep{beddhu2004hypothesis,mafra2008body}.

On the other hand, race ($X_{2}$) and insurance coverage ($X_{4}$) have significant impacts on survival. 
It has been widely accepted that race and insurance
coverage are highly correlated with patients' socioeconomic status, which plays an
crucial role in the choice of chronic kidney disease treatment, especially for the end-stage patients\citep{lewis2010racial,muntner2012racial,nicholas2013racial,webster2017chronic}.
The other significant variables are also known risk factors for the ESRD mortality  in the literature  \citep{kauffman2005transplant,mehdi2009anemia,kayler2011impact,pyram2012chronic,lim2015peak}.

\begin{figure}
	\centering
        \includegraphics[width=4.5cm]{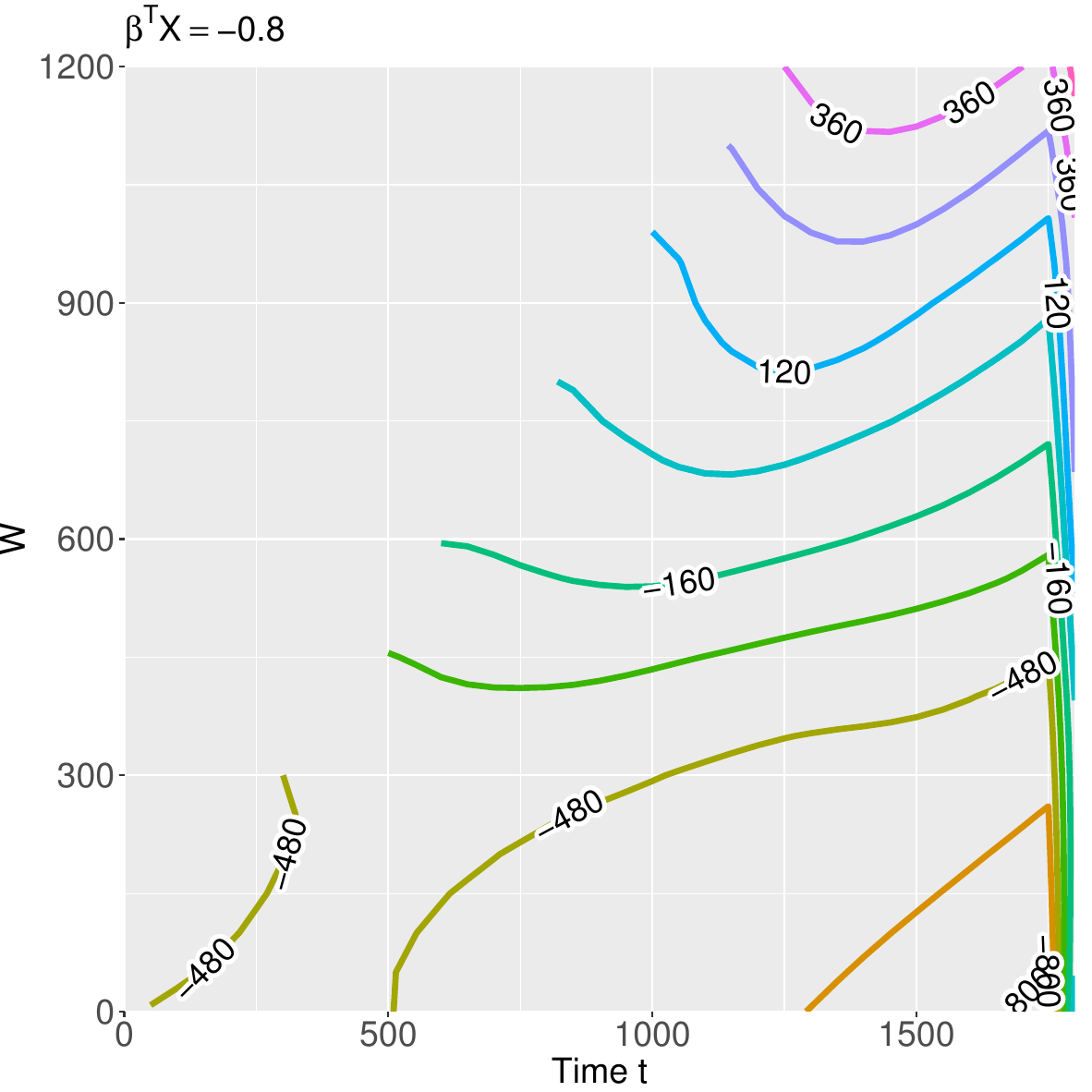}
        \includegraphics[width=4.5cm]{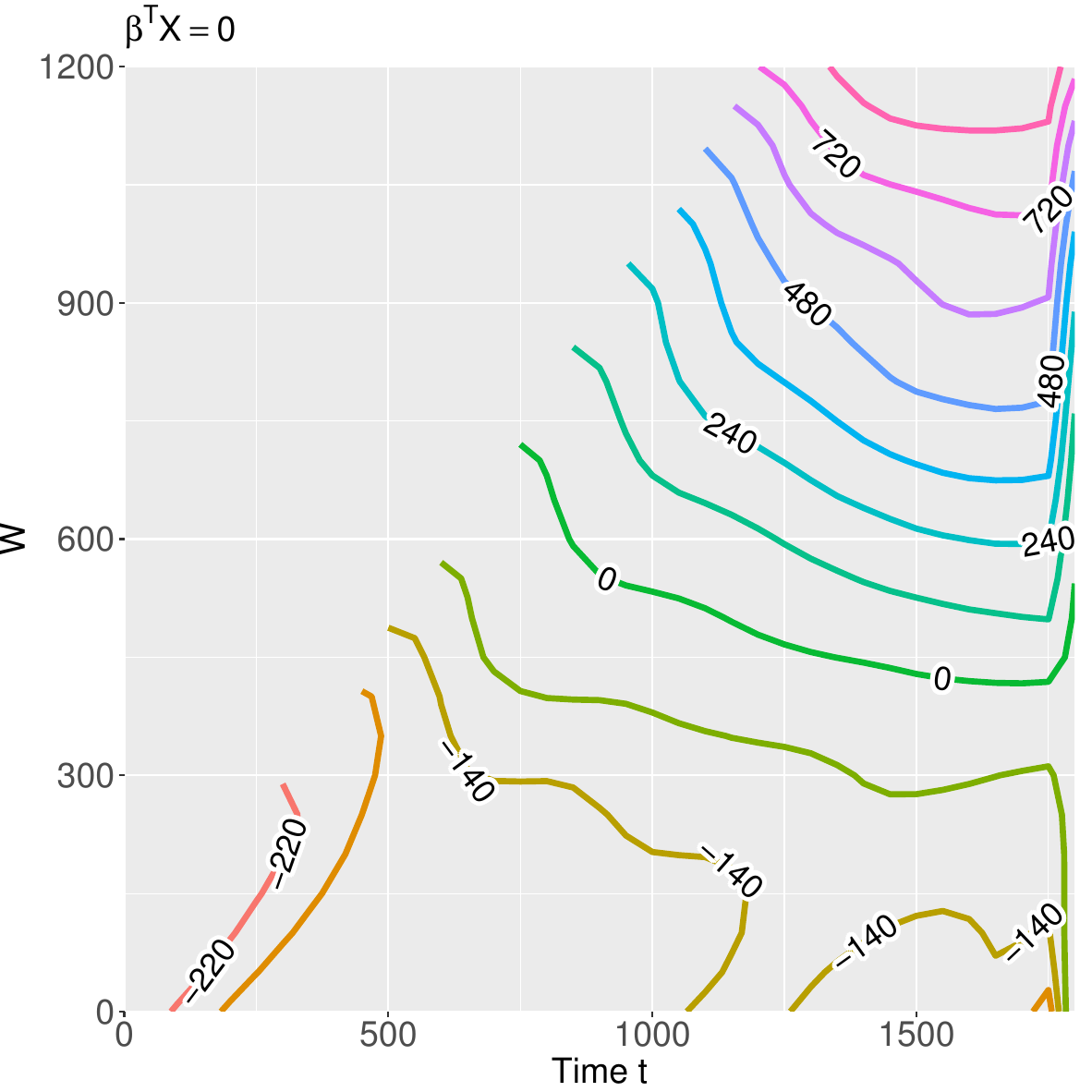}
        \includegraphics[width=4.5cm]{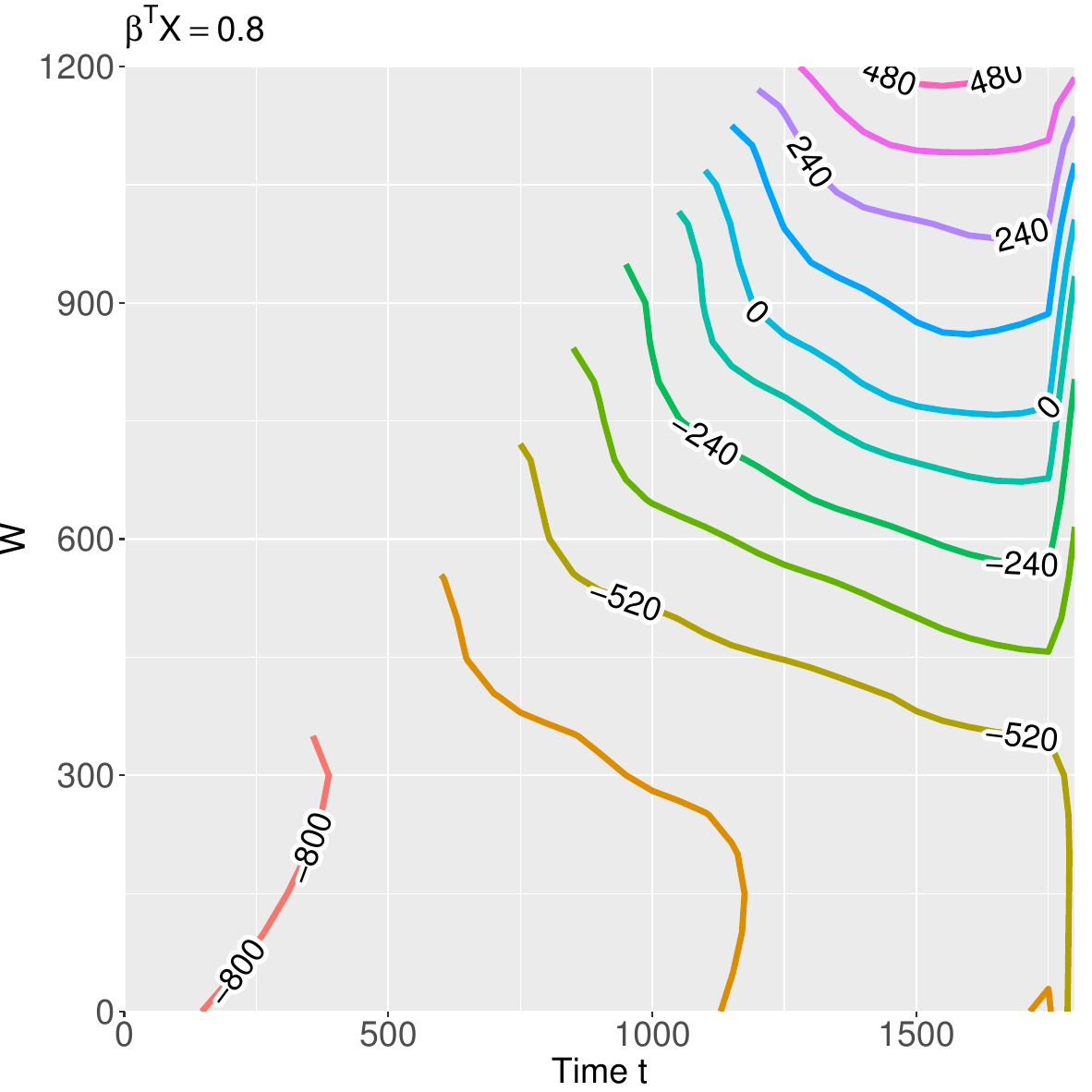}
	\caption{Mean residual life improvement from UNOS/OPTN data. Panels from left to right: $\bb\trans\x=-0.8,0,0.8$.}
	\label{fig:appDiffContour}
\end{figure}

Given a patient with characteristics
$\x$, alive at time $t$, and waiting time $w$,
 $\wh m_{T}(t-w, \wh\bb\trans\x,w)-\wh
m_{N}(t,\wh\bb\trans\x)$
provides an estimate of the patient's mean residual life improvement after receiving a kidney transplant at $w$. Because the
difference is a function of  $t$,
$\wh\bb\trans\x$ and $w$, we present
the
difference using various
plots. Figure \ref{fig:appDiffContour} plots contours that
change with $t$ and $\wh\bb\trans\x$ at several fixed
$w$ values.

Several important observations can be made. 
First, with the waiting time being close to 0 in each panel of Figure \ref{fig:appDiffContour}, kidney transplant led to less survival gains  compared to dialysis treatment, possibly
because  patients transplanted without waiting were likely to be high-risk patients and 
postoperative complications, such as cardiovascular and urological complications, increase mortality risk among them \citep{rahnemai2015independent,den2020predictors}. Second, 
as the waiting time $w$ increases, kidney transplant could result in a reasonably larger improvement compared to dialysis. This is because these patients tended to be more stable, allowing kidney transplant to provide a notable survival advantage  \citep{ingsathit2013survival,schold2014association,bui2019functional}. Moreover, with $t$ and $w$ fixed, complex relationships existed between the patient's index value $\bb\trans\x$ and the survival improvement. The improvement is larger at $\bb\trans\x=0$ than that at $\bb\trans\x=-0.8,0.8$. It is very likely that large or small values of $\bb\trans\x$ were resulted by extreme health conditions which led to the worse improvement. Thus, this index in general measured patients' overall health condition.

\begin{figure}
	\centering
	\captionsetup[subfigure]{labelformat=empty}
	\begin{tikzpicture}
		\node (img11) {\includegraphics[width=3cm]{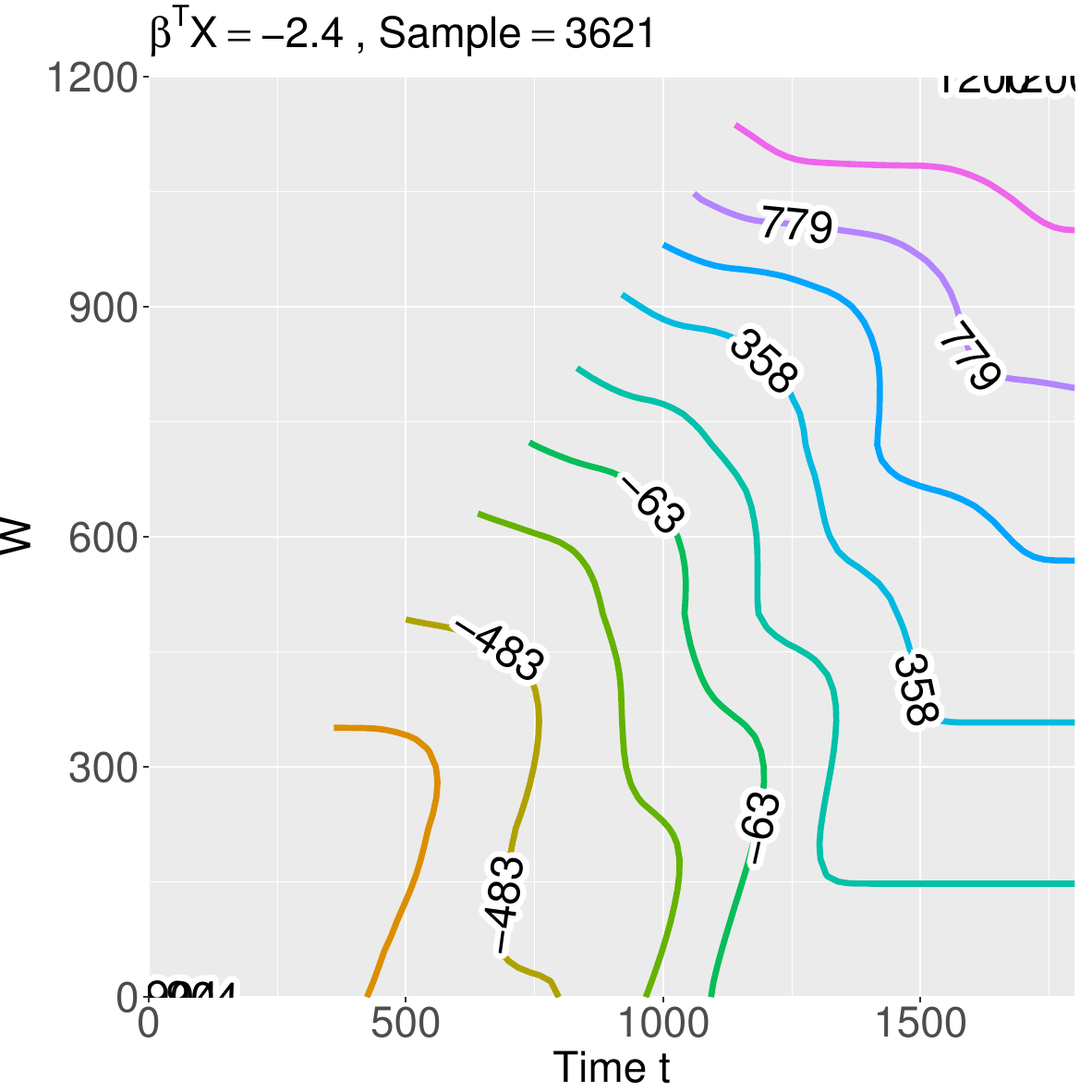}};
		\node[left=of img11, node distance=0cm, rotate=90, anchor=center,font=\tiny,xshift=0.2cm,yshift=-1cm] {African American};
		\node[above=of img11, node distance=0cm, anchor=center,font=\tiny,xshift=0.3cm,yshift=-1cm] {Female,Public};
	\end{tikzpicture}
	\begin{tikzpicture}
		\node (img12) {\includegraphics[width=3cm]{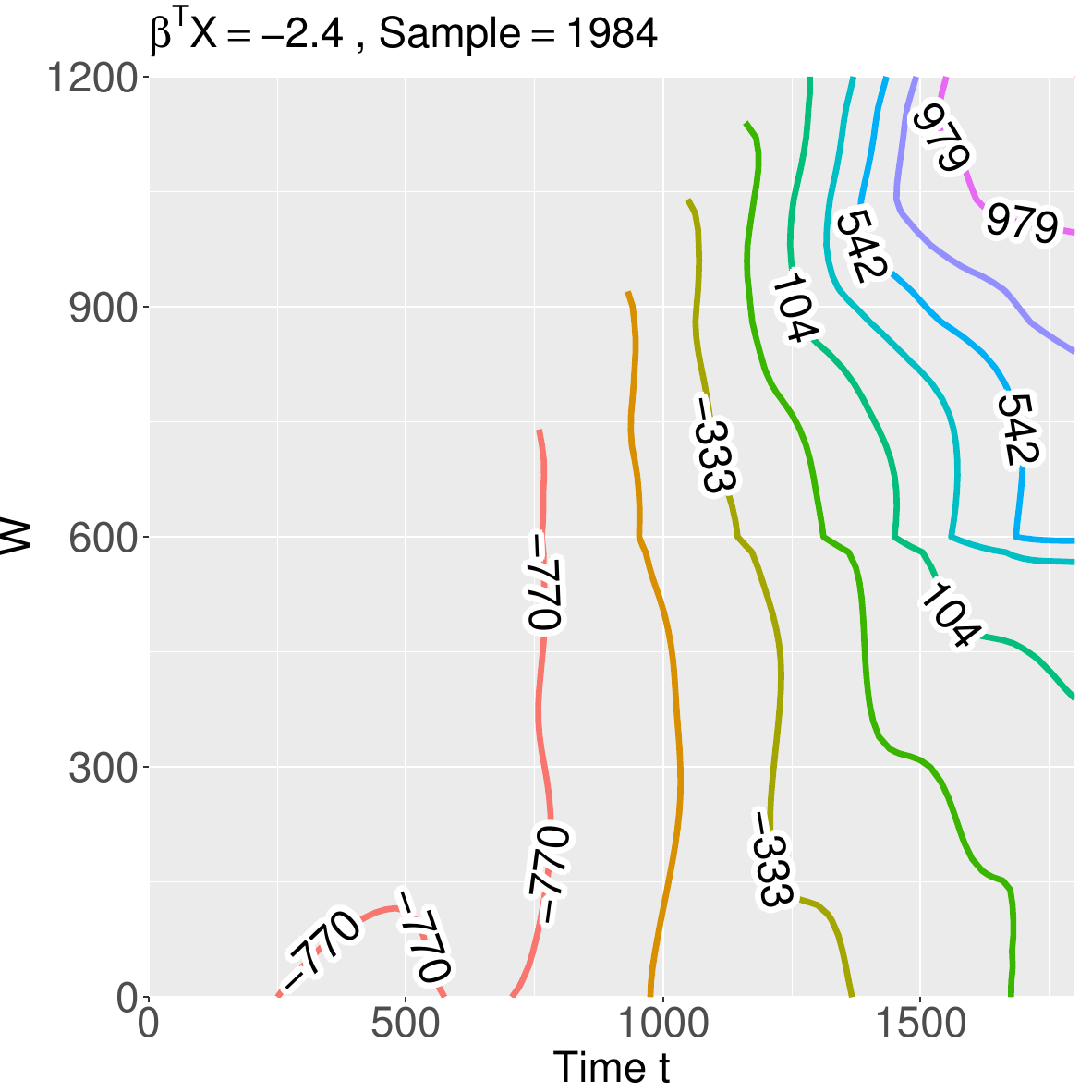}};
		\node[above=of img12, node distance=0cm, anchor=center,font=\tiny,xshift=0.3cm,yshift=-1cm] {Female,Private};
	\end{tikzpicture}
	\begin{tikzpicture}
		\node (img13) {\includegraphics[width=3cm]{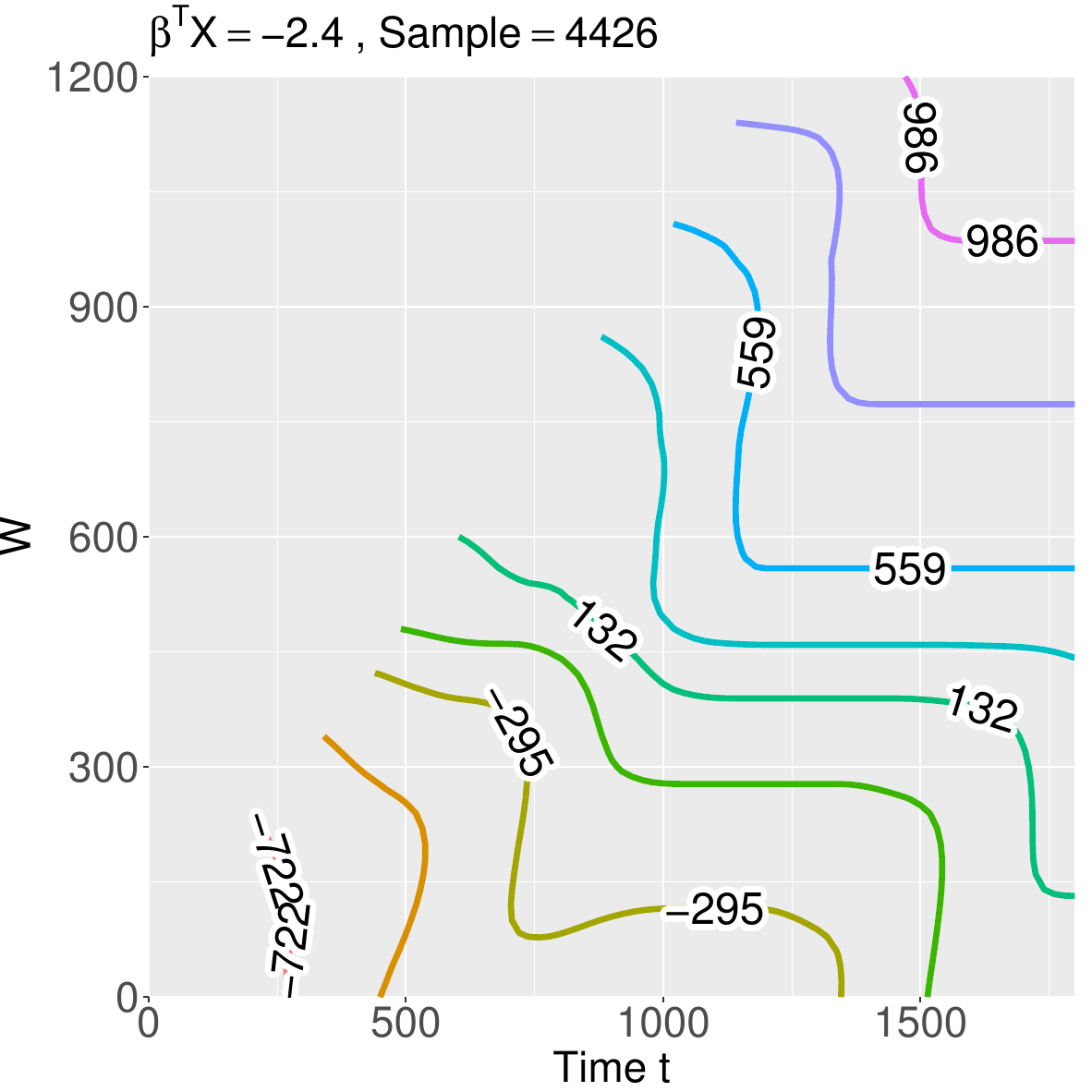}};
		\node[above=of img13, node distance=0cm, anchor=center,font=\tiny,xshift=0.3cm,yshift=-1cm] {Male,Public};
	\end{tikzpicture}
	\begin{tikzpicture}
		\node (img14) {\includegraphics[width=3cm]{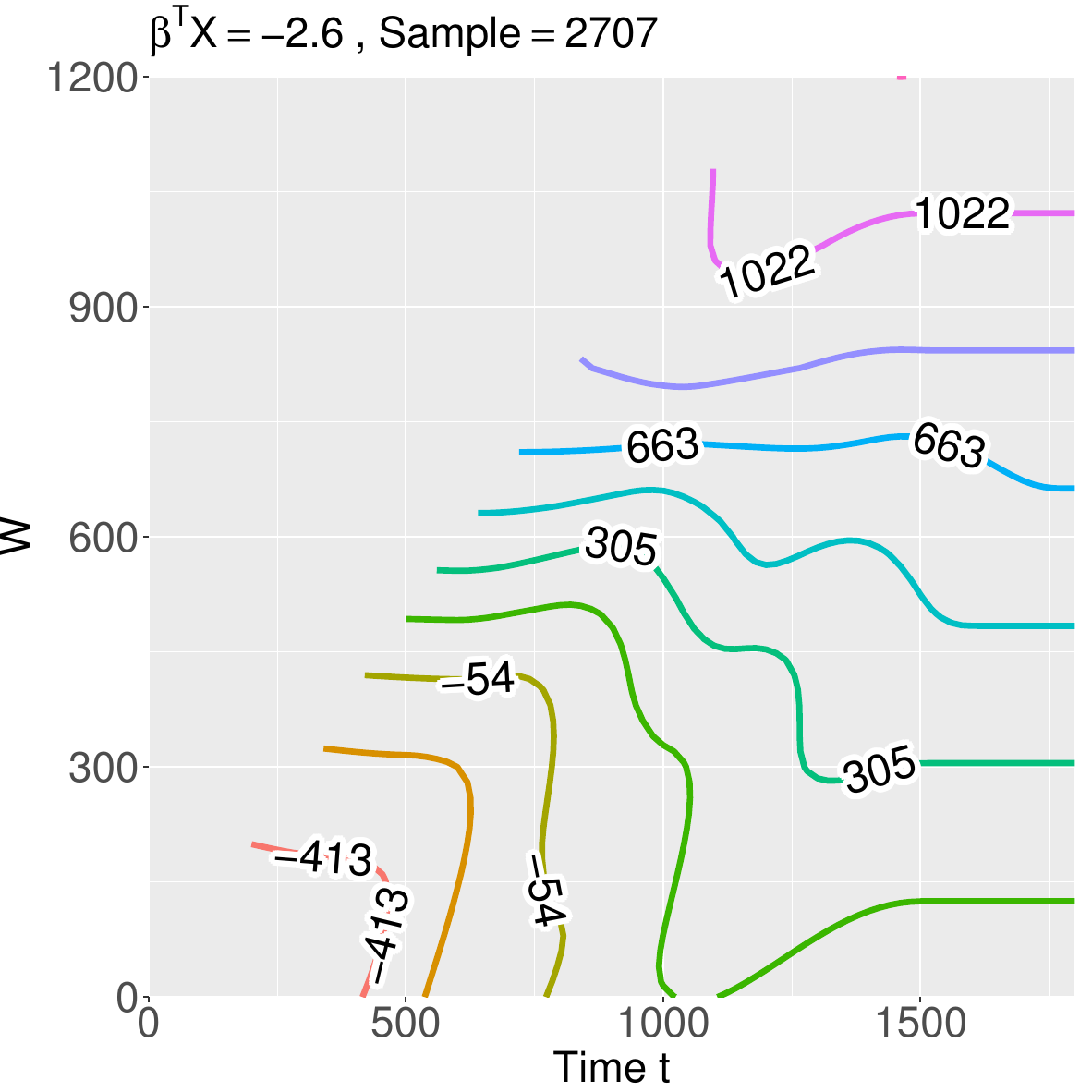}};
		\node[above=of img14, node distance=0cm, anchor=center,font=\tiny,xshift=0.3cm,yshift=-1cm] {Male,Private};
	\end{tikzpicture}\\
	\begin{tikzpicture}
		\node (img21) {\includegraphics[width=3cm]{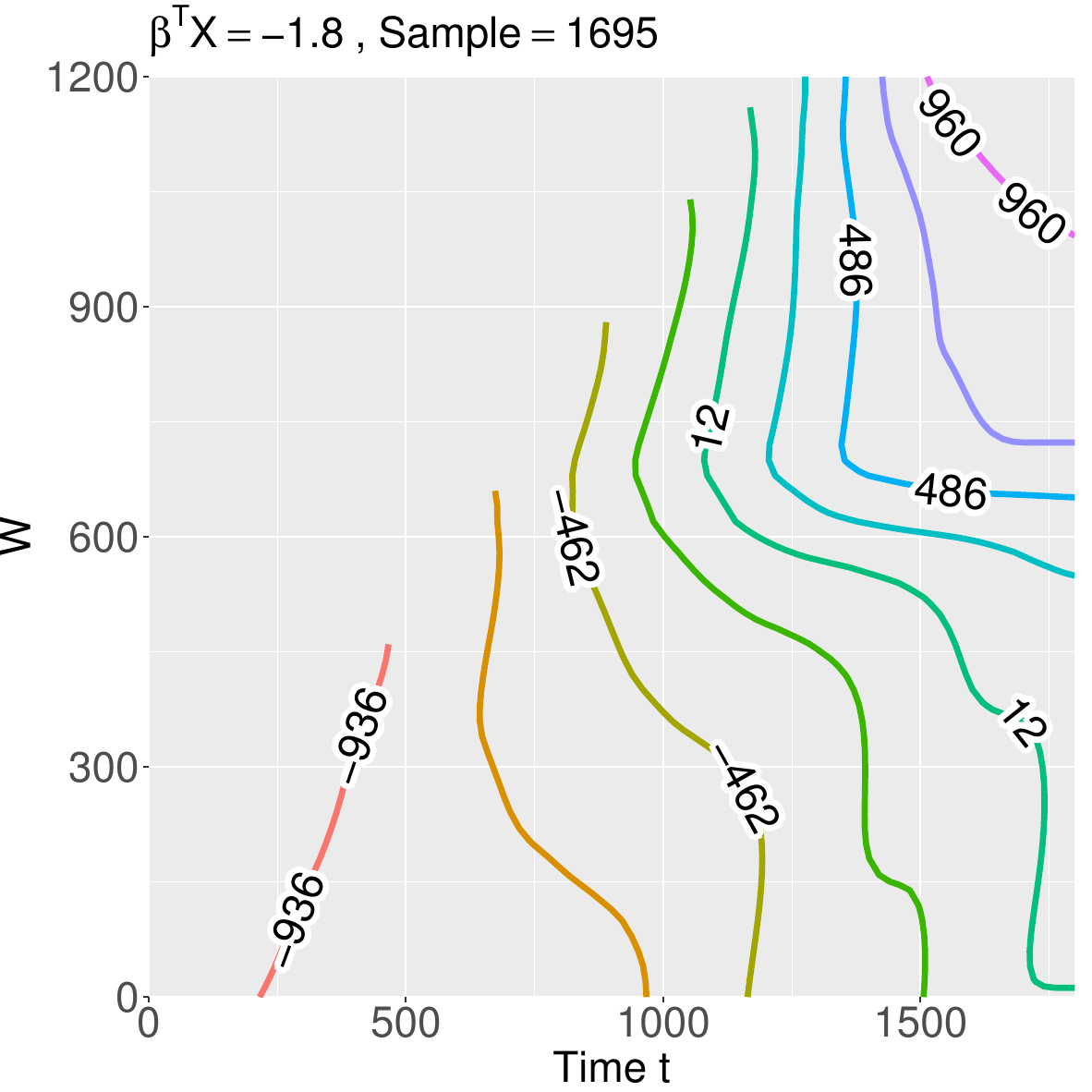}};
		\node[left=of img11, node distance=0cm, rotate=90, anchor=center,font=\tiny,xshift=0.2cm,yshift=-1cm] {Hispanic};
	\end{tikzpicture}
	\begin{tikzpicture}
		\node (img22) {\includegraphics[width=3cm]{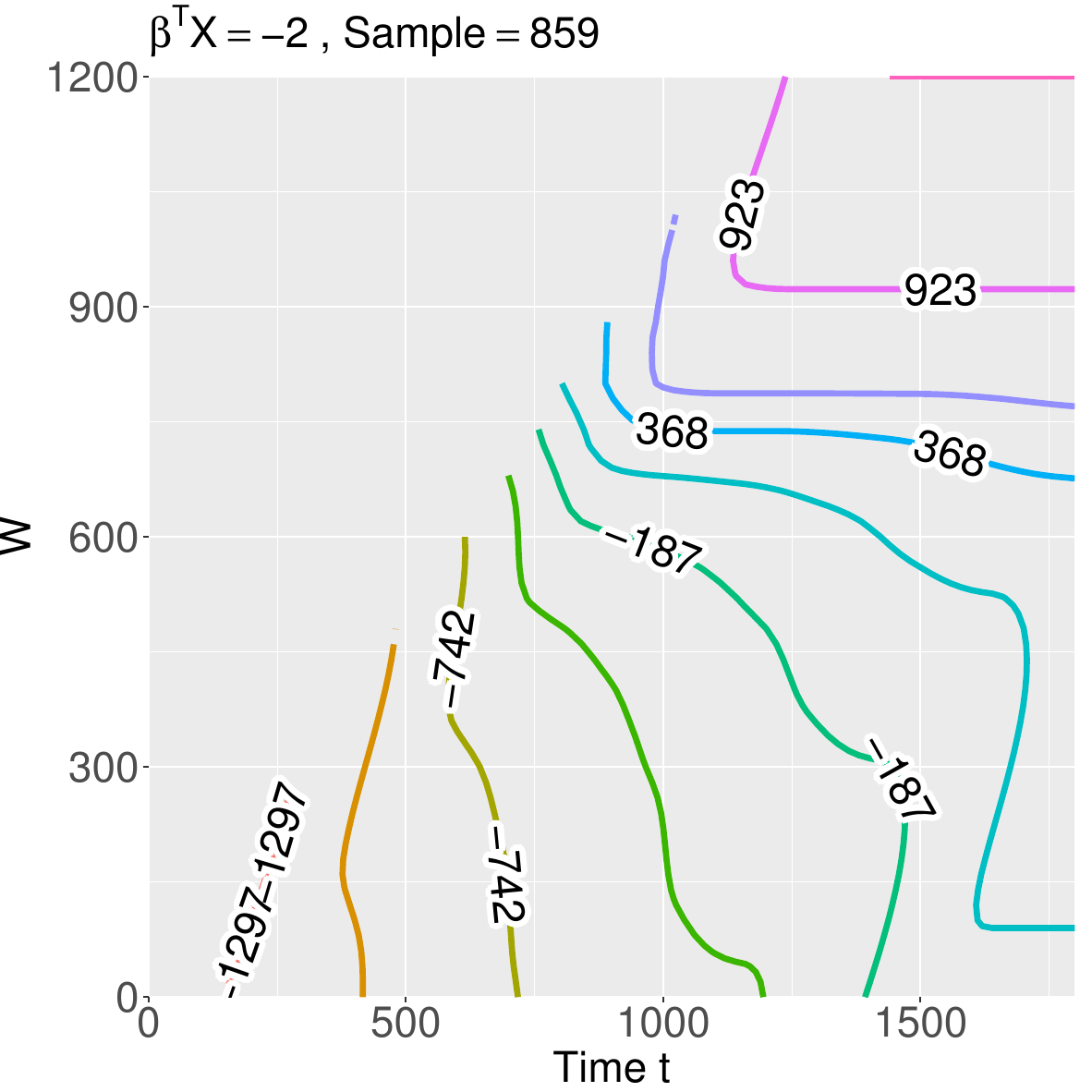}};
	\end{tikzpicture}
	\begin{tikzpicture}
		\node (img23) {\includegraphics[width=3cm]{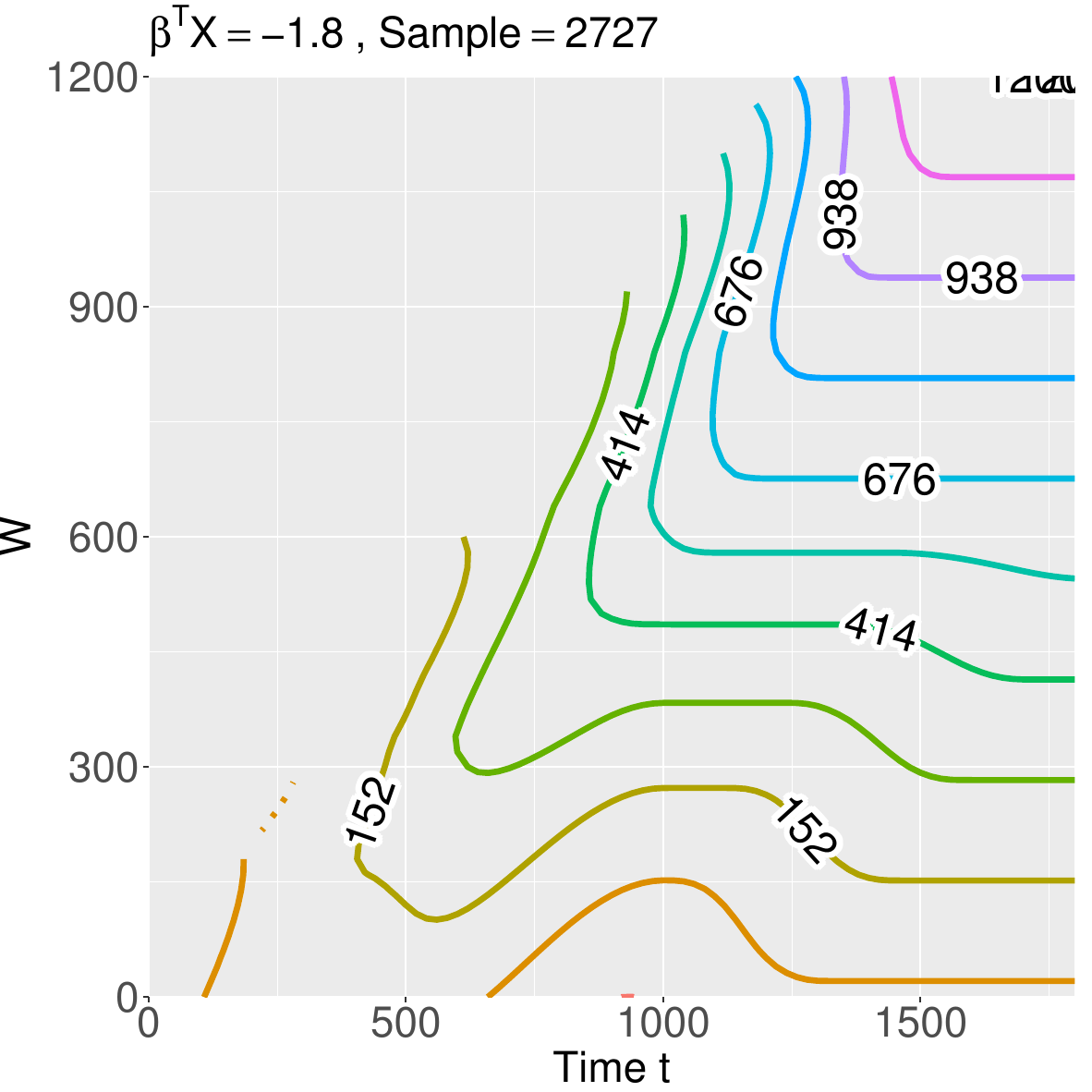}};
	\end{tikzpicture}
	\begin{tikzpicture}
		\node (img24) {\includegraphics[width=3cm]{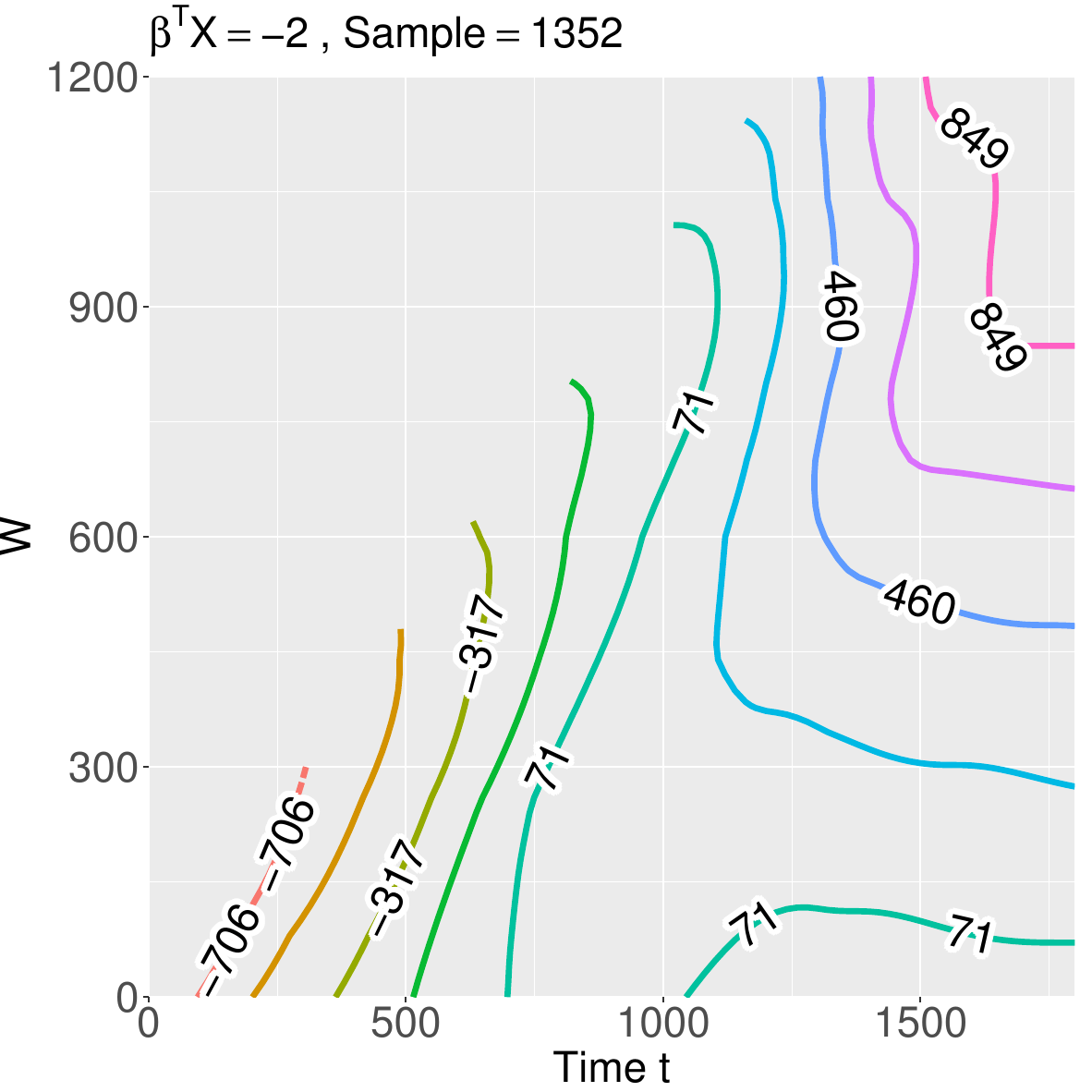}};
	\end{tikzpicture}\\
	\begin{tikzpicture}
		\node (img31) {\includegraphics[width=3cm]{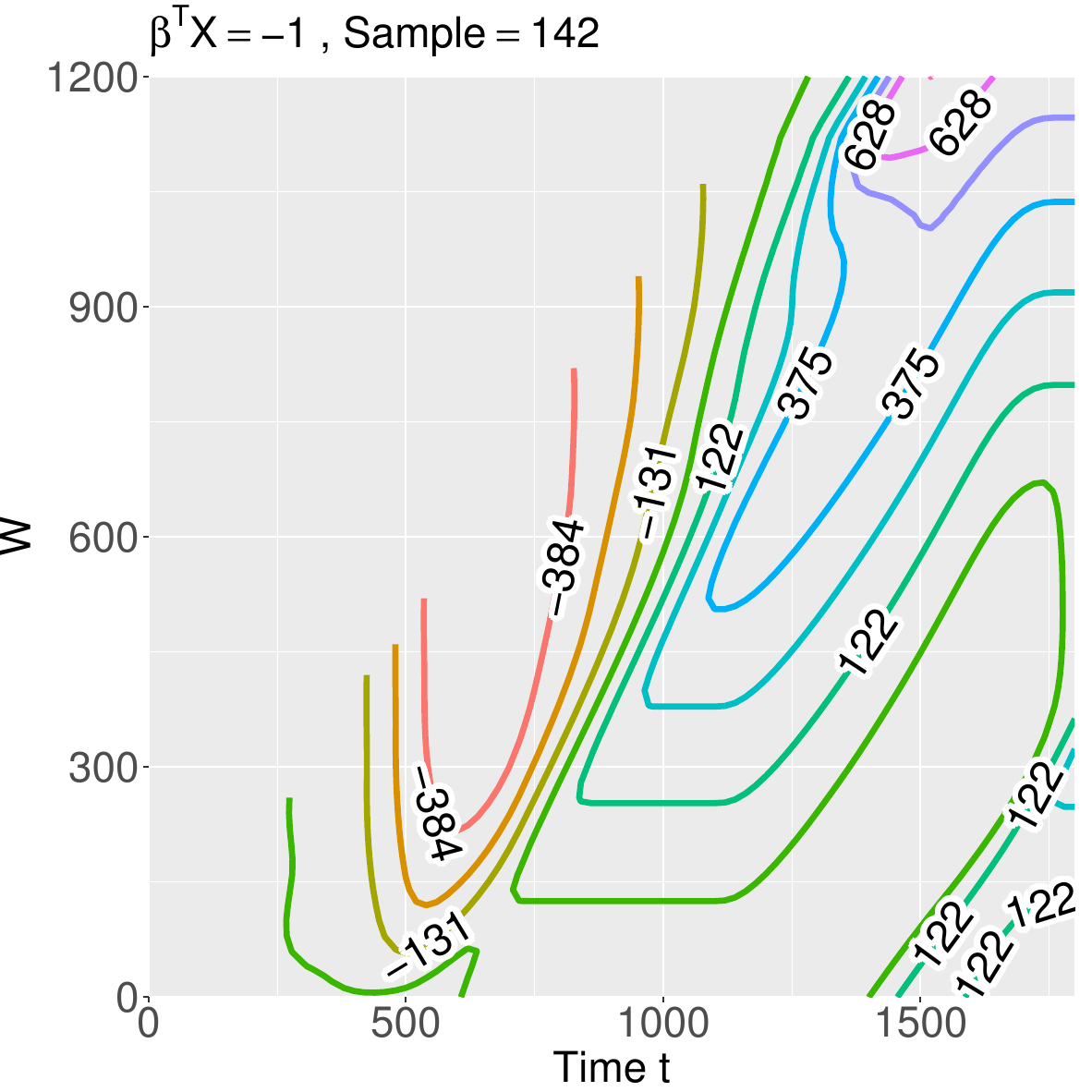}};
		\node[left=of img11, node distance=0cm, rotate=90, anchor=center,font=\tiny,xshift=0.2cm,yshift=-1cm] {Native};
	\end{tikzpicture}
	\begin{tikzpicture}
		\node (img32) {\includegraphics[width=3cm]{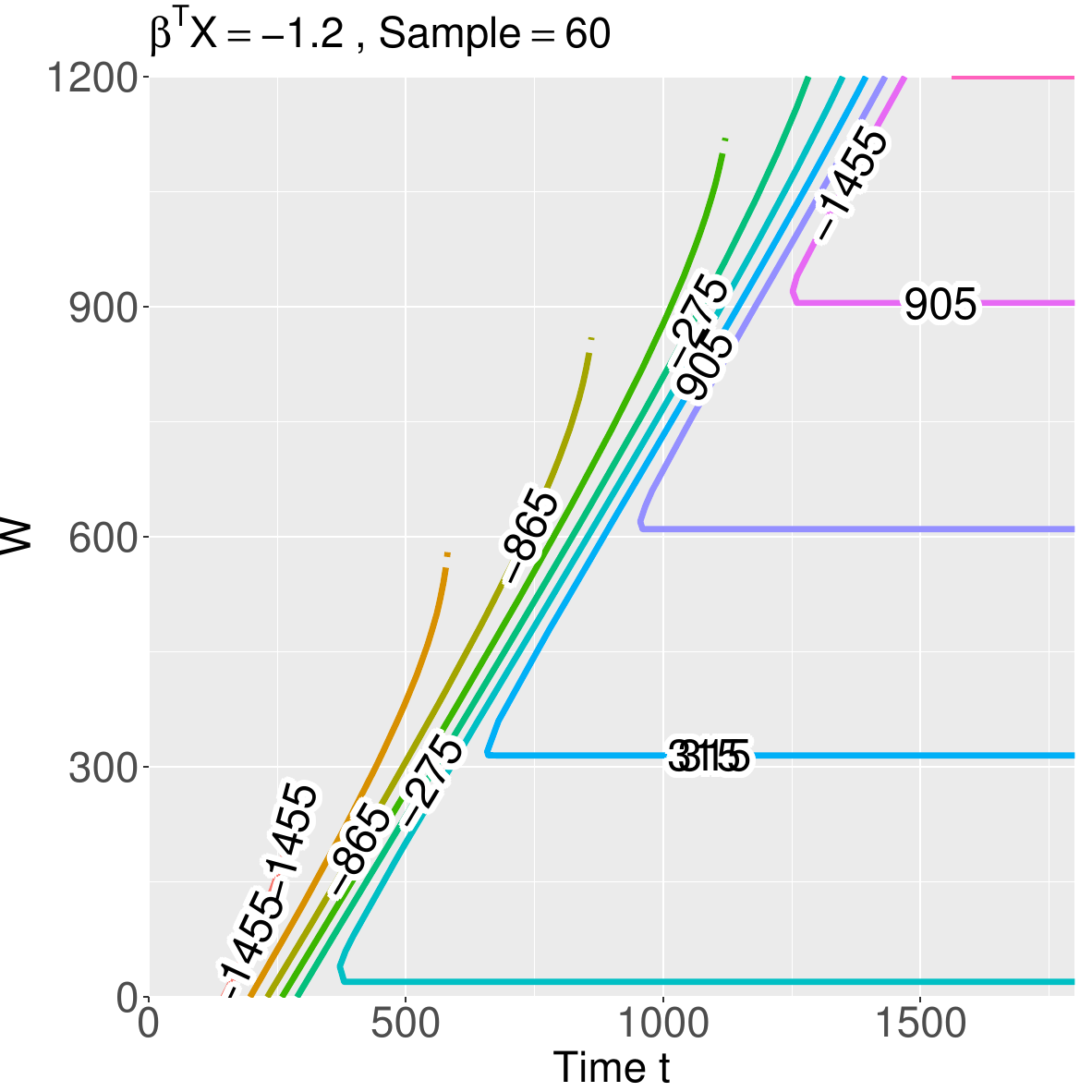}};
	\end{tikzpicture}
	\begin{tikzpicture}
		\node (img33) {\includegraphics[width=3cm]{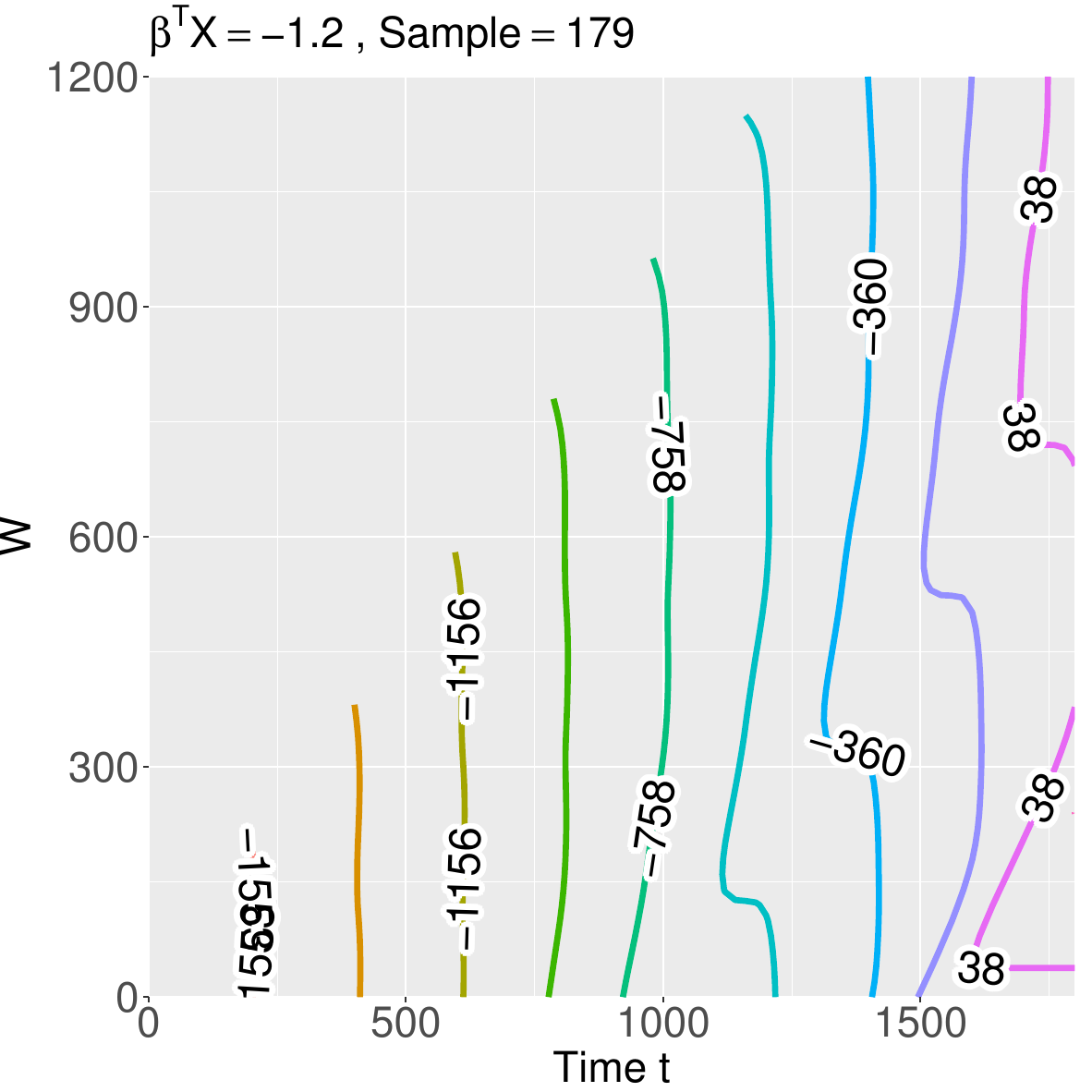}};
	\end{tikzpicture}
	\begin{tikzpicture}
		\node (img34) {\includegraphics[width=3cm]{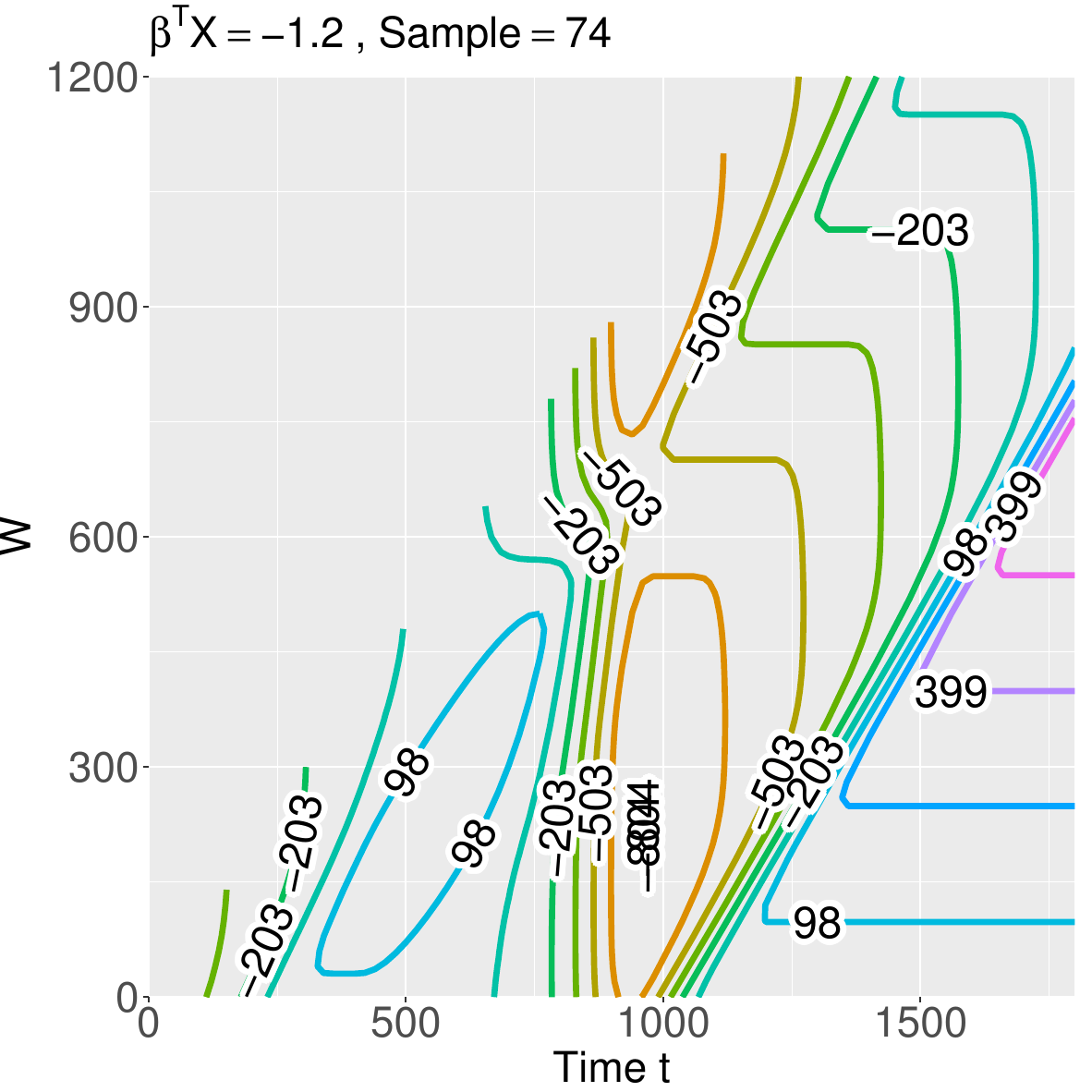}};
	\end{tikzpicture}\\
	\begin{tikzpicture}
		\node (img41) {\includegraphics[width=3cm]{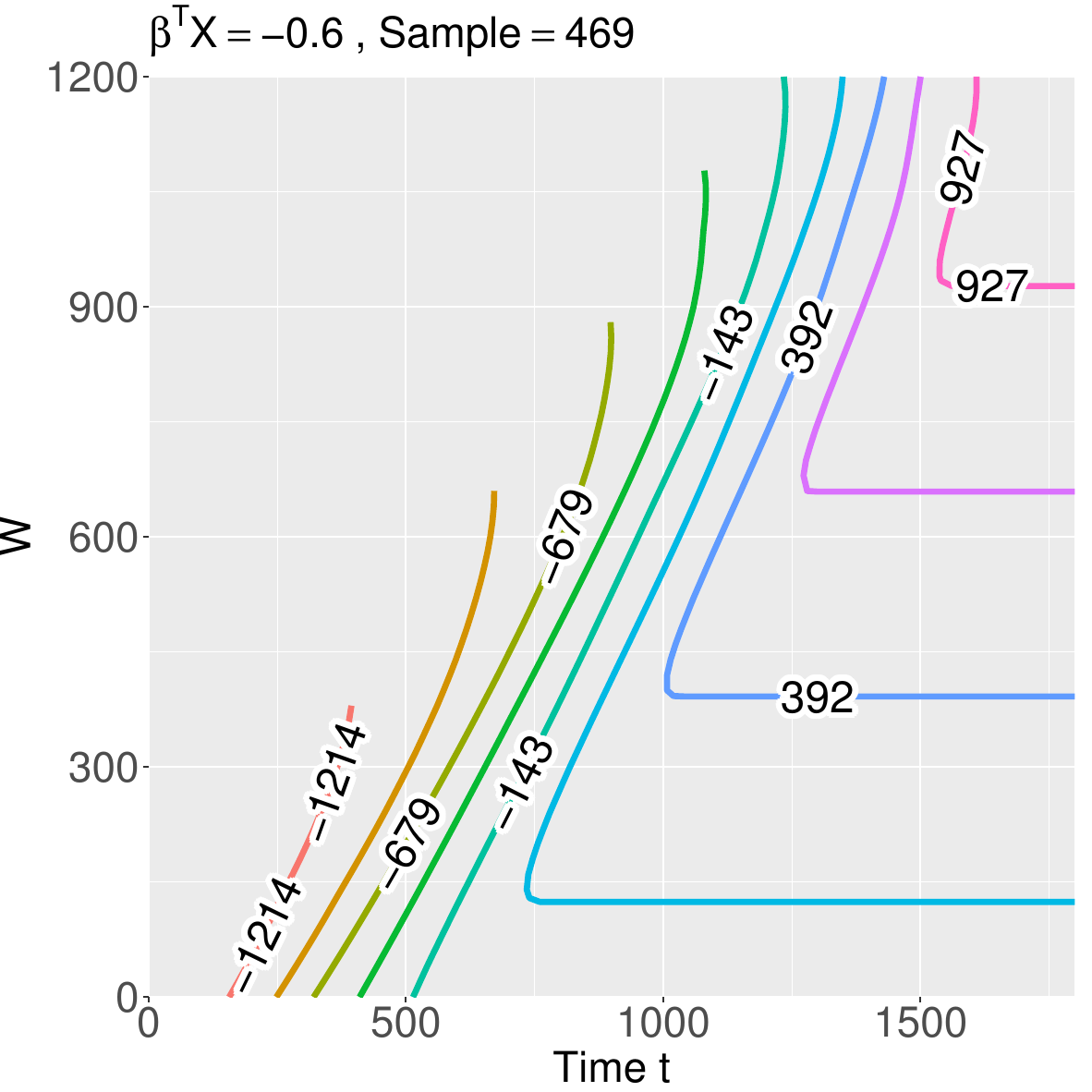}};
		\node[left=of img11, node distance=0cm, rotate=90, anchor=center,font=\tiny,xshift=0.2cm,yshift=-1cm] {Asian};
	\end{tikzpicture}
	\begin{tikzpicture}
		\node (img42) {\includegraphics[width=3cm]{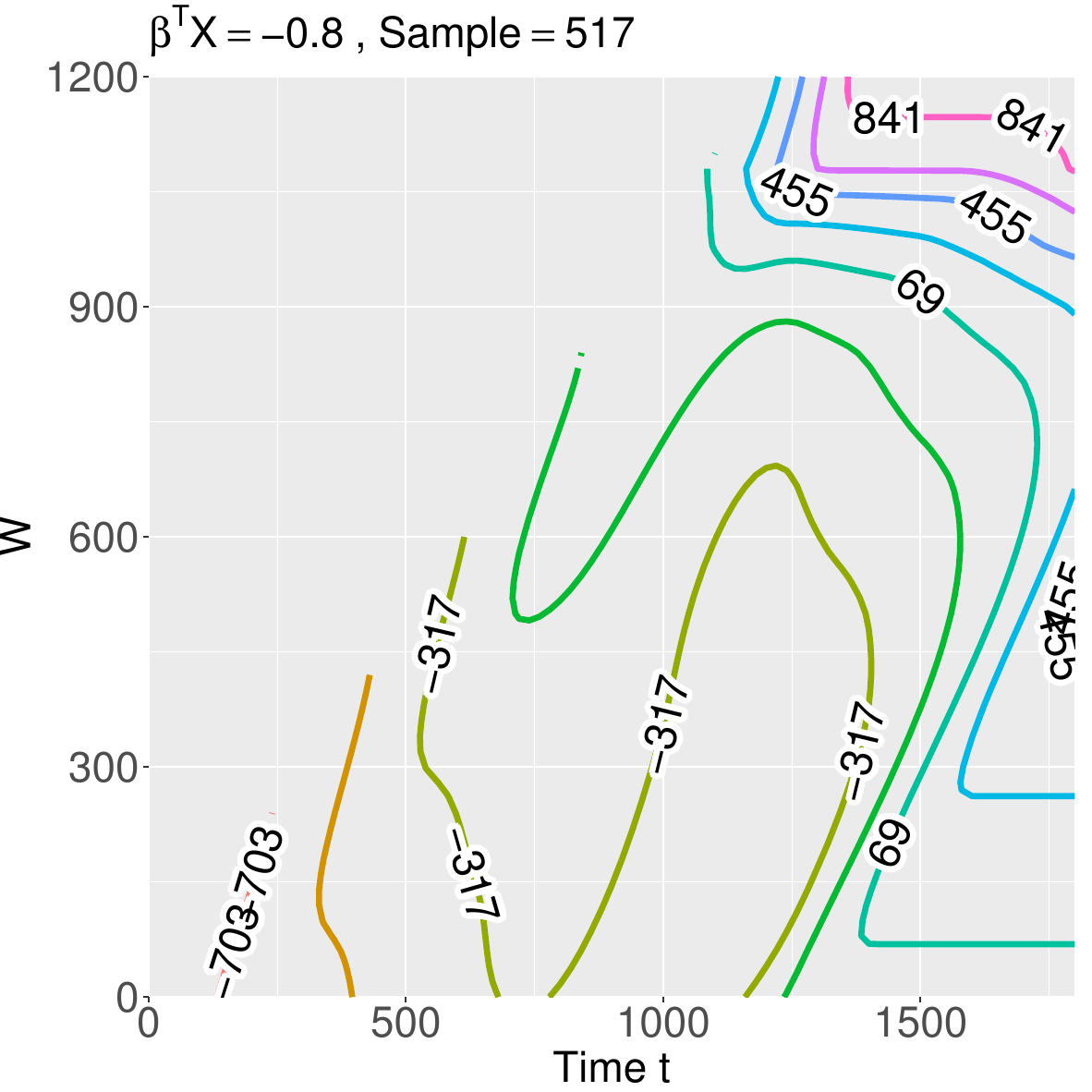}};
	\end{tikzpicture}
	\begin{tikzpicture}
		\node (img43) {\includegraphics[width=3cm]{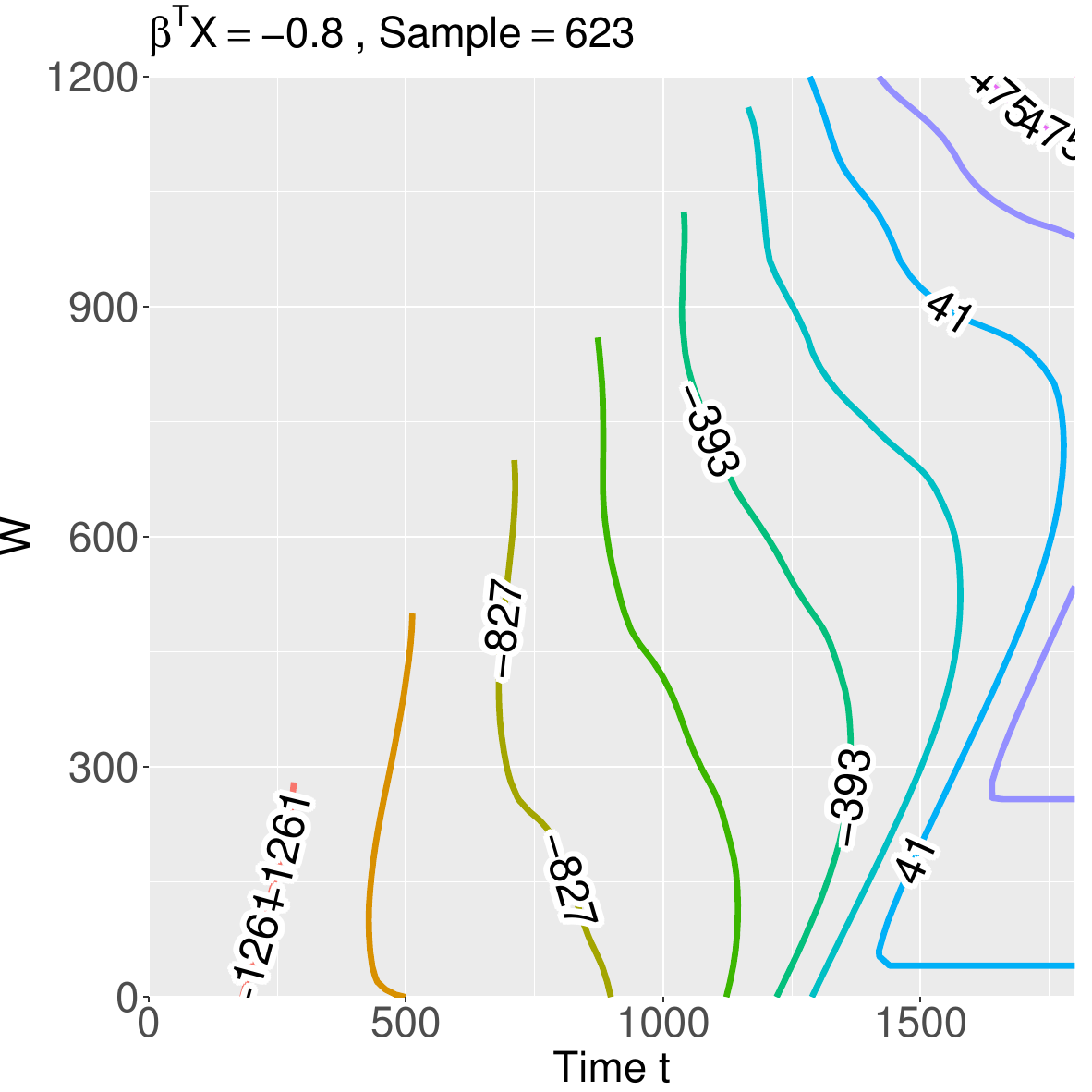}};
	\end{tikzpicture}
	\begin{tikzpicture}
		\node (img44) {\includegraphics[width=3cm]{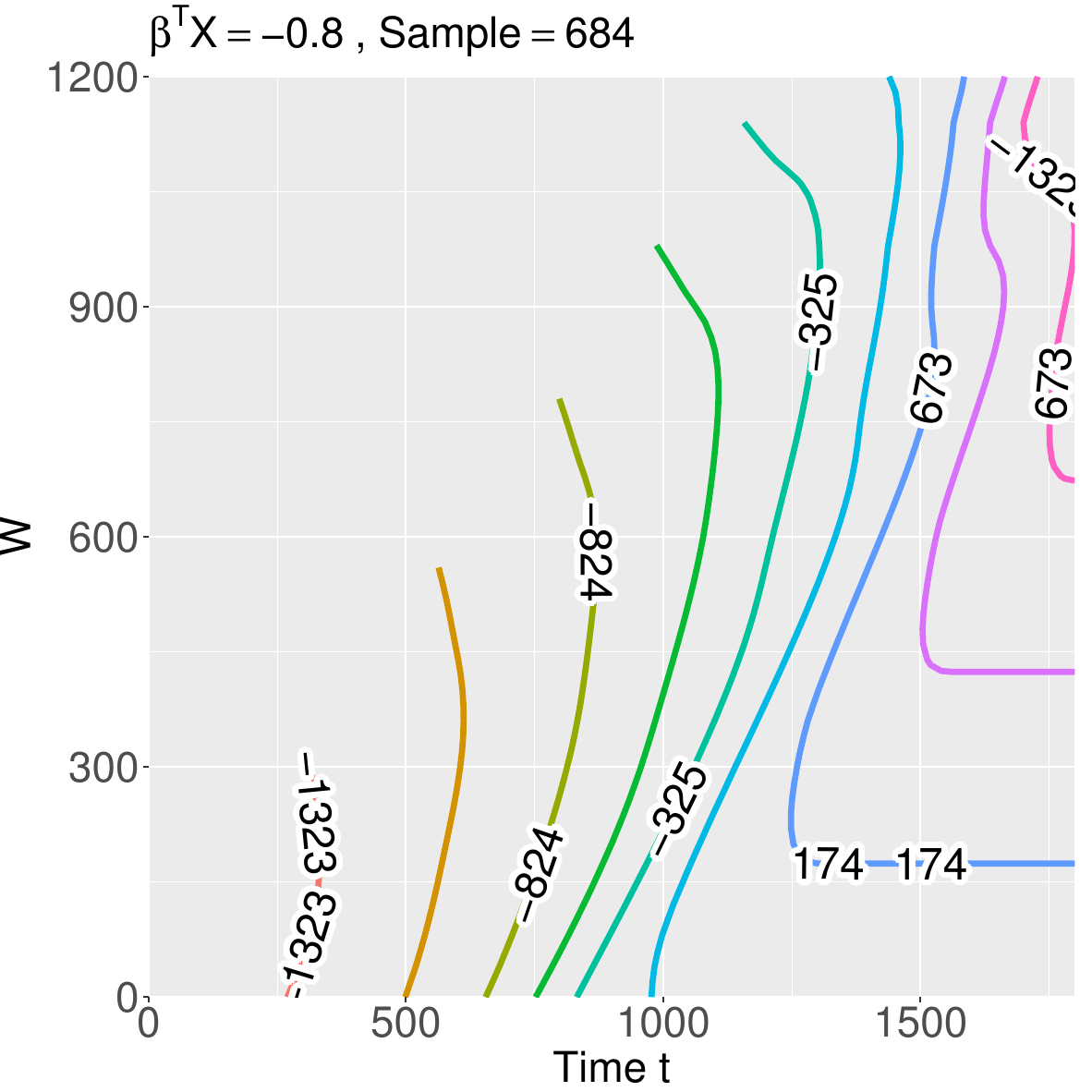}};
	\end{tikzpicture}\\
	\begin{tikzpicture}
		\node (img51) {\includegraphics[width=3cm]{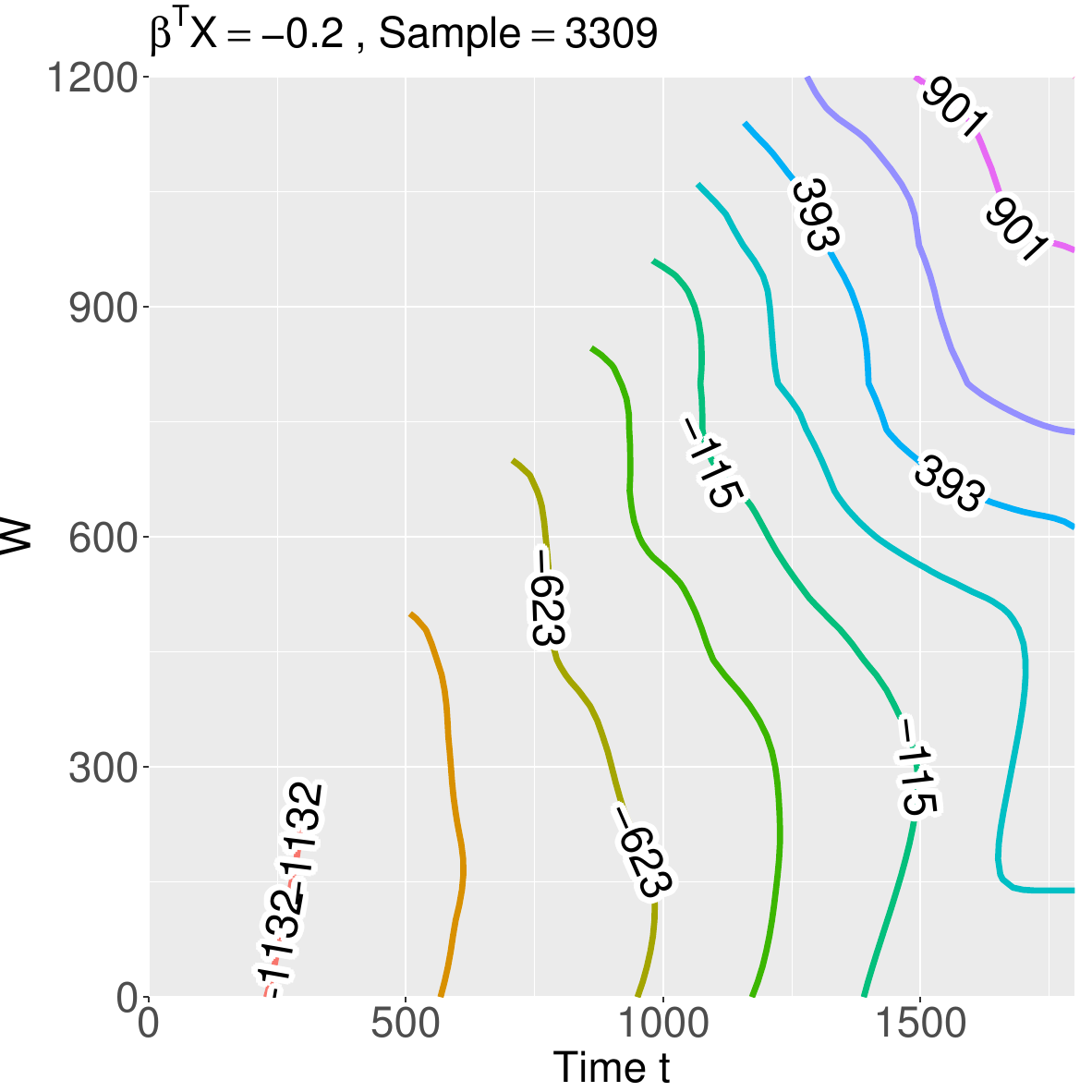}};
		\node[left=of img11, node distance=0cm, rotate=90, anchor=center,font=\tiny,xshift=0.2cm,yshift=-1cm] {White};
	\end{tikzpicture}
	\begin{tikzpicture}
		\node (img52) {\includegraphics[width=3cm]{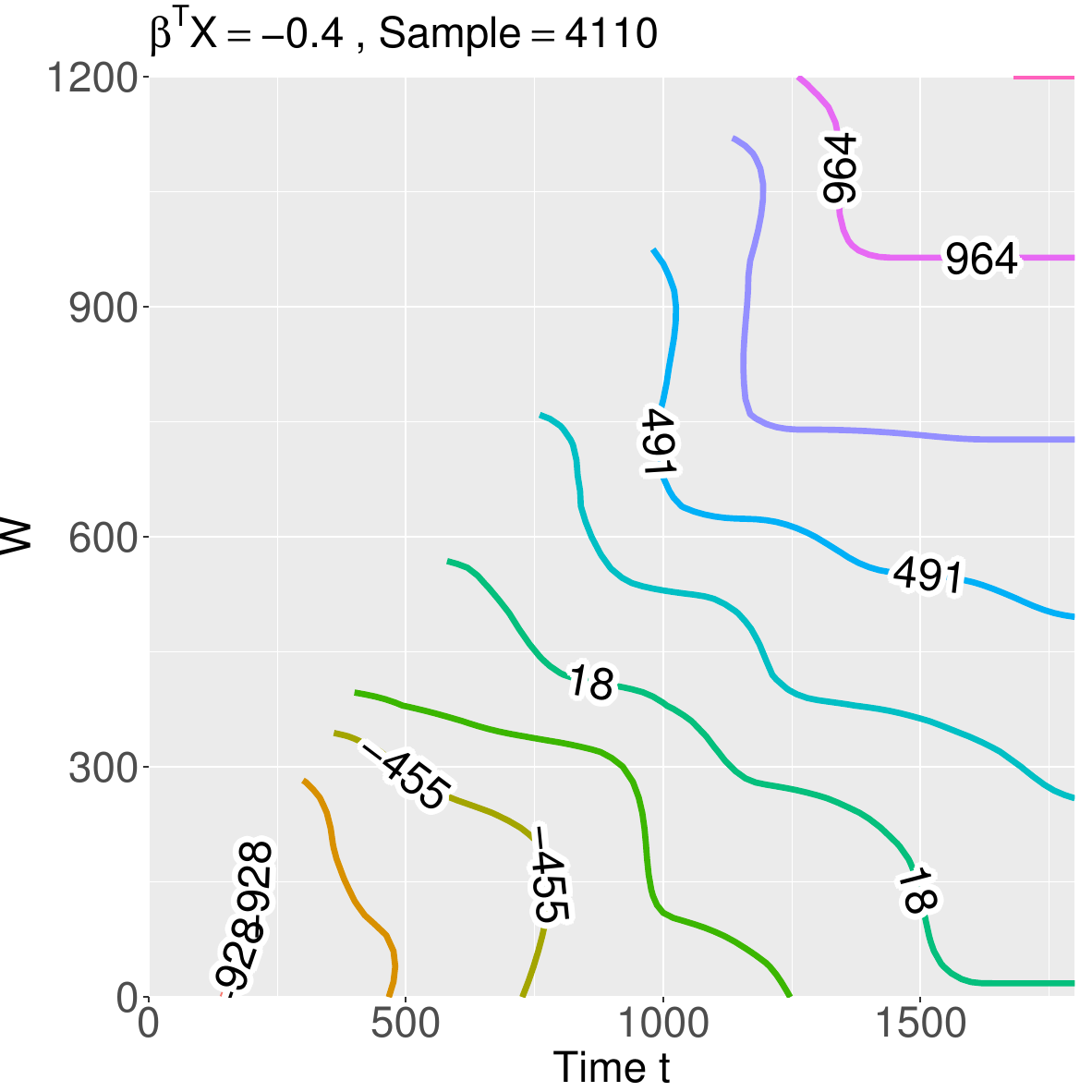}};
	\end{tikzpicture}
	\begin{tikzpicture}
		\node (img53) {\includegraphics[width=3cm]{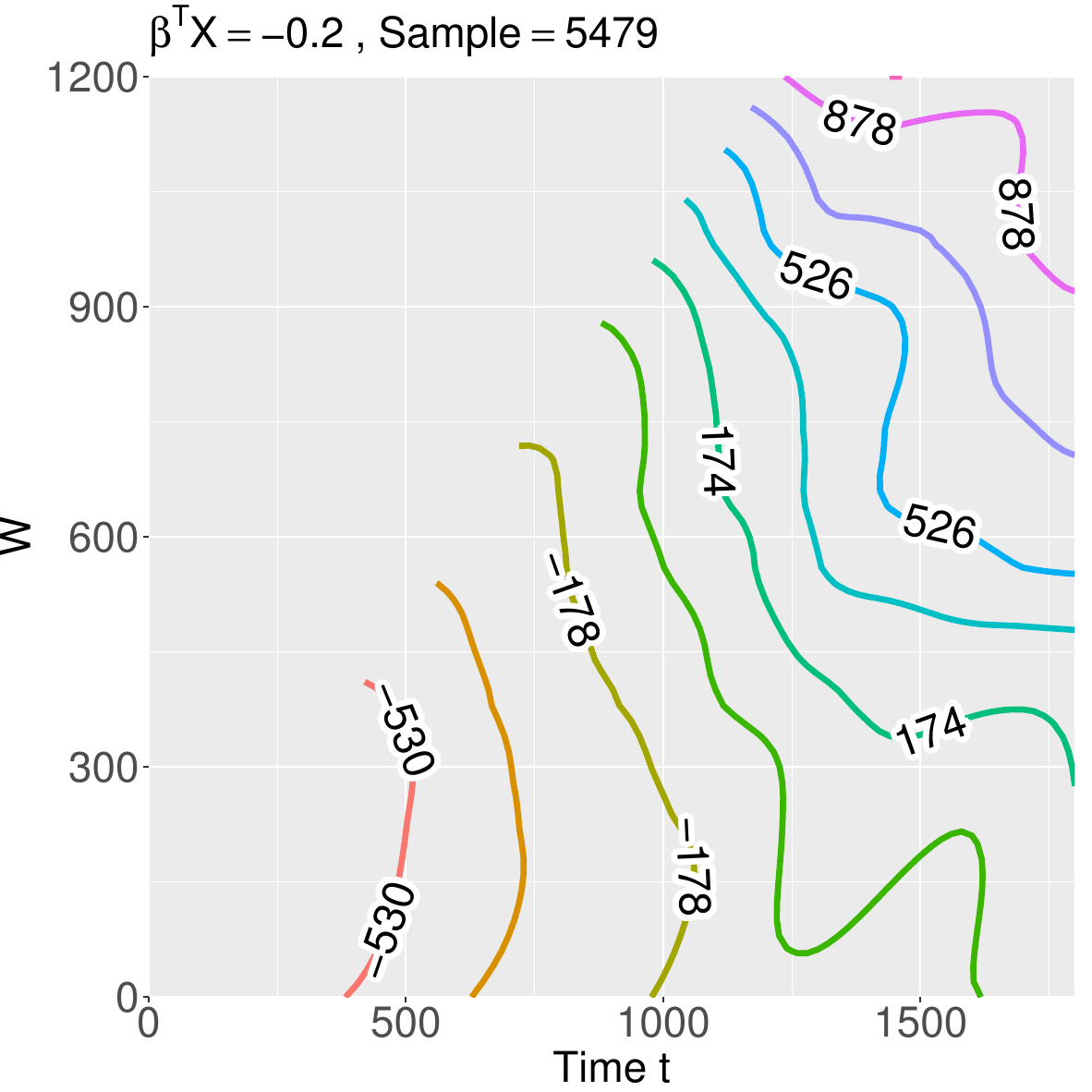}};
	\end{tikzpicture}
	\begin{tikzpicture}
		\node (img54) {\includegraphics[width=3cm]{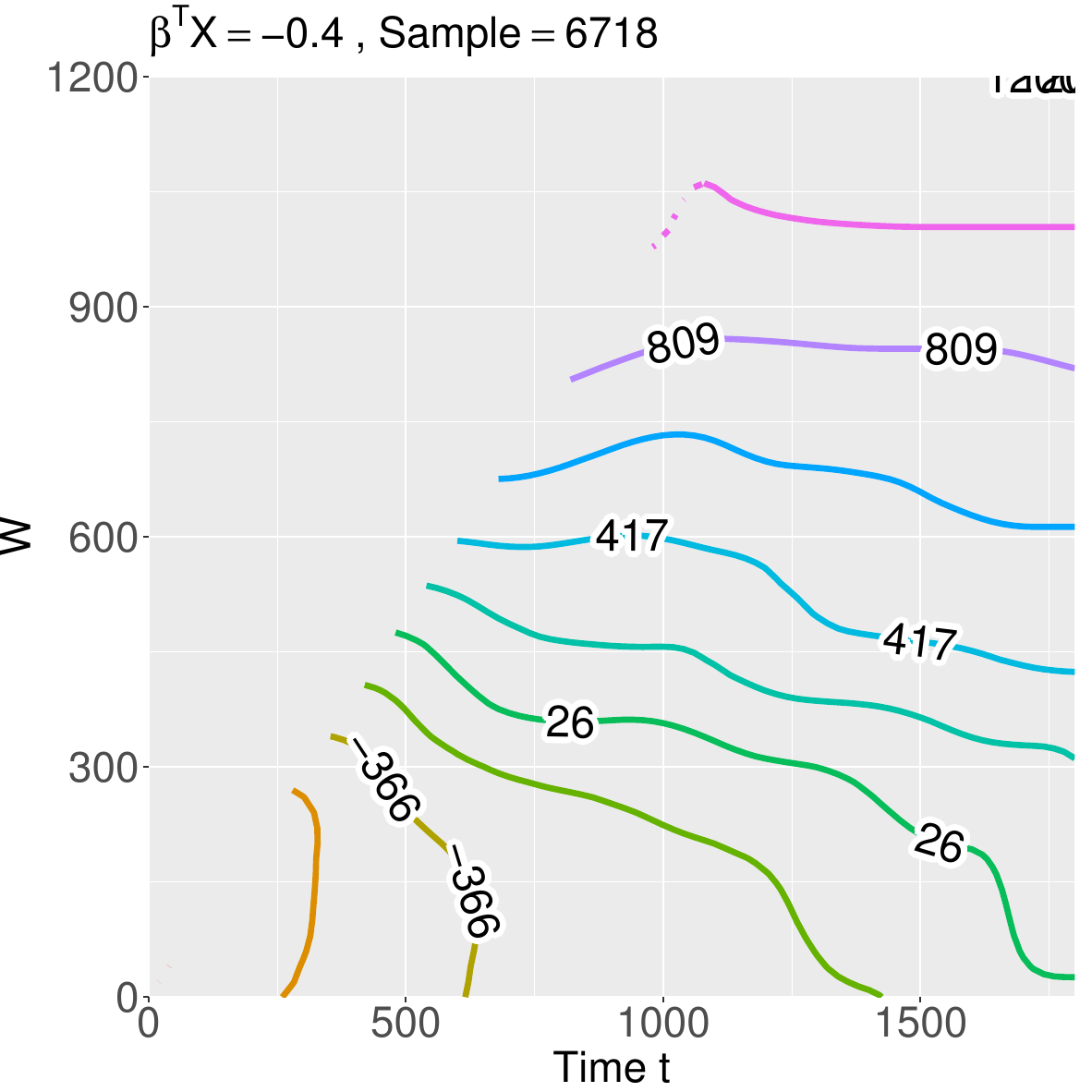}};
	\end{tikzpicture}\\
	\caption{Mean residual life improvement from UNOS/OPTN data. Stratified by Race, Gender, and Insurance Status with minimum $\bb\trans\x$ per strata.}
	\label{fig:appDiffContourCloser1}
\end{figure}
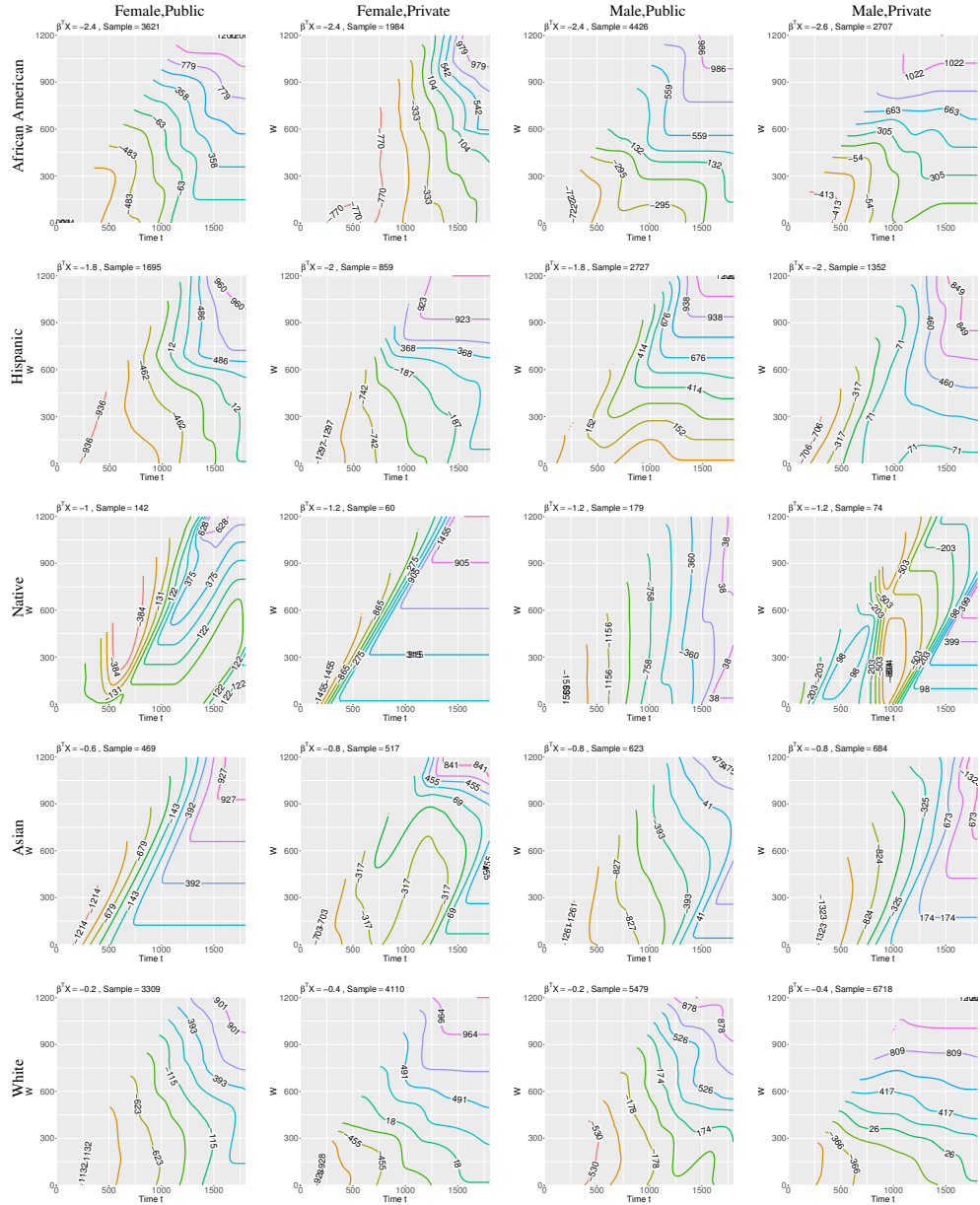
\begin{figure}
	\centering
	\captionsetup[subfigure]{labelformat=empty}
	\begin{tikzpicture}
		\node (img11) {\includegraphics[width=3cm]{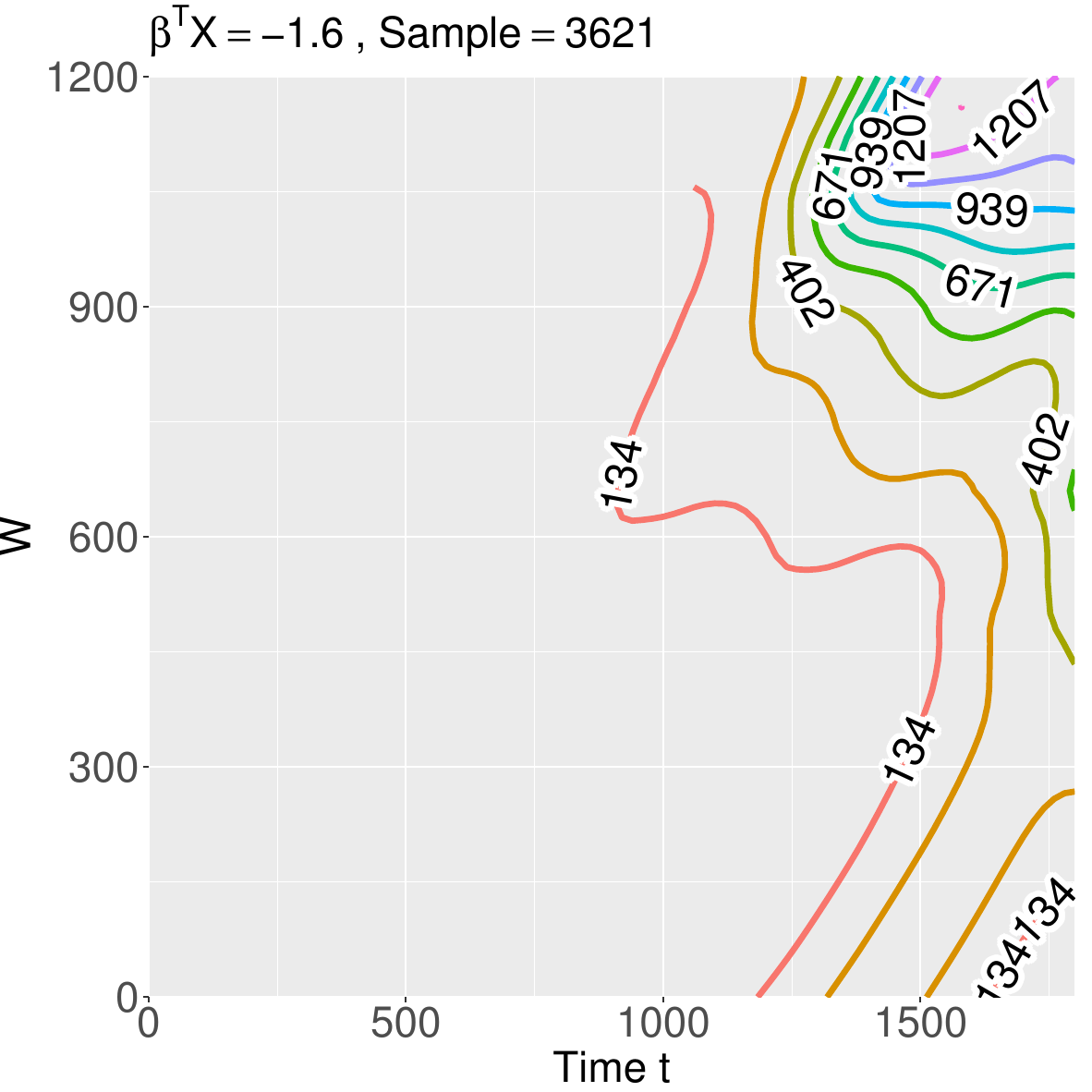}};
		\node[left=of img11, node distance=0cm, rotate=90, anchor=center,font=\tiny,xshift=0.2cm,yshift=-1cm] {African American};
		\node[above=of img11, node distance=0cm, anchor=center,font=\tiny,xshift=0.3cm,yshift=-1cm] {Female,Public};
	\end{tikzpicture}
	\begin{tikzpicture}
		\node (img12) {\includegraphics[width=3cm]{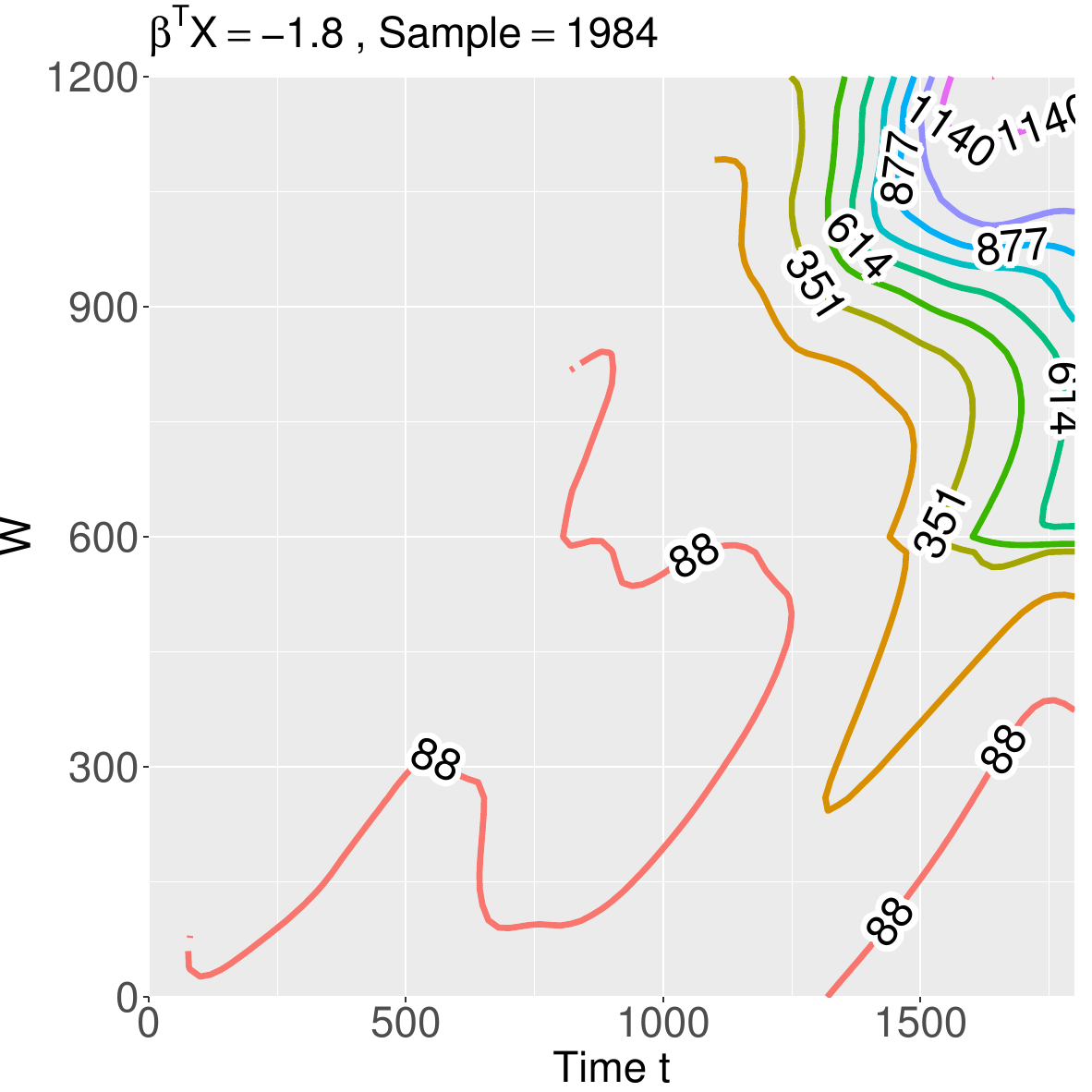}};
		\node[above=of img12, node distance=0cm, anchor=center,font=\tiny,xshift=0.3cm,yshift=-1cm] {Female,Private};
	\end{tikzpicture}
	\begin{tikzpicture}
		\node (img13) {\includegraphics[width=3cm]{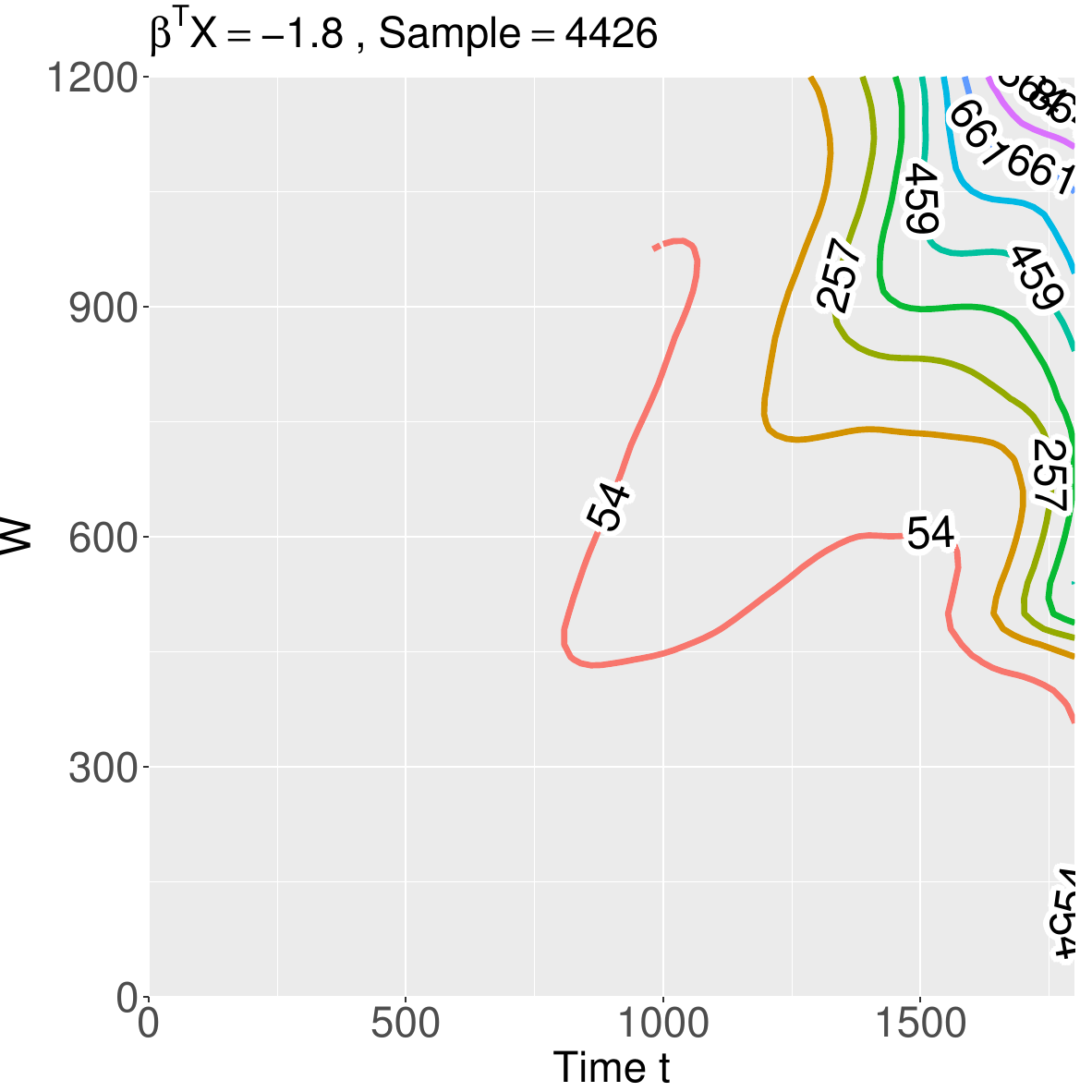}};
		\node[above=of img13, node distance=0cm, anchor=center,font=\tiny,xshift=0.3cm,yshift=-1cm] {Male,Public};
	\end{tikzpicture}
	\begin{tikzpicture}
		\node (img14) {\includegraphics[width=3cm]{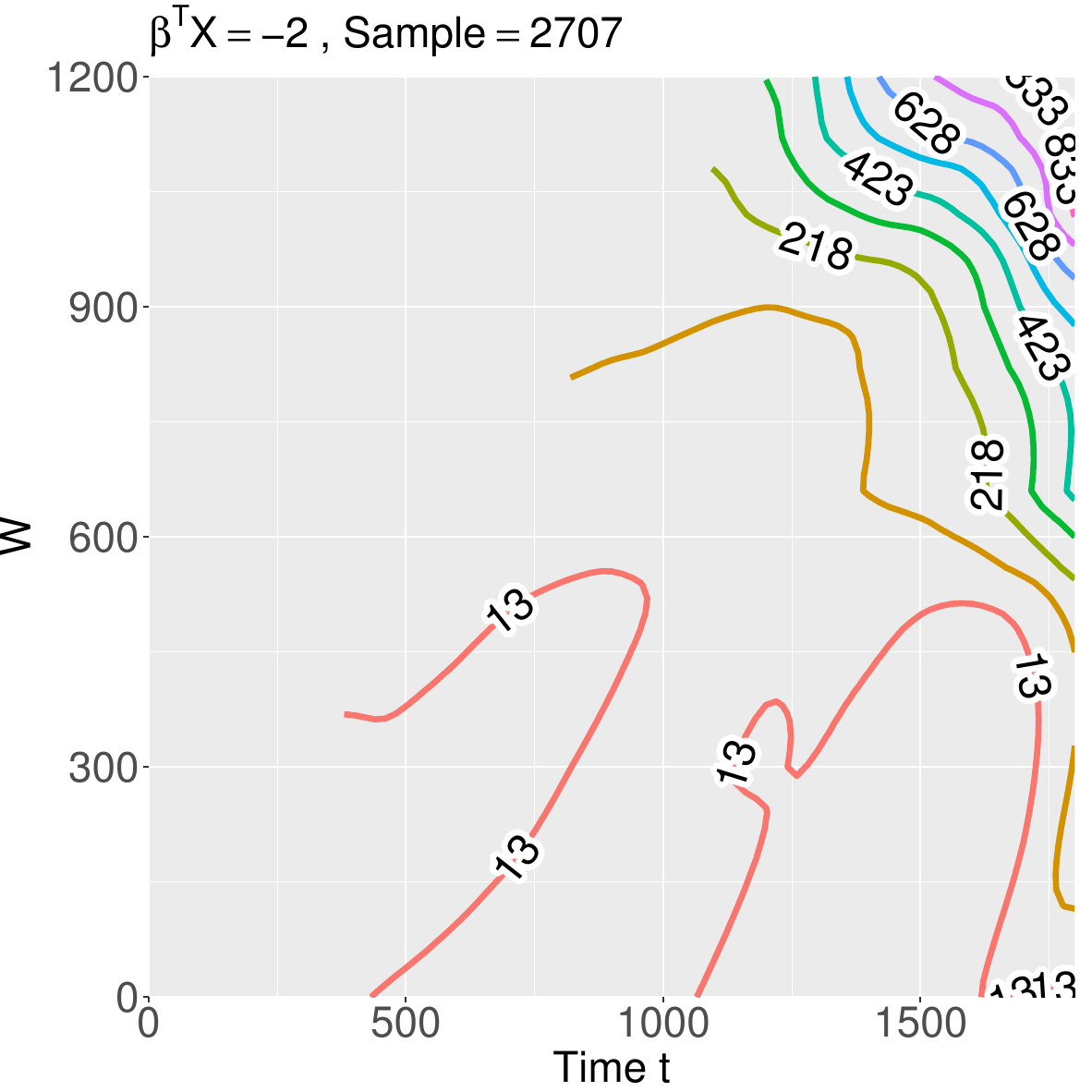}};
		\node[above=of img14, node distance=0cm, anchor=center,font=\tiny,xshift=0.3cm,yshift=-1cm] {Male,Private};
	\end{tikzpicture}\\
        \begin{tikzpicture}
		\node (img21) {\includegraphics[width=3cm]{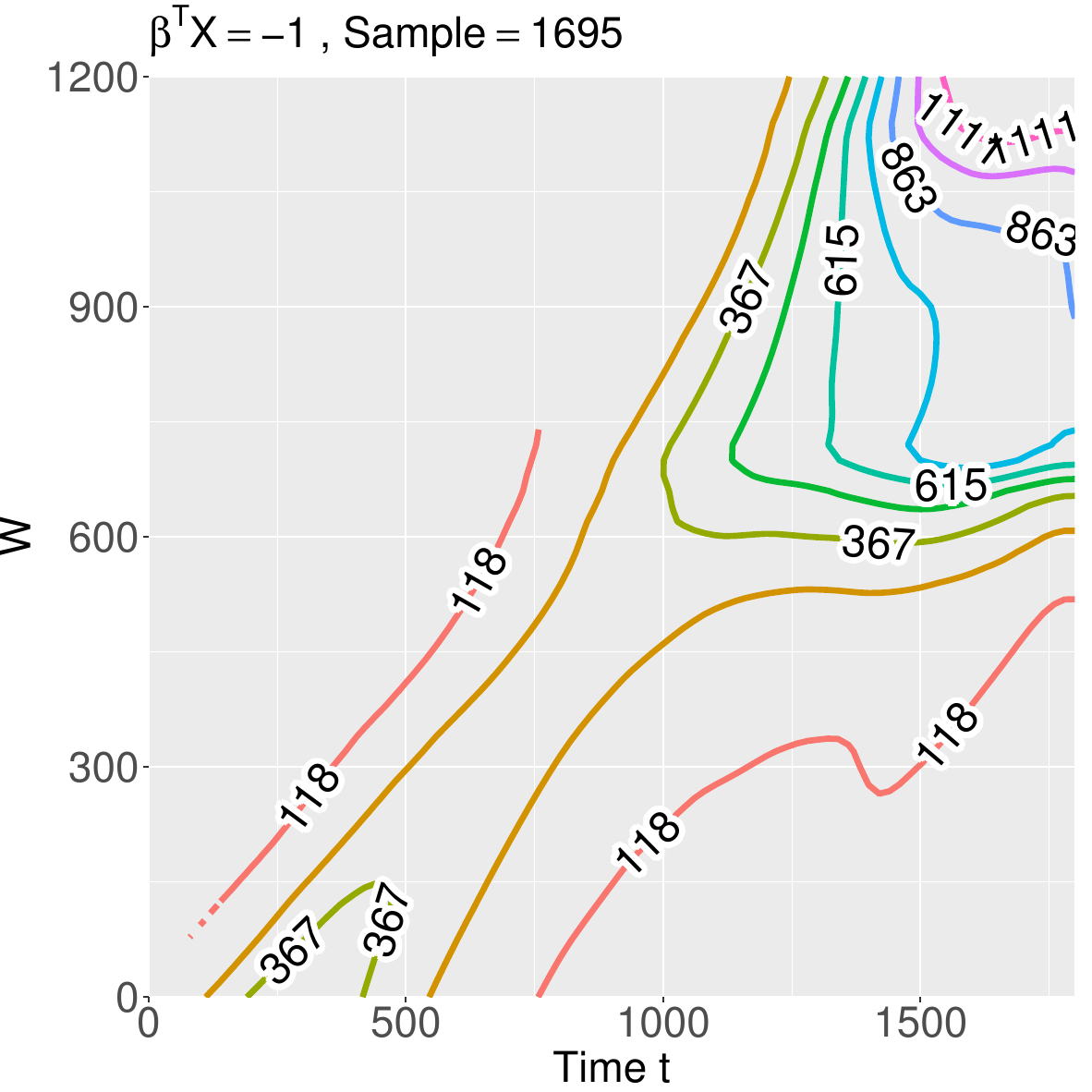}};
		\node[left=of img11, node distance=0cm, rotate=90, anchor=center,font=\tiny,xshift=0.2cm,yshift=-1cm] {Hispanic};
	\end{tikzpicture}
	\begin{tikzpicture}
		\node (img22) {\includegraphics[width=3cm]{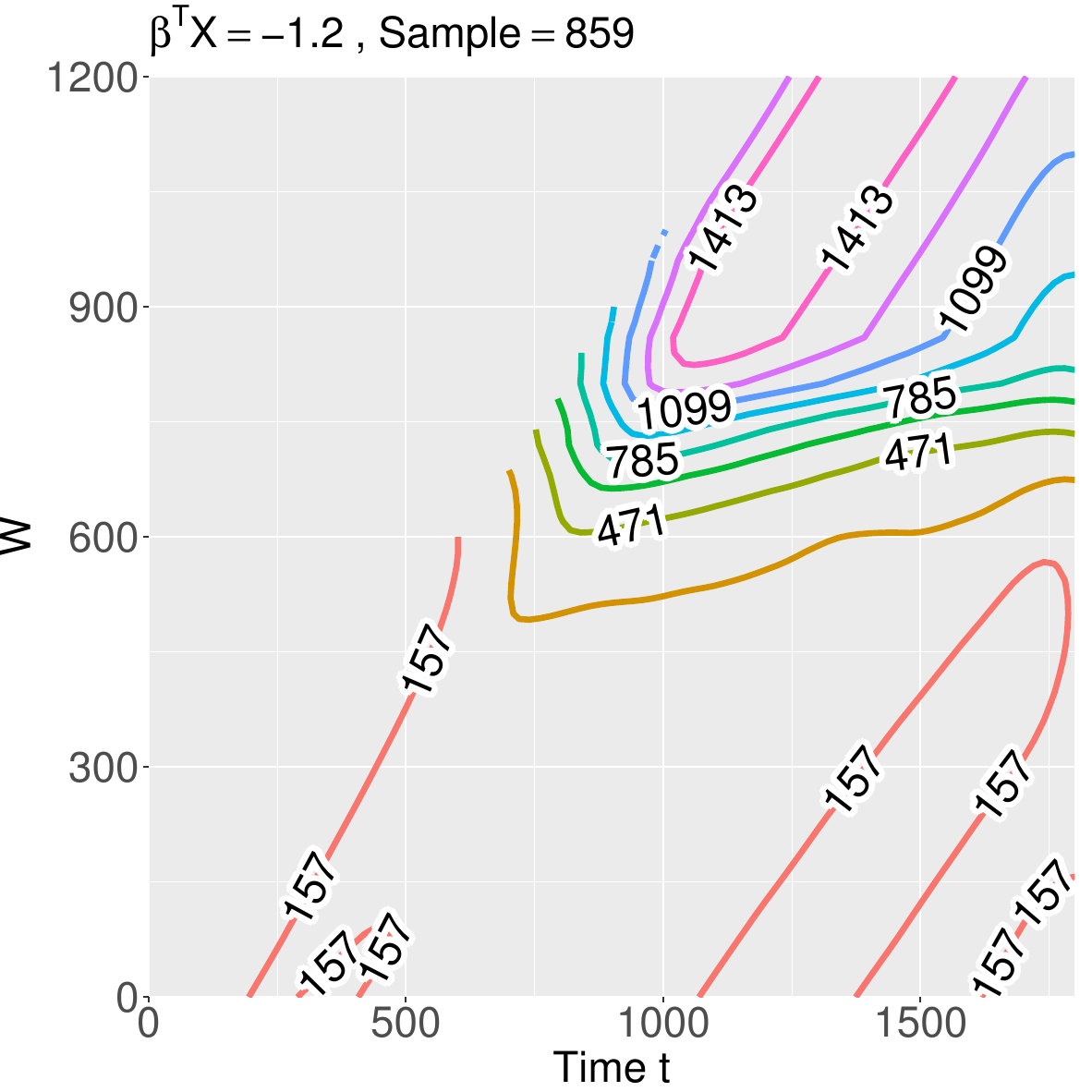}};
	\end{tikzpicture}
	\begin{tikzpicture}
		\node (img23) {\includegraphics[width=3cm]{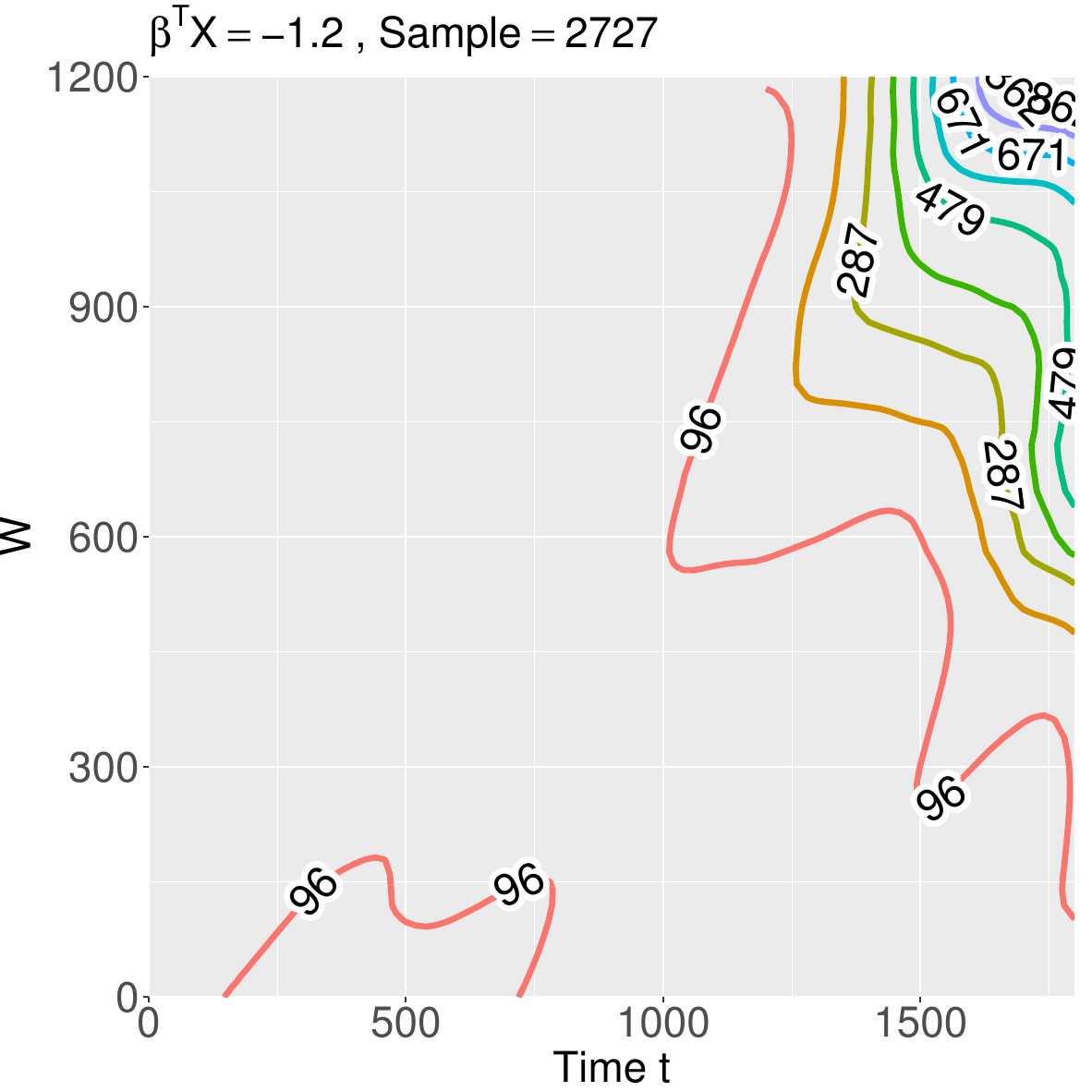}};
	\end{tikzpicture}
	\begin{tikzpicture}
		\node (img24) {\includegraphics[width=3cm]{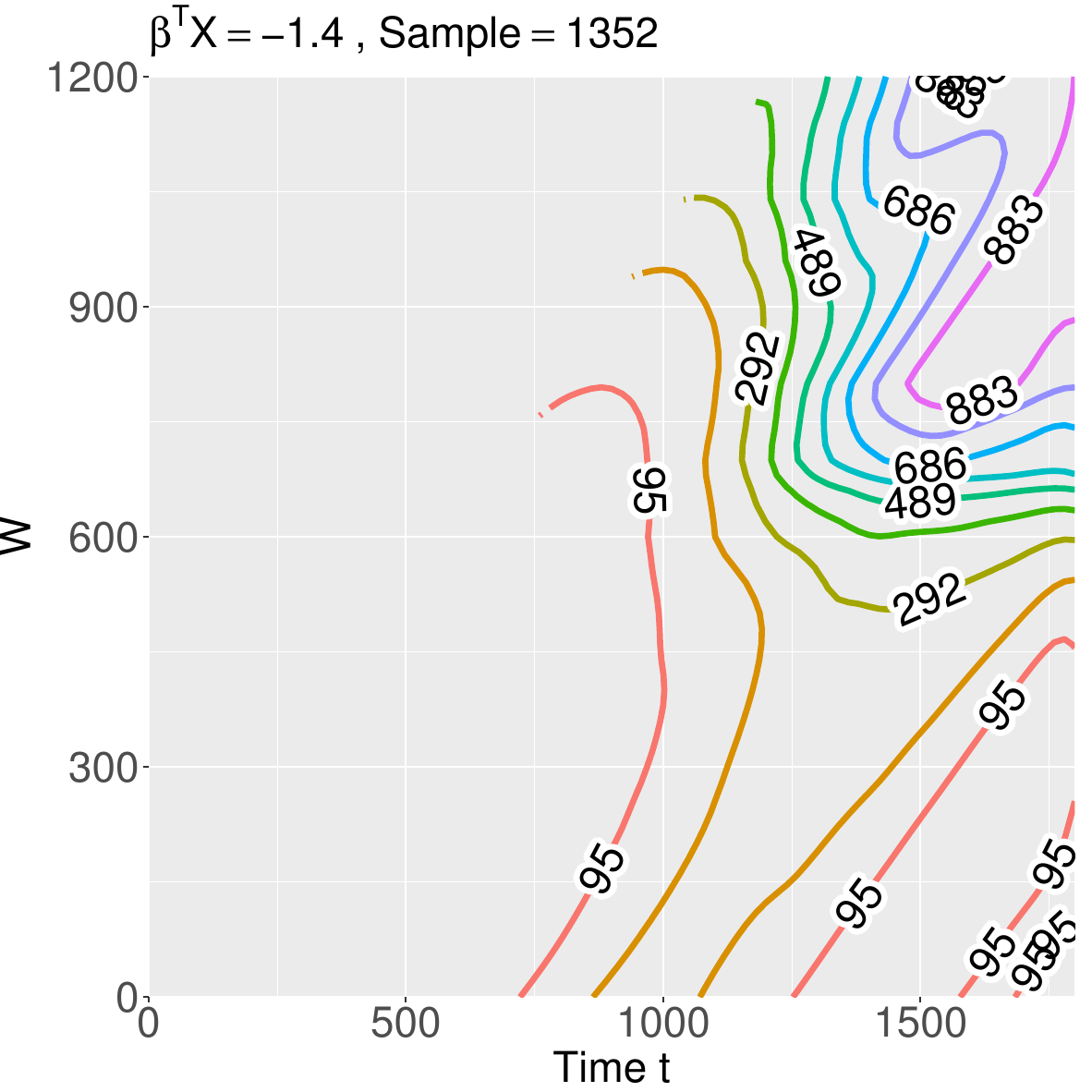}};
	\end{tikzpicture}\\
	\begin{tikzpicture}
		\node (img31) {\includegraphics[width=3cm]{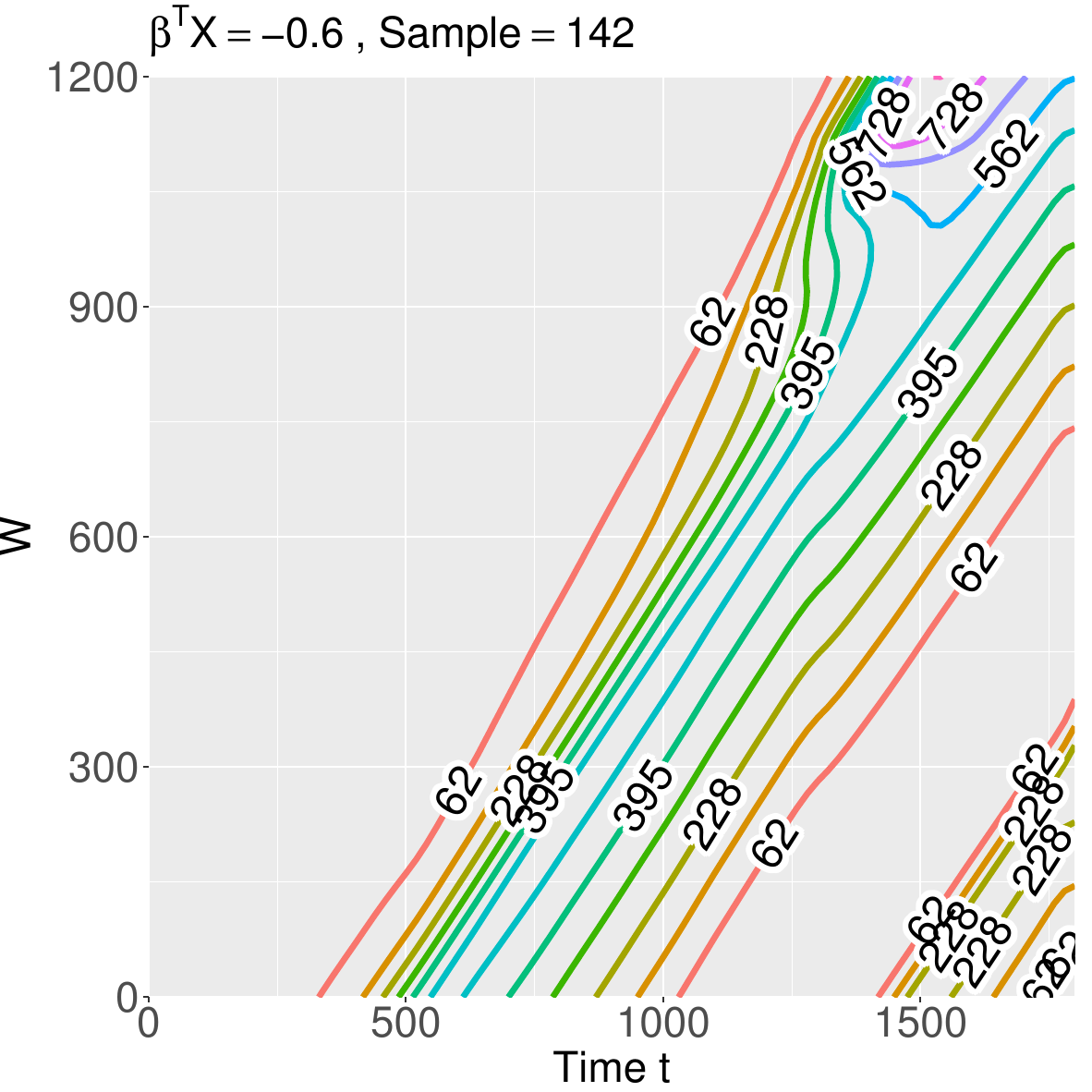}};
		\node[left=of img11, node distance=0cm, rotate=90, anchor=center,font=\tiny,xshift=0.2cm,yshift=-1cm] {Native};
	\end{tikzpicture}
	\begin{tikzpicture}
		\node (img32) {\includegraphics[width=3cm]{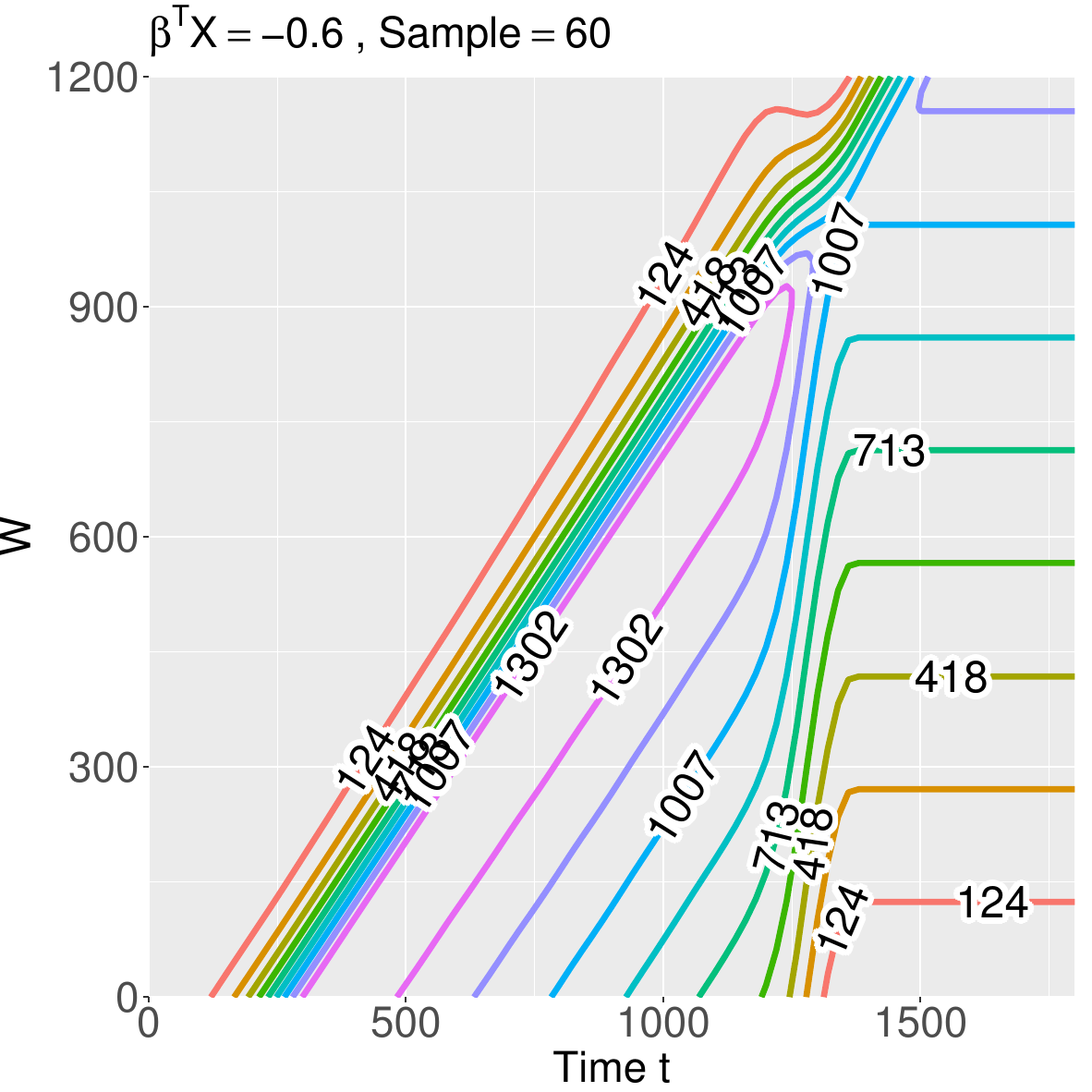}};
	\end{tikzpicture}
	\begin{tikzpicture}
		\node (img33) {\includegraphics[width=3cm]{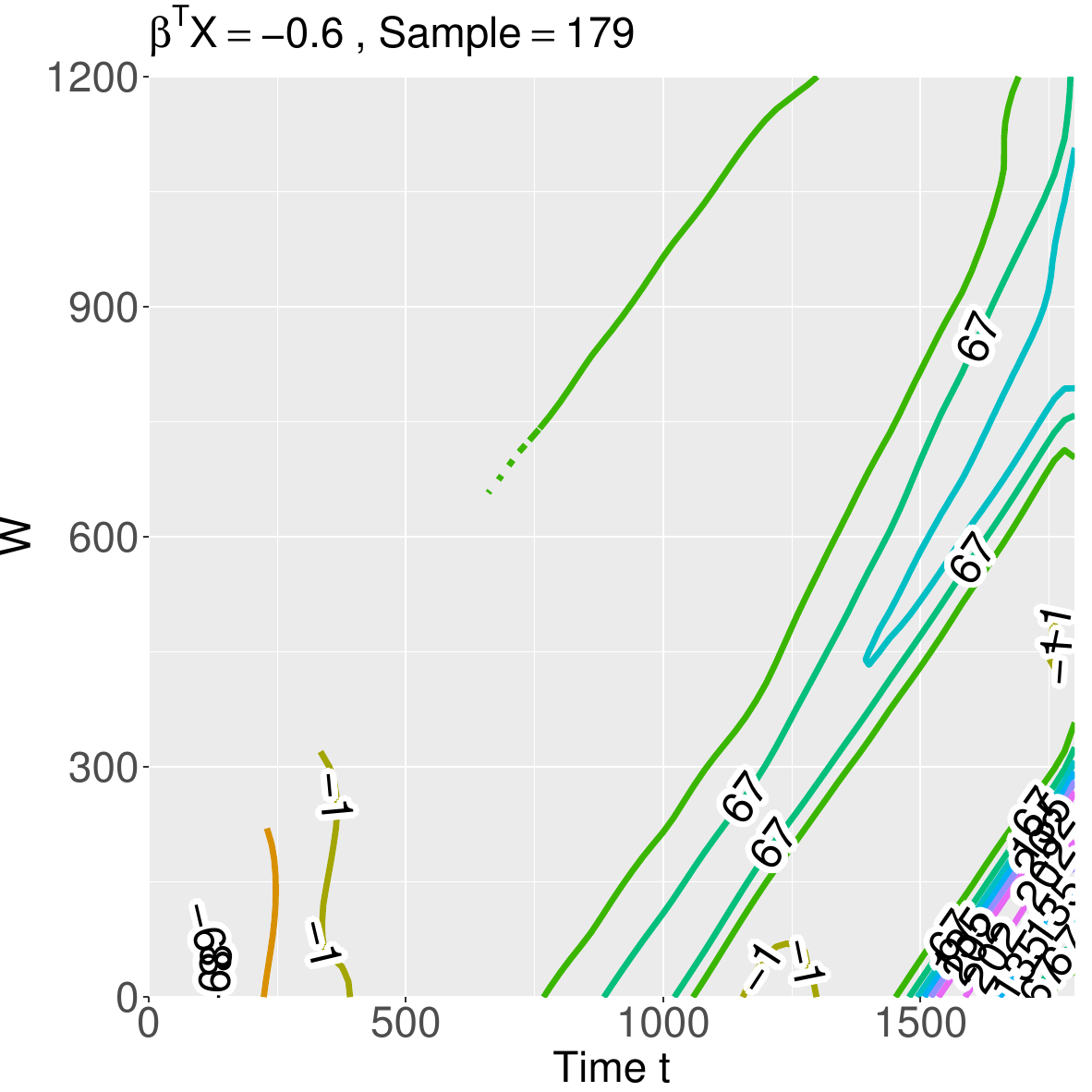}};
	\end{tikzpicture}
	\begin{tikzpicture}
		\node (img34) {\includegraphics[width=3cm]{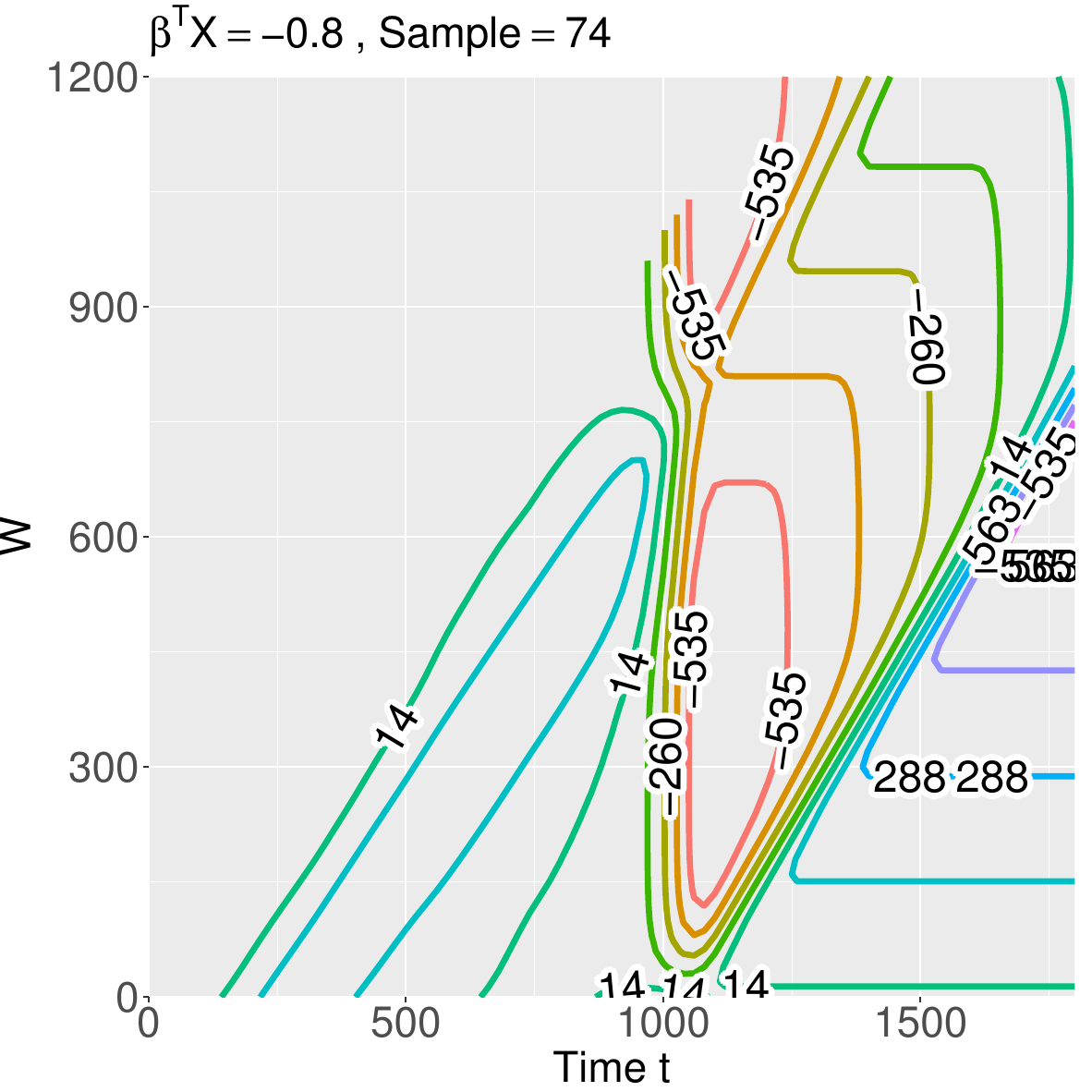}};
	\end{tikzpicture}\\
	\begin{tikzpicture}
		\node (img41) {\includegraphics[width=3cm]{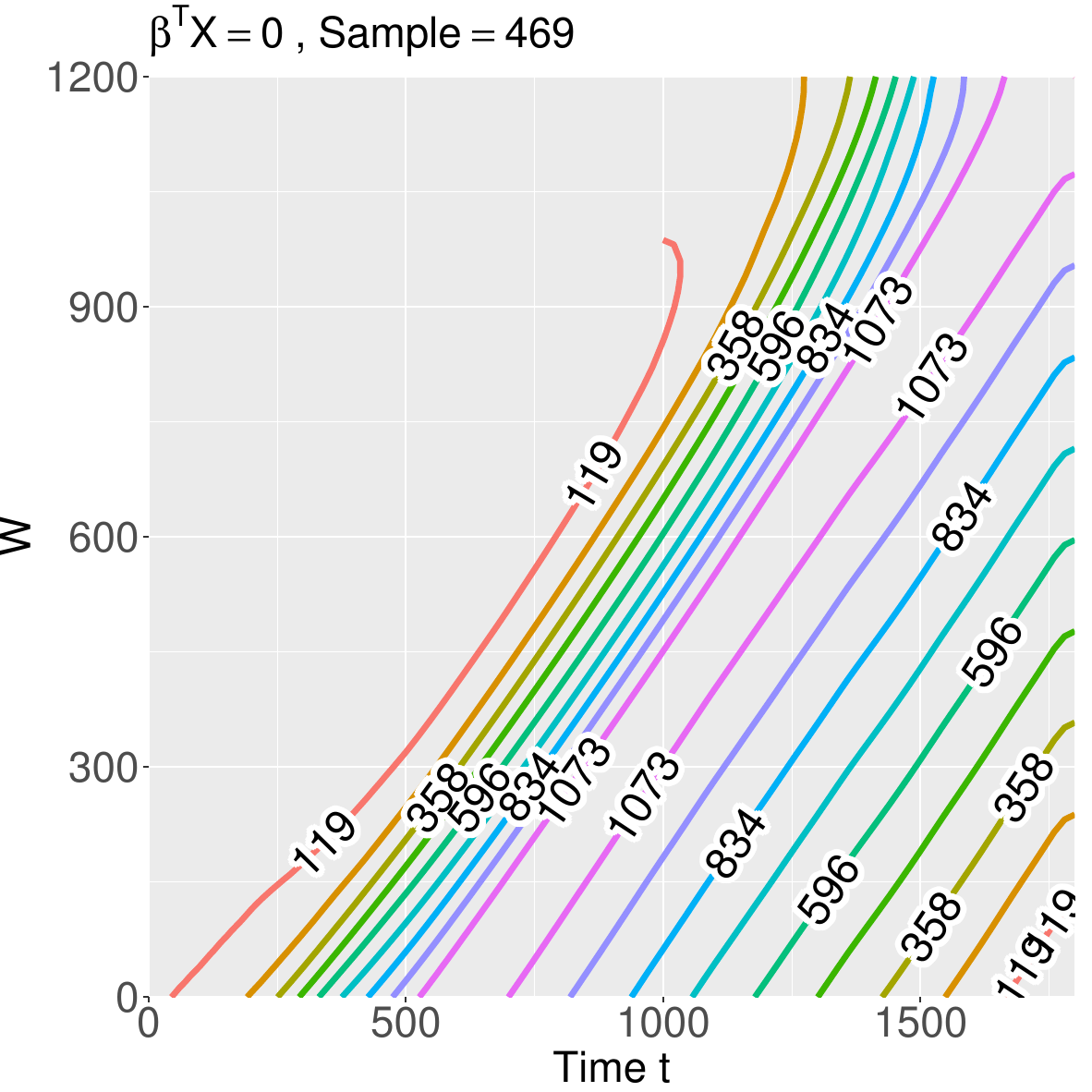}};
		\node[left=of img11, node distance=0cm, rotate=90, anchor=center,font=\tiny,xshift=0.2cm,yshift=-1cm] {Asian};
	\end{tikzpicture}
	\begin{tikzpicture}
		\node (img42) {\includegraphics[width=3cm]{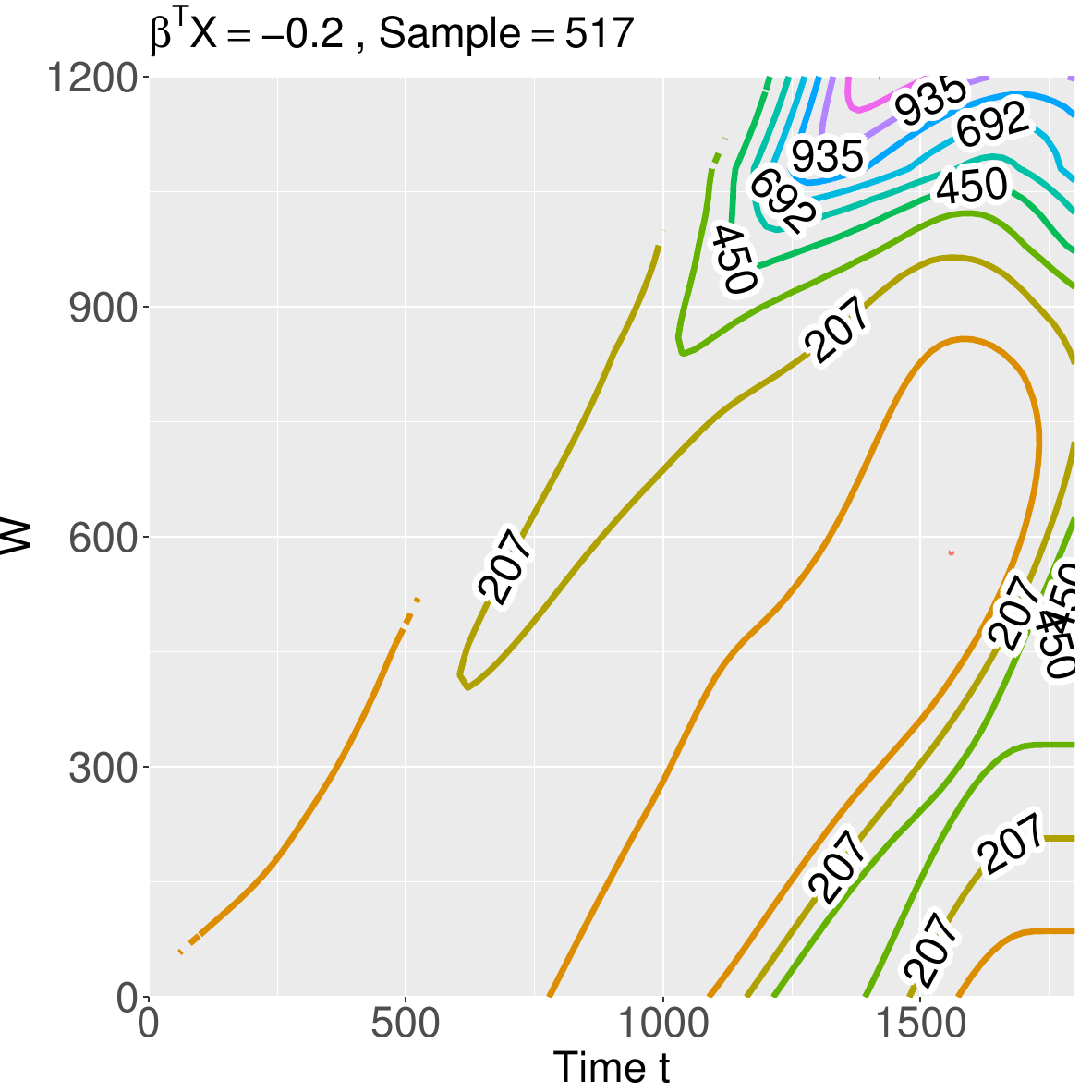}};
	\end{tikzpicture}
	\begin{tikzpicture}
		\node (img43) {\includegraphics[width=3cm]{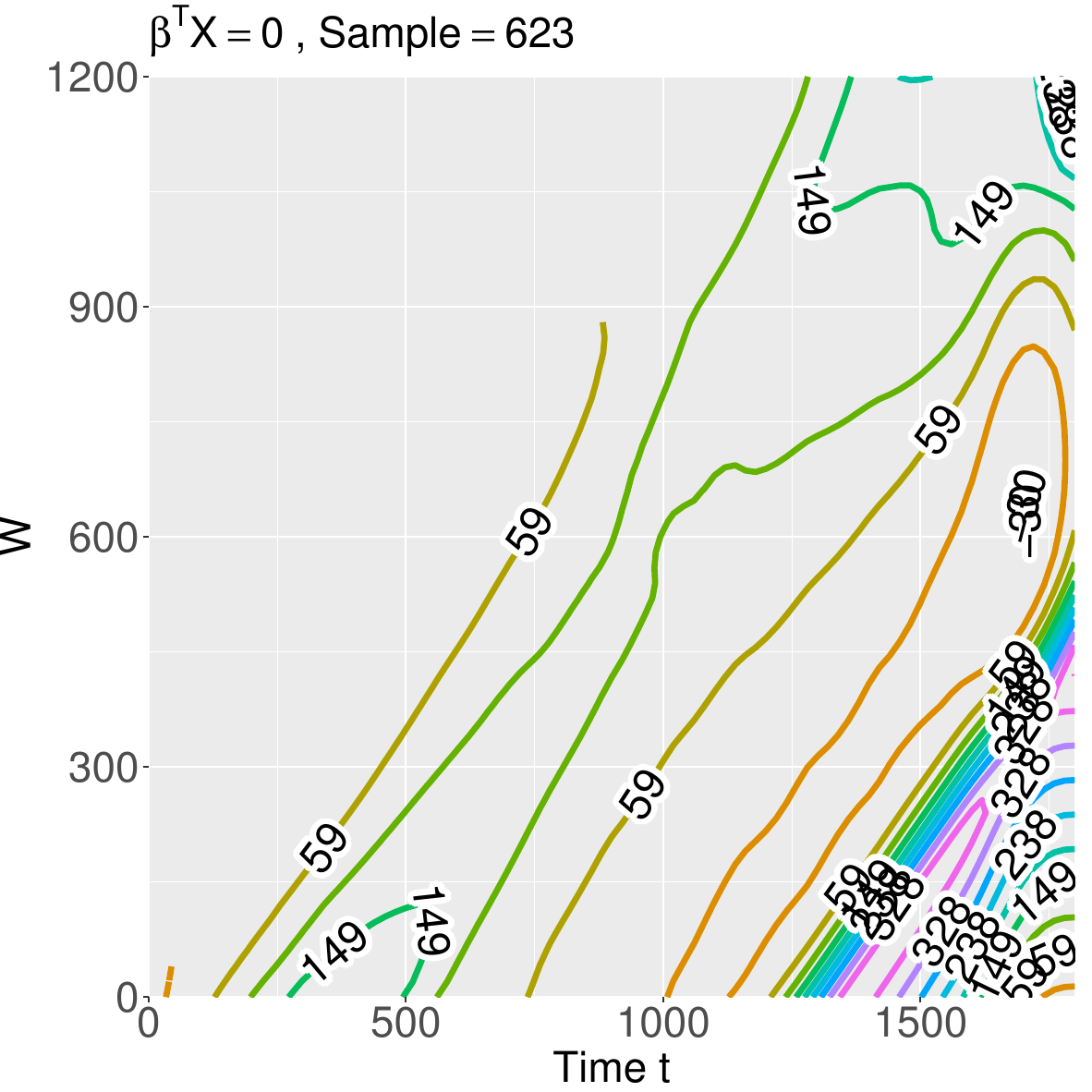}};
	\end{tikzpicture}
	\begin{tikzpicture}
		\node (img44) {\includegraphics[width=3cm]{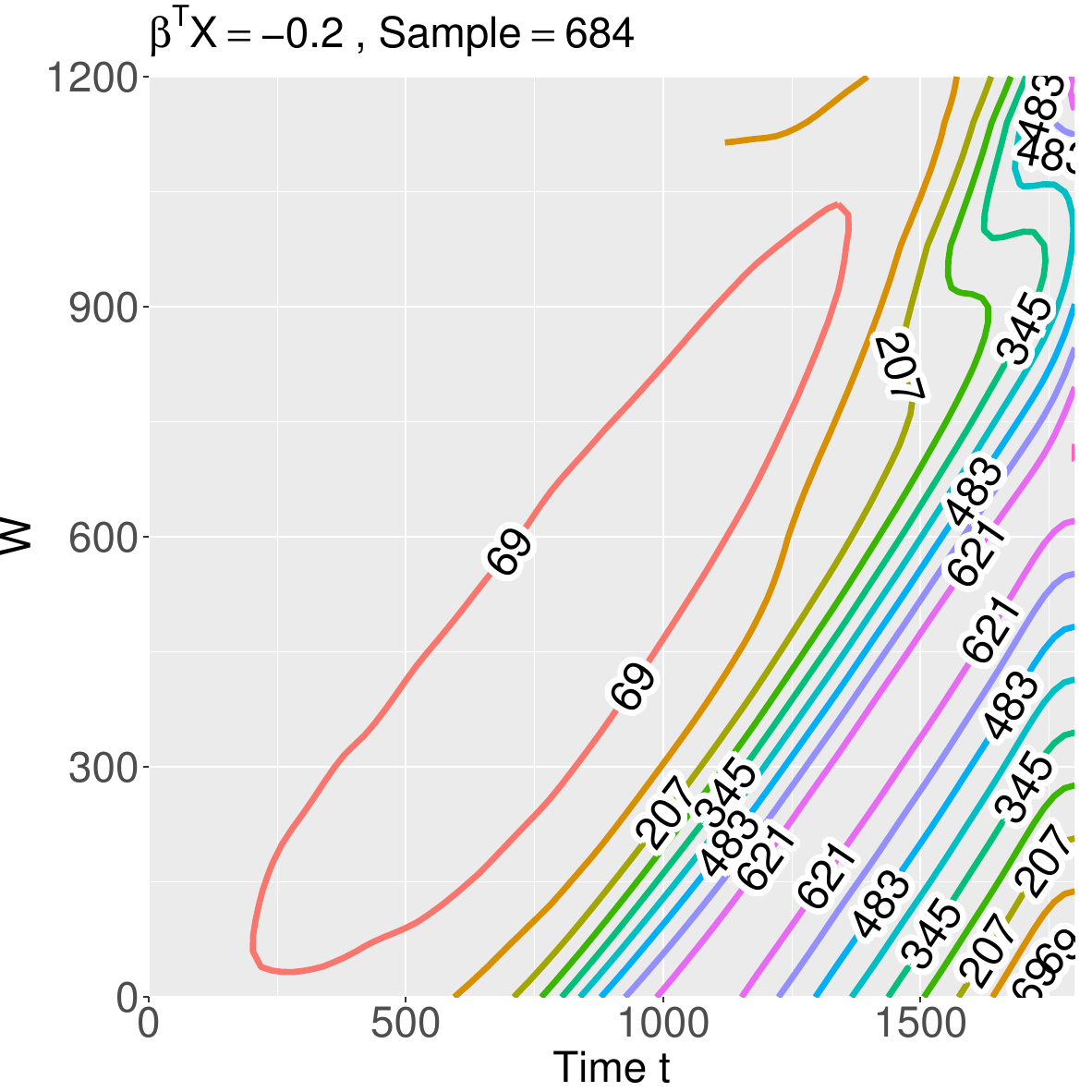}};
	\end{tikzpicture}\\
	\begin{tikzpicture}
		\node (img51) {\includegraphics[width=3cm]{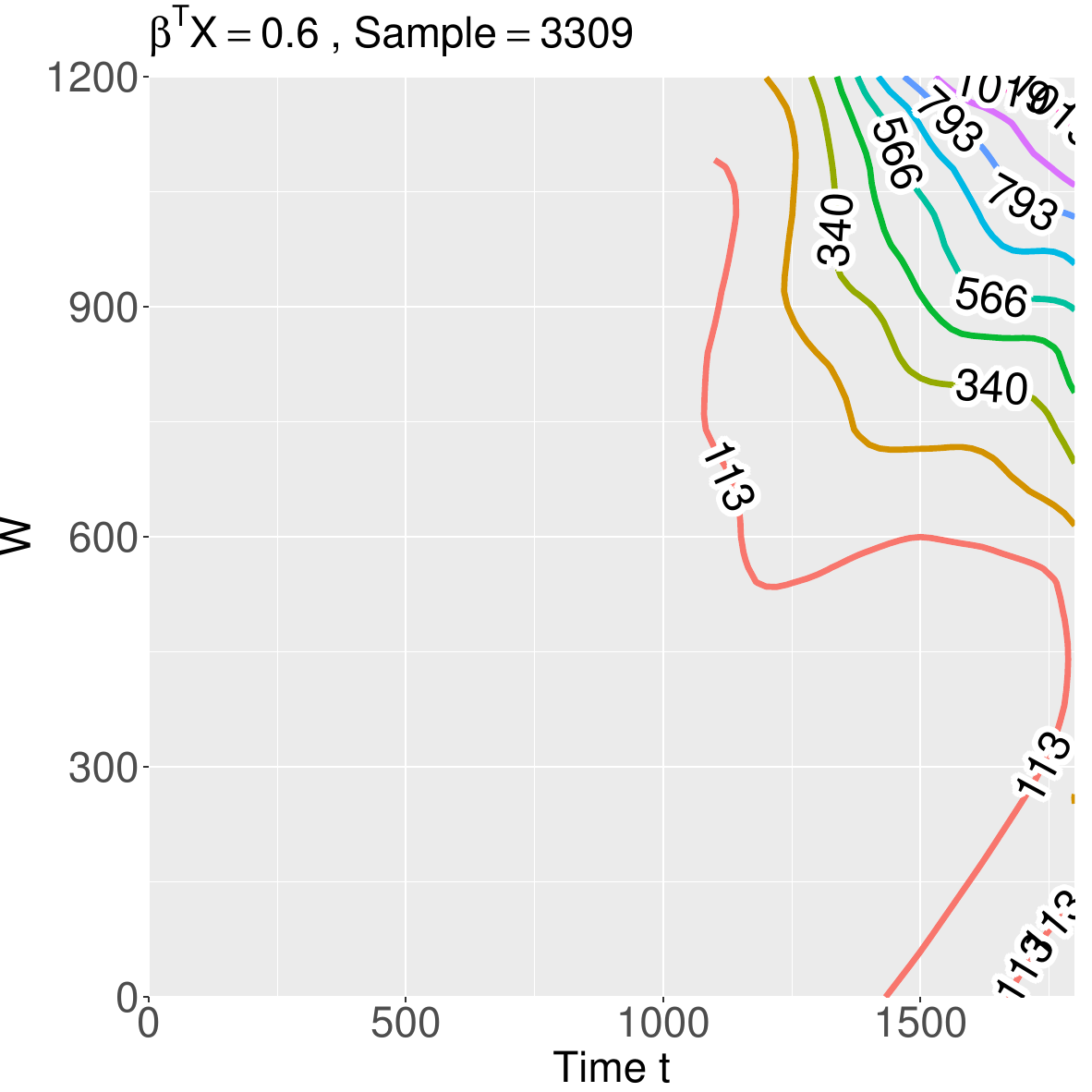}};
		\node[left=of img11, node distance=0cm, rotate=90, anchor=center,font=\tiny,xshift=0.2cm,yshift=-1cm] {White};
	\end{tikzpicture}
	\begin{tikzpicture}
		\node (img52) {\includegraphics[width=3cm]{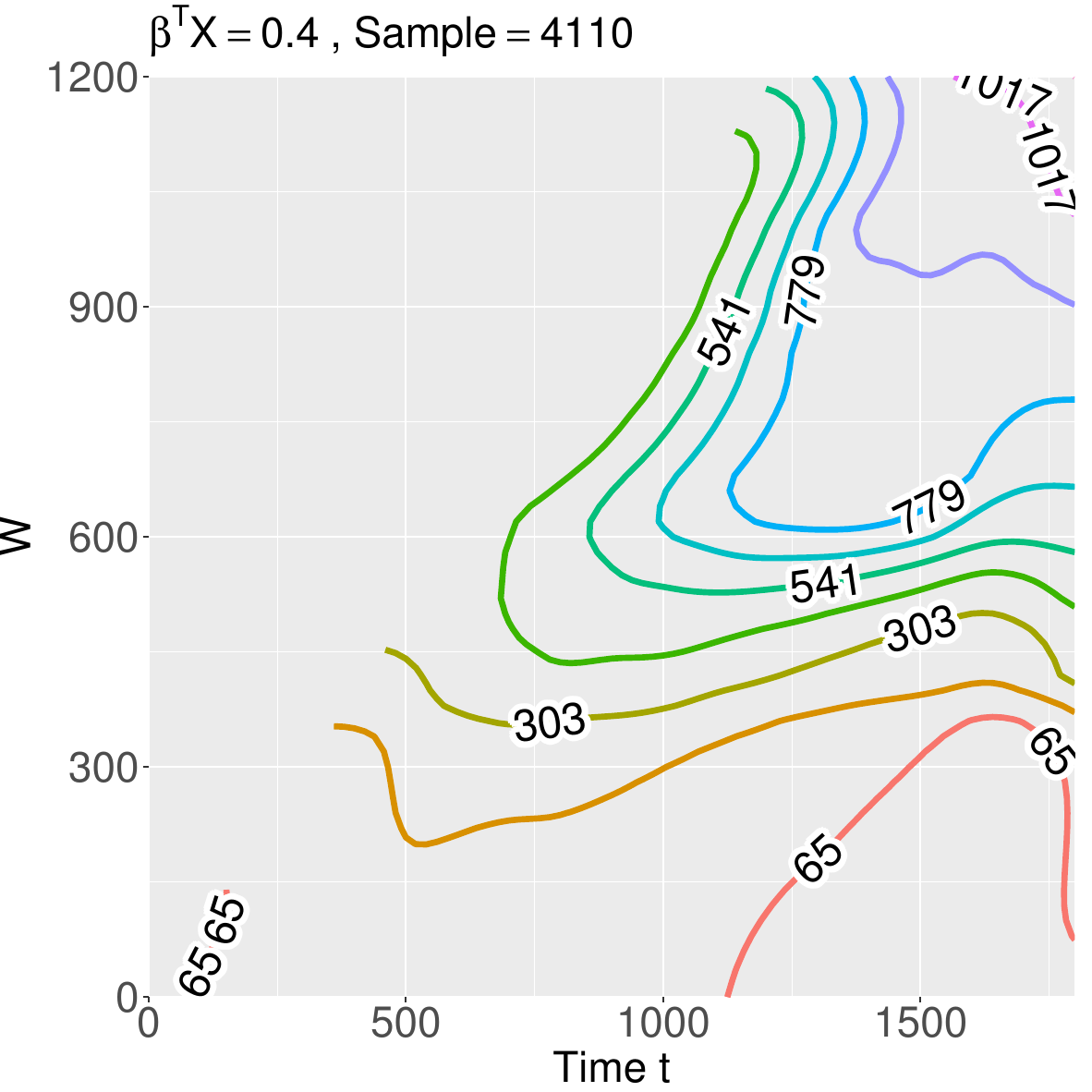}};
	\end{tikzpicture}
	\begin{tikzpicture}
		\node (img53) {\includegraphics[width=3cm]{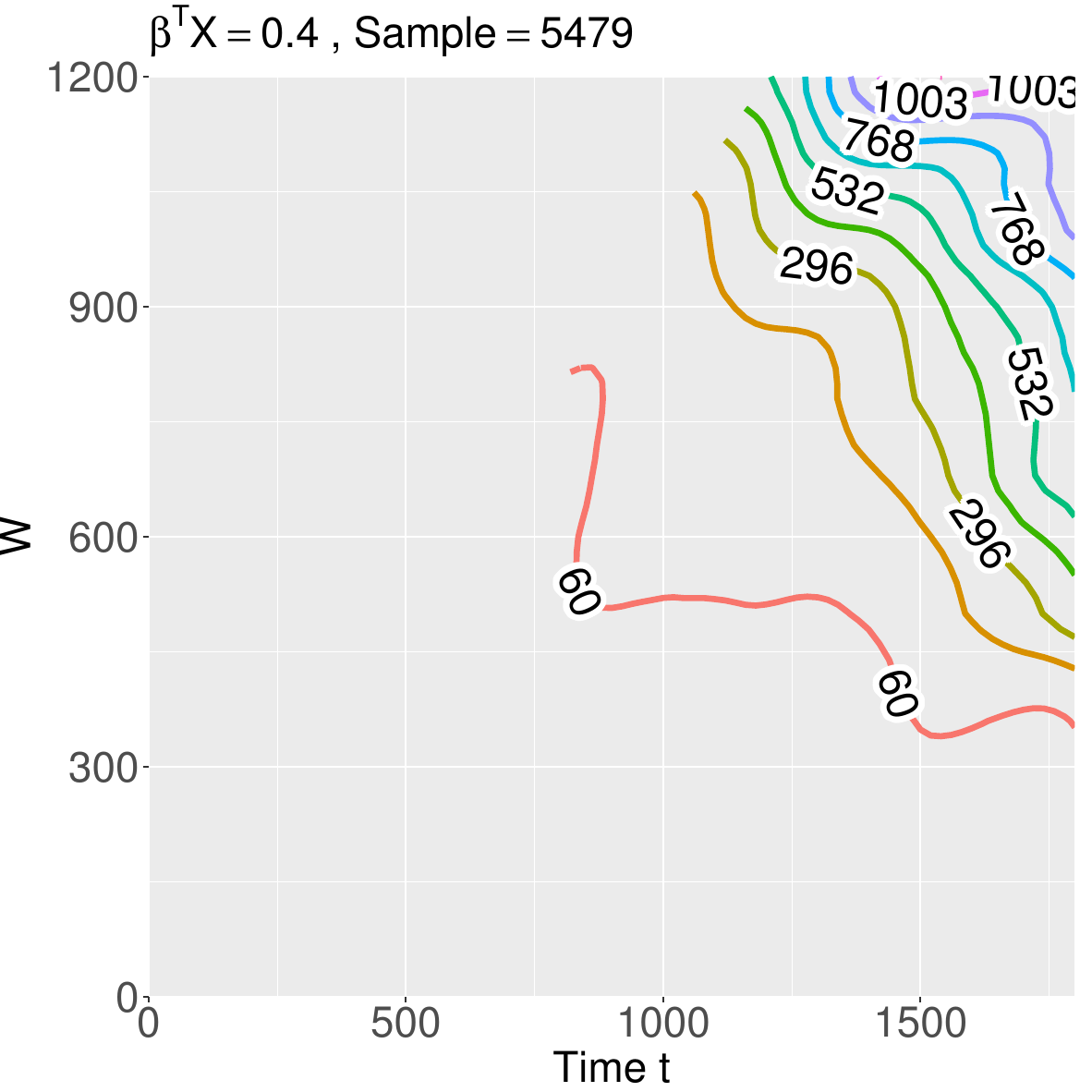}};
	\end{tikzpicture}
	\begin{tikzpicture}
		\node (img54) {\includegraphics[width=3cm]{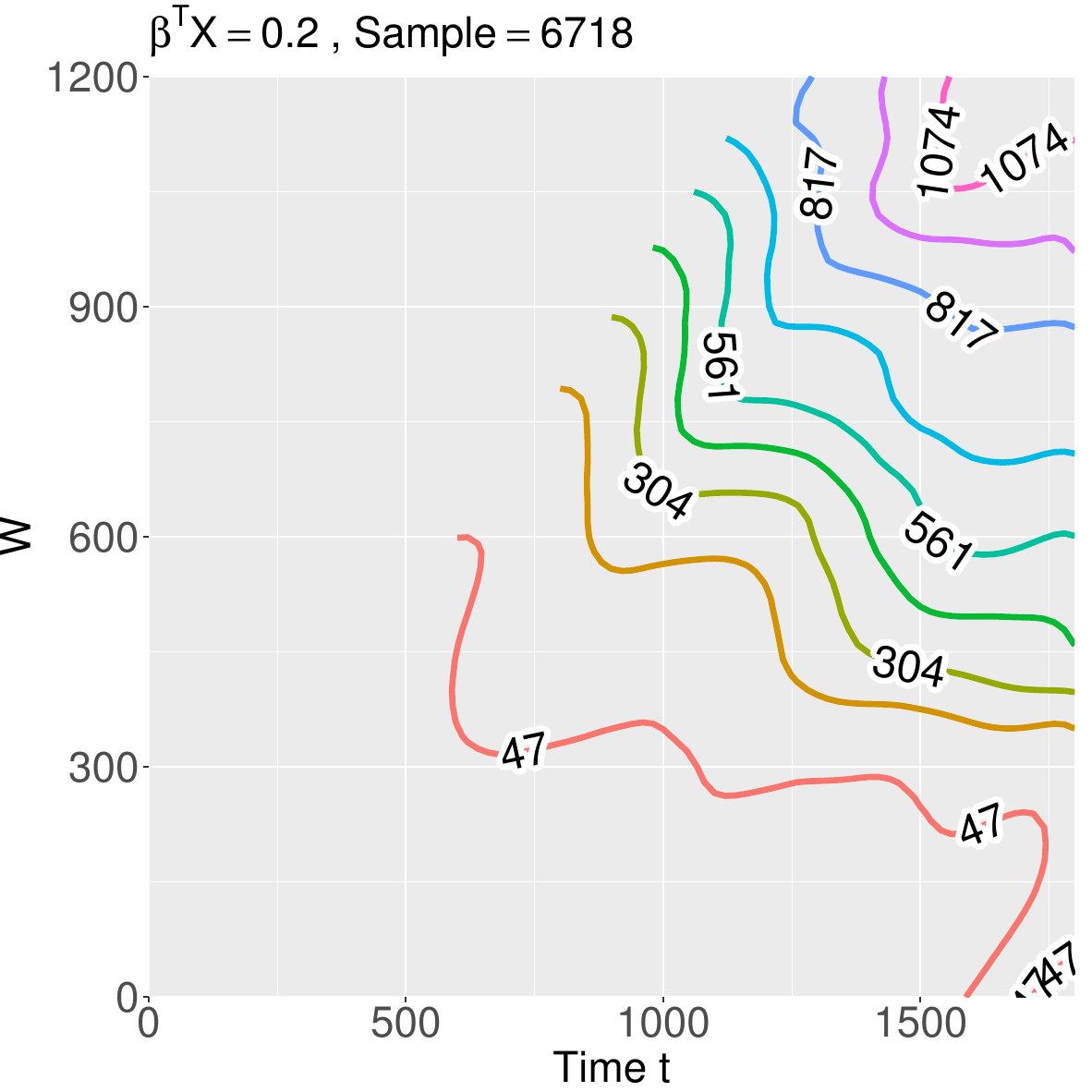}};
	\end{tikzpicture}\\
	\caption{Mean residual life improvement from UNOS/OPTN data. Stratified by Race, Gender, and Insurance Status with median $\bb\trans\x$ per strata.}
	\label{fig:appDiffContourCloser2}
\end{figure}
\begin{figure}
	\centering
	\captionsetup[subfigure]{labelformat=empty}
	\begin{tikzpicture}
		\node (img11) {\includegraphics[width=3cm]{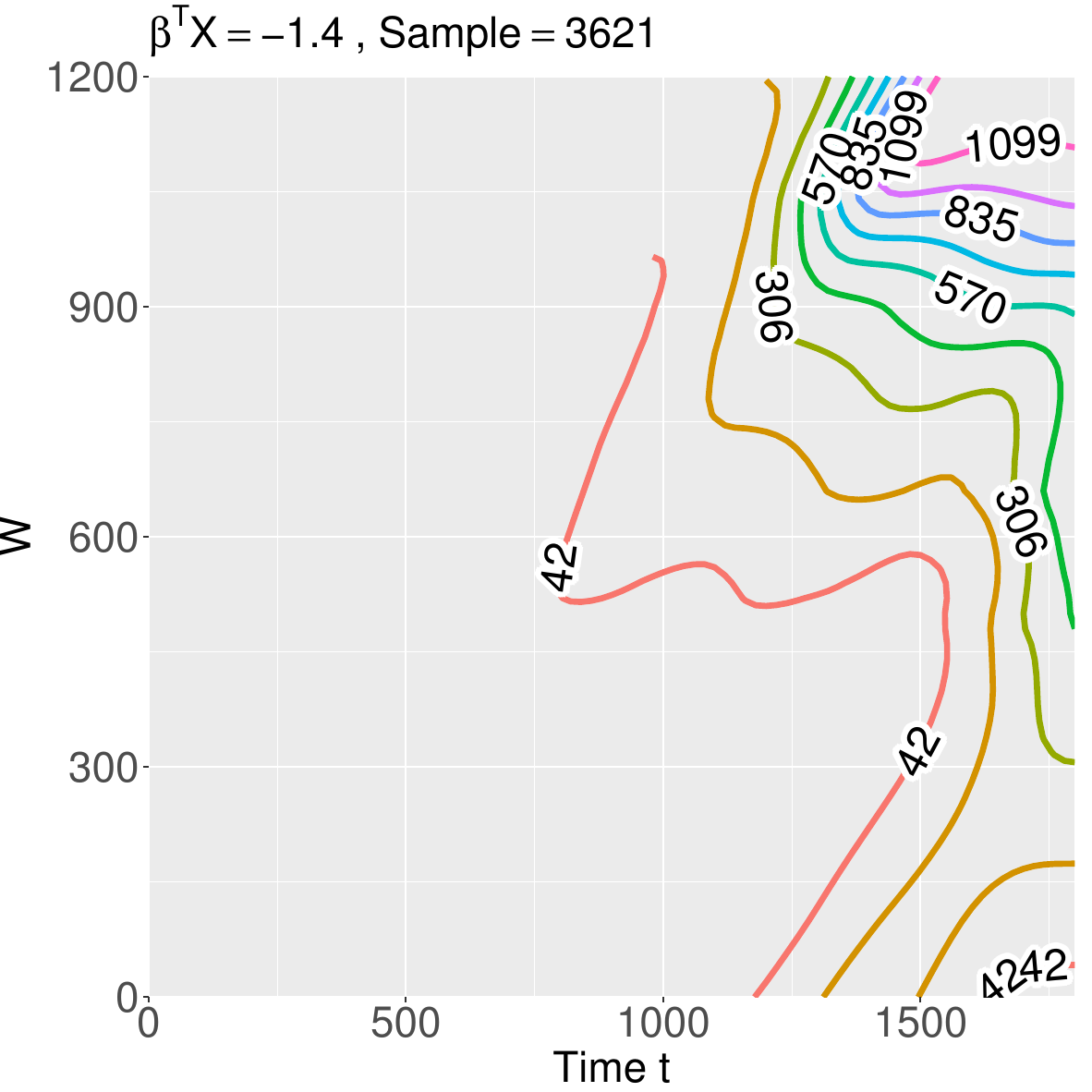}};
		\node[left=of img11, node distance=0cm, rotate=90, anchor=center,font=\tiny,xshift=0.2cm,yshift=-1cm] {African American};
		\node[above=of img11, node distance=0cm, anchor=center,font=\tiny,xshift=0.3cm,yshift=-1cm] {Female,Public};
	\end{tikzpicture}
	\begin{tikzpicture}
		\node (img12) {\includegraphics[width=3cm]{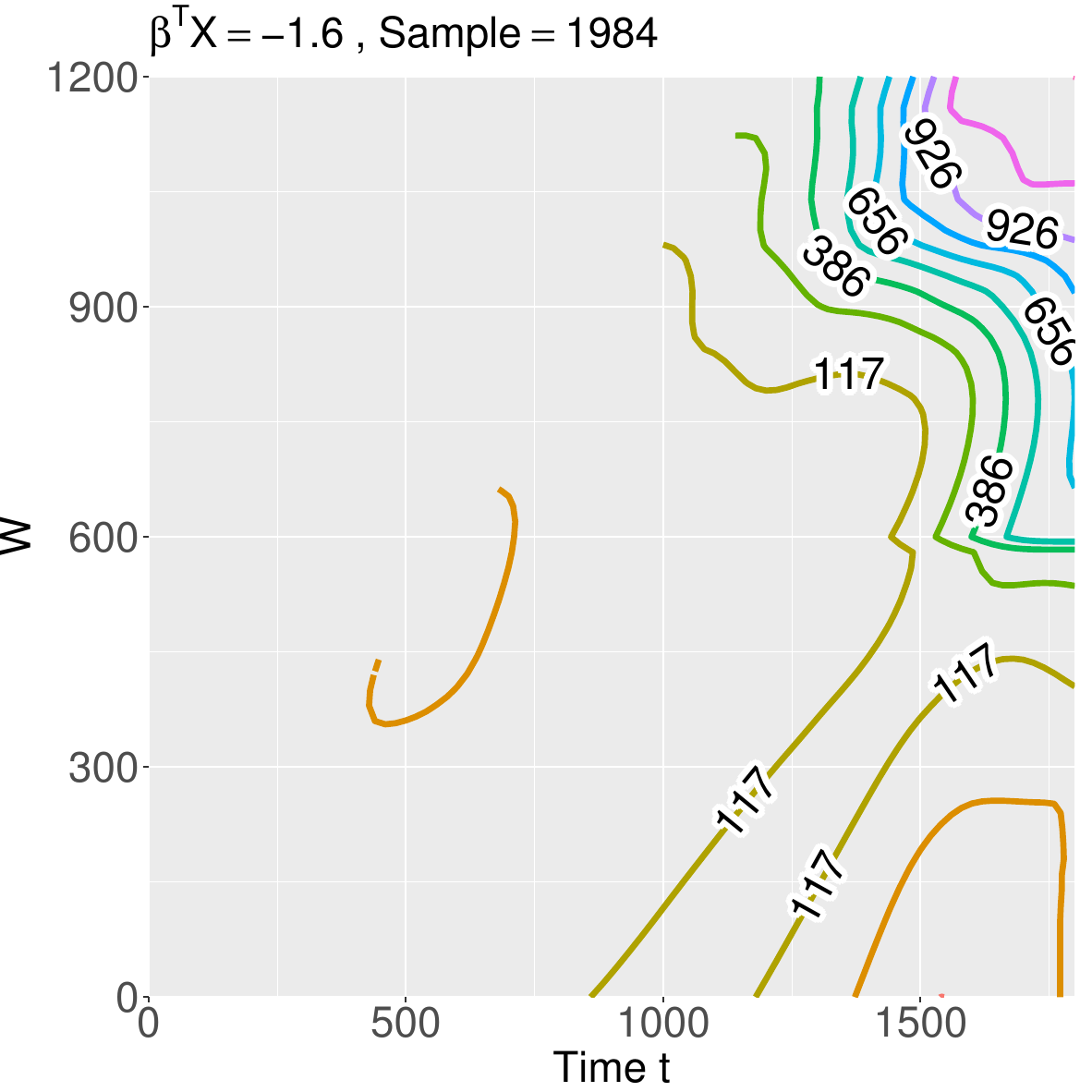}};
		\node[above=of img12, node distance=0cm, anchor=center,font=\tiny,xshift=0.3cm,yshift=-1cm] {Female,Private};
	\end{tikzpicture}
	\begin{tikzpicture}
		\node (img13) {\includegraphics[width=3cm]{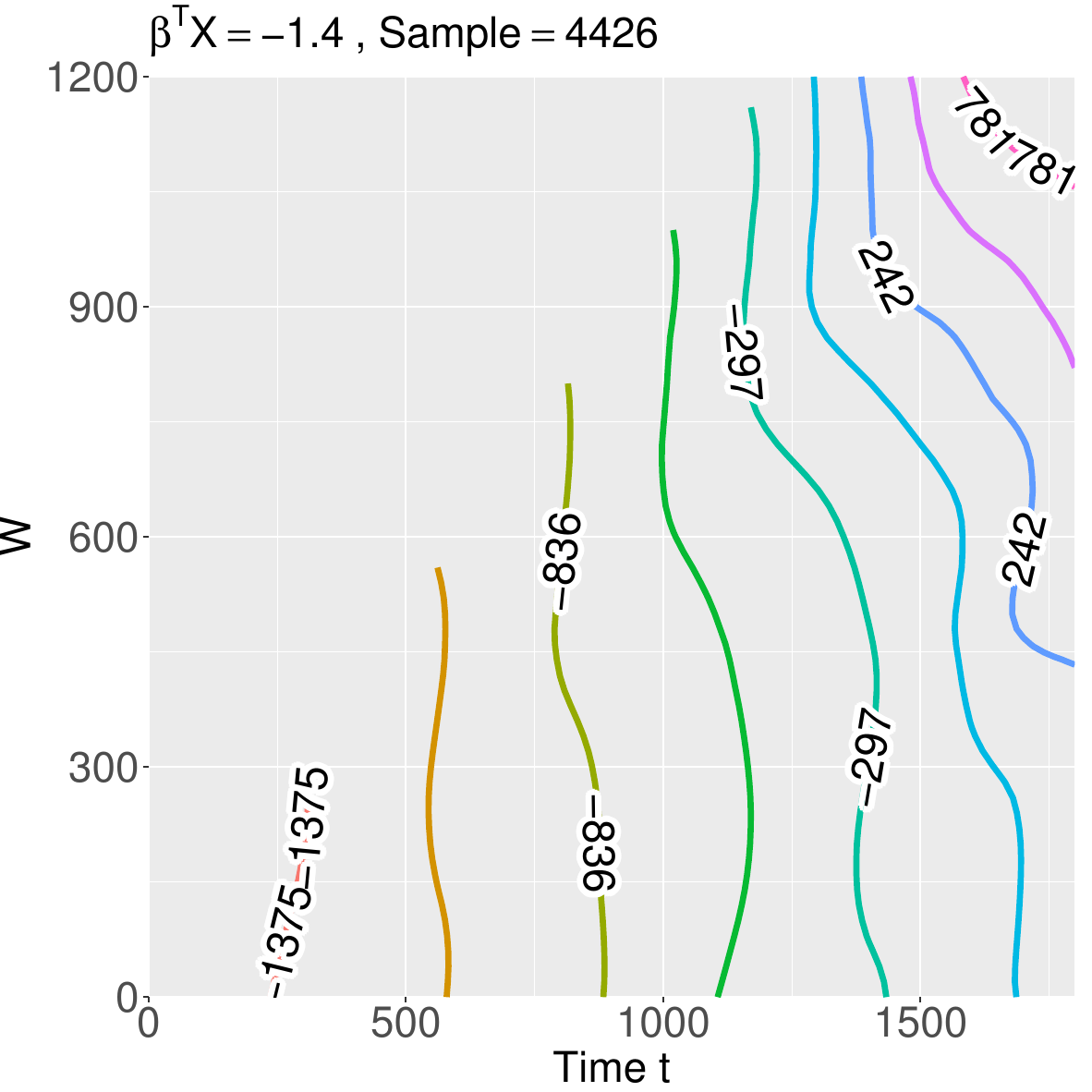}};
		\node[above=of img13, node distance=0cm, anchor=center,font=\tiny,xshift=0.3cm,yshift=-1cm] {Male,Public};
	\end{tikzpicture}
	\begin{tikzpicture}
		\node (img14) {\includegraphics[width=3cm]{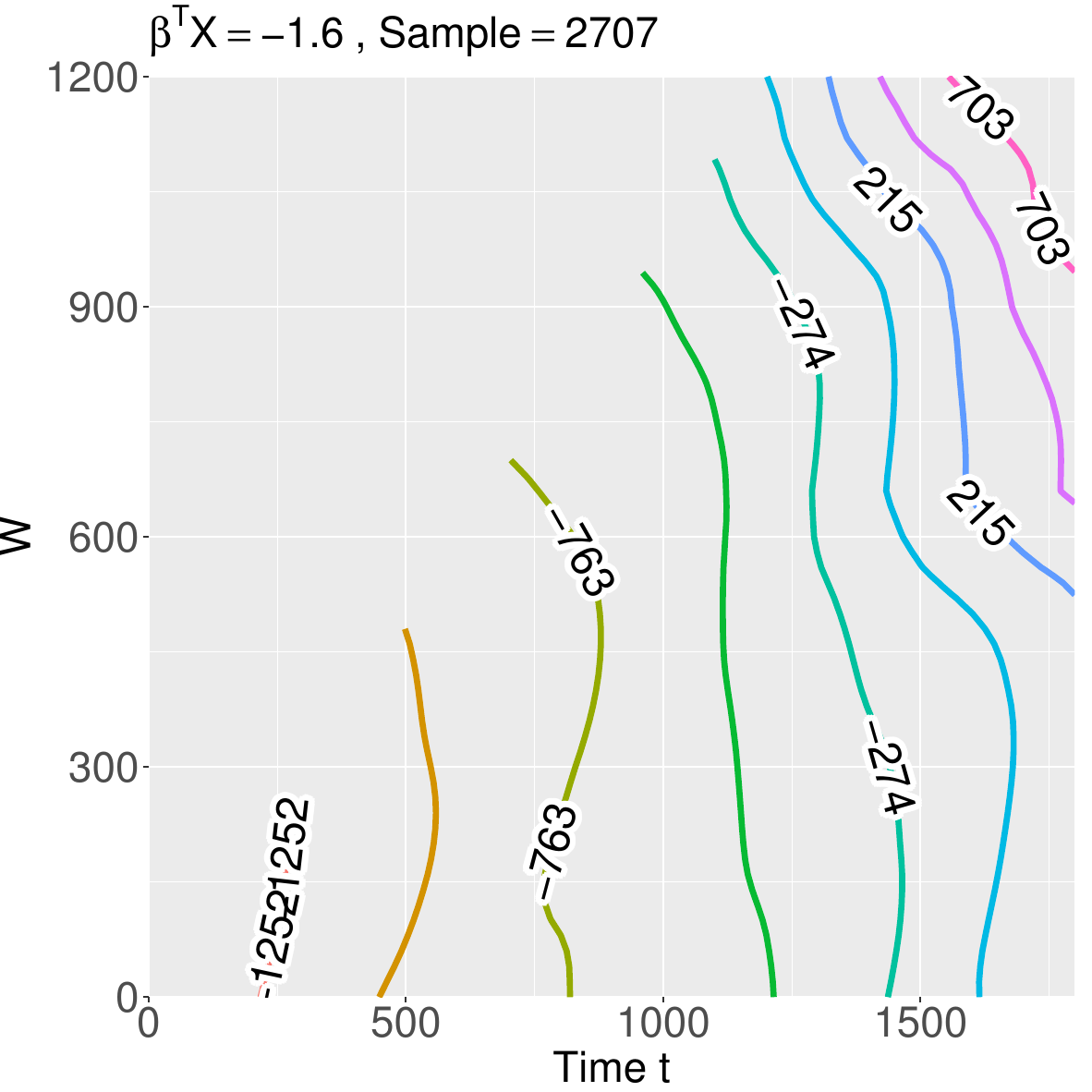}};
		\node[above=of img14, node distance=0cm, anchor=center,font=\tiny,xshift=0.3cm,yshift=-1cm] {Male,Private};
	\end{tikzpicture}\\
	\begin{tikzpicture}
		\node (img21) {\includegraphics[width=3cm]{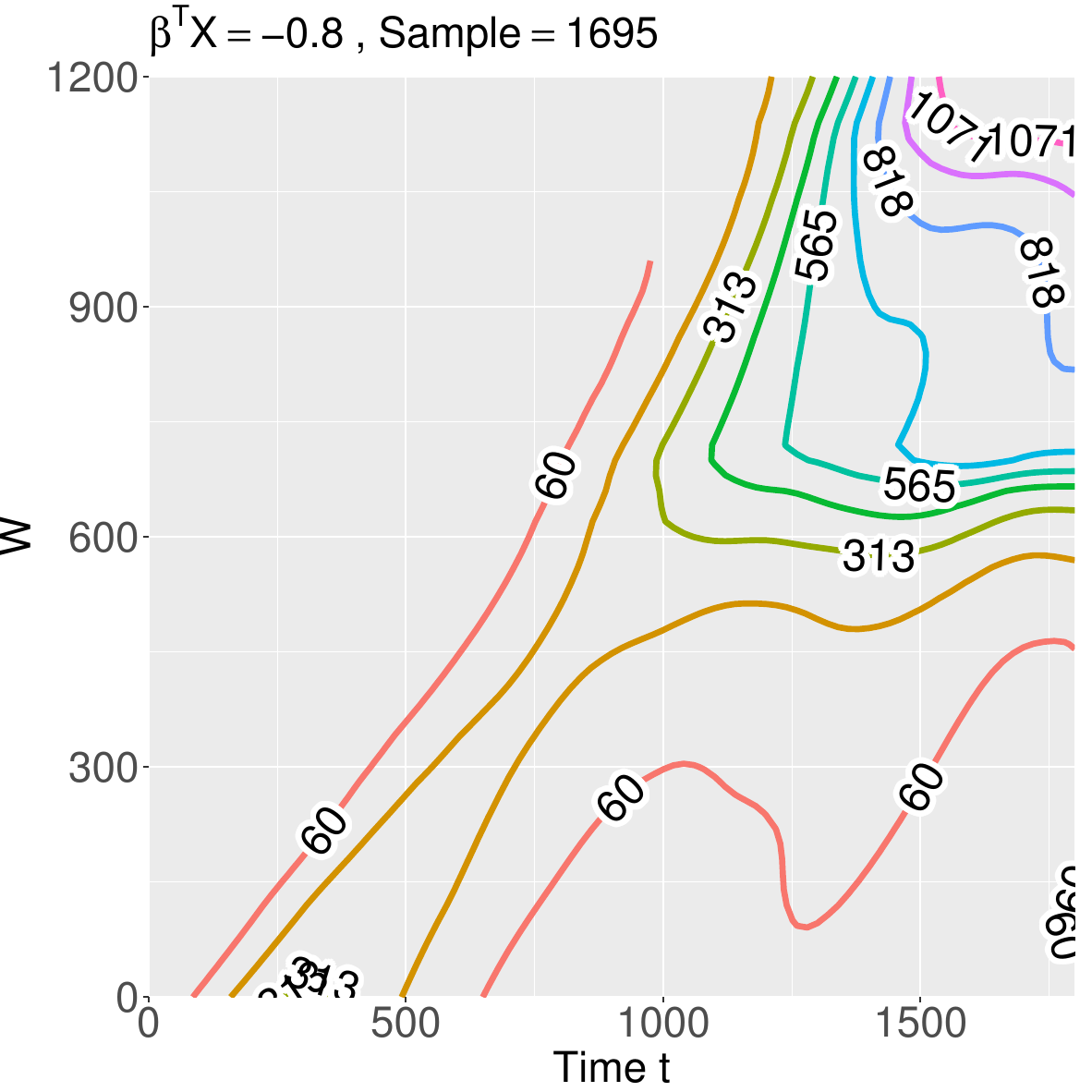}};
		\node[left=of img11, node distance=0cm, rotate=90, anchor=center,font=\tiny,xshift=0.2cm,yshift=-1cm] {Hispanic};
	\end{tikzpicture}
	\begin{tikzpicture}
		\node (img22) {\includegraphics[width=3cm]{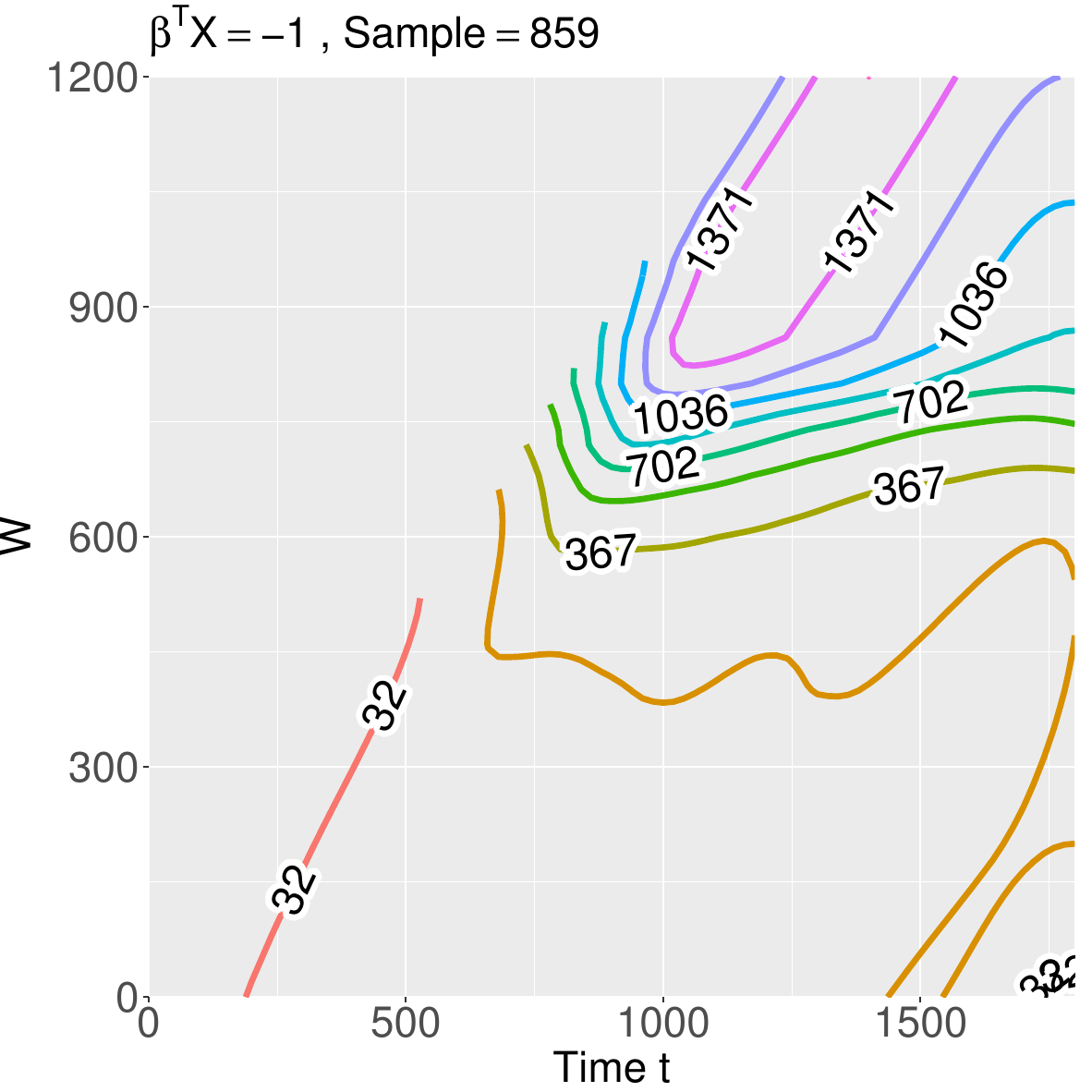}};
	\end{tikzpicture}
	\begin{tikzpicture}
		\node (img23) {\includegraphics[width=3cm]{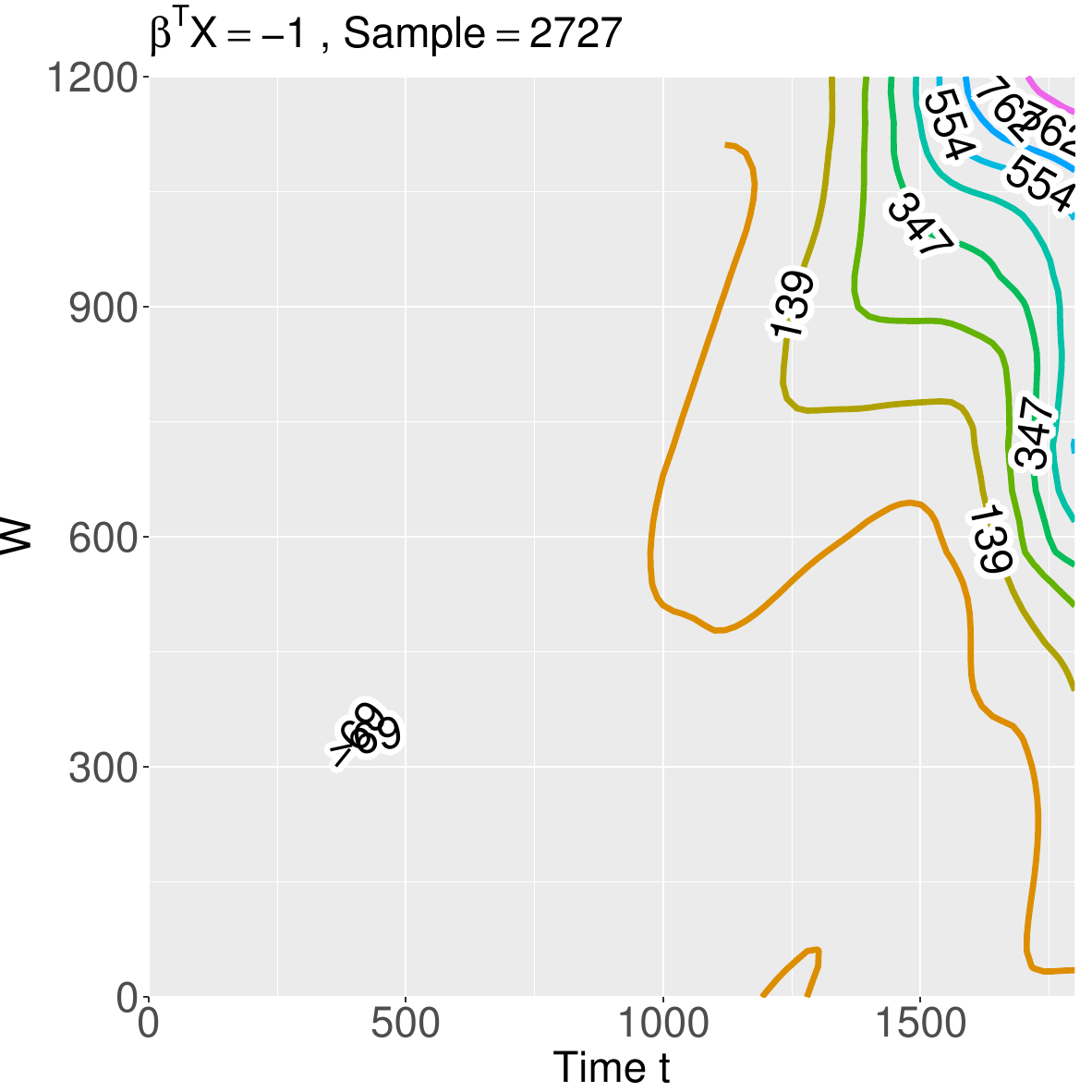}};
	\end{tikzpicture}
	\begin{tikzpicture}
		\node (img24) {\includegraphics[width=3cm]{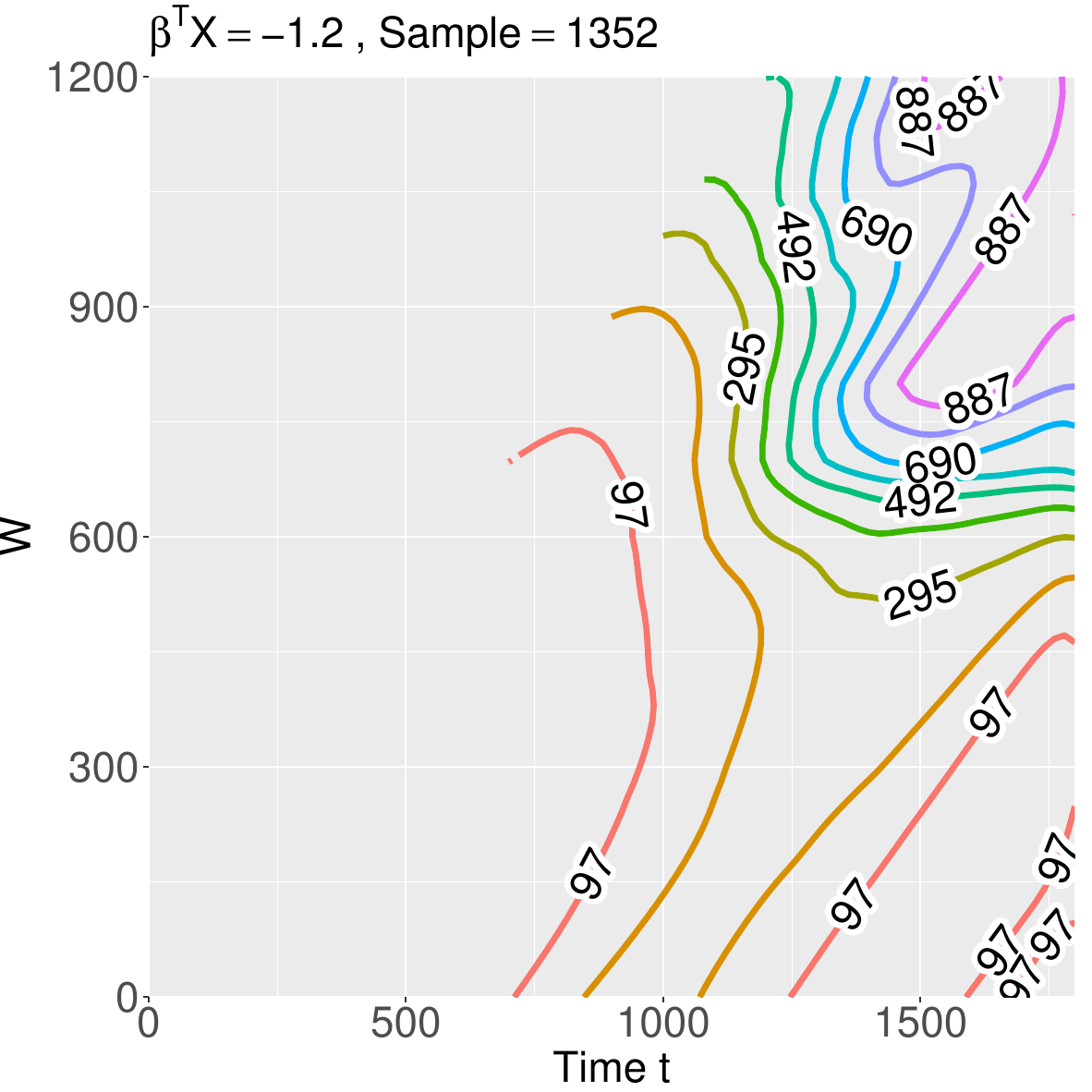}};
	\end{tikzpicture}\\
	\begin{tikzpicture}
		\node (img31) {\includegraphics[width=3cm]{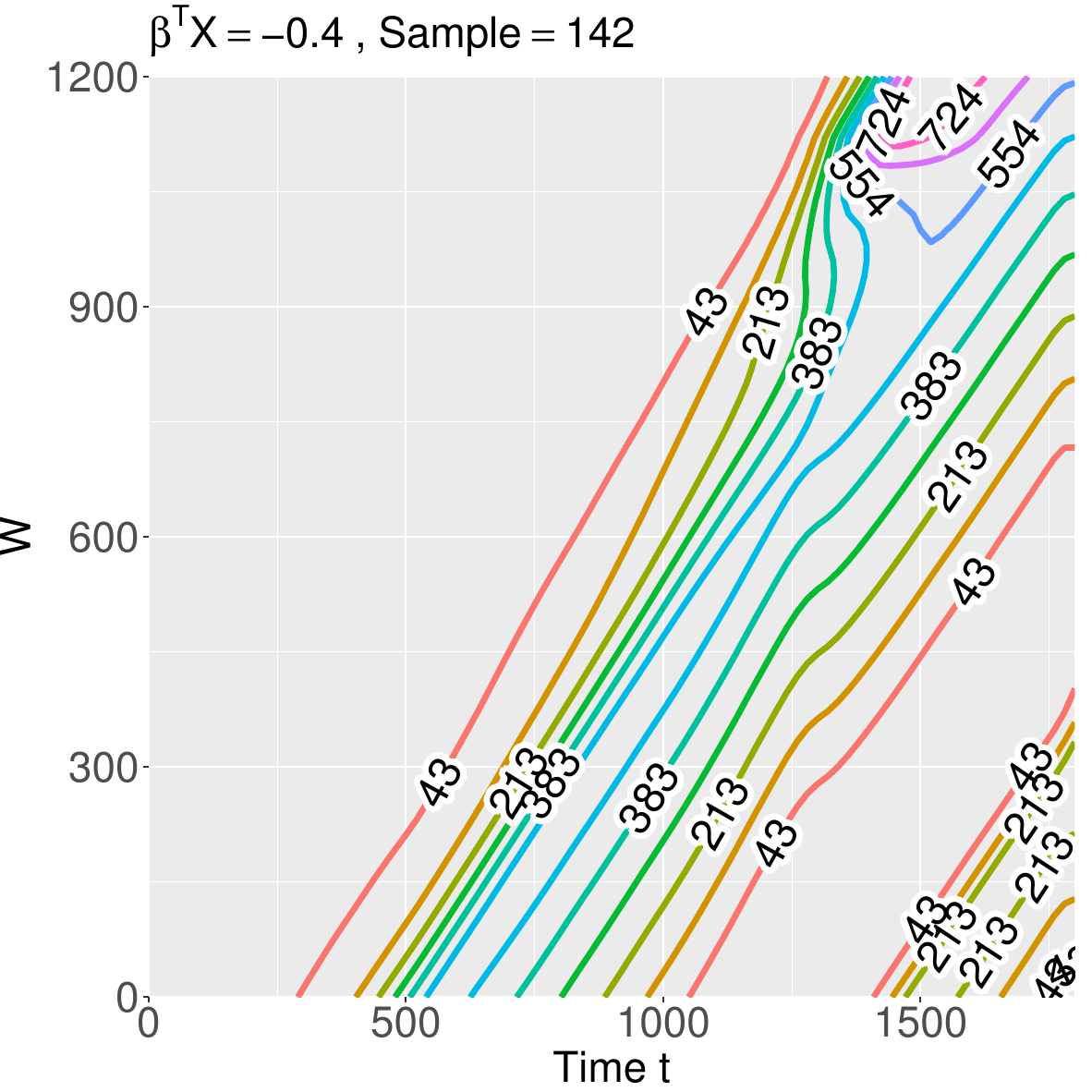}};
		\node[left=of img11, node distance=0cm, rotate=90, anchor=center,font=\tiny,xshift=0.2cm,yshift=-1cm] {Native};
	\end{tikzpicture}
	\begin{tikzpicture}
		\node (img32) {\includegraphics[width=3cm]{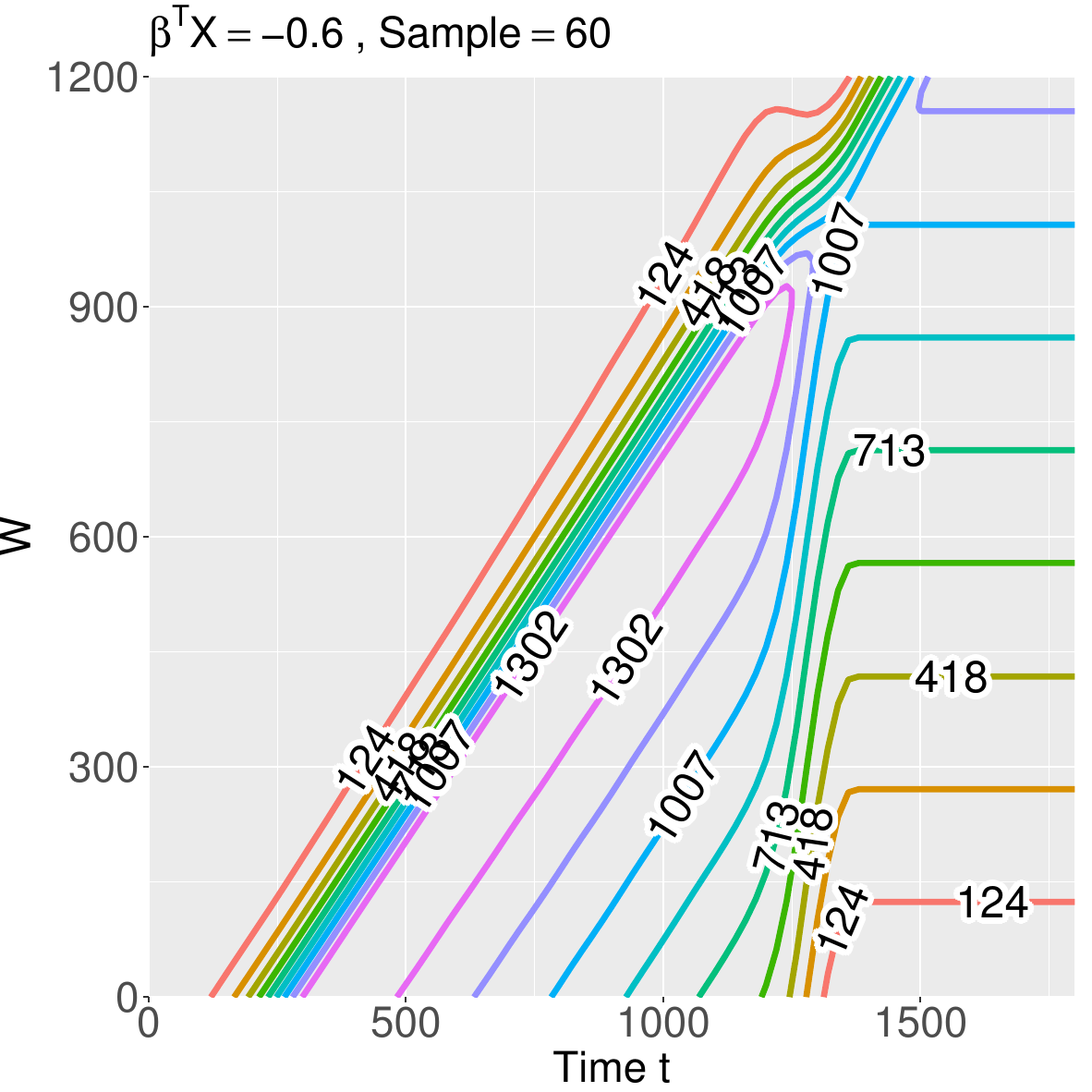}};
	\end{tikzpicture}
	\begin{tikzpicture}
		\node (img33) {\includegraphics[width=3cm]{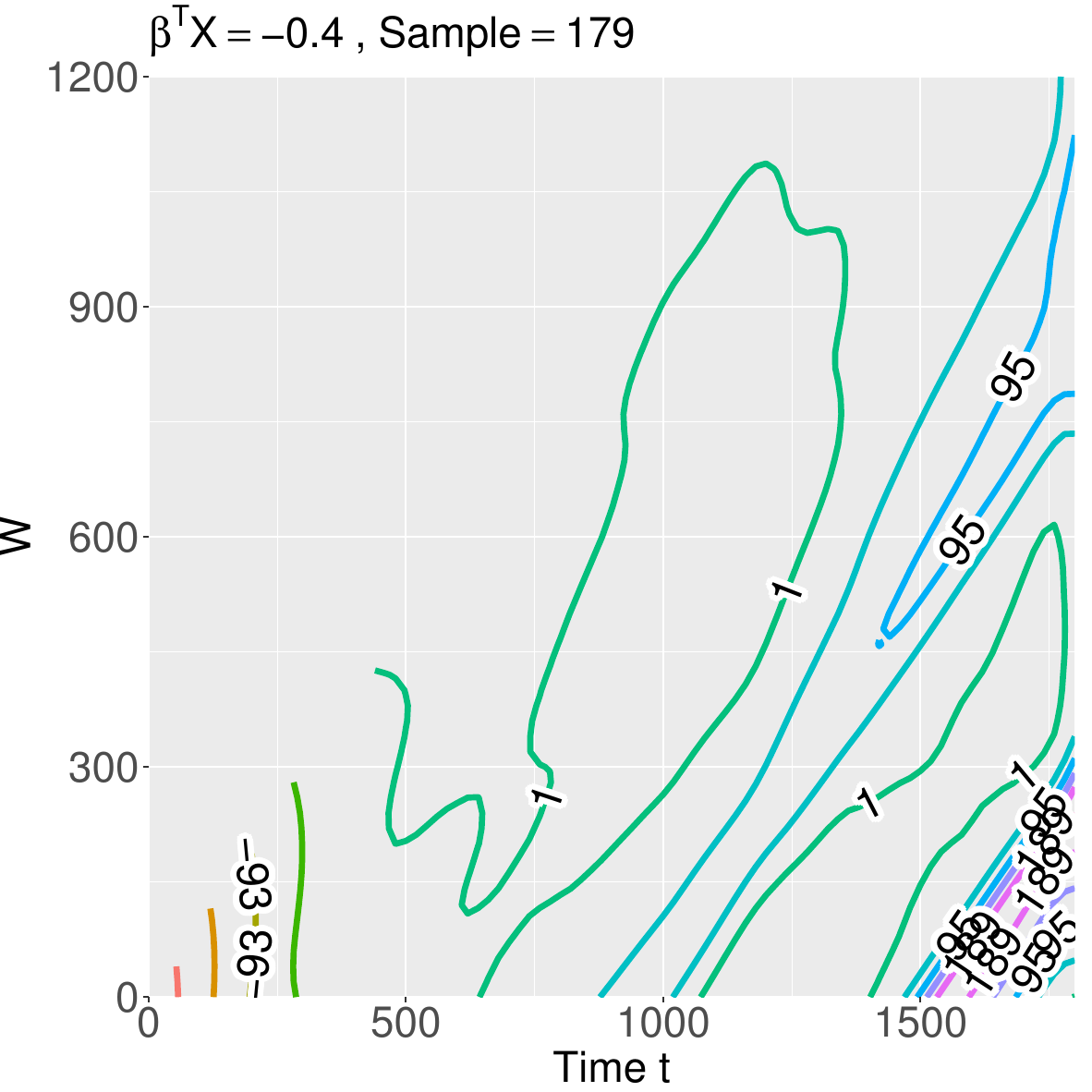}};
	\end{tikzpicture}
	\begin{tikzpicture}
		\node (img34) {\includegraphics[width=3cm]{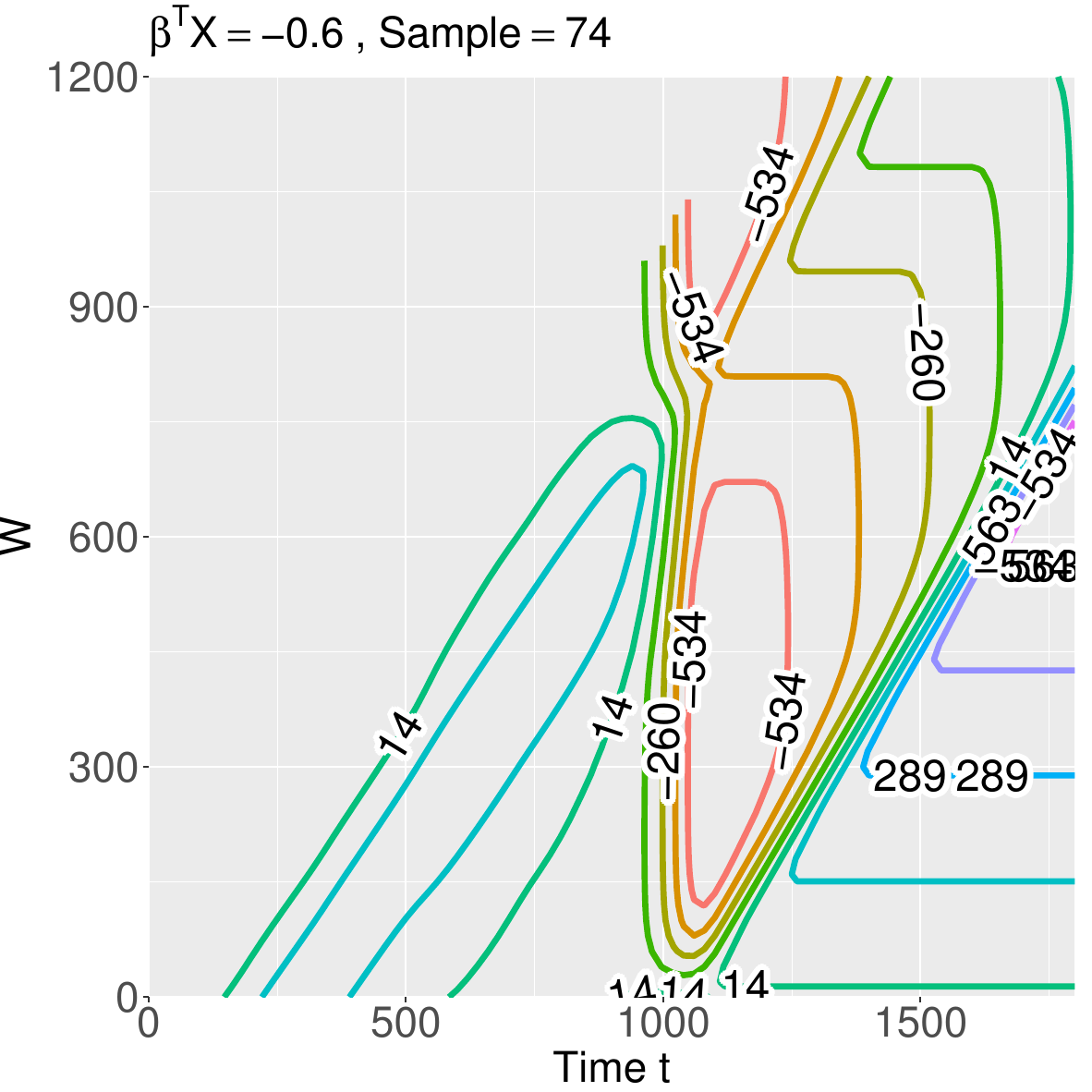}};
	\end{tikzpicture}\\
	\begin{tikzpicture}
		\node (img41) {\includegraphics[width=3cm]{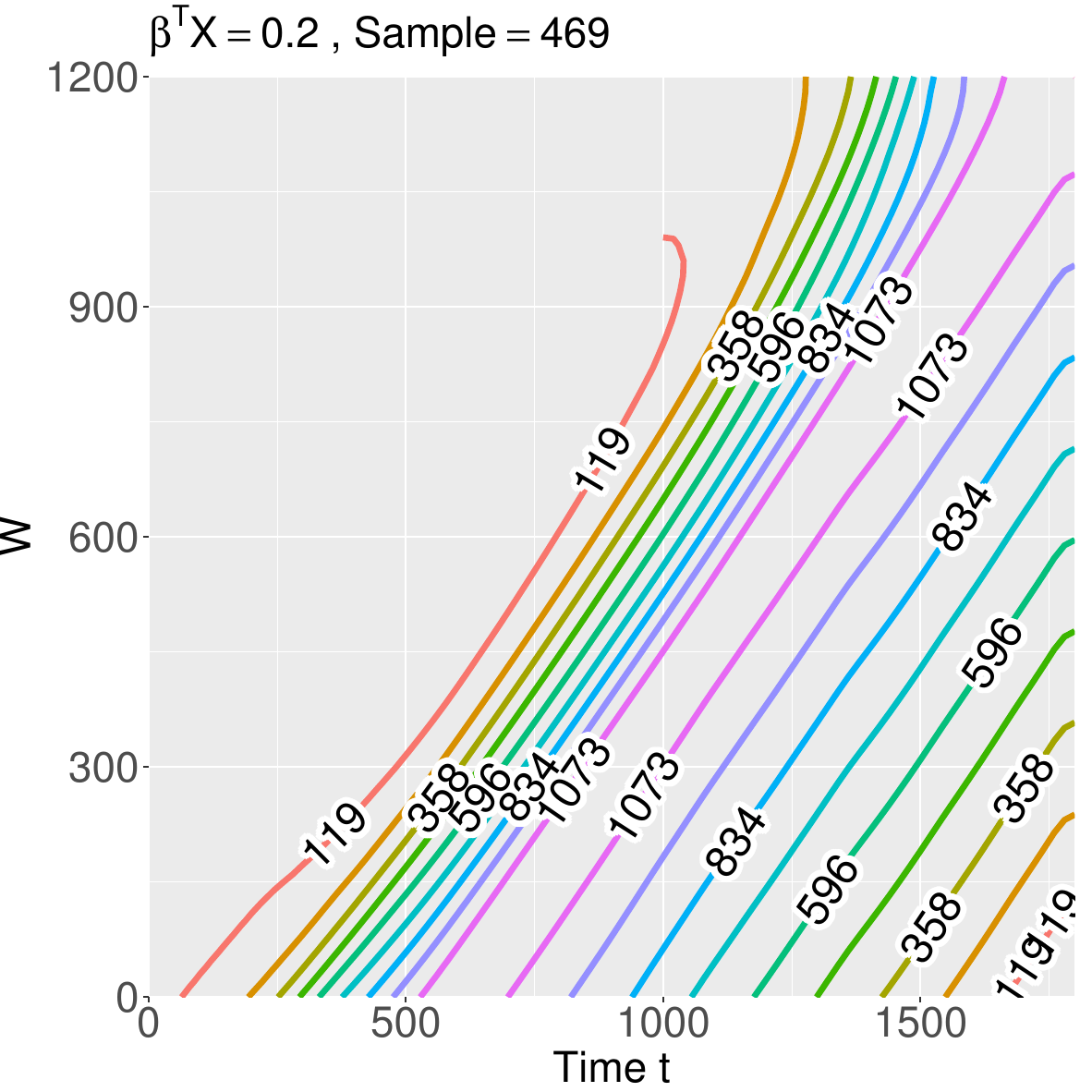}};
		\node[left=of img11, node distance=0cm, rotate=90, anchor=center,font=\tiny,xshift=0.2cm,yshift=-1cm] {Asian};
	\end{tikzpicture}
	\begin{tikzpicture}
		\node (img42) {\includegraphics[width=3cm]{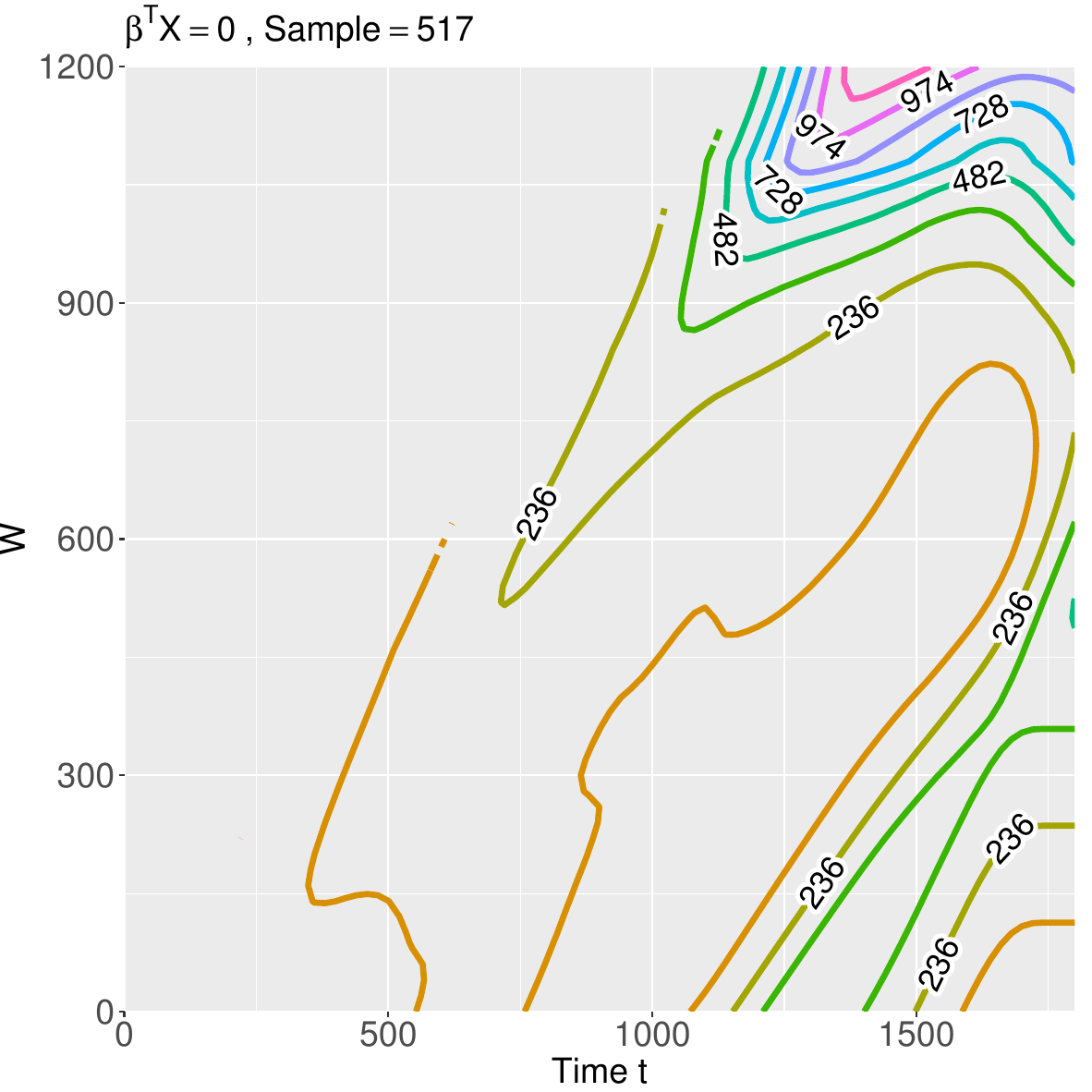}};
	\end{tikzpicture}
	\begin{tikzpicture}
		\node (img43) {\includegraphics[width=3cm]{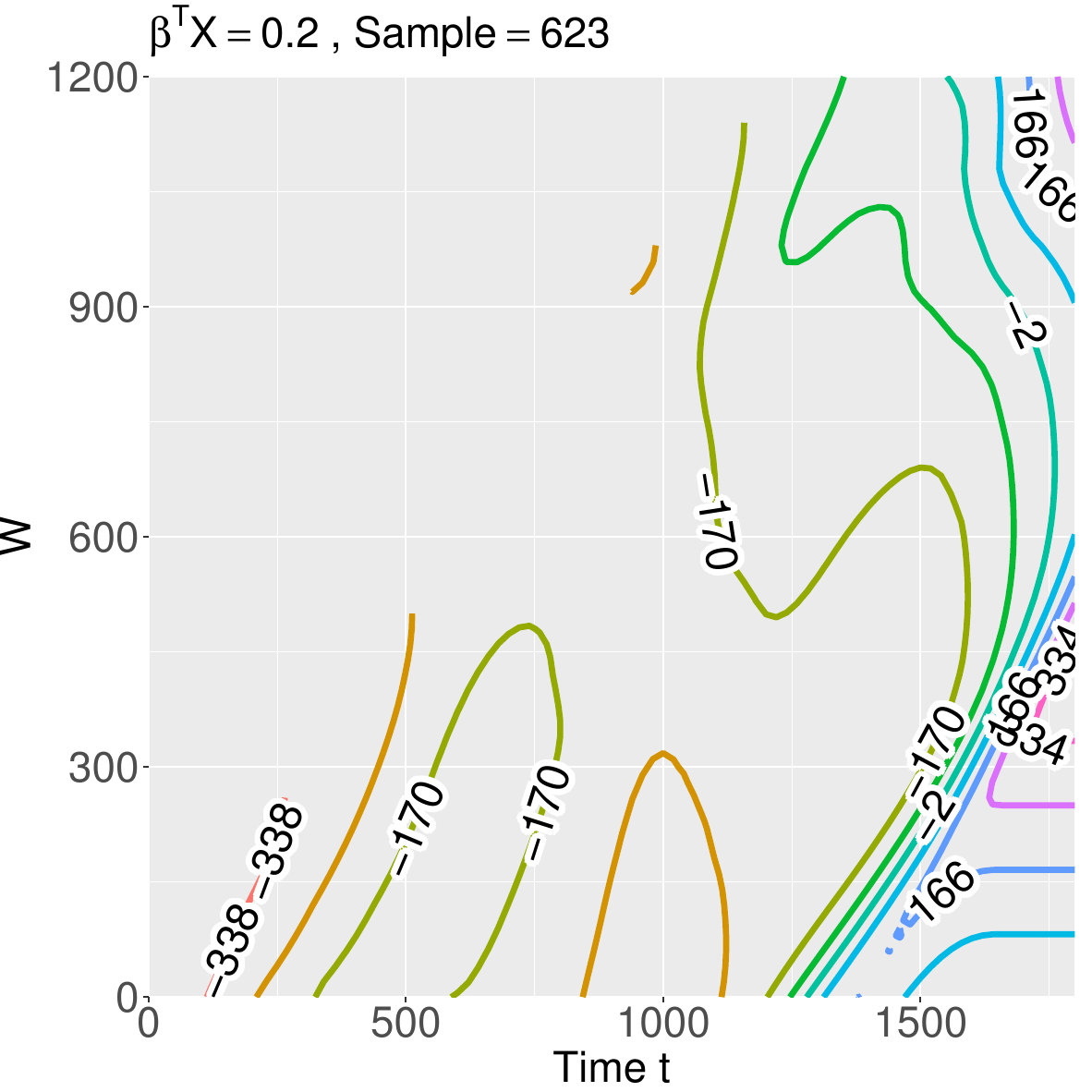}};
	\end{tikzpicture}
	\begin{tikzpicture}
		\node (img44) {\includegraphics[width=3cm]{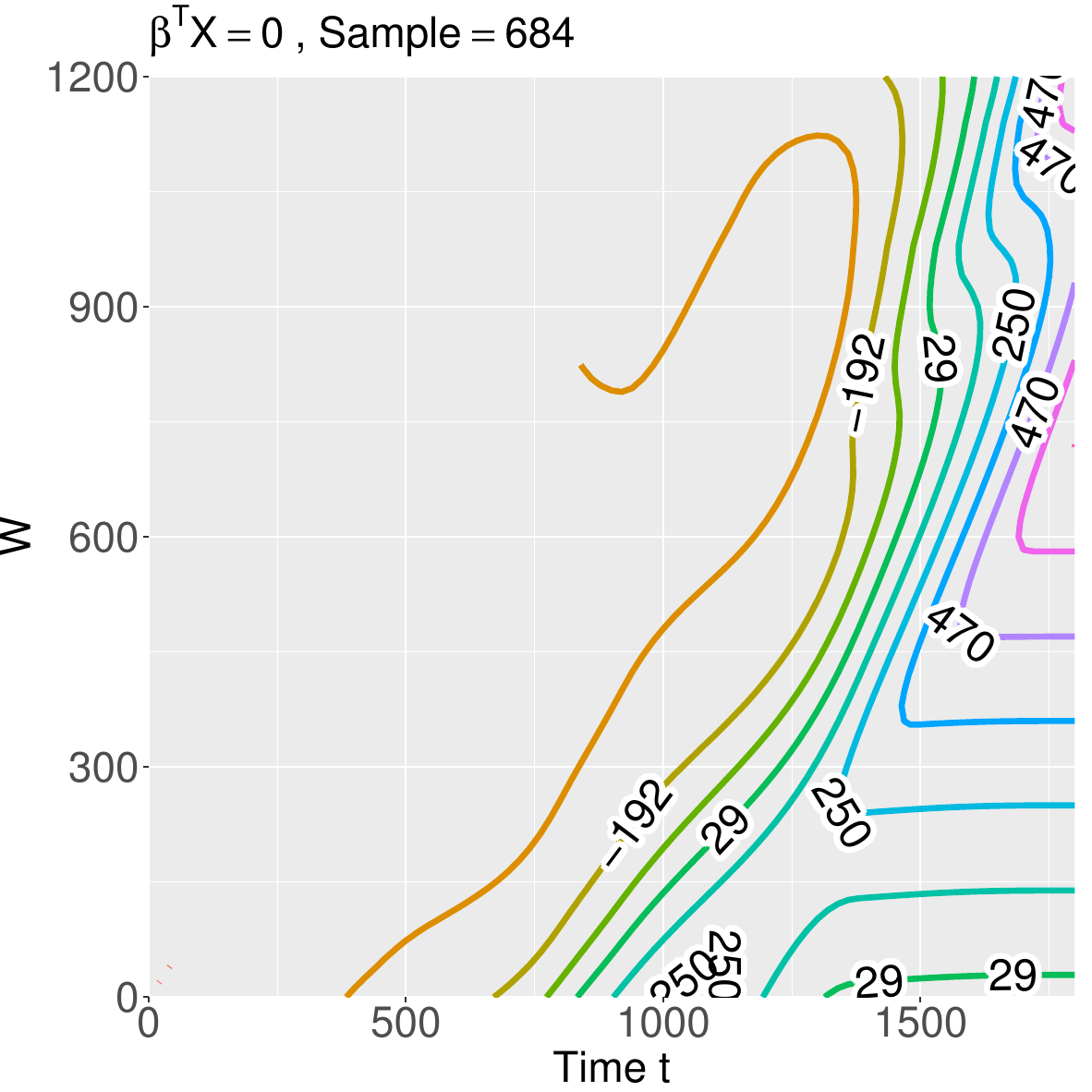}};
	\end{tikzpicture}\\
	\begin{tikzpicture}
		\node (img51) {\includegraphics[width=3cm]{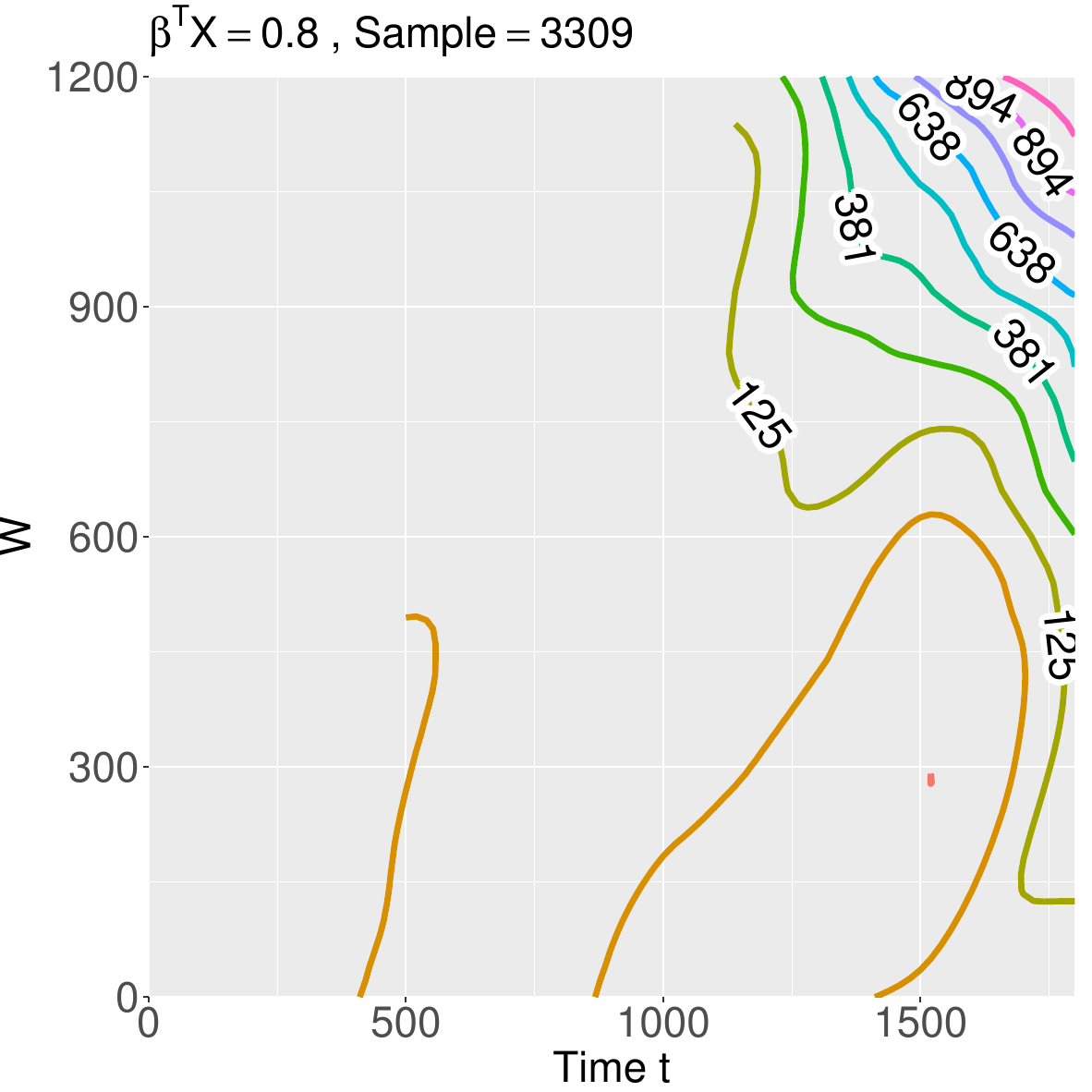}};
		\node[left=of img11, node distance=0cm, rotate=90, anchor=center,font=\tiny,xshift=0.2cm,yshift=-1cm] {White};
	\end{tikzpicture}
	\begin{tikzpicture}
		\node (img52) {\includegraphics[width=3cm]{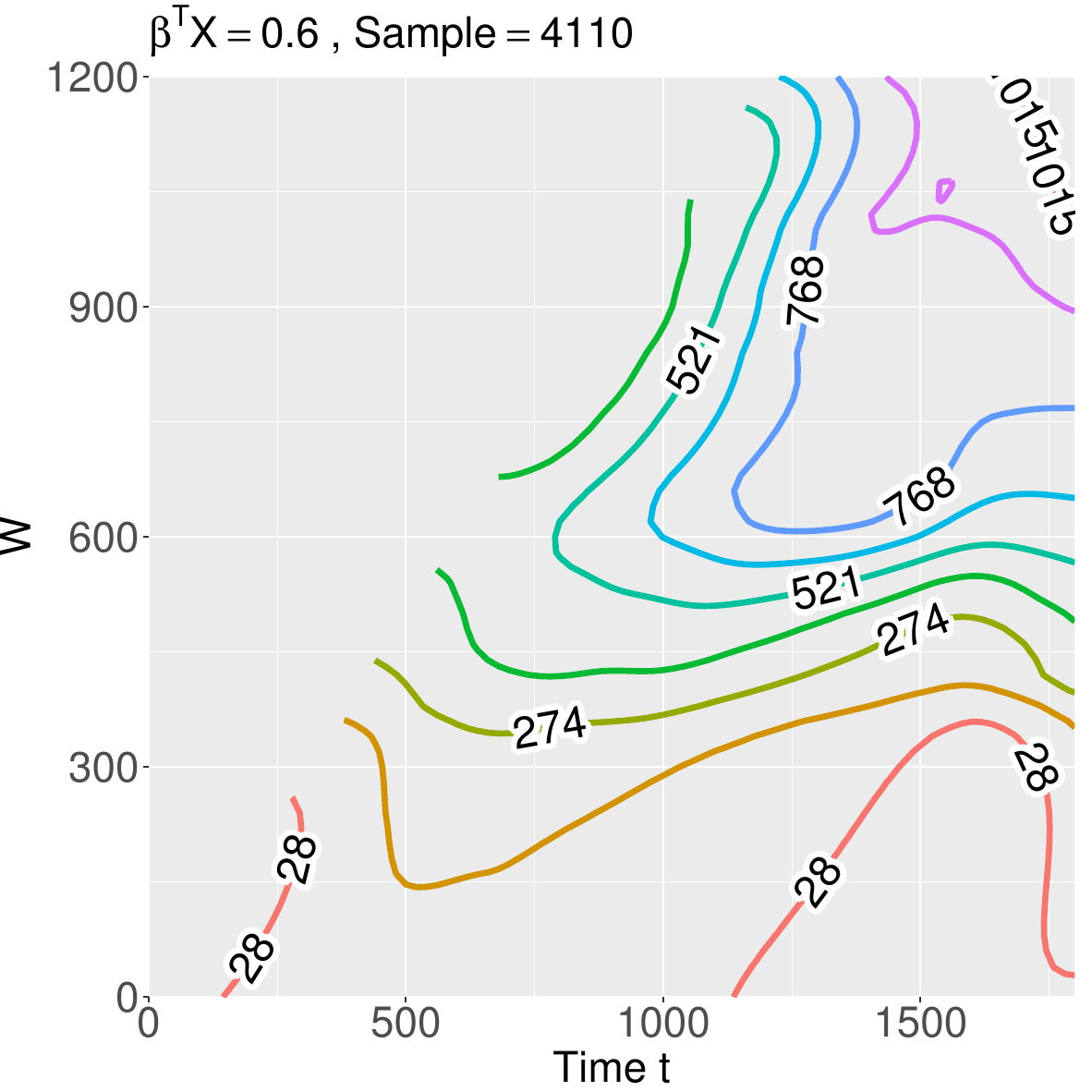}};
	\end{tikzpicture}
	\begin{tikzpicture}
		\node (img53) {\includegraphics[width=3cm]{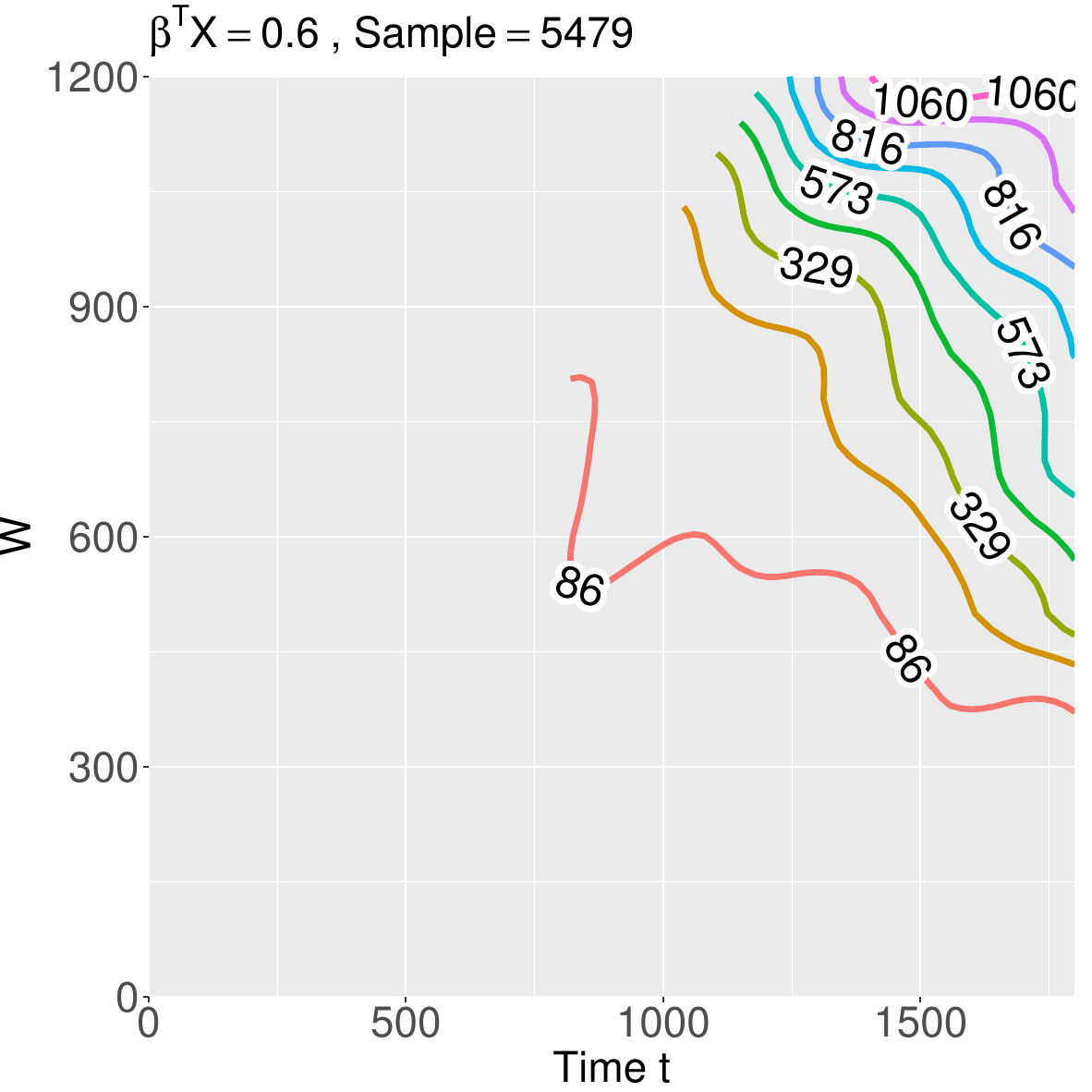}};
	\end{tikzpicture}
	\begin{tikzpicture}
		\node (img54) {\includegraphics[width=3cm]{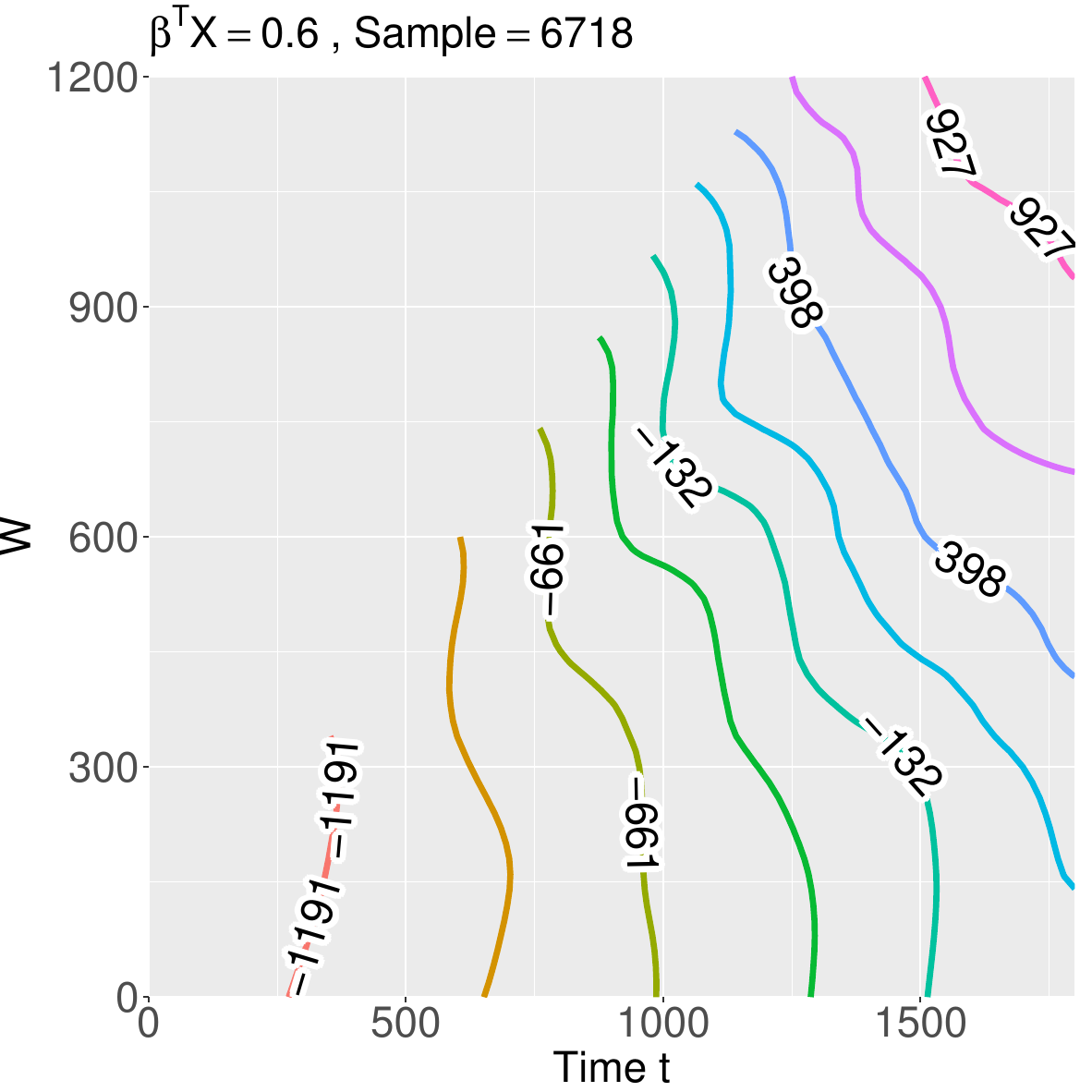}};
	\end{tikzpicture}\\
	\caption{Mean residual life improvement from UNOS/OPTN data. Stratified by Race, Gender, and Insurance Status with maximum $\bb\trans\x$ per strata.}
	\label{fig:appDiffContourCloser3}
\end{figure}
Figure \ref{fig:appDiffContourCloser1}-\ref{fig:appDiffContourCloser3} further reveal the mean residual life improvement stratified by gender ($X_{1}$), race ($X_{2}$), and insurance coverage ($X_{4}$) at different values of $\bb\trans\x$. Several inequalities are noteworthy. 
First, patients with private insurance performed better than those with  public insurance in most of cases,  possibly due to socioeconomic status differences and the affordability for disease maintenance and treatment \citep{goldfarb2006role,nicholas2015socioeconomic}. Second, the life gains   were not very similar across male and female patients, especially the patterns differed between them. Females tended to have  life gains positively related to the index; in contrast, males tended to have  higher residual life gains at small absolute index. It is likely that  heterogeneous kidney disease progression rates and lifestyles might lead to the pattern discrepancy \citep{baylis2009sexual,okada2014sex,pscheidt2015sex}, though the negligible difference in quantity between genders exemplified the ``canceled survival advantage between genders'' phenomenon \citep{oien2006gender,carrero2010gender,cobo2016sex}.	

	Survival gains of African Americans with public insurance were not monotonically related to the index regardless of gender given $t$; further, it was observed that African American females experienced greater life expansion compared to their male counterparts. Interestingly, we found the survival gains of African Americans were comparable with those of non-Hispanic Whites, even though these two racial groups had very different mortality among the general population \citep{lewis2010racial}. However, the African American groups exhibited markedly different indices compared to the non-Hispanic white groups, possibly due to lifestyles and lack of access to healthcare among those groups \citep{kasiske1998race,nicholas2013racial,fedewa2014association}. 
        
	The Hispanic patients displayed consistent patterns in relation to their insurance type. Notably, there was minimal improvement in life gains when $w$ was less than 500 and when $\bb\trans\x>-1.2$. However, at $\bb\trans\x=-1.8$, life gains showed a notable increase as $w$ increased. The most significant decrease in life gains was observed under specific conditions: when $t$ and $w$ were small, and $\bb\trans\x<-1.5$. These trends held true for both genders and across different insurance types.
	
	Among Asian patients, a distinct pattern emerged in life gains with respect to the variables $t$ and $w$. Initially, life gains exhibited a rising trend, followed by a subsequent decline, with fluctuations observed along these dimensions. However, when $\bb\trans\x<0$, the alterations in life gains were less pronounced. Remarkably, the impact of insurance type on survival outcomes varied between genders. For female patients, the choice between private and public insurance did not yield significantly divergent survival gains. Conversely, among male patients, private insurance demonstrated a better outcomes when compared to public insurance.

\section{Discussion}\label{sec:conclusion}

Addressing
a severe shortage of organs that are needed to sustain ESRD patients' life, this work aims to design a feasible strategy to increase
the potential efficiency brought by each available kidney. Instead of
evaluating the patients' expected survival time,  as is done
in the literature, we
 consider
the potential residual life prolonged by kidney transplant. By
comparing  patients' expected residual life with and without
transplant, we use their difference to gauge the potential benefit
gained from the transplant; patients with larger
differences may  have a higher
priority for organ allocations than those
 with smaller  values.

 A natural extension of our model is to compute the causal effect between two groups. In the absence of a strong confounder ``age", it is impossible to draw causal conclusion in this study. However, our comparison between two groups has a capability to analyze the causality as long as all confounders are included. On the other hand, our model compares the transplant cohort to a special case of the nontransplant cohort in which the transplantation would never happen in the future. A more general model will be studied in the future that transplant occurred at any time. Therefore, excepting comparing the mean residual life, many other quantities such as all-cause survivals and hazard ratio will be considered \citep{aalen2015does,andersen2017causal,syriopoulou2020marginal}.

Our semiparametric regression model of mean
 residual life  relaxes the parametric assumptions on the
 dependence of mean residual life on covariates and
	how long a patient has lived.
	To strike a balance
between interpretation and flexibility,  our procedure enables one to reduce the covariate dimensions from $p$ to $d$:
when $d=1$, the  model falls to the  single index model, while $d=p$ corresponds to a completely
nonparametric model. We suggest to use the Validated
Information Criterion  \citep{ma2015validated} to choose $d$,
 which seems to fare well in practice.

 In our model, both the transplant and non-transplant groups are characterized by the same set of indices denoted as $\bb$ throughout the study. An alternative would be to conduct separate estimation processes to distinguish the effects of these indices on different transplantation statuses. While our method can accommodate this suggestion by applying the model separately to each group, this separation approach may imply independent progressions for the same patient before and after the transplant. This may be beyond the scope of our primary objective of consistently quantifying survival improvements. To estimate the mean residual life enhancement attributable to inherent factors which reflect patient functional status, we intend to treat each patient's progression as a cohesive whole. This seems reasonable in the realm of kidney transplantation as studies showed that transplant does not interact significantly with patients' pre-operative functional status and is ``associated with substantial improvement in all stages of functional capacity" \citep{ali2021impact}.

To ensure estimability, we have assumed the complete follow-up
condition, which is reasonable in clinical studies with  high
	event rates and long followup \citep{sun2012mean}, such as in
        studies with advanced stage cancer patients
        \citep{chenetal2005} and ESRD patients
        \citep{mansourvar2016additive}. Our data example features
        renal failure patients with a long followup, which may satisfy
        the  assumption. We also acknowledge that, while the complete follow-up condition is a
        common assumption \citep{chenetal2005,chencheng2005,tsiatis1990estimating,sun2009class},
	it incurs some limitations. For example,
	\cite{ying1993large} pointed out that this assumption
	implies that the knowledge of the support is obtained in
	advance to assure a reasonable maximum followup time $\tau$.
	\cite{sun2012mean}, \cite{chen2006linear} and
	\cite{mansourvar2015semiparametric}
	proposed various ways of selecting a reasonable $\tau$, all requiring
	certain pre-knowledge.  Due to these limitations,
          in Supplement \ref{sec:relax},
          we further relax the complete
          followup condition, where we allow an unbounded support for the
          event time and only require a tail condition on the
          distribution
          similar to but weaker than the sub-Gaussian type. Finally, we
          are aware that the kidney transplant data
          from the U.S. SRTR may represent a biased sample, that is, the included patients were those with access to transplantation. In order to make results generalizable to  a more general population, it is vital to take the probability of accessing  transplantation into account. Estimation of this probability, however, is challenging because of  many tangible and intangible factors involved in the process \citep{axelrod2008rates,weng2010barriers,kucirka2015improving,carrero2018sex}. More research is warranted.

\begin{funding}
	G. Zhao was supported by Faculty Development Program, Portland State University (FEAGXZ). H. Lin was supported by Natural Science Foundation of China (No. 11931014 and 11829101).
\end{funding}

\bibliographystyle{imsart-nameyear} 
\bibliography{meanres}       

\newpage

\title{Supplement to ``Evaluation of transplant benefits with the U.S. Scientific Registry of Transplant Recipients by semiparametric regression of mean residual life"}
		Section 1 derives
the estimators for the finite dimensional parameters and nonparametric functions.
Section 2 presents the asymptotic results of these estimators and lists
the sufficient regularity conditions. Section 3 proves the main theorems for
the asymptotic results. Section 4 considers a possible relaxation of an assumption made
on event times. 
\setcounter{section}{0}

\section{Derivations of the Estimators}

\subsection{Derivation of an efficient score function}\label{sec:score}

Let $Y(t) = I(Z\ge
t)$ be the at risk process and
$N(t) = I(Z\le
t)\Delta$ be the counting process.
Define the filtration ${\cal F}_t$ to be $\sigma\{
N(u), Y(u), \X, I(W \le u), WI(W\le u), 0\le u < t\}$,  and, thus,  
$M(t)
=N(t)-\int_0^t
Y(s)\lambda\{s,\bb\trans\X, I(W \le s), WI(W\le s)\}ds$ is
a martingale with respect to ${\cal F}_t$.  
The nuisance tangent space, which will be utilized for deriving our
estimator, is obtained as follows.

\begin{Pro}\label{pro:nuisance}
	The nuisance tangent space is
	${\cal T}={\cal T}_1\oplus{\cal T}_2\oplus{\cal T}_3$, where each component corresponds to $f_{\X, W}$, $m$ defined in (\ref{eq:meanResidualLife}), and $\lambda_c$,
	respectively. Specifically,	\bse
	&&{\cal T}_1\\
	&=&[\a(\X,W): E\{\a(\X,W)\}=\0, \a(\X,W)\in{\cal
		R}^{(p-d)d}, \var\{\a(\X,W)\}<\infty],\\
	&&{\cal T}_2\\
	&=&\left(\int_0^\infty
	\left[\frac{\h_1\{s,\bb\trans\X,I(W \le s),WI(W\le s)\}}{m_1\{s,\bb\trans\X,I(W\le s),WI(W\le s)\}+1}
	-\frac{\h\{s,\bb\trans\X,I(W \le s),WI(W\le s)\}}{m\{s,\bb\trans\X,I(W \le s),WI(W\le s)\}}\right]\right.\\
	&&\times dM\{s,\bb\trans\X,I(W \le s),WI(W\le s)\}:\forall\h\{z,\bb\trans\X,I(W \le z),WI(W\le z)\}\in{\cal
		R}^{(p-d)d},\\
	&&\var\{z,\bb\trans\X,I(W \le z),WI(W\le z)\}<\infty\bigg),\\
	&&{\cal T}_3\\
	&=&\left[\int_0^\infty\h(s,\X)dM_c(s,\X):\forall\h(z,\X)\in{\cal
		R}^{(p-d)d},\var\{\h(z,\X)\}<\infty\right].
	\ese
\end{Pro}
The derivation of Proposition \ref{pro:nuisance} is provided in
Supplement \ref{app:nuisance}.
Taking the derivative of the logarithm of (\ref{eq:pdf})
with
respect to  $\bb$, we obtain
the score function
\bse
&&\bS_{\bb}\{\Delta,Z,\X,I(W\le Z), WI(W\le Z)\}\\
&=&\int_0^\infty
\left[\frac{\m_{12}\{s,\bb\trans\X,I(W \le s),WI(W\le s)\}}{m_1\{s,\bb\trans\X,I(W \le s),WI(W\le s)\}+1}
-\frac{\m_2\{s,\bb\trans\X,I(W \le s),WI(W\le s)\}
}{m\{s,\bb\trans\X,I(W \le s),WI(W\le s)\}}\right]\\
&&\otimes\X_l
dM\{s,\bb\trans\X,I(W \le s),WI(W\le s)\},
\ese
where $m_{1}\{s,\v, \cdot,\cdot)\}\equiv
\partial m\{s,\v,\cdot,\cdot\}/\partial s =  \partial m_T(s-W,\v)/\partial s I(W\le s)+\partial m_N(s,\v)/\partial s \{1-I(W\le s)\}$, $\m_{2}\{s,\v,\cdot,\cdot\}\equiv
\partial m\{s,\v,\cdot,\cdot\}/\partial \v=\partial m_T(s-W,\v)/\partial \v I(W\le s)+\partial m_N(s,\v)\allowbreak/\partial \v \{1-I(W\le s)\}$, $\m_{12}\{s,\v,\cdot,\cdot\}\equiv
\partial\m_2\{s,\v,\cdot,\cdot\}/\partial s = \partial^2 m_T(s-W,\v)/\partial s\partial \v I(W\le s)+\partial ^2 m_N(s,\v)$

\noindent$/\partial s\partial \v \{1-I(W\le s)\}$.
We can verify that, at $\bb_0$, $\bS_{\bb}\{\Delta,Z,\X,I(W\le Z), WI(W\le Z)\}\perp{\cal T}_1$ and
$\bS_{\bb}\{\Delta,Z,\X,I(W\le Z), WI(W\le Z)\}\perp{\cal T}_3$ due to the martingale
properties.
To look for an efficient score by projecting
$\bS_{\bb}\{\Delta,Z,\X,I(W\le Z), WI(W\le Z)\}$ at $\bb_0$ to ${\cal T}_2$, we search for
$\h^*\{s,\bb\trans\X,I(W \le s),WI(W\le s)\}$ such that
\bse
&&\bS\eff\{\Delta,Z,\X,I(W\le Z), WI(W\le Z)\}\\
&=&\bS_{\bb}\{\Delta,Z,\X,I(W\le Z), WI(W\le Z)\}
-\int_0^\infty
\left[\frac{\h_1^*\{s,\bb_0\trans\X,I(W \le s),WI(W\le s)\}}{m_1\{s,\bb_0\trans\X,I(W \le s),WI(W\le s)\}+1}\right.\\
&&-\left.\frac{\h^*\{s,\bb_0\trans\X,I(W \le s),WI(W\le s)\}}{m\{s,\bb_0\trans\X,I(W \le s),WI(W\le s)\}}\right]
dM\{s,\bb_0\trans\X,I(W \le s),WI(W\le s)\}\\
&=&\int_0^\infty
\left[\frac{\m_{12}\{s,\bb_0\trans\X,I(W \le s),WI(W\le s)\}\otimes\X_l-\h_1^*\{s,\bb_0\trans\X,I(W \le s),WI(W\le s)\}}{m_1\{s,\bb_0\trans\X,I(W \le s),WI(W\le s)\}+1}\right.\\
&&\quad\left.-\frac{\m_2\{s,\bb_0\trans\X,I(W \le s),WI(W\le s)\}
	\otimes\X_l-\h^*\{s,\bb_0\trans\X,I(W \le s),WI(W\le s)\}}{m\{s,\bb_0\trans\X,I(W \le s),WI(W\le s)\}}\right]\\
&&\times dM\{s,\bb_0\trans\X,I(W \le s),WI(W\le s)\},
\ese
where $\h_1^*\{s,\v,\cdot,\cdot\}=\partial \h^*\{s,\v,\cdot,\cdot\}/\partial s$, is orthogonal to ${\cal T}_2$. 
It implies that, for any $\h\{s,\bb_0\trans\X,$

\noindent$I(W \le s),WI(W\le s)\}$, the following must hold
\be\label{eq:must}
0
&=&E\left(\int_0^\infty \a\{s,\bb_0\trans\X,I(W \le s),WI(W\le s)\}\trans
\left[\frac{\h_1\{s,\bb_0\trans\X,I(W \le s),WI(W\le s)\}}{m_1\{s,\bb_0\trans\X,I(W \le s),WI(W\le s)\}+1}\right.\right.\n\\
&&\left.\left.-\frac{\h\{s,\bb_0\trans\X,I(W \le s),WI(W\le s)\}}{m\{s,\bb_0\trans\X,I(W \le s),WI(W\le s)\}}\right]
ds\right),
\ee
where
\be\label{eq:a}
&&\a\{s,\bb_0\trans\X,I(W \le s),WI(W\le s)\}\n\\
&\equiv&
E\left(\left[\frac{\m_{12}\{s,\bb_0\trans\X,I(W \le s),WI(W\le s)\}\otimes\X_l-\h_1^*\{s,\bb_0\trans\X,I(W \le s),WI(W\le s)\}}{m_1\{s,\bb_0\trans\X,I(W \le s),WI(W\le s)\}+1}\right.\right.\n\\
&&\quad\left.-\frac{\m_2\{s,\bb_0\trans\X,I(W \le s),WI(W\le s)\}
	\otimes\X_l-\h^*\{s,\bb_0\trans\X,I(W \le s),WI(W\le s)\}}{m\{s,\bb_0\trans\X,I(W \le s),WI(W\le s)\}}\right]\n\\
&&\times S_c(s,\X)\mid\bb_0\trans\X\bigg)
S\{s,\bb_0\trans\X,I(W \le s),WI(W\le s)\}\n\\
&&\times\frac{m_1\{s,\bb_0\trans\X,I(W \le s),WI(W\le s)\}+1}{m\{s,\bb_0\trans\X,I(W \le s),WI(W\le s)\}},
\ee
and $\h_1\{s,\v,\cdot,\cdot\}=\partial \h\{s,\v,\cdot,\cdot\}/\partial s$.
We can choose any $\h\{s,\bb_0\trans\X,I(W \le s),WI(W\le s)\}$ function. Specifically, by letting $\h\{s,\bb_0\trans\X,I(W \le s),WI(W\le s)\}=0$ for $s<t$ and
$\h\{s,\bb_0\trans\X,I(W \le s),WI(W\le s)\}=\c(\bb_0\trans\X)$ for $s\ge t$ with an
arbitrary
function $\c(\bb_0\trans\X)$, we obtain
$
\frac{\a\{t,\bb_0\trans\X,I(W \le t),WI(W\le t)\}}{m_1\{t,\bb_0\trans\X,I(W \le t),WI(W\le t)\}+1}-\int_t^\infty\frac{\a\{s,\bb_0\trans\X,I(W \le s),WI(W\le s)\}}{m\{s,\bb_0\trans\X,I(W \le s),WI(W\le s)\}}ds=\0.
$
Solving this integral equation leads to
\bse
&&\a\{t,\bb_0\trans\X,I(W \le t),WI(W\le t)\}\\
&=&\{m_1\{t,\bb_0\trans\X,I(W \le t),WI(W\le t)\}+1\}\\
&&\times\exp\left[-\int_0^t\frac{m_1\{s,\bb_0\trans\X,I(W \le s),WI(W\le s)\}+1}{m\{s,\bb_0\trans\X,I(W \le s),WI(W\le s)\}}ds\right]\c(\bb_0\trans\X),
\ese
for  function $\c(\cdot)$. Thus, reusing (\ref{eq:must}), we require that for all
$\h\{t,\bb_0\trans\X,I(W \le t),WI(W\le t)\}$,
\bse
0&=&E\left(\int_0^\infty [m_1\{t,\bb_0\trans\X,I(W \le t),WI(W\le t)\}+1]\right.\\
&&\left.\times\exp\left[-\int_0^t\frac{m_1\{s,\bb_0\trans\X,I(W \le s),WI(W\le s)\}+1}{m\{s,\bb_0\trans\X,I(W \le s),WI(W\le s)\}}ds\right]\c(\bb_0\trans\X)\trans\right.\\
&&\left.\times\left[\frac{\h_1\{t,\bb_0\trans\X,I(W \le t),WI(W\le t)\}}{m_1\{t,\bb_0\trans\X,I(W \le t),WI(W\le t)\}+1}
-\frac{\h\{t,\bb_0\trans\X,I(W \le t),WI(W\le t)\}}{m\{t,\bb_0\trans\X,I(W \le t),WI(W\le t)\}}\right]dt\right)\\
&=&-E\left[\c(\bb_0\trans\X)\trans
\h\{0,\bb_0\trans\X,I(W\le 0),WI(W\le 0)\}\right].
\ese
Letting $\h\{0,\bb_0\trans\X,I(W\le 0),WI(W\le 0)\}=\c(\bb_0\trans\X)$ yields the only
possibility of $\c(\bb_0\trans\X)=\0$, hence
$\a\{t,\bb_0\trans\X,I(W \le t),WI(W\le t)\}=\0$.
Inserting the expression of
$\a\{t,\bb_0\trans\X,I(W \le t),WI(W\le t)\}$ into (\ref{eq:a}), we have
\bse
&&\frac{\h_1^*\{t,\bb_0\trans\X,I(W \le t),WI(W\le t)\}}{m_1\{t,\bb_0\trans\X,I(W \le t),WI(W\le t)\}+1}
-\frac{\h^*\{t,\bb_0\trans\X,I(W \le t),WI(W\le t)\}}{m\{t,\bb_0\trans\X,I(W \le t),WI(W\le t)\}}\\
&=&\b\{t,\bb_0\trans\X,I(W \le t),WI(W\le t)\},
\ese
where
\bse
&&\b\{t,\bb_0\trans\X,I(W \le t),WI(W\le t)\}
=\left\{\frac{\m_{12}\{t,\bb_0\trans\X,I(W \le t),WI(W\le t)\}}{m_1\{t,\bb_0\trans\X,I(W \le t),WI(W\le t)\}+1}\right.\\
&&\quad\left.-\frac{\m_2\{t,\bb_0\trans\X,I(W \le t),WI(W\le t)\}
}{m\{t,\bb_0\trans\X,I(W \le t),WI(W\le t)\}}\right\} \otimes \frac{E\left\{\X_l
	S_c(t,\X)\mid\bb_0\trans\X,W\right\}}{E\left\{S_c(t,\X)\mid\bb_0\trans\X,W\right\}}.
\ese
Thus an efficient score is
\be\label{eq:effscore}
&&\bS\eff\{\Delta,Z,\X,I(W \le Z),WI(W\le Z)\}\n\\
&=&\int_0^\infty
\left[\frac{\m_{12}\{s,\bb_0\trans\X,I(W\le s),WI(W\le s)\}}{m_1\{s,\bb_0\trans\X,I(W\le s),WI(W\le s)\}+1}
-\frac{\m_2\{s,\bb_0\trans\X,I(W\le s),WI(W\le s)\} }{m\{s,\bb_0\trans\X,I(W\le s),WI(W\le s)\}}\right]\n\\
&&\otimes
\left[\X_l-
\frac{E\left\{\X_l
	S_c(s,\X)\mid\bb_0\trans\X,W\right\}}{E\left\{S_c(s,\X)\mid\bb_0\trans\X,W\right\}}\right]
dM\{s,\bb_0\trans\X,I(W\le s),WI(W\le s)\}.\n
\ee

\subsubsection{Proof of Proposition \ref{pro:nuisance}  \label{app:nuisance}}
\noindent Proof:
Let ${\cal T}_1$, ${\cal T}_2$ and ${\cal T}_3$ be the nuisance
tangent
spaces  corresponding to $f_{\X,W}$, $m$ and $\lambda_c$
respectively.
The result of ${\cal T}_1$ follows obviously.
To obtain ${\cal T}_2$, let
$m\{z,\bb\trans\X,I(W\le z),WI(W\le z)\}+\bg\trans\h\{z,\bb\trans\X,I(W\le z),WI(W\le z)\}$ be a sub model of
$m\{z,\bb\trans\X,I(W\le z),WI(W\le z)\}$, where $\h\{z,\bb\trans\X,I(W\le z),WI(W\le z)\}\in{\cal
	R}^{(p-d)d}$ with $\var[\h\{z,\bb\trans\X,$

\noindent$I(W\le z),WI(W\le z)\}]<\infty,$
differentiate the log of  (\ref{eq:pdf}) with
respect  to $\bg$ and evaluate it at
$\bg=0$. Then, ${\cal T}_2$ is
\bse
&&\Delta \left\{\frac{\h_1\{z,\bb\trans\X,I(W\le z),WI(W\le z)\}}{m_1\{z,\bb\trans\X,I(W\le z),WI(W\le z)\}+1}
-\frac{\h\{z,\bb\trans\X,I(W\le z),WI(W\le z)\}}{m\{z,\bb\trans\X,I(W\le z),WI(W\le z)\}}\right\}\\
&&-\int_0^z
\frac{\h_1\{s,\bb\trans\X,I(W\le s),WI(W\le s)\}m\{s,\bb\trans\X,I(W\le s),WI(W\le s)\}}{m^2\{s,\bb\trans\X,I(W\le s),WI(W\le s)\}}\\
&&\quad-\frac{\{m_1\{s,\bb\trans\X,I(W\le s),WI(W\le s)\}+1\}\h\{s,\bb\trans\X,I(W\le s),WI(W\le s)\}
}{m^2\{s,\bb\trans\X,I(W\le s),WI(W\le s)\}}dt\\
&=&\Delta \left\{\frac{\h_1\{z,\bb\trans\X,I(W\le z),WI(W\le z)\}}{m_1\{z,\bb\trans\X,I(W\le z),WI(W\le z)\}+1}
-\frac{\h\{z,\bb\trans\X,I(W\le z),WI(W\le z)\}}{m\{z,\bb\trans\X,I(W\le z),WI(W\le z)\}}\right\}\\
&&-\int_0^z
\left[\frac{\h_1\{s,\bb\trans\X,I(W\le s),WI(W\le s)\}}{m_1\{s,\bb\trans\X,I(W\le s),WI(W\le s)\}+1}-\frac{\h\{s,\bb\trans\X,I(W\le s),WI(W\le s)\}
}{m\{s,\bb\trans\X,I(W\le s),WI(W\le s)\}}\right]\\
&&\quad\lambda\{s,\bb\trans\X,I(W\le s),WI(W\le s)\}ds\\
&=&\int_0^\infty
\left[\frac{\h_1\{s,\bb\trans\X,I(W\le s),WI(W\le s)\}}{m_1\{s,\bb\trans\X,I(W\le s),WI(W\le s)\}+1}-\frac{\h\{s,\bb\trans\X,I(W\le s),WI(W\le s)\}
}{m\{s,\bb\trans\X,I(W\le s),WI(W\le s)\}}\right]\\
&&dM\{s,\bb\trans\X,I(W\le s),WI(W\le s)\}.
\ese

To obtain ${\cal T}_3$,
let
$\lambda_c(t,\X)\{1+\bg\trans\h(t,\X)\}$ be
a submodel of $\lambda_{c}(t,\X)$,
where $\h(z,\X)\in{\cal
	R}^{(p-d)d},$ with $\var\{\h(z,\X)\}<\infty$.
We then obtain ${\cal T}_3$ as follows
\bse
\frac{\partial\log f(\X,Z,\Delta)}{\partial \bg} | _{\bg=0}
&=&(1-\Delta)\h(Z,\X)-\int_0^Z\h(s,\X)\lambda_c(s,\X)ds\\
&=&\int_0^\infty\h(s,\X)dM_c(s,\X),
\ese
where $M_c(t,\X)=N_c(t)-\int_0^tI(Z\ge
s)\lambda_c(s,\X)ds$
is a martingale by Theorem 1.3.2 in \cite{fh1991}.
Because
$\lambda_c(t,\X)$ can be any positive function,
$\h(s,\X)$ can be any function. This leads to the form of ${\cal T}_3$.

By taking  conditional expectations given $\X$ and $W$,
it follows that ${\cal T}_1\perp{\cal T}_2$ and
${\cal T}_1\perp{\cal T}_3$.
Further, ${\cal T}_2\perp{\cal T}_3$ because the martingale
integrals  associated with $M\{s,\bb\trans\X,I(W\le s),WI(W\le s)\}$ and $M_c(s,\X)$ are
independent
conditional on $\X$ due to $T\indep C\mid W, \X$.
\qed

\subsubsection{Proof of Equations (\ref{eq:useful}) and
	(\ref{eq:expecteff})}\label{app:eff}
Proof:
First, we note that, when $t \le \tau$,
\bse
E\left\{ \X_lY(t)\mid\bb\trans\X, W\right\}
&=&E[E\left\{\X_l
I(T \ge t)I(C \ge t)\mid\bb\trans\X, \X, W\right\} \mid \bb\trans\X, W]
\\
&=&E[E\left\{\X_l
I(T \ge t)I(C \ge t)\mid\X, W\right\} \mid \bb\trans\X, W]
\\
&=&E\left[\X_l S\{t, \bb\trans\X,I(W\le t),WI(W\le t)\}
S_c(t, \X) \mid\bb\trans\X, W\right]
\\
&=& S\{t, \bb\trans\X,I(W\le t),WI(W\le t)\} E\left\{\X_l
S_c(t, \X) \mid\bb\trans\X\right\},
\ese
where
the second to last equality holds because of $T\indep C\mid \X,W$
and that $\pr(T \ge t|\X)$ is a function of $\bb\trans\X$ and $W$.
Similarly, for $t \le \tau$, we obtain that
$$ E\left\{ Y(t)\mid\bb\trans\X,W\right\}
= S\{t, \bb\trans\X,I(W\le t),WI(W\le t)\} E\left\{
S_c(t, \X) \mid\bb\trans\X\right\}.
$$
Hence, when $t\le \tau$,
\be \label{junk}
\frac{E\left\{ \X_lY(t)\mid\bb\trans\X,W\right\}}
{E\left\{Y(t)\mid\bb\trans\X,W\right\}}
&=&\frac{E\left\{\X_lS_c(t,\X)\mid\bb\trans\X\right\}}
{E\left\{S_c(t,\X)\mid\bb\trans\X\right\}}.
\ee

Second, when $t>\tau$, $Y(t)=0$ and $S_c( t,\X)=0$. Hence, we have a $0/0$ scenario in (\ref{junk}), in which case,
we define (\ref{junk}) to be \bse
\frac{E\left\{\X_lS_c(\tau,\X)\mid\bb\trans\X\right\}}
{E\left\{S_c(\tau, \X)\mid\bb\trans\X\right\}}
= \frac{E\left\{\X_l p(\X)\mid\bb\trans\X\right\}}
{E\left\{p(\X)\mid\bb\trans\X\right\}},
\ese
a time-invariant constant.
Here, $p(\X)$ is defined in Section \ref{sec:estimation}.
Hence, (\ref{eq:useful}) holds  over $[0, \infty)$, with the truth of $\bb$ being $\bb_0$.

With a generic $\bb$, (\ref{junk}) leads to
\bse
&&E\left(\int_0^\infty\g\{s,\bb\trans\X,I(W \le s),WI(W\le s)\}
\otimes\left[\X_l-
\frac{E\left\{\X_l
	S_c(s,\X)\mid\bb\trans\X\right\}}
{E\left\{S_c(s,\X)\mid\bb\trans\X\right\}}\right]\right.\\
&&\times \left.Y(s)\lambda_0\{s,\bb\trans\X,I(W\le s),WI(W\le s)\}ds\right)\\
&=&E\left(\int_0^\infty\g\{s,\bb\trans\X,I(W \le s),WI(W\le s)\}
\otimes\left[E\{\X_lY(s)\mid\bb\trans\X,W\}\right.\right.\\
&&\left.\left.-\frac{E\left\{\X_l
	S_c(s,\X)\mid\bb\trans\X\right\}}
{E\left\{S_c(s,\X)\mid\bb\trans\X\right\}}
E\{Y(s)\mid\bb\trans\X,I(W \le s),WI(W\le s)\}\right]\lambda_0(s,\bb\trans\X)ds\right)\\
&=&\0,
\ese
as the quantity inside the square bracket is zero.
In addition,
\bse
&&E\left(\int_0^\infty\g\{s,\bb\trans\X,I(W \le s),WI(W\le s)\}\right.\\
&&\quad\left.\otimes\left[\X_l-
\frac{E\left\{\X_l
	S_c(s,\X)\mid\bb\trans\X\right\}}
{E\left\{S_c(s,\X)\mid\bb\trans\X\right\}}\right]dM\{s,\bb\trans\X,I(W \le s),WI(W\le s)\}\right)=\0,
\ese
because  $dM\{s,\bb\trans\X,I(W \le s),WI(W\le s)\}=dN(s)-Y(s)\lambda_0\{s,\bb\trans\X,I(W \le s),W\allowbreak I(W\le s)\}ds$ is a martingale.
Therefore, we have
\bse
E\left(\int_0^\infty\g\{s,\bb\trans\X,I(W \le s),WI(W\le s)\}
\otimes\left[\X_l-
\frac{E\left\{\X_l
	S_c(s,\X)\mid\bb\trans\X\right\}}
{E\left\{S_c(s,\X)\mid\bb\trans\X\right\}}\right]dN(s)\right)=\0.
\ese
Hence,  (\ref{eq:expecteff}) holds with the truth of $\bb$ being $\bb_0$.
\qed

\subsection{ Explicit Expressions of Nonparametric Estimators}

\subsubsection{Nonparametric Estimators of Hazard and Mean Residual Life Functions and Their Derivatives}\label{sec:nonpara}

The term $W$ within the estimator of $\Lambda_T$ is equivalent to $\bb\trans\X$ when conducting the kernel estimation therefore to be omitted thereafter for the sake of ease notation. Thus the estimators of $\Lambda_T$ and $\Lambda_N$ have a similar format and only differ in the samples utilized. We propose a general form of the nonparametric estimators of
$\bLam_2(t,\bb\trans\X)$,
$\lambda(t,\bb\trans\X)$, $\blam_2(t,\bb\trans\X)$, where $\bLam_2(t,\v)\equiv\partial \Lambda(t,\v)/\partial \v$ and $\blam_2(t,\v)\equiv\partial \lambda(t,\v)/\partial \v$.
\be
&&\wh\bLam_2(t,\bb\trans\X)\n\\
&=&-\sum_{i=1}^n \frac{I(Z_i\le
	t)\Delta_i \K_h'(\bb\trans\X_i-\bb\trans\X)}{\sumj
	I(Z_j\ge
	Z_i)K_h(\bb\trans\X_j-\bb\trans\X)}\n\\
&&+\sum_{i=1}^nI(Z_i\le t)\Delta_iK_h(\bb\trans\X_i-\bb\trans\X)
\frac{\sumj
	I(Z_j\ge Z_i) \K_h'(\bb\trans\X_j-\bb\trans\X)}{\{\sumj
	I(Z_j\ge
	Z_i)K_h(\bb\trans\X_j-\bb\trans\X)\}^2},\label{eq:Lambda2}\\
&&\wh{\lambda}(t,\bb\trans\X) \n\\
&=& \sum_{i=1}^n
\frac{K_b(Z_i-t)\Delta_iK_h(\bb\trans\X_i-\bb\trans\X)}{\sumj I(Z_j\ge
	Z_i)K_h(\bb\trans\X_j-\bb\trans\X)},\label{eq:lambda}\\
&&\wh\blam_2(t,\bb\trans\X)\n\\
&=&-\sum_{i=1}^n
\frac{K_b(Z_i-t)\Delta_i\K_h'(\bb\trans\X_i-\bb\trans\X)}{\sumj I(Z_j\ge
	Z_i)K_h(\bb\trans\X_j-\bb\trans\X)}\n\\
&&+\sum_{i=1}^nK_b(Z_i-t)\Delta_iK_h(\bb\trans\X_i-\bb\trans\X)
\frac{\sumj
	I(Z_j\ge
	Z_i)\K_h'(\bb\trans\X_j-\bb\trans\X)}{\{\sumj I(Z_j\ge
	Z_i)K_h(\bb\trans\X_j-\bb\trans\X)\}^2},\label{eq:lambda1}
\ee
The estimators of ${m}_N(t,\bb\trans\x)$, ${m}_T(t,\bb\trans\x)$, $m_1\{t,\bb\trans\X,I(W \le t),WI(W\le t)\}$, $\m_2\{t,\bb\trans\X,$

\noindent$I(W \le t),WI(W\le t)\}$, and $\m_{12}\{t,\bb\trans\X,I(W \le t),WI(W\le t)\}$ are
\bse
&& \wh{m}_N(t,\bb\trans\x)=e^{{\wh\Lambda_N}(t,\bb\trans\x)}\int_{t}^{\tau}
e^{-{\wh\Lambda_N}(s,\bb\trans\x)}ds,\\
&& \wh{m}_T(t,\bb\trans\x)=e^{{\wh\Lambda_T}(t,\bb\trans\x)}\int_{t}^{\tau}
e^{-{\wh\Lambda_T}(s,\bb\trans\x)}ds,\\
&&\wh m_1\{t,\bb\trans\X,I(W \le t),WI(W\le t)\}\\
&=&\wh\lambda_T(t-W,\bb\trans\X)e^{\wh\Lambda_T(t-W,\bb\trans\X)}
\int_t^\infty
e^{-\wh\Lambda_T(s-W,\bb\trans\X)}ds I(W\le t)\n\\
&+&\wh\lambda_N(t,\bb\trans\X)e^{\wh\Lambda_N(t,\bb\trans\X)}
\int_t^\infty
e^{-\wh\Lambda_N(s,\bb\trans\X)}ds \{1-I(W\le t)\}-1,\\
&& \wh\m_2\{t,\bb\trans\X,I(W \le t),WI(W\le t)\}\n\\
&=&\left\{\wh\bLam_{T2}(t-W,\bb\trans\X)e^{\wh\Lambda_T(t-W,\bb\trans\X)}
\int_t^\infty
e^{-\wh\Lambda_T(s-W,\bb\trans\X)}ds\right.\n\\
&&\left.-e^{\wh\Lambda_T(t-W,\bb\trans\X)}
\int_t^\infty
\wh\bLam_{T2}(s-W,\bb\trans\X)e^{-\wh\Lambda_T(s-W,\bb\trans\X)}ds\right\}I(W\le t)\n\\
&+&\left\{\wh\bLam_{N2}(t,\bb\trans\X)e^{\wh\Lambda_N(t,\bb\trans\X)}
\int_t^\infty
e^{-\wh\Lambda_N(s,\bb\trans\X)}ds\right.\n\\
&&\left.-e^{\wh\Lambda_N(t,\bb\trans\X)}
\int_t^\infty
\wh\bLam_{N2}(s,\bb\trans\X)e^{-\wh\Lambda_N(s,\bb\trans\X)}ds\right\}\{1-I(W\le t)\},\\
&& \wh\m_{12}\{t,\bb\trans\X,I(W \le t),WI(W\le t)\}\n\\
&=&\left\{\wh\blam_{T2}(t-W,\bb\trans\X)e^{\wh\Lambda_T(t-W,\bb\trans\X)}
\int_t^\infty
e^{-\wh\Lambda_T(s-W,\bb\trans\X)}ds\right.\n\\
&&-\wh\lambda_T(t-W,\bb\trans\X)e^{\wh\Lambda_T(t-W,\bb\trans\X)}
\int_t^\infty\wh\bLam_{T2}(s-W,\bb\trans\X)
e^{-\wh\Lambda_T(s-W,\bb\trans\X)}ds\n\\
&&\left.+\wh\lambda_T(t-W,\bb\trans\X)\wh\bLam_{T2}(t-W,\bb\trans\X)e^{\wh\Lambda_T(t-W,\bb\trans\X)}
\int_t^\infty
e^{-\wh\Lambda_T(s-W,\bb\trans\X)}ds\right\}I(W\le t)\n\\
&+&\left\{\wh\blam_{N2}(t,\bb\trans\X)e^{\wh\Lambda_N(t,\bb\trans\X)}
\int_t^\infty
e^{-\wh\Lambda_N(s,\bb\trans\X)}ds\right.\n\\
&&-\wh\lambda_N\{t,\bb\trans\X,I(W \le t),WI(W\le t)\}e^{\wh\Lambda_N(t,\bb\trans\X)}
\int_t^\infty\wh\bLam_{N2}(s,\bb\trans\X)
e^{-\wh\Lambda_N(s,\bb\trans\X)}ds.\n\\
&&\left.+\wh\lambda_N(t,\bb\trans\X)\wh\bLam_{N2}(t,\bb\trans\X)e^{\wh\Lambda_N(t,\bb\trans\X)}
\int_t^\infty
e^{-\wh\Lambda_N(s,\bb\trans\X)}ds\right\}\{1-I(W\le t)\}.
\ese

\section{Asymptotic properties and semiparametric efficiency}\label{sec:asymp}

We list the  regularity conditions for the results of
root-$n$ consistency and  asymptotic normality of the estimators proposed
in Section \ref{sec:estimation}.
We  also establish semiparametric
efficiency of the
estimator obtained by solving (\ref{eq:eff}).

\begin{enumerate}[label=C\arabic*]
	
	\item\label{assum:kernel} ({\itshape kernel function})
	The kernel function $K_h(\cdot)=h^{-d}K(\cdot/h)$ where
	$K(\a)=\prod_{j=1}^{d}K(a_j)$ for
	$\a=(a_1,...,a_d)\trans$ is symmetric on each individual
	entry and $K(a_j)$ is differentiable,
	decreasing when $x\ge0$, and $\int K(x)dx=1$,
	$\int x^j K(x)dx=0$, for $1\le j<\nu$, $0<\int x^\nu
	K(x)dx<\infty$, and
	$\int K^2(x)dx$,
	$\int x^2 K^2(x)dx$,
	$\int K'^2(x)dx$,
	$\int x^2K'^2(x)dx$,
	$\int K''^2(x)dx$,
	$\int x^2K''^2(x)dx$ are all bounded. When there is no confusion,
	we use the same $K$ for both
	univariate and multivariate kernel functions for simplicity.

	\item\label{assum:bandwidth}({\itshape bandwidths})
	The bandwidths $h$ and $b$ satisfy $h\to0$,	$nh^{2\nu}\to0$, $b\to0$ and
	$nh^{d+2}b\to\infty$, where $2\nu> d+1$.

	\item\label{assum:fbeta}({\itshape density functions of
		covariates})
	For all $\bb\in{\cal B}$, the
	parameter space,
	the probability density function of $\bb\trans\X$,
	$f_{\bb\trans\X}(\bb\trans\x)$, has a compact support and is
	bounded away from zero and $\infty$. Furthermore,
	the first and second derivatives of $f_{\bb\trans\X}(\bb\trans\x)$ exist and are Lipschitz continuous.
	
	\item\label{assum:exi}({\itshape smoothness})
	For all $\bb\in{\cal B}$ and $t>0$,
	
	$E\{\X_jI(Z_j\ge
	t)\mid\bb\trans\X_j=\bb\trans\x\}$
	is bounded and its first derivative is a Lipschitz continuous function of $\bb\trans\x$.
	$E\{\X_j\X_j\trans I(Z_j\ge
	t)\mid\bb\trans\X_j=\bb\trans\x\}$	
	is a bounded and Lipschitz continuous function of $\bb\trans\x$.

	\item\label{assum:survivalfunction}({\itshape survival
		function})
	
	For all $\bb\in{\cal B}$ and $t\in(0,\tau)$,
	$E\{S_c(t,\X)\mid \bb\trans\x\}$ and $f(t,\bb\trans\x)$ are
	bounded away from zero. Their first derivatives with respect
	to $t$ and first and second derivatives with respect to
	$\bb\trans\x$ exist and are Lipschitz continuous. In
	addition, $S_c(\tau,\X)$ is bounded
	way from zero.
	
	\item\label{assum:bounded}({\itshape boundedness})
	The true parameter $\bb_0$ is an interior point in  $\cal B$
	and ${\cal B}$ is bounded.

	\item\label{assum:unique}({\itshape uniqueness})
	The equation
	\bse
	&&E\left( \Delta\left[\frac{\m_{12}\{Z,\bb\trans\X,I(W \le Z),WI(W\le Z)\}}{
		m_1\{Z,\bb\trans\X,I(W \le Z),WI(W\le Z)\}+1}\right.\right.\\
	&&\left.\quad-\frac{\m_2\{Z,\bb\trans\X,I(W \le Z),WI(W\le Z)\}}{m\{Z,\bb\trans\X,I(W \le Z),WI(W\le Z)\}}
	\right]\\
	&&\quad\left.\otimes\left[\X_{l}-
	\frac{ E\left\{\X_{l}
		Y(Z)\mid\bb\trans\X,I(W \le Z),WI(W\le Z)\right\}}
	{E\left\{Y(Z)\mid\bb\trans\X,I(W \le Z),WI(W\le Z)\right\}}\right]\right)
	=\0
	\ese
	has a unique solution in $\cal B$.

\end{enumerate}

Conditions \ref{assum:kernel} and
\ref{assum:bandwidth} are commonly assumed in kernel
regression analysis \citep{silverman1986density,ma2013efficient}. Conditions
\ref{assum:fbeta}-\ref{assum:survivalfunction} assume boundedness
of event time, censoring time, covariates and their
expectations, which hold for real datasets. The smoothness of several functions is imposed by constraining
their derivatives, which are  common conditions \citep{silverman1978weak}. It is natural to make a boundedness assumption on
the parameter  space
$\cal B$ as in Condition \ref{assum:bounded} in practical
problems \citep{hardle1997semiparametric}.
Condition \ref{assum:unique} precludes
that the estimating equation
is degenerate.

Theorems \ref{th:consistency} and \ref{th:eff} demonstrate
the root-$n$ consistency and asymptotical normality of the
profile parameter estimator $\wh\bb$.  The proofs  are given in Supplement \ref{app:consistofb} and
\ref{app:asympofb}.

\begin{Th}\label{th:consistency}
	Under Conditions
	\ref{assum:kernel}-\ref{assum:unique},
	the estimator, $\wh\bb$, obtained by solving (\ref{eq:general}) or
	(\ref{eq:eff})
	is
	consistent,
	i.e.
	$\wh\bb-\bb\to\0$ in probability when $n\to\infty$.
\end{Th}

\begin{Th}\label{th:eff}
	Under Conditions
	\ref{assum:kernel}-\ref{assum:unique},
	the estimator, $\wh\bb$, obtained by solving (\ref{eq:general}) or
	(\ref{eq:eff})
	satisfies
	$
	\sqrt{n}(\wh\bb-\bb)\to N(\0, \A^{-1}\B{\A^{-1}}\trans)
	$
	in distribution when $n\to\infty$, where
	\bse
	\A&=&E\left\{\frac{\partial}{\partial\vecl(\bb)\trans}
	\vecl\left(\Delta\g\{Z,\bb\trans\X,I(W\le Z), WI(W\le Z)\}\right.\right.\\
	&&\left.\left.\otimes\left[\a(\X_{l})-
	\frac{E\left\{\a(\X_{l})
		Y(Z)\mid\bb\trans\X,I(W \le Z),WI(W\le Z)\right\}}
	{E\left\{Y(Z)\mid\bb\trans\X,I(W \le Z),WI(W\le Z)\right\}}\right]\right)\right\},\\
	\B&=&E\left\{\vecl\left(\Delta\g\{Z,\bb\trans\X,I(W\le Z), WI(W\le Z)\}\right.\right.\\
	&&\left.\left.\otimes\left[\a(\X_{l})-
	\frac{E\left\{\a(\X_{l})
		Y(Z)\mid\bb\trans\X,I(W \le Z),WI(W\le Z)\right\}}
	{E\left\{Y(Z)\mid\bb\trans\X,I(W \le Z),WI(W\le Z)\right\}}\right]\right)^{\otimes2}\right\}.
	\ese
	Here $\vecl(\A)$ represents the vectorization of the lower
	$(p-d)\times d$ block of a generic matrix $\A$ and
	$\A^{\otimes2}=\A\A\trans$ for any matrix or vector $\A$.
	Note that in (\ref{eq:general}), $\a(\X_l)=\X_l$ and in
	(\ref{eq:eff}), $\a(\X_l)=\X_l$,
	$g\{Z,\bb\trans\X,I(W\le Z), WI(W\le Z)\}=\wh\m_{12}\{Z,\bb\trans\X,I(W\le Z), WI(W\le Z)\}/\{\wh
	m_1\{Z,\bb\trans\X,I(W\le Z), WI(W\le Z)\}+1\}
	-\wh\m_{2}\{Z,\bb\trans\X,I(W\le Z), WI(W\le Z)\}/\wh
	m\{Z,\bb\trans\X,I(W\le Z), WI(W\le Z)\}$.
	Further, the estimator, $\wh\bb$, obtained from solving
	(\ref{eq:eff}) is semiparametrically efficient and  satisfies
	\bse
	\sqrt{n}(\wh\bb-\bb)\to N\{\0, (E[\bS\eff ^{\otimes2}
	\{\Delta,
	Z,\X,I(W\le Z), WI(W\le Z)\})^{-1}\}
	\ese
	in distribution, where $\bS\eff\{Z,\bb\trans\X,I(W\le Z), WI(W\le Z)\}$ is given
	in (\ref{eq:effscore}).
\end{Th}

With $\bS\eff$ being a martingale,
\bse
&&E[\bS\eff ^{\otimes2} \{Z,\bb\trans\X,I(W\le Z), WI(W\le Z)\}]\\
&=&E\left\{\int_0^\infty
\left(\left[\frac{\m_{12}\{s,\bb\trans\X,I(W \le s),WI(W\le s)\}}{
	m_1\{s,\bb\trans\X,I(W \le s),WI(W\le s)\}+1}\right.\right.\right.\\
&&\left.-\frac{\m_2\{s,\bb\trans\X,I(W \le s),WI(W\le s)\} }{m\{s,\bb\trans\X,I(W \le s),WI(W\le s)\}}\right]\\
&&\left.\left.\otimes\left[\X_l-
\frac{E\left\{\X_l
	S_c(s,\X)\mid\bb\trans\X\right\}}
{E\left\{S_c(s,\X)\mid\bb\trans\X\right\}}\right]\right)^{\otimes2}
dN(s)\right\},
\ese
which leads to a consistent estimator of $E[\bS\eff ^{\otimes2} \{s,\bb\trans\X,I(W \le s),WI(W\le s)\}]$ as follows
\bse
&&\frac{1}{n}\sumi\delta_i
\left(\frac{\wh\blam_{2}\{z_i,\wh\bb\trans\x_i,I(w_i\le z_i),w_i I(w_i\le z_i)\}}{\wh\lambda\{z_i,\wh\bb\trans\x_i,I(w_i\le z_i),w_i I(w_i\le z_i)\}}\right.\\
&&\left.\otimes\left[\x_{il}-
\frac{\wh E\left\{\X_{l}
	Y(z_i)\mid\wh\bb\trans\x_i,I(w_i \le z_i),w_i I(w_i\le z_i)\right\}}
{\wh
	E\left\{Y(z_i)\mid\wh\bb\trans\x_i,I(w_i \le z_i),w_i I(w_i\le z_i)\right\}}\right]\right)^{\otimes2}.
\ese
Here $\wh
E\left\{Y(z_i)\mid\wh\bb\trans\x_i,I(w_i \le z_i),w_iI(w_i\le z_i)\right\}$,
$\wh E\left\{\X_{l}
Y(z_i)\mid\wh\bb\trans\x_i,I(w_i \le z_i)\allowbreak,w_iI(w_i\le z_i)\right\}$,
$\wh\lambda\{z_i,\wh\bb\trans\x_i,I(w_i\le z_i),w_iI(w_i\le z_i)\}$
and $\wh\blam_{2}\{z_i,\wh\bb\trans\x_i,I(w_i\le z_i),w_iI(w_i\le z_i)\}$
are
given in (\ref{eq:expectY}), (\ref{eq:expectXY}),
(\ref{eq:lambda}) and (\ref{eq:lambda1}) respectively.

\begin{Th}\label{th:m}
	Under Conditions
	\ref{assum:kernel}-\ref{assum:unique}, the
	nonparametric estimators $\wh m_N(t,\wh\bb\trans\x)$ and $\wh m_T(t-w,\wh\bb\trans\x,w)$ 
	satisfy
	\bse
	&&\sqrt{nh}\left\{\wh{m}_N(t,\wh\bb\trans\x)-{m}_N(t,\bb\trans\x)\right\}\to
	N\{0,\sigma^2_N(t,\bb\trans\x)\}\\
	&&\sqrt{nh}\left\{\wh{m}_T(t-w,\wh\bb\trans\x,w)-{m}_T(t-w,\bb\trans\x,w)\right\}\to
	N\{0,\sigma^2_T(t-w,\bb\trans\x,w)\}
	\ese
	in distribution for all $t$, $w<t$ and
	$\x$, where
	\bse
	&&\sigma^2_N(t,\bb\trans\x)=e^{2\Lambda_N(t,\bb\trans\x)}\frac{\int
		K^2(u)du}{f_{\bb\trans\X}(\bb\trans\x)}
	\int_{0}^{\tau}\frac{\lambda_N(r,\bb\trans\x)}
	{E\{I(Z\ge r)\mid\bb\trans\x\}}\\
	&&\quad\times\left\{I(r<t)\int_{t}^{\tau}
	e^{-\Lambda_N(s,\bb\trans\x)}ds+\int_{\max(r,t)}^{\tau}
	e^{-\Lambda_N(s,\bb\trans\x)}ds\right\}^2dr,\\
	&&\sigma^2_T(t-w,\bb\trans\x,w)=e^{2\Lambda_T(t-w,\bb\trans\x,w)}\frac{\int
		K^2(u)du}{f_{\bb\trans\X,W}(\bb\trans\x,w)}
	\int_{0}^{\tau}\frac{\lambda_T(r,\bb\trans\x,w)}
	{E\{I(Z\ge r)\mid\bb\trans\x,w\}}\\
	&&\quad\times\left\{I(r<t-w)\int_{t-w}^{\tau}
	e^{-\Lambda_T(s-w,\bb\trans\x,w)}ds+\int_{\max(r,t-w)}^{\tau}
	e^{-\Lambda_T(s-w,\bb\trans\x,w)}ds\right\}^2dr.
	\ese
\end{Th}
The proof  is provided in
Supplement \ref{app:m}. Theorem
\ref{th:m} implies that
we can even
estimate the variance $\sigma_T^2(t-w,\bb\trans\x,w)$, $\sigma^2_N(t,\bb\trans\x)$, without estimating
$\lambda$ or $f_{\bb\trans\X}(\bb\trans\x)$, by using
\bse
&&\wh\sigma_T^2(t-w,\wh\bb\trans\x,w)\\
&=&I(w\le t)e^{2\wh\Lambda_T(t-w,\wh\bb\trans\x,w)}\int K^2(u)du
\sum_{i=1}^{n}I(w_i\le t)\\
&&\quad\left(\frac{\wh\Lambda_T(t_{(i)},	
	\wh\bb\trans\x,w)-\wh\Lambda_T(t_{(i-1)},\wh\bb\trans\x,w)}	
{1/n\sumj
	Y_j(t_{(i-1)})K_h(\wh\bb\trans\x_j-\wh\bb\trans\x,w_j-w)}\right.\\
&&\left[I(t_{(i-1)}<t-w)\sum_{j=1}^{n}I(t_{(j)}>
t-w)e^{-\wh\Lambda_T(t_{(j-1)}-w,\wh\bb\trans\x,w)}\{t_{(j)}-\max(t-w,t_{(j-1)})\}
\right.\\
&&\left.\left.+\sum_{j=1}^{n}I\{t_{(j)}>\max(t-w,t_{(i-1)})\}
e^{-\wh\Lambda_T(t_{(j-1)}-w,\wh\bb\trans\x,w)}\{t_{(j)}-\max(t_{(i-1)},t_{(j-1)})\}\right]^2
\right),\\
&&\wh\sigma_N^2(t,\wh\bb\trans\x)\\
&=&\{1-I(w\le t)\}e^{2\wh\Lambda_N(t,\wh\bb\trans\x)}\int K^2(u)du
\sum_{i=1}^{n}\{1-I(w_i\le t)\}\\
&&\times\left(\frac{\wh\Lambda_N(t_{(i)},	
	\wh\bb\trans\x)-\wh\Lambda_N(t_{(i-1)},\wh\bb\trans\x)}	
{1/n\sumj
	Y_j(t_{(i-1)})K_h(\wh\bb\trans\x_j-\wh\bb\trans\x)}\right.\\
&&\left[I(t_{(i-1)}<t)\sum_{j=1}^{n}I(t_{(j)}>
t)e^{-\wh\Lambda_N(t_{(j-1)},\wh\bb\trans\x)}\{t_{(j)}-\max(t,t_{(j-1)})\}
\right.\\
&&\left.\left.+\sum_{j=1}^{n}I\{t_{(j)}>\max(t,t_{(i-1)})\}
e^{-\wh\Lambda_N(t_{(j-1)},\wh\bb\trans\x)}\{t_{(j)}-\max(t_{(i-1)},t_{(j-1)})\}\right]^2
\right),
\ese
where $t_{(1)}< t_{(2)}<\dots < t_{(n_T)}$ are the observed
event times and $t_{(0)} =0$.

\section{Proofs of  Theorems 1--3}

\subsection{Two useful lemmas}\label{sec:prepare}

\subsubsection{Lemma \ref{lem:pre}}\label{sec:prooflem}
\begin{Lem}\label{lem:pre} 
	Under the regularity conditions
	\ref{assum:kernel}-\ref{assum:survivalfunction}  listed
	above,
	\be
	\wh E\left\{Y(Z)\mid\bb\trans\X\right\}
	&=&E\{Y(Z)\mid\bb\trans\X\}+O_p\{(nh)^{-1/2}+h^2\},\label{eq:lemeq1}\\
	\wh E\left\{\X Y(Z)\mid\bb\trans\X\right\}
	&=&E\{\X Y(Z)\mid\bb\trans\X\}+O_p\{(nh)^{-1/2}+h^2\},
	\label{eq:lemeq2}\\
	\wh{\lambda}(z,\bb\trans\X)&=&\lambda(z,\bb\trans\X)+O_p\{(nhb)^{-1/2}+h^2+b^2\}\label{eq:lemeq5}\\
	\wh\blam_2(z,\bb\trans\X)&=&\blam_2(z,\bb\trans\X)+O_p\{(nbh^3)^{-1/2}+h^2+b^2\}\label{eq:lemeq6}\\
	\wh{\Lambda}(z,\bb\trans\X)&=&\Lambda(z,\bb\trans\X)+O_p\{(nh)^{-1/2}+h^2\}\label{eq:lemeq7}\\
	\wh\bLam_2(z,\bb\trans\X)&=&\bLam_2(z,\bb\trans\X)+O_p\{(nh^3)^{-1/2}+h^2\}\label{eq:lemeq8}
	\ee
	uniformly for all $z,\bb\trans\X$.
\end{Lem}

\noindent Proof: For notation convenience, we prove the results for $d=1$. We prove
\bse
\wh E\left\{\X Y(Z)\mid\bb\trans\X\right\}
&=&E\{\X Y(Z)\mid\bb\trans\X\}+O_p\{(nh)^{-1/2}+h^2\}
\ese
and
\bse
\wh\bLam_2(z,\bb\trans\X)&=&\bLam_2(z,\bb\trans\X)+O_p\{(nh^3)^{-1/2}+h^2\}.
\ese
and skip the remaining results because of the similar arguments.

First, for any $\X$ and $\bb$ in a local neighborhood of $\bb_0$,
\be
\frac{1}{n}\sumj
K_h(\bb\trans\X_j-\bb\trans\X)=f_{\bb\trans\X}(\bb\trans\X)+
O_p(n^{-1/2}h^{-1/2}+h^2),\label{eq:fbeta}
\ee
To see this, the absolute bias of the left hand
side of
(\ref{eq:fbeta}) is
\bse
&&\left|E\left\{\frac{1}{n}\sumj
K_h(\bb\trans\X_j-\bb\trans\X)\right\}-f_{\bb\trans\X}(\bb\trans\X)\right|\\
&=&\left|\int\frac{1}{h}K\left(\frac{\bb\trans\x_j-\bb\trans\X}{h}\right)
f_{\bb\trans\X}(\bb\trans\x_j)d\bb\trans\x_j
-f_{\bb\trans\X}(\bb\trans\X)\right|\\
&=&\left|\int K(u)f_{\bb\trans\X}(\bb\trans\X+hu)du
-f_{\bb\trans\X}(\bb\trans\X)\right|\\
&=&\left|\int
K(u)\left\{f_{\bb\trans\X}(\bb\trans\X)+f_{\bb\trans\X}'
(\bb\trans\X)hu+O(h^2)u^2\right\}du
-f_{\bb\trans\X}(\bb\trans\X)\right|\\
&=& O(h^2)
\ese
under Conditions \ref{assum:kernel}-\ref{assum:fbeta}.
The variance is
\bse
&&\var\left\{\frac{1}{n}\sumj
K_h(\bb\trans\X_j-\bb\trans\X)\right\}\\
&=&\frac{1}{n}\var K_h(\bb\trans\X_j-\bb\trans\X)\\
&=&\frac{1}{n}\left[EK_h^2(\bb\trans\X_j-\bb\trans\X)-
\left\{EK_h(\bb\trans\X_j-\bb\trans\X)\right\}^2\right]\\
&=&\frac{1}{n}\left[\int\frac{1}{h^2}
K^2\{(\bb\trans\x_j-\bb\trans\X)/h\}f_{\bb\trans\X}
(\bb\trans\x_j)d\bb\trans\x_j-f_{\bb\trans\X}^2(\bb\trans\X)+O(h^2)\right]\\
&=&\frac{1}{nh}\int
K^2(u)f_{\bb\trans\X}(\bb\trans\X+hu)du-\frac{1}{n}f_{\bb\trans\X}^2
(\bb\trans\X)+O(h^2/n)
\\
&\le&\frac{1}{nh}f_{\bb\trans\X}(\bb\trans\X)\int
K^2(u)du
+O(h/n)\int u^2K^2(u)du+\frac{1}{n}|f_{\bb\trans\X}^2(\bb\trans\X)|+O(h^2/n).
\ese
Therefore, applying the central limit theorem, we have that
\bse
\frac{1}{n}\sumj
K_h(\bb\trans\X_j-\bb\trans\X)=f_{\bb\trans\X}(\bb\trans\X)+O_p(n^{-1/2}h^{-1/2}+h^2)
\ese
for all
$\bb$ under conditions \ref{assum:kernel}-\ref{assum:fbeta}. Condition \ref{assum:fbeta}
also holds for any $\bb$ in a local neighborhood of $\bb_0$ due to the continuity.
Similarly, We have
\be
\label{eq:fbeta1}
-\frac{1}{n}\sumj
K_h'(\bb\trans\X_j-\bb\trans\X)=f_{\bb\trans\X}'(\bb\trans\X)+
O_p(n^{-1/2}h^{-3/2}+h^2).
\ee

To show (\ref{eq:lemeq2}), the absolute bias is
\bse
&&\left|
E\left\{\frac{1}{n}\sumj \X_j I(Z_j\ge z)
K_h(\bb\trans\X_j-\bb\trans\X)\right\}
-f_{\bb\trans\X}(\bb\trans\X)E\{\X_j(Z_j\ge
z)\mid\bb\trans\X\}\right|\\
&=&\left|\int
f_{\bb\trans\X}(\bb\trans\X) E
\left\{\X_jI(Z_j\ge z)\mid\bb\trans\X\right\}
K(u)du-f_{\bb\trans\X}(\bb\trans\X)E\{\X_j(Z_j\ge
z)\mid\bb\trans\X\}\right.\\
&&\left.+h\int
\frac{\partial}{\partial(\bb\trans\X)}
f_{\bb\trans\X}(\bb\trans\X) E
\left\{\X_jI(Z_j\ge z)\mid\bb\trans\X\right\}
uK(u)du + O(h^2)\right|\\
&=& O(h^2)
\ese
under Conditions \ref{assum:kernel}-\ref{assum:exi}. The variance is
\bse
&&\var\left\{-\frac{1}{n}\sumj \X_j I(Z_j\ge z)
K_h(\bb\trans\X_j-\bb\trans\X)\right\}\\
&=&\frac{1}{n^2}\sumj\left( E \left\{ \X_j\X_j\trans I(Z_j\ge z)
K^2_h(\bb\trans\X_j-\bb\trans\X)\right\}\right.\\
&&\left.-\left[E\left\{ \X_j I(Z_j\ge z)
K_h(\bb\trans\X_j-\bb\trans\X)\right\}\right]^2\right)\\
&\le&\frac{1}{nh}
\sup_{\bb\trans\X}\left|f_{\bb\trans\X}(\bb\trans\X)E\{\X_j\X_j\trans
I(Z_j\ge z)\mid\bb\trans\X\}\right|\int K^2(u)du+O(1/n)
\ese
under conditions \ref{assum:kernel}-\ref{assum:exi}.
So
\be\label{eq:fbetaxi}
\frac{1}{n}\sumj \X_j I(Z_j\ge z)  K_h(\bb\trans\X_j-\bb\trans\X)
&=&f_{\bb\trans\X}(\bb\trans\X)E\{\X_j I(Z_j\ge
z)\mid\bb\trans\X\}\n\\
&&+O_p(n^{-1/2}h^{-1/2}+h^2)
\ee
under conditions \ref{assum:kernel}-\ref{assum:exi}.

To show (\ref{eq:lemeq8}), let
\bse
\wh\bLam_{21}(z,\bb\trans\X)&=&
-\sum_{i=1}^n \frac{I(Z_i\le
	z)\Delta_iK_h'(\bb\trans\X_i-\bb\trans\X)}{\sumj
	I(Z_j\ge
	Z_i)K_h(\bb\trans\X_j-\bb\trans\X)}\\
\wh\bLam_{22}(z,\bb\trans\X)&=&\sum_{i=1}^nI(Z_i\le
z)\Delta_iK_h(\bb\trans\X_i-\bb\trans\X) \frac{\sumj I(Z_j\ge
	Z_i)K_h'(\bb\trans\X_j-\bb\trans\X)}{\{\sumj I(Z_j\ge
	Z_i)K_h(\bb\trans\X_j-\bb\trans\X)\}^2}.
\ese
Then
$\wh\bLam_2(z,\bb\trans\X)=\wh\bLam_{21}(z,\bb\trans\X)+\wh\bLam_{22}(z,\bb\trans\X)$.
To analyze $\wh\bLam_{21}$,
\bse
&&E\wh\bLam_{21}(z,\bb\trans\X)\\
&=&E\left[ \frac{-I(Z_i\le
	z)\Delta_iK_h'(\bb\trans\X_i-\bb\trans\X)}{f_{\bb\trans\X}(\bb\trans\X)S(Z_i,\bb\trans\X)E\{S_c(Z_i,\X_j)\mid\bb\trans\X_j=\bb\trans\X,Z_i\}}\right]\\
&&+E\left[ \frac{1}{n}\sumi\frac{-I(Z_i\le
	z)\Delta_iK_h'(\bb\trans\X_i-\bb\trans\X)}{f_{\bb\trans\X}(\bb\trans\X)S(Z_i,\bb\trans\X)E\{S_c(Z_i,\X_j)\mid\bb\trans\X_j=\bb\trans\X,Z_i\}}O_p(A)\right].
\ese
The first term is
\be
&&E\left[ \frac{-I(Z_i\le
	z)\Delta_iK_h'(\bb\trans\X_i-\bb\trans\X)}{f_{\bb\trans\X}(\bb\trans\X)S(Z_i,\bb\trans\X)E\{S_c(Z_i,\X_j)\mid\bb\trans\X_j=\bb\trans\X,Z_i\}}\right]\n\\
&=&\int I(z_i\le
z)\frac{\partial\left[f(z_i,\bb\trans\X)E\{S_c(z_i,\X_i)\mid\bb\trans\X,z_i\}f_{\bb\trans\X}(\bb\trans\X)\right]/\partial\bb\trans\X}{f_{\bb\trans\X}(\bb\trans\X)
	S(z_i,\bb\trans\X)E\{S_c(
	z_i,\X_j)\mid\bb\trans\X,z_i\}}dz_i\n\\
&&+\iint I(z_i\le
z)O(h^2)u^3K'(u)dudz_i\label{eq:expectationOf1}
\ee
under Condition \ref{assum:kernel} and \ref{assum:fbeta}-\ref{assum:survivalfunction}.
Hence
\bse
&&\left|E\left[ \frac{-I(Z_i\le
	z)\Delta_iK_h'(\bb\trans\X_i-\bb\trans\X)}
{f_{\bb\trans\X}(\bb\trans\X)S(Z_i,\bb\trans\X)E\{S_c(Z_i,\X_j)
	\mid\bb\trans\X_j=\bb\trans\X,Z_i\}}\right]\right.\\
&&\left.-\int I(z_i\le z)
\frac{\partial\left[f(z_i,\bb\trans\X)E\{S_c(z_i,\X_i)
	\mid\bb\trans\X,z_i\}f_{\bb\trans\X}(\bb\trans\X)\right]
	/\partial\bb\trans\X}{f_{\bb\trans\X}(\bb\trans\X)S(z_i,\bb\trans\X)
	E\{S_c(z_i,\X_j)\mid\bb\trans\X,z_i\}}dz_i\right|\\
&=&O(h^2)
\ese

under Condition
\ref{assum:kernel} and \ref{assum:fbeta}-\ref{assum:survivalfunction}. Similarly, we conclude that
\bse
\frac{1}{n}\sumi\frac{-I(Z_i\le
	z)\Delta_iK_h'(\bb\trans\X_i-\bb\trans\X)}{f_{\bb\trans\X}(\bb\trans\X)S(Z_i,\bb\trans\X)E\{S_c(Z_i,\X_j)\mid\bb\trans\X_j=\bb\trans\X,Z_i\}}O_p(A)=O_p\{h^2+(nh)^{-1/2}\}
\ese
under conditions
\ref{assum:kernel}-\ref{assum:survivalfunction} due
to $A=O_p\{h^2+(nh)^{-1/2}\}$. Therefore
\bse
E\wh\bLam_{21}(z,\bb\trans\X)&=&\int I(z_i\le
z)\frac{\partial\left[f(z_i,\bb\trans\X)E\{S_c(z_i,\X_i)\mid\bb\trans\X,z_i\}f_{\bb\trans\X}(\bb\trans\X)\right]/\partial\bb\trans\X}{f_{\bb\trans\X}(\bb\trans\X)
	S(z_i,\bb\trans\X)E\{S_c(
	z_i,\X_j)\mid\bb\trans\X,z_i\}}dz_i\\
&&+O\{(nh)^{-1/2}+h^2\}.
\ese
For $\wh\bLam_{22}$, let
$B=	-1/n\sumj I(Z_j\ge Z_i)  K_h'(\bb\trans\X_j-\bb\trans\X)
-\partial
f_{\bb\trans\X}(\bb\trans\X)E\{I(Z_j\ge
Z_i)\mid\bb\trans\X\}/\partial
\bb\trans\X$,
then
\bse
&&\wh\bLam_{22}(z,\bb\trans\X)\\
&=&-\frac{1}{n}\sumi I(Z_i\le
z)\Delta_iK_h(\bb\trans\X_i-\bb\trans\X)\\
&&\times\frac{\partial\left[
	f_{\bb\trans\X}(\bb\trans\X)S(z_i,\bb\trans\X)E\{S_c(z_i,\X_j)\mid\bb\trans\X,z_i\}\right]/
	\partial
	\bb\trans\X}{f_{\bb\trans\X}^2(\bb\trans\X)S^2(z_i,\bb\trans\X)E^2\{S_c(z_i,\X_j)\mid\bb\trans\X,z_i\}}\\
&&\times\left\{1+O_p(B)+O_p(A)\right\}.
\ese
We have
\be
&&E\left[-\frac{1}{n}\sumi I(Z_i\le
z)\Delta_iK_h(\bb\trans\X_i-\bb\trans\X)\right.\n\\
&&\quad\left.\frac{\partial\left[
	f_{\bb\trans\X}(\bb\trans\X)E\{I(Z_j\ge
	Z_i)\mid\bb\trans\X,Z_i\}\right]/
	\partial
	\bb\trans\X}{f_{\bb\trans\X}^2(\bb\trans\X)E^2\{I(Z_j\ge
	Z_i)\mid\bb\trans\X,Z_i\}}\right]\n\\
&=&-\iint I(z_i\le z)K_h(\bb\trans\X_i-\bb\trans\X)\n\\
&&\times\frac{\partial
	\left[f_{\bb\trans\X}(\bb\trans\X)S(z_i,\bb\trans\X)E\{S_c(z_i,\X_j)\mid\bb\trans\X,z_i\}\right]/
	\partial
	\bb\trans\X}{f_{\bb\trans\X}^2(\bb\trans\X)S^2(z_i,\bb\trans\X)E^2\{S_c(
	z_i,\X_j)\mid\bb\trans\X,z_i\}}\n\\
&&\times
f(z_i,\bb\trans\X_i)E\{S_c(z_i,\X_i)\mid\bb\trans\X_i,z_i\}f_{\bb\trans\X}(\bb\trans\X_i)dz_id\bb\trans\X_i\n\\
&=&-\int I(z_i\le z)\frac{\partial
	\left[f_{\bb\trans\X}(\bb\trans\X)S(z_i,\bb\trans\X)E\{S_c(
	z_i,\X_j)\mid\bb\trans\X,z_i\}\right]/\partial\bb\trans\X}
{f_{\bb\trans\X}(\bb\trans\X)
	S^2(z_i,\bb\trans\X)E\{S_c(z_i,\X_j)\mid\bb\trans\X,z_i\}}
f(z_i,\bb\trans\X)dz_i\n\\
&&-\iint  I(z_i\le z)\frac{\partial
	\left[f_{\bb\trans\X}(\bb\trans\X)S(z_i,\bb\trans\X)E\{S_c(
	z_i,\X_j)\mid\bb\trans\X,z_i\}\right]/ \partial
	\bb\trans\X}{f_{\bb\trans\X}^2(\bb\trans\X)S^2(z_i,\bb\trans\X)E^2\{S_c(
	z_i,\X_j)\mid\bb\trans\X,z_i\}}\n\\
&&\times O(h^2)u^2K(u)dz_idu,\label{eq:expectationOf2}\n\\
\ee
therefore
\bse
&&\left|E\left[-\frac{1}{n}\sumi I(Z_i\le
z)\Delta_iK_h(\bb\trans\X_i-\bb\trans\X) \right.\right.\\
&&\left.\left.\frac{\partial\left[
	f_{\bb\trans\X}(\bb\trans\z)E\{I(Z_j\ge
	Z_i)\mid\bb\trans\X,Z_i\}\right]/
	\partial
	\bb\trans\X}{f_{\bb\trans\X}^2(\bb\trans\X)E^2\{I(Z_j\ge
	Z_i)\mid\bb\trans\X,Z_i\}}\right]\right.\\
&&\left.+\int I(z_i\le z)\frac{\partial
	\left[f_{\bb\trans\X}(\bb\trans\X)S(z_i,\bb\trans\X)E\{S_c(
	z_i,\X_j)\mid\bb\trans\X,z_i\}\right]/ \partial
	\bb\trans\X}{f_{\bb\trans\X}(\bb\trans\X)S^2(z_i,\bb\trans\X)E\{S_c(
	z_i,\X_j)\mid\bb\trans\X,z_i\}}f(z,\bb\trans\X)dz_i\right|\\
&=&O(h^2)
\ese
under conditions
\ref{assum:kernel}-\ref{assum:survivalfunction}.
Recall $B=O_p(n^{-1/2}h^{-3/2}+h^2)$, then
\bse
&&-\frac{1}{n}\sumi I(Z_i\le
z)\Delta_iK_h(\bb\trans\X_i-\bb\trans\X)\\
&&\quad\times\frac{\partial\left[
	f_{\bb\trans\X}(\bb\trans\X)E\{I(Z_j\ge
	Z_i)\mid\bb\trans\X,Z_i\}\right]/ \partial
	\bb\trans\X}{f_{\bb\trans\X}^2(\bb\trans\X)E^2\{I(Z_j\ge
	Z_i)\mid\bb\trans\X,Z_i\}}O_p(B)\\
&=&O_p(n^{-1/2}h^{-3/2}+h^2),\\
&&-\frac{1}{n}\sumi I(Z_i\le
z)\Delta_iK_h(\bb\trans\X_i-\bb\trans\X)\\
&&\quad\times\frac{\partial\left[
	f_{\bb\trans\X}(\bb\trans\X)E\{I(Z_j\ge
	Z_i)\mid\bb\trans\X,Z_i\}\right]/ \partial
	\bb\trans\X}{f_{\bb\trans\X}^2(\bb\trans\X)E^2\{I(Z_j\ge
	Z_i)\mid\bb\trans\X,Z_i\}}O_p(A)\\
&=&O_p(n^{-1/2}h^{-1/2}+h^2)
\ese
under condition
\ref{assum:kernel}-\ref{assum:survivalfunction}.
Therefore
\bse
E\wh\bLam_{22}&=&-\int \frac{\partial
	\left[f_{\bb\trans\X}(\bb\trans\X)S(z_i,\bb\trans\X)E\{S_c(
	z_i,\X_j)\mid\bb\trans\X,z_i\}\right]/ \partial
	\bb\trans\X}{f_{\bb\trans\X}(\bb\trans\X)S^2(z_i,\bb\trans\X)E\{S_c(
	z_i,\X_j)\mid\bb\trans\X,z_i\}}\\
&&\quad\times I(z_i\le z)f(z_i,\bb\trans\X)dz_i+O(n^{-1/2}h^{-3/2}+h^2).
\ese
In addition
\bse
&&\int \left(I(z_i\le
z)\frac{\partial\left[f(z_i,\bb\trans\X)E\{S_c(z_i,\X_j)\mid\bb\trans\X,z_i\}f_{\bb\trans\X}(\bb\trans\X)\right]/\partial\bb\trans\X}{f_{\bb\trans\X}(\bb\trans\X)
	S(z_i,\bb\trans\X)E\{S_c(
	z_i,\X_j)\mid\bb\trans\X,z_i\}}\right.\\
&&-\left.I(z_i\le z)\frac{\partial
	\left[f_{\bb\trans\X}(\bb\trans\X)S(z_i,\bb\trans\X)E\{S_c(
	z_i,\X_j)\mid\bb\trans\X,z_i\}\right]/ \partial
	\bb\trans\X}{f_{\bb\trans\X}(\bb\trans\X)S^2(z_i,\bb\trans\X)E\{S_c(
	z_i,\X_j)\mid\bb\trans\X,z_i\}}f(z_i,\bb\trans\X)\right)dz_i\\
&=&\bLam_2(z,\bb\trans\X).
\ese
Combining $E\wh\bLam_{21}(z,\bb\trans\X)$ and
$E\wh\bLam_{22}(z,\bb\trans\X)$ gives
\bse
\left|E\wh\bLam_2(z,\bb\trans\X)-\bLam_2(z,\bb\trans\X)\right|
=O(n^{-1/2}h^{-3/2}+h^2)
\ese
under conditions
\ref{assum:kernel}-\ref{assum:survivalfunction}.

The variance of $\wh\bLam_2(z,\bb\trans\X)$ is
\bse
\var\left\{\wh\bLam_2(z,\bb\trans\X)\right\}
&=&\var\left\{\wh\bLam_{21}(z,\bb\trans\X)+\wh\bLam_{22}(z,\bb\trans\X)\right\}\\
&\le&2\var\{\wh\bLam_{21}(z,\bb\trans\X)\}+2\var\{\wh\bLam_{22}(z,\bb\trans\X)\}.
\ese
The first term
\bse
&&\var\{\wh\bLam_{21}(z,\bb\trans\X)\}\\
&\le&\frac{2}{n}\var\left[\frac{-I(Z_i\le
	z)\Delta_iK_h'(\bb\trans\X_i-\bb\trans\X)}{f_{\bb\trans\X}(\bb\trans\X)S(Z_i,\bb\trans\X)E\{S_c(Z_i,\X_j)\mid\bb\trans\X_i=\bb\trans\X,Z_i\}}\right]\\
&&+2\var\left[\frac{1}{n}\sumi\frac{-I(Z_i\le
	z)\Delta_iK_h'(\bb\trans\X_i-\bb\trans\X)}{f_{\bb\trans\X}(\bb\trans\X)S(Z_i,\bb\trans\X)E\{S_c(Z_i,\X_j)\mid\bb\trans\X_i=\bb\trans\X,Z_i\}}O_p(A)\right]\\
\ese
The first part is

\be
&&\frac{2}{n}\var\left[\frac{-I(Z_i\le
	z)\Delta_iK_h'(\bb\trans\X_i-\bb\trans\X)}{f_{\bb\trans\X}(\bb\trans\X)S(Z_i,\bb\trans\X)E\{S_c(Z_i,\X_j)\mid\bb\trans\X_i=\bb\trans\X,Z_i\}}\right]\n\\
&=&\frac{2}{n}E\left[\frac{-I(Z_i\le
	z)\Delta_iK_h'(\bb\trans\X_i-\bb\trans\X)}{f_{\bb\trans\X}(\bb\trans\X)S(Z_i,\bb\trans\X)E\{S_c(Z_i,\X_j)\mid\bb\trans\X_i=\bb\trans\X,Z_i\}}\right]^2\n\\
&&-\frac{2}{n}\left(E\left[\frac{-I(Z_i\le
	z)\Delta_iK_h'(\bb\trans\X_i-\bb\trans\X)}{f_{\bb\trans\X}(\bb\trans\X)S(Z_i,\bb\trans\X)E\{S_c(Z_i,\X_j)\mid\bb\trans\X_i=\bb\trans\X,Z_i\}}\right]\right)^2\n\\
&\le&\frac{2}{nh^3}\int \frac{I(z_i\le z)f(z_i,\bb\trans\X)}
{f_{\bb\trans\X}(\bb\trans\X)S^2(z_i,\bb\trans\X)E\{S_c(z_i,\X_j)\mid\bb\trans\X,z_i\}}dz_i\left\{\int
K'^2(u)du\right\}\n\\
&&+\frac{1}{nh^3}\int O(h^2) u^2K'^2(u)du+O(1/n)\n\\
&=&O\{1/(nh^3)\}\label{eq:varianceOf1}
\ee
under condition
\ref{assum:kernel}-\ref{assum:survivalfunction}.
The second part is
\bse
&&2\var\left[\frac{1}{n}\sumi\frac{-I(Z_i\le
	z)\Delta_iK_h'(\bb\trans\X_i-\bb\trans\X)}{f_{\bb\trans\X}(\bb\trans\X)S(Z_i,\bb\trans\X)E\{S_c(Z_i,\X_j)\mid\bb\trans\X_i=\bb\trans\X,Z_i\}}O_p(A)\right]\\
&\le&2E\left[\frac{1}{n}\sumi\frac{-I(Z_i\le
	z)\Delta_iK_h'(\bb\trans\X_i-\bb\trans\X)}{f_{\bb\trans\X}(\bb\trans\X)S(Z_i,\bb\trans\X)E\{S_c(Z_i,\X_j)\mid\bb\trans\X_i=\bb\trans\X,Z_i\}}O_p(A)\right]^2\\
&=&\Bigg\{\left(E\left[\frac{-I(Z_i\le
	z)\Delta_iK_h'(\bb\trans\X_i-\bb\trans\X)}{f_{\bb\trans\X}(\bb\trans\X)S(Z_i,\bb\trans\X)E\{S_c(Z_i,\X_j)\mid\bb\trans\X_i=\bb\trans\X,Z_i\}}\right]\right)^2\\
&&+\frac{1}{n}\var\left[\frac{-I(Z_i\le
	z)\Delta_iK_h'(\bb\trans\X_i-\bb\trans\X)}{f_{\bb\trans\X}(\bb\trans\X)S(Z_i,\bb\trans\X)E\{S_c(Z_i,\X_j)\mid\bb\trans\X_i=\bb\trans\X,Z_i\}}\right]\Bigg\}O\{1/(nh)+h^4\}\\
&=& O\{1/(nh)+h^4\}
\ese
under conditions
\ref{assum:kernel}-\ref{assum:survivalfunction}, where the second last equation is because of
(\ref{eq:expectationOf1}) and
(\ref{eq:varianceOf1}).
Therefore
$
\var\{\wh\bLam_{21}(z,\bb\trans\X)\}=O\{1/(nh^3)\}
$
under conditions
\ref{assum:kernel}-\ref{assum:survivalfunction}.

For $\wh\bLam_{22}(z,\bb\trans\X)$,
\bse
&&\var\{\wh\bLam_{22}(z,\bb\trans\X)\}\\
&\le&\frac{2}{n}\var\left[I(Z_i\le
z)\Delta_iK_h(\bb\trans\X_i-\bb\trans\X)
\frac{\partial\left[
	f_{\bb\trans\X}(\bb\trans\X)E\{I(Z_j\ge
	Z_i)\mid\bb\trans\X\}\right]/
	\partial
	\bb\trans\X}{f_{\bb\trans\X}^2(\bb\trans\X)E^2\{I(Z_j\ge
	Z_i)\mid\bb\trans\X\}}\right]\\
&&+4\var\left[\frac{1}{n}\sumi I(Z_i\le
z)\Delta_iK_h(\bb\trans\X_i-\bb\trans\X) \right.\\
&&\quad\times\left.\frac{\partial\left[
	f_{\bb\trans\X}(\bb\trans\X)E\{I(Z_j\ge
	Z_i)\mid\bb\trans\X\}\right]/
	\partial
	\bb\trans\X}{f_{\bb\trans\X}^2(\bb\trans\X)E^2\{I(Z_j\ge
	Z_i)\mid\bb\trans\X\}}O_p(B)\right]\\
&&+4\var\left[\frac{1}{n}\sumi I(Z_i\le
z)\Delta_iK_h(\bb\trans\X_i-\bb\trans\X)\right.\\
&&\quad\times\left. \frac{\partial\left[
	f_{\bb\trans\X}(\bb\trans\X)E\{I(Z_j\ge
	Z_i)\mid\bb\trans\X\}\right]/
	\partial
	\bb\trans\X}{f_{\bb\trans\X}^2(\bb\trans\X)E^2\{I(Z_j\ge
	Z_i)\mid\bb\trans\X\}}O_p(A)\right].
\ese
The first part is
\be
&&\frac{2}{n}\var\left[I(Z_i\le
z)\Delta_iK_h(\bb\trans\X_i-\bb\trans\X)
\frac{\partial\left[
	f_{\bb\trans\X}(\bb\trans\X)E\{I(Z_j\ge
	Z_i)\mid\bb\trans\X,Z_i\}\right]/
	\partial
	\bb\trans\X}{f_{\bb\trans\X}^2(\bb\trans\X)E^2\{I(Z_j\ge
	Z_i)\mid\bb\trans\X,Z_i\}}\right]\n\\
&\le&\frac{2}{nh}\int I(z_i\le
z)f(z_i,\bb\trans\X)\frac{\left(\partial\left[
	f_{\bb\trans\X}(\bb\trans\X)S(z_i,\bb\trans\X)E\{S_c(z_i,\X_j)\mid\bb\trans\X\}\right]/
	\partial
	\bb\trans\X\right)^2}{f_{\bb\trans\X}^3(\bb\trans\X)S^4(z_i,\bb\trans\X)E^3\{S_c(z_i,\X_j)\mid\bb\trans\X,z_i\}}dz_i\n\\
&&\times \left\{\int K^2(u) du\right\}\n\\
&&+\frac{1}{nh}\sup_{z,z_i,\bb\trans\X}
\left|\frac{\left(\partial\left[
	f_{\bb\trans\X}(\bb\trans\X)S(z_i,\bb\trans\X)E\{S_c(z_i,\X_i)\mid\bb\trans\X\}\right]/
	\partial
	\bb\trans\X\right)^2}{f_{\bb\trans\X}^4(\bb\trans\X)S^4(z_i,\bb\trans\X)E^4\{S_c(z_i,\X_i)\mid\bb\trans\X,z_i\}}\right|\n\\
&&\times \left\{\int
O(h^2)u^2K^2(u)du\right\}+O(1/n)\n\\
&=&O\{1/(nh)\}\n\\\label{eq:varianceOf2}
\ee
under conditions
\ref{assum:kernel}-\ref{assum:survivalfunction}.
The second part is
\bse
&&4\var\left[\frac{1}{n}\sumi I(Z_i\le
z)\Delta_iK_h(\bb\trans\X_i-\bb\trans\X)\right.\\
&&\quad\left.\frac{\partial\left[
	f_{\bb\trans\X}(\bb\trans\X)E\{I(Z_j\ge
	Z_i)\mid\bb\trans\X\}\right]/
	\partial
	\bb\trans\X}{f_{\bb\trans\X}^2(\bb\trans\X)E^2\{I(Z_j\ge
	Z_i)\mid\bb\trans\X,Z_i\}}O_p(B)\right]\\
&\le&4E\left(\left[\frac{1}{n}\sumi I(Z_i\le
z)\Delta_iK_h(\bb\trans\X_i-\bb\trans\X)\right.\right.\\
&&\quad\times\left.\left. \frac{\partial\left[
	f_{\bb\trans\X}(\bb\trans\X)E\{I(Z_j\ge
	Z_i)\mid\bb\trans\X\}\right]/
	\partial
	\bb\trans\X}{f_{\bb\trans\X}^2(\bb\trans\X)E^2\{I(Z_j\ge
	Z_i)\mid\bb\trans\X,Z_i\}}\right]^2\right)\\
&&\times O\{1/(nh^3)+h^4\}\\
&=&O\{1/(nh^3)+h^4\}
\ese
under conditions \ref{assum:kernel}-\ref{assum:survivalfunction}, where the second last equation is because of
(\ref{eq:expectationOf2}) and
(\ref{eq:varianceOf2}). The last part is
\bse
&&4\var\left[\frac{1}{n}\sumi I(Z_i\le
z)\Delta_iK_h(\bb\trans\X_i-\bb\trans\X)\right.\\
&&\quad\times\left.\frac{\partial\left[
	f_{\bb\trans\X}(\bb\trans\X)E\{I(Z_j\ge
	Z_i)\mid\bb\trans\X\}\right]/
	\partial
	\bb\trans\X}{f_{\bb\trans\X}^2(\bb\trans\X)E^2\{I(Z_j\ge
	Z_i)\mid\bb\trans\X,Z_i\}}O_p(A)\right]\\
&\le&4E\left(\left[\frac{1}{n}\sumi I(Z_i\le
z)\Delta_iK_h(\bb\trans\X_i-\bb\trans\X)\right.\right.\\
&&\quad\times\left.\left.\frac{\partial\left[
	f_{\bb\trans\X}(\bb\trans\X)E\{I(Z_j\ge
	Z_i)\mid\bb\trans\X\}\right]/
	\partial
	\bb\trans\X}{f_{\bb\trans\X}^2(\bb\trans\X)E^2\{I(Z_j\ge
	Z_i)\mid\bb\trans\X,Z_i\}}\right]^2\right)\\
&&\times O\{(nh)^{-1}+h^4\}\\
&=&O\{(nh)^{-1}+h^4\}
\ese
under conditions \ref{assum:kernel}-\ref{assum:survivalfunction}
where the second last equation is because of
(\ref{eq:expectationOf2}) and
(\ref{eq:varianceOf2}).
Therefore
$
\var\{\wh\bLam_{22}(z,\bb\trans\X)\}=O\{1/(nh^3)\}
$
under conditions
\ref{assum:kernel}-\ref{assum:survivalfunction}.

Summarizing the results above,
$\var\{\wh\bLam_2(z,\bb\trans\X)\}=O\{1/(nh^3)\}$.
Hence
the estimator $\wh\bLam_2(z,\bb\trans\X)$ satisfies
\be
\wh\bLam_2(z,\bb\trans\X)=\bLam_2(z,\bb\trans\X)+O_p\{(nh^3)^{-1/2}+h^2\}\n
\ee
under conditions
\ref{assum:kernel}-\ref{assum:survivalfunction}.

We give detailed proof for uniform result of (\ref{eq:lemeq2}) only. Because
the domain of $\bb\trans\X$ is compact, we divide it into rectangular
regions. In each region, the distance between a point $\bb\trans\x$ in
this region and the nearest grid point is less than $n^{-2}$. We need only $N\le Cn^2$ grid points, where $C$ is a constant. Let the grid points be
$\bkappa_1, \dots, \bkappa_N$. Let
$\wh\brho(\bb\trans\X)=\wh E\{\X
Y(Z)\mid\bb\trans\X\}$ and
$\brho(\bb\trans\X)=E\{\X
Y(Z)\mid\bb\trans\X\}$. Then for any
$(\bb\trans\X)$, there exists a $\bkappa_i$, $1 \le i \le N$, such that \bse
|\wh\brho(\bb\trans\X) -\brho(\bb\trans\X)| &\le&|\wh\brho(\bkappa_i)
-\brho(\bkappa_i)| +|\wh\brho(\bb\trans\X) -\wh\brho(\bkappa_i)|+
|\brho(\bb\trans\X) -\brho(\bkappa_i)|\\ &\le&|\wh\brho(\bkappa_i)
-\brho(\bkappa_i)| +D_1n^{-2},
\ese
for an absolute constant $D_1$ under
Conditions \ref{assum:kernel} and \ref{assum:survivalfunction}. Thus, for
any $D\ge D_1$,
\bse &&\pr(\sup_{\bb\trans\x}|\wh\brho(\bb\trans\X)-\brho(\bb\trans\X)|>2D[h^2+\{\log n (nh)^{-1}\}^{1/2}])\\
&\le&\pr(\sup_{\bkappa_i}|\wh\brho(\bkappa_i )-\brho(\bkappa_i)|>2D[h^2+\{\log n
(nh)^{-1}\}^{1/2}] -D_1n^{-2})\\
&\le&\pr(\sup_{\bkappa_i}|\wh\brho(\bkappa_i )-\brho(\bkappa_i)|>D[h^2+\{\log n
(nh)^{-1}\}^{1/2}])
\ese
under Condition
\ref{assum:bandwidth}. Using
Bernstein's inequality on $\wh\brho(\bkappa_i)$,  under Conditions
\ref{assum:kernel}-\ref{assum:survivalfunction}, we have
\bse
\pr
[|\wh\brho(\bkappa_i )-\brho(\bkappa_i)|\ge A\{\log
n/(nh)\}^{1/2}]
&\le&
2\exp\left\{\frac{-n A^2\log n/(nh)}{2D_2h^{-1}+2/3AD_3(\log
	n)^{1/2}(nh)^{-1/2}}\right\}\\
&=&2\exp\left\{\frac{- A^2\log
	n}{2D_2+2/3AD_3(\log
	nh/n)^{1/2}}\right\}\\
&\le&2\exp\left(\frac{- A^2\log
	n}{2D_2+AD_3}\right)
\ese
for all $A>D_3+\sqrt{D_3^2+4D_2}$, where $D_2$ and $D_3$ are constants satisfying
\bse
\var\{\wh\brho(\bb\trans\X) -\brho(\bb\trans\X)\}&\le& \frac{D_2}{nh},\\
\left|\frac{K_h(\bb\trans\X_i-\bb\trans\X)\X_iI(Z_i\ge Z)}{1/n\sumj K_h(\bb\trans\X_j-\bb\trans\X)}-\brho(\bb\trans\X)\right|&\le& D_3 \text{ with probability 1}.
\ese
This leads to
\bse \pr
[\sup_{\bkappa_i}|\wh\brho(\bkappa_i )-E\wh\brho(\bkappa_i)|\ge A\{\log
n/(nh)\}^{1/2}] &\le& 2Cn^{2}\exp\left(\frac{- A^2\log n}{2D_2+AD_3}\right)\\
&=&2C\exp\left[\{2- A^2/(2D_2+AD_3)\}\log n\right]\to0
\ese
because $A>D_3+\sqrt{D_3^2+4D_2}$.
Combining the above results, for $A_1=\max(A,D)$, \bse
&&\pr(\sup_{\bb\trans\x}|\wh\brho(\bb\trans\X,
\bb)-\brho(\bb\trans\X)|>2A_1[h^2+\{\log n (nh)^{-1}\}^{1/2}])\\ &\le&
\pr(\sup_{\bkappa_i}|\wh\brho(\bkappa_i )-\brho(\bkappa_i)|>A_1[h^2+\{\log n
(nh)^{-1}\}^{1/2}])\\ &\le&\pr\{\sup_{\bkappa_i}|\wh\brho(\bkappa_i
)-\brho(\bkappa_i)|>A_1h^2\} +\pr([\sup_{\bkappa_i}|\wh\brho(\bkappa_i
)-\brho(\bkappa_i)|\ge A_1\{\log n (nh)^{-1}\}^{1/2}])\\ &\to&0. \ese
The uniform convergence results concerning (\ref{eq:lemeq5})-(\ref{eq:lemeq8})
are slightly different because these functions contain the additional component
$Z$. Nevertheless, under Condition \ref{assum:survivalfunction}, the support of
$(\bb\trans\X_i,Z_i)$ or $(\bb\trans\X_i, Z_j)$  is also bounded so we
can similarly divide the region using $N\le
Cn^{2+2}$ grid points while the
distance of a point to the nearest grid point is less than $n^{-2}$. The rest of
the analysis can then be similarly carried out  as above, then the uniform convergence is established. \qed

\subsubsection{Lemma \ref{th:Lambda}}
\begin{Lem}\label{th:Lambda}
	The estimator $\wh{\Lambda}(t,\wh\bb\trans\X)$ has the
	expansion
	\bse
	&&\sqrt{nh}\left\{\wh{\Lambda}(t,\wh\bb\trans\X)-
	{\Lambda}(t,\bb\trans\X)\right\}\\
	&=&
	\sqrt{\frac{h}{n}}\sumi\int_{0}^{t}\frac{I\left\{
		\sumj Y_j(s)K_h(\bb\trans\X_j-\bb\trans\X)>0\right\}}
	{f_{\bb\trans\X}(\bb\trans\X)E\{I(z\ge s)\mid\bb\trans\X\}}
	K_h(\bb\trans\X_i-\bb\trans\X)\\
	&& \times dM_i\{s,\bb\trans\X,I(W\le s),WI(W\le s)\}+o_p(1),
	\ese
	and
	satisfies
	\bse
	\sqrt{nh}\left\{\wh{\Lambda}(t,\wh\bb\trans\X)-{\Lambda}(t,\bb\trans\X)\right\}\to
	N\{0,\sigma^2(t,\bb\trans\X)\}
	\ese
	in distribution when $n\to\infty$ for all $t,\bb\trans\X$
	under Conditions
	\ref{assum:kernel}-\ref{assum:survivalfunction}, where
	\bse
	\sigma^2(t,\bb\trans\X)=\int
	K^2(u)du\int_{0}^{t}\frac{\lambda(s,\bb\trans\X)}
	{f_{\bb\trans\X}(\bb\trans\X)E\{I(Z\ge
		s)\mid\bb\trans\X\}}ds.
	\ese
\end{Lem}
\noindent Proof:

We only need to prove
$\sqrt{nh}\left\{\wh{\Lambda}(t,\bb\trans\X)-{\Lambda}(t,\bb\trans\X)\right\}\to
N\{0,\sigma^2(t,\bb\trans\X)\}$ because
$\sqrt{nh}\left\{\wh{\Lambda}(t,\wh\bb\trans\X)-{\wh\Lambda}(t,\bb\trans\X)\right\}=O_p(\sqrt{h})=o_p(1)$
due to Theorems \ref{th:consistency} and
\ref{th:eff}.
For notational convenience, let $d=1$ and
$\nu=2$.
For any
$t$ and
$\bb\trans\X$, define
\bse
\phi_n(s,\bb\trans\X)&\equiv&\sumj
Y_j(s)K_h(\bb\trans\X_j-\bb\trans\X),\\
Q_n(t,\bb\trans\X)
&\equiv&\wh\bLam(t,\bb\trans\X)-\int_{0}^{t}I\left\{
\phi_n(s,\bb\trans\X)>0\right\}\lambda(s,\bb\trans\X)ds,\\
D_n(t,\bb\trans\X)&\equiv&\int_{0}^{t}\lambda(s,\bb\trans\X)
\left[1-I\left\{\phi_n(s,\bb\trans\X)>0\right\}\right]ds.
\ese
Then
\bse
\sqrt{nh}\left\{\wh{\Lambda}(t,\bb\trans\X)-
{\Lambda}(t,\bb\trans\X)\right\}=\sqrt{nh}Q_n(t,\bb\trans\X)
-\sqrt{nh}D_n(t,\bb\trans\X).
\ese
We first show that $\sqrt{nh}D_n(t,\bb\trans\X)\to 0$ in
probability uniformly. It suffices to show that
\bse
\sqrt{nh}\left[1-I\left\{\phi_n(s,\bb\trans\X)>0\right\}\right]\overset{p}{\to}
0,
\ese
which is equivalent to show that for any $\epsilon>0$,
\bse
\Pr\left(\sqrt{nh}\left[1-I\left\{\phi_n(s,\bb\trans\X)>0\right\}\right]>\epsilon\right){\to}
0.
\ese
Now for $\epsilon\ge\sqrt{nh}$, the above automatically holds.
For $\epsilon<\sqrt{nh}$, this is equivalent to show
\be
\Pr\left\{n^{-1}\phi_n(s,\bb\trans\X)\le 0\right\}{\to}
0.\label{eq:sumto0}
\ee
Because
$n^{-1}\phi_n(s,\bb\trans\X)=f_{\bb\trans\X}(\bb\trans\X)
E\{I(Z_j\ge s)\mid\bb\trans\X\}+O_p\{h^2+(nh)^{-1/2}\}$,
(\ref{eq:sumto0}) is equivalent to
\bse
\Pr\left[f_{\bb\trans\X}(\bb\trans\X)E\{I(Z_j\ge
s)\mid\bb\trans\X\}+O_p\{h^2+(nh)^{-1/2}\}
\le 0\right]{\to} 0,
\ese
which automatically holds under Condition \ref{assum:fbeta} and
\ref{assum:survivalfunction}.
Hence $\sqrt{nh}D_n(t,\bb\trans\X)\to 0$ in
probability uniformly.
Second we inspect the asymptotic property of
$\sqrt{nh}Q_n(t,\bb\trans\X)
$.
Recall $M_i(s,\bb\trans\X)$ is the martingale corresponding
to the
counting process $N_i(s)$ and satisfies
$dM_i(s,\bb\trans\X)=dN_i(s)-Y_i(s)\lambda(s,\bb\trans\X)ds$.
\be
&&\sqrt{nh}Q_n(t,\bb\trans\X)\n\\
&=&\int_{0}^{t}\sqrt{nh}\frac{1}{\sumj 	
	Y_j(s)K_h(\bb\trans\X_j-\bb\trans\X)}\sumi
K_h(\bb\trans\X_i-\bb\trans\X)dN_i(s)\n\\
&&-\int_{0}^{t}\sqrt{nh}I\left\{\phi_n(s,\bb\trans\X)>0\right\}\lambda(s,\bb\trans\X)ds\n\\
&=&\int_{0}^{t}\sqrt{nh}\frac{I\{\phi_n(s,\bb\trans\X)\le
	0\}}{\sumj
	Y_j(s)K_h(\bb\trans\X_j-\bb\trans\X)}\sumi
K_h(\bb\trans\X_i-\bb\trans\X)dN_i(s)\n\\
&&+\int_{0}^{t}\sqrt{nh}\frac{I\{\phi_n(s,\bb\trans\X)>
	0\}}{\sumj Y_j(s)K_h(\bb\trans\X_j-\bb\trans\X)}\sumi
K_h(\bb\trans\X_i-\bb\trans\X)dN_i(s)\n\\
&&-\int_{0}^{t}\sqrt{nh}\frac{I\{\phi_n(s,\bb\trans\X)>
	0\}}{\sumj Y_j(s)K_h(\bb\trans\X_j-\bb\trans\X)}\sumi
K_h(\bb\trans\X_i-\bb\trans\X)Y_i(s)\lambda(s,\bb\trans\X)ds\n\\
&=&\int_{0}^{t}\sqrt{nh}\frac{I\{\phi_n(s,\bb\trans\X)>0\}}{\sumj Y_j(s)K_h(\bb\trans\X_j-\bb\trans\X)}\n\\
&&\times\sumi
K_h(\bb\trans\X_i-\bb\trans\X)dM_i\{s,\bb\trans\X,I(W \le s),WI(W\le s)\}\n\\\label{eq:Qn1}\\
&&+\int_{0}^{t}\sqrt{nh}\frac{I\{\phi_n(s,\bb\trans\X)\le
	0\}}{\sumj Y_j(s)K_h(\bb\trans\X_j-\bb\trans\X)}\sumi
K_h(\bb\trans\X_i-\bb\trans\X)dN_i(s).\label{eq:Qn2}
\ee
In (\ref{eq:Qn1}), it leads to the same martingale $dM_i\{s,\bb\trans\X,I(W \le s),WI(W\le s)\}=dN_i(t)-
Y_i(s)[\lambda_T(t-W,\bb\trans\X)I(W\le t)+\lambda_N(t,\bb\trans\X)\{1-I(W\le t)\}]ds$ in  either group of transplant or nontransplant because one of $\lambda_T(t,\bb\trans\X)$ and $\lambda_N(t,\bb\trans\X)$ is zero.

We decompose (\ref{eq:Qn1}) as
\be
&&\int_{0}^{t}\sqrt{nh}\frac{I\{\phi_n(s,\bb\trans\X)>0\}}
{\sumj Y_j(s)K_h(\bb\trans\X_j-\bb\trans\X)}\n\\
&&\quad\sumi
K_h(\bb\trans\X_i-\bb\trans\X)dM_i\{s,\bb\trans\X,I(W \le s),WI(W\le s)\}\n\\
&=&\sqrt{\frac{h}{n}}\sumi\int_{0}^{t}
\frac{I\{\phi_n(s,\bb\trans\X)>0\}}{1/n\sumj
	Y_j(s)K_h(\bb\trans\X_j-\bb\trans\X)}\n\\
&&\quad K_h(\bb\trans\X_i-\bb\trans\X)dM_i\{s,\bb\trans\X,I(W \le s),WI(W\le s)\}\n\\
&=&\sqrt{\frac{h}{n}}\sumi\int_{0}^{t}I\left\{\phi_n(s,\bb\trans\X)>0\right\}\left[\frac{1}
{f_{\bb\trans\X}(\bb\trans\X)E\{I(Z_j\ge
	s)\mid\bb\trans\X\}}\right.\n\\
&&-\frac{1/n\sumj Y_j(s)K_h(\bb\trans\X_j-\bb\trans\X)}
{f^2_{\bb\trans\X}(\bb\trans\X)E^2\{I(Z_j\ge
	s)\mid\bb\trans\X\}}
+\frac{1}{f_{\bb\trans\X}(\bb\trans\X)E\{I(Z_j\ge 	
	s)\mid\bb\trans\X\}}\n\\
&& \left.+O_p\{h^4+(nh)^{-1}\}\right]\times K_h(\bb\trans\X_i-\bb\trans\X)dM_i\{s,\bb\trans\X,I(W \le s),WI(W\le s)\}\n\\
&=&Q_{n1}-Q_{n2}+o_p(1),\label{eq:Qn1decompose}
\ee
where
\bse
Q_{n1}&=&\sqrt{\frac{h}{n}}\sumi\int_{0}^{t}\frac{I\left\{\phi_n(s,\bb\trans\X)>0\right\}K_h(\bb\trans\X_i-\bb\trans\X)}
{f_{\bb\trans\X}(\bb\trans\X)E\{I(Z\ge
	s)\mid\bb\trans\X\}}\\
&&\quad\times dM_i\{s,\bb\trans\X,I(W \le s),WI(W\le s)\},\\
Q_{n2}&=&\sqrt{\frac{h}{n}}\sumi\int_{0}^{t}
\frac{I\left\{\phi_n(s,\bb\trans\X)>0\right\}}
{f^2_{\bb\trans\X}(\bb\trans\X)E^2\{I(Z_j\ge
	s)\mid\bb\trans\X\}}\n\\
&&\times\left[\frac{1}{n}\sumj
Y_j(s)K_h(\bb\trans\X_j-\bb\trans\X)
-f_{\bb\trans\X}(\bb\trans\X)
E\{I(Z\ge s)\mid\bb\trans\X\}\right]\n\\
&&	\times K_h(\bb\trans\X_i-\bb\trans\X)dM_i\{s,\bb\trans\X,I(W \le s),WI(W\le s)\}\n\\
&=&\sqrt{\frac{h}{n^3}}\sumi\sumj\int_{0}^{t}
\frac{I\left\{\phi_n(s,\bb\trans\X)>0\right\}}
{f^2_{\bb\trans\X}(\bb\trans\X)E^2\{I(Z_j\ge
	s)\mid\bb\trans\X\}}\n\\
&&\times\left[Y_j(s)K_h(\bb\trans\X_j-\bb\trans\X)-
f_{\bb\trans\X}(\bb\trans\X)E\{I(Z\ge s)\mid\bb\trans\X\}\right]
\\
&&\times K_h(\bb\trans\X_i-\bb\trans\X)dM_i\{s,\bb\trans\X,I(W \le s),WI(W\le s)\}\n,
\ese
and the remaining term in (\ref{eq:Qn1decompose}) is $o_p(1)$
because
$\sqrt{n/h} O_p\{h^4+(nh)^{-1}\}
=O_p\{n^{1/2}h^{7/2}+(nh^3)^{-1/2}\}=o_p(1)$ by Condition
\ref{assum:bandwidth}.

Using the U-statistic property, $Q_{n2}$ has leading order terms
$Q_{n21}+Q_{n22}-Q_{n23}$, where
\bse
Q_{n21}&=&\sqrt{\frac{h}{n}}E\Bigg(\sumi\int_{0}^{t}
\frac{I\left\{\phi_n(s,\bb\trans\X)>0\right\}}
{f^2_{\bb\trans\X}(\bb\trans\X)E^2\{I(Z_j\ge
	s)\mid\bb\trans\X\}}\n\\
&&\times\left[Y_j(s)K_h(\bb\trans\X_j-\bb\trans\X)-
f_{\bb\trans\X}(\bb\trans\X)E\{I(Z\ge s)\mid\bb\trans\X\}\right]K_h(\bb\trans\X_i-\bb\trans\X)
\\
&&\times
dM_i\{s,\bb\trans\X,I(W \le s),WI(W\le s)\}
\mid\Delta_i,\bb\trans\X_i,Z_i,I(W \le s),WI(W\le s)\Bigg),\\
Q_{n22}&=&\sqrt{\frac{h}{n}}E\Bigg(\sumj
\int_{0}^{t}\frac{I\left\{\phi_n(s,\bb\trans\X)>0\right\}}
{f^2_{\bb\trans\X}(\bb\trans\X)E^2\{I(Z_j\ge
	s)\mid\bb\trans\X\}}\n\\
&&\times\left[Y_j(s)K_h(\bb\trans\X_j-\bb\trans\X)-
f_{\bb\trans\X}(\bb\trans\X)E\{I(Z\ge s)\mid\bb\trans\X\}\right]K_h(\bb\trans\X_i-\bb\trans\X)
\\
&&\times
dM_i\{s,\bb\trans\X,I(W \le s),WI(W\le s)\}
\mid\Delta_j,\bb\trans\X_j,Z_j,I(W \le s),WI(W\le s)\Bigg),\\
Q_{n23}&=&\sqrt{nh}E\Bigg(\int_{0}^{t}
\frac{I\left\{\phi_n(s,\bb\trans\X)>0\right\}}
{f^2_{\bb\trans\X}(\bb\trans\X)E^2\{I(Z_j\ge
	s)\mid\bb\trans\X\}}\n\\
&&\times\left[Y_j(s)K_h(\bb\trans\X_j-\bb\trans\X)-
f_{\bb\trans\X}(\bb\trans\X)E\{I(Z\ge
s)\mid\bb\trans\X\}\right]\\
&&\times
E\left\{K_h(\bb\trans\X_i-\bb\trans\X)dM_i\{s,\bb\trans\X,I(W \le s),WI(W\le s)\}\right\}\Bigg).
\ese
$I\left\{\phi_n(s,\bb\trans\X)>0\right\}
=I\left[f_{\bb\trans\X}(\bb\trans\X)E\{I(Z_j\ge
s)\mid\bb\trans\X\}+O_p\{h^2+(nh)^{-1}\}>0\right]=1
$
almost surely. Thus, almost surely,
\bse
Q_{n21}
&=&\sqrt{\frac{h}{n}}\sumi\Bigg(\int_{0}^{t}
\frac{1}{f_{\bb\trans\X}(\bb\trans\X)E\{I(Z_j\ge
	s)\mid\bb\trans\X\}}\n\\
&&\times E\left[Y_j(s)K_h(\bb\trans\X_j-\bb\trans\X)-
f_{\bb\trans\X}(\bb\trans\X)E\{I(Z\ge s)\mid\bb\trans\X\}\right]
\\
&&\times
K_h(\bb\trans\X_i-\bb\trans\X)dM_i\{s,\bb\trans\X,I(W \le s),WI(W\le s)\}\Bigg)\\
&=&\sqrt{\frac{h}{n}}\sumi
\int_{0}^{t}\frac{O(h^2)K_h(\bb\trans\X_i-\bb\trans\X)}{f_{\bb\trans\X}(\bb\trans\X)E\{I(Z_j\ge
	s)\mid\bb\trans\X\}}
dM_i\{s,\bb\trans\X,I(W \le s),WI(W\le s)\}\\
&\to& 0.
\ese
uniformly as $h\to 0$.
Similarly, almost surely,
\bse
Q_{n22}
&=&\sqrt{\frac{h}{n}}\sumj
\int_{0}^{t}\frac{1}{f^2_{\bb\trans\X}(\bb\trans\X)E^2\{I(Z_j\ge
	s)\mid\bb\trans\X\}}\n\\
&&\times\left[Y_j(s)K_h(\bb\trans\X_j-\bb\trans\X)-
f_{\bb\trans\X}(\bb\trans\X)E\{I(Z\ge s)\mid\bb\trans\X\}\right]
\\
&&\times
E[K_h(\bb\trans\X_i-\bb\trans\X)dM_i\{s,\bb\trans\X,I(W \le s),WI(W\le s)\}]\\
&=&0.
\ese
Obviously, $Q_{n23}=E(Q_{n22})=0$,
hence $Q_{n2}\to0$ in probability uniformly as $n\to\infty$.

For (\ref{eq:Qn2})
\bse
&&\int_{0}^{t}\sqrt{nh}\frac{I\{\phi_n(s,\bb\trans\X)\le
	0\}}{\sumj Y_j(s)K_h(\bb\trans\X_j-\bb\trans\X)}\sumi
K_h(\bb\trans\X_i-\bb\trans\X)dN_i(s)
\to 0
\ese
in probability uniformly.
We have obtained
\be
&&\sqrt{nh}Q_n(t,\bb\trans\X)\n\\
&=&\sqrt{\frac{h}{n}}\sumi\int_{0}^{t}\frac{I\left\{\phi_n(s,\bb\trans\X)>0\right\}}
{f_{\bb\trans\X}(\bb\trans\X)E\{I(Z\ge
	s)\mid\bb\trans\X\}}K_h(\bb\trans\X_i-\bb\trans\X)\n\\
&&\quad\times dM_i\{s,\bb\trans\X,I(W \le s),WI(W\le s)\}\n\\
\label{eq:Qn1a}\\
&&+o_p(1).\n
\ee
Applying martingale
central limit theorem on (\ref{eq:Qn1a}),
we have
\be
&&\frac{h}{n}\sumi\int_{0}^{t}\frac{\lambda(s,\bb\trans\X)
	I\{\phi_n(s,\bb\trans\X)>0\}}{[f_{\bb\trans\X}(\bb\trans\X)E\{I(Z\ge
	s)\mid\bb\trans\X\}]^2}
K^2_h(\bb\trans\X_i-\bb\trans\X)Y_i(s)ds\n\\
&=&\int_{0}^{t}\frac{\lambda(s,\bb\trans\X)I\{\phi_n(s,\bb\trans\X)>0\}}
{[f_{\bb\trans\X}(\bb\trans\X)E\{I(Z\ge
	s)\mid\bb\trans\X\}]^2}\frac{1}{n}\sumi
hK^2_h(\bb\trans\X_i-\bb\trans\X)Y_i(s)ds\n\\
&=&\int_{0}^{t}\frac{\lambda(s,\bb\trans\X)
	I\{\phi_n(s,\bb\trans\X)>0\}}{[f_{\bb\trans\X}(\bb\trans\X)
	E\{I(Z\ge s)\mid\bb\trans\X\}]^2}\n\\
&&\times \left[f_{\bb\trans\X}(\bb\trans\X)E\{I(Z_i\ge
s)\mid\bb\trans\X\}\int
K^2(u)du+O_p(n^{-1/2}h^{-1/2}+h^2)\right]ds\n\\
&\overset{p}{\to}&\int
K^2(u)du\int_{0}^{t}\frac{\lambda(s,\bb\trans\X)}
{f_{\bb\trans\X}(\bb\trans\X)E\{I(Z_j\ge
	s)\mid\bb\trans\X\}}ds\n\\
&=&\sigma^2(t,\bb\trans\X)\label{eq:MCLTa}.
\ee
Next we inspect the following integration for any $\epsilon>0$.
\bse
&&\sumi\int_{0}^{t}\frac{h}{n}\frac{I\{\phi_n(s,\bb\trans\X)>0\}}
{[f_{\bb\trans\X}(\bb\trans\X)E\{I(Z\ge s)\mid\bb\trans\X\}]^2}
K^2_h(\bb\trans\X_i-\bb\trans\X)Y_i(s)\\
&&\times
I\left[\sqrt{\frac{h}{n}}\left|\frac{I\{\phi_n(s,\bb\trans\X)>0\}}
{f_{\bb\trans\X}(\bb\trans\X)E\{I(Z\ge	s)\mid\bb\trans\X\}}
K_h(\bb\trans\X_i-\bb\trans\X)\right|>\epsilon\right]
\lambda(s,\bb\trans\X)ds\\
&=&\int_{0}^{t}\frac{I\{\phi_n(s,\bb\trans\X)>0\}}
{[f_{\bb\trans\X}(\bb\trans\X)E\{I(Z\ge s)\mid\bb\trans\X\}]^2}
\frac{1}{n}\sumi \Bigg(hK^2_h(\bb\trans\X_i-\bb\trans\X)Y_i(s)\\
&&\times
I\left[\sqrt{\frac{h}{n}}\left|\frac{I\{\phi_n(s,\bb\trans\X)>0\}}
{f_{\bb\trans\X}(\bb\trans\X)E\{I(Z\ge s)\mid\bb\trans\X\}}
K_h(\bb\trans\X_i-\bb\trans\X)\right|>\epsilon\right]\Bigg)\lambda(s,\bb\trans\X)ds\\
&\le&\int_{0}^{t}\lambda(s,\bb\trans\X)
\frac{I\{\phi_n(s,\bb\trans\X)>0\}}{[f_{\bb\trans\X}(\bb\trans\X)
	E\{I(Z\ge s)\mid\bb\trans\X\}]^2}
\left\{\frac{1}{n}\sumi
hK^2_h(\bb\trans\X_i-\bb\trans\X)Y_i(s)\right\}\\
&&\times \sup_{1\le i\le n}
I\left[\left|f_{\bb\trans\X}(\bb\trans\X)E\{I(Z\ge
s)\mid\bb\trans\X\}\right|<\frac{\xi_i}{\epsilon\sqrt{nh}}\right]ds,
\ese
where
$\xi_i=|I\{\phi_n(s,\bb\trans\X)>0\}K\{(\bb\trans\X_i-\bb\trans\X)/h\}|$,
which is bounded following Condition \ref{assum:kernel}.
In the above display,
\bse
\sup_{1\le i\le
	n}I\left[\left|f_{\bb\trans\X}(\bb\trans\X)E\{I(Z_j\ge
s)\mid\bb\trans\X\}\right|<\frac{\xi_i}{\epsilon\sqrt{nh}}\right]=0
\ese
as long as $n$ is large enough because the right hand side
converges to 0 by
Condition \ref{assum:bandwidth} but the left hand side will be
always
larger than 0 by conditions \ref{assum:fbeta} and
\ref{assum:survivalfunction}.
On the other hand,
\bse
&&\frac{1}{n}\sumi hK^2_h(\bb\trans\X_i-\bb\trans\X)Y_i(s)\\
&=&f_{\bb\trans\X}(\bb\trans\X)E\{I(Z_i\ge
s)\mid\bb\trans\X\}\int
K^2(u)du+O_p(n^{-1/2}h^{-1/2}+h^2)\\
&\to&0
\ese
in probability uniformly.
Hence
\be
&&\lim_{n\to\infty}\sumi\int_{0}^{t}\frac{h}{n}
\frac{I\{\phi_n(s,\bb\trans\X)>0\}}{[f_{\bb\trans\X}(\bb\trans\X)
	E\{I(Z\ge s)\mid\bb\trans\X\}]^2}
K^2_h(\bb\trans\X_i-\bb\trans\X)Y_i(s)\n\\
&&\times
I\left[\sqrt{\frac{h}{n}}\left|\frac{I\{\phi_n(s,\bb\trans\X)>0\}}
{f_{\bb\trans\X}(\bb\trans\X)E\{I(Z\ge s)\mid\bb\trans\X\}}
K_h(\bb\trans\X_i-\bb\trans\X)\right|>\epsilon\right]\lambda(s,\bb\trans\X)ds=0\n\\
\label{eq:MCLTb}
\ee
with probability 1 uniformly for any $\epsilon>0$.

In summary
\bse
\sqrt{nh}\left\{\wh{\Lambda}(t,\bb\trans\X)-{\Lambda}(t,\bb\trans\X)\right\}\to
N\{0,\sigma^2(t,\bb\trans\X)\}
\ese
uniformly.
\qed

\subsection{Proof of Theorem \ref{th:consistency}}\label{app:consistofb}
Because the result regarding (\ref{eq:eff}) is the most
difficult to
establish, we provide only the proof concerning (\ref{eq:eff}),
the result concerning (\ref{eq:general}) 
is based on a similar proof.

For each $n$, let $\wh\bb_n$ satisfy
\bse
&&\frac{1}{n}\sumi
\Delta_i\frac{\wh\blam_2\{Z_i,\wh\bb_n\trans\X_i,I(W_i\le Z_i),W_iI(W_i\le Z_i)\}}{\wh\lambda\{Z_i,\wh\bb_n\trans\X_i,I(W_i\le Z_i),W_iI(W_i\le Z_i)\}}\\
&&\quad\otimes\left[\X_{li}-
\frac{\wh E\left\{\X_{li}
	Y_i(Z_i)\mid\wh\bb_n\trans\X_i,I(W_i\le Z_i),W_iI(W_i\le Z_i)\right\}}
{\wh E\left\{Y_i(Z_i)\mid\wh\bb_n\trans\X_i,I(W_i\le Z_i),W_iI(W_i\le Z_i)\right\}}\right]=\0.
\ese
Under condition \ref{assum:bounded}, there exists a subsequence
of
$\wh\bb_n, n=1, 2,\dots$,
that converges. For notational simplicity,  we still write
$\wh\bb_n,
n=1, 2, \dots, $
as the subsequence that
converges and let the limit be $\bb^*$.

From the uniform convergence in (\ref{eq:lemeq1}),
(\ref{eq:lemeq2}),
(\ref{eq:lemeq5}), (\ref{eq:lemeq6}) given in
Lemma \ref{lem:pre},
\bse
&&\frac{1}{n}\sumi
\Delta_i\frac{\wh\blam_2\{Z_i,\wh\bb_n\trans\X_i,I(W_i\le Z_i),W_iI(W_i\le Z_i)\}}{\wh\lambda\{Z_i,\wh\bb_n\trans\X_i,I(W_i\le Z_i),W_iI(W_i\le Z_i)\}}\\
&&\quad\otimes\left[\X_{li}-
\frac{\wh E\left\{\X_{li}
	Y_i(Z_i)\mid\wh\bb_n\trans\X_i,I(W_i\le Z_i),W_iI(W_i\le Z_i)\right\}}
{\wh E\left\{Y_i(Z_i)\mid\wh\bb_n\trans\X_i,I(W_i\le Z_i),W_iI(W_i\le Z_i)\right\}}\right]\\
&=&\frac{1}{n}\sumi
\Delta_i\frac{\blam_2\{Z_i,\wh\bb_n\trans\X_i,I(W_i\le Z_i),W_iI(W_i\le Z_i)\}+O_p\{(nbh^3)^{-1/2}+h^2+b^2\}}{\lambda\{Z_i,\wh\bb_n\trans\X_i,I(W_i\le Z_i),W_iI(W_i\le Z_i)\}+O_p\{(nbh)^{-1/2}+h^2+b^2\}}\\
&&
\otimes\left[\X_{li}-
\frac{E\left\{\X_{li}
	Y_i(Z_i)\mid\wh\bb_n\trans\X_i,I(W_i\le Z_i),W_iI(W_i\le Z_i)\right\}+O_p\{(nh)^{-1/2}+h^2\}}
{E\left\{Y_i(Z_i)\mid\wh\bb_n\trans\X_i,I(W_i\le Z_i),W_iI(W_i\le Z_i)\right\}+O_p\{(nh)^{-1/2}+h^2\}}\right]\\
&=&\frac{1}{n}\sumi
\Delta_i\left[\frac{\blam_2\{Z_i,\wh\bb_n\trans\X_i,I(W_i\le Z_i),W_iI(W_i\le Z_i)\}}{\lambda\{Z_i,\wh\bb_n\trans\X_i,I(W_i\le Z_i),W_iI(W_i\le Z_i)\}}+O_p\{(nbh^3)^{-1/2}+h^2+b^2\}\right]\\
&&
\otimes\left[\X_{li}-
\frac{E\left\{\X_{li}
	Y_i(Z_i)\mid\wh\bb_n\trans\X_i,I(W_i\le Z_i),W_iI(W_i\le Z_i)\right\}}
{E\left\{Y_i(Z_i)\mid\wh\bb_n\trans\X_i,I(W_i\le Z_i),W_iI(W_i\le Z_i)\right\}}+O_p\{(nh)^{-1/2}+h^2\}\right]\\
&=&\frac{1}{n}\sumi
\Delta_i\frac{\blam_2\{Z_i,\wh\bb_n\trans\X_i,I(W_i\le Z_i),W_iI(W_i\le Z_i)\}}{\lambda\{Z_i,\wh\bb_n\trans\X_i,I(W_i\le Z_i),W_iI(W_i\le Z_i)\}}\\
&& \quad\otimes\left[\X_{li}-
\frac{E\left\{\X_{li}
	Y_i(Z_i)\mid\wh\bb_n\trans\X_i,I(W_i\le Z_i),W_iI(W_i\le Z_i)\right\}}
{E\left\{Y_i(Z_i)\mid\wh\bb_n\trans\X_i,I(W_i\le Z_i),W_iI(W_i\le Z_i)\right\}}\right]+o_p(1).
\ese
Thus,  for sufficiently large $n$,
\bse
&&\frac{1}{n}\sumi
\Delta_i\frac{\blam_2\{Z_i,\wh\bb_n\trans\X_i,I(W_i\le Z_i),W_iI(W_i\le Z_i)\}}{\lambda\{Z_i,\wh\bb_n\trans\X_i,I(W_i\le Z_i),W_iI(W_i\le Z_i)\}}\\
&&\quad\otimes\left[\X_{li}-
\frac{E\left\{\X_{li}
	Y_i(Z_i)\mid\wh\bb_n\trans\X_i,I(W_i\le Z_i),W_iI(W_i\le Z_i)\right\}}
{E\left\{Y_i(Z_i)\mid\wh\bb_n\trans\X_i,I(W_i\le Z_i),W_iI(W_i\le Z_i)\right\}}\right]\\
&=&\frac{1}{n}\sumi
\Delta_i\frac{\blam_2\{Z_i,{\bb^*}\trans\X_i,I(W_i\le Z_i),W_iI(W_i\le Z_i)\}}{\lambda\{Z_i,{\bb^*}\trans\X_i,I(W_i\le Z_i),W_iI(W_i\le Z_i)\}}\\
&&\quad\otimes\left[\X_{li}-
\frac{E\left\{\X_{li}
	Y_i(Z_i)\mid{\bb^*}\trans\X_i,I(W_i\le Z_i),W_iI(W_i\le Z_i)\right\}}
{E\left\{Y_i(Z_i)\mid{\bb^*}\trans\X_i,I(W_i\le Z_i),W_iI(W_i\le Z_i)\right\}}\right]+O_p(\wh\bb_n-\bb^*)\\
&=&\frac{1}{n}\sumi
\Delta_i\frac{\blam_2\{Z_i,{\bb^*}\trans\X_i,I(W_i\le Z_i),W_iI(W_i\le Z_i)\}}{\lambda\{Z_i,{\bb^*}\trans\X_i,I(W_i\le Z_i),W_iI(W_i\le Z_i)\}}\\
&&\quad\otimes\left[\X_{li}-
\frac{E\left\{\X_{li}
	Y_i(Z_i)\mid{\bb^*}\trans\X_i,I(W_i\le Z_i),W_iI(W_i\le Z_i)\right\}}
{E\left\{Y_i(Z_i)\mid{\bb^*}\trans\X_i,I(W_i\le Z_i),W_iI(W_i\le Z_i)\right\}}\right]+o_p(1),
\ese
under Condition
\ref{assum:kernel}-\ref{assum:bandwidth}, where
the last equality holds because $\wh\bb_n$ converges to $\bb^*$.
In addition,
\bse
&&\frac{1}{n}\sumi
\Delta_i\frac{\blam_2\{Z_i,{\bb^*}\trans\X_i,I(W_i\le Z_i),W_iI(W_i\le Z_i)\}}{\lambda\{Z_i,{\bb^*}\trans\X_i,I(W_i\le Z_i),W_iI(W_i\le Z_i)\}}\\
&&\quad\otimes\left[\X_{li}-
\frac{E\left\{\X_{li}
	Y_i(Z_i)\mid{\bb^*}\trans\X_i,I(W_i\le Z_i),W_iI(W_i\le Z_i)\right\}}
{E\left\{Y_i(Z_i)\mid{\bb^*}\trans\X_i,I(W_i\le Z_i),W_iI(W_i\le Z_i)\right\}}\right]\\
&=&E\left(\Delta\frac{\blam_2\{Z,{\bb^*}\trans\X,I(W\le Z),WI(W\le Z)\}}{\lambda\{Z,{\bb^*}\trans\X,I(W\le Z),WI(W\le Z)\}}\right.\\
&&\quad\left.\otimes\left[\X_{l}-
\frac{E\left\{\X_{l}
	Y(Z)\mid{\bb^*}\trans\X,I(W_i\le Z_i),W_iI(W_i\le Z_i)\right\}}
{E\left\{Y(Z)\mid{\bb^*}\trans\X,I(W\le Z),WI(W\le Z)\right\}}\right]\right)\\
&&+o_p(1)
\ese
under Condition \ref{assum:kernel}-\ref{assum:bandwidth}. Thus,
for sufficient
large $n$
\bse
\0&=&\frac{1}{n}\sumi
\Delta_i\frac{\wh\blam_2\{Z_i,\wh\bb_n\trans\X_i,I(W_i\le Z_i),W_iI(W_i\le Z_i)\}}{\wh\lambda\{Z_i,\wh\bb_n\trans\X_i,I(W_i\le Z_i),W_iI(W_i\le Z_i)\}}\\
&&\quad\otimes\left[\X_{li}-
\frac{\wh E\left\{\X_{li}
	Y_i(Z_i)\mid\wh\bb_n\trans\X_i,I(W_i\le Z_i),W_iI(W_i\le Z_i)\right\}}
{\wh E\left\{Y_i(Z_i)\mid\wh\bb_n\trans\X_i,I(W_i\le Z_i),W_iI(W_i\le Z_i)\right\}}\right]\\
&=&\frac{1}{n}\sumi
\Delta_i\frac{\blam_2\{Z_i,\wh\bb_n\trans\X_i,I(W_i\le Z_i),W_iI(W_i\le Z_i)\}}{\lambda\{Z_i,\wh\bb_n\trans\X_i,I(W_i\le Z_i),W_iI(W_i\le Z_i)\}}\\
&&\quad\otimes\left[\X_{li}-
\frac{E\left\{\X_{li}
	Y_i(Z_i)\mid\wh\bb_n\trans\X_i,I(W_i\le Z_i),W_iI(W_i\le Z_i)\right\}}
{E\left\{Y_i(Z_i)\mid\wh\bb_n\trans\X_i,I(W_i\le Z_i),W_iI(W_i\le Z_i)\right\}}\right]+o_p(1)\\
&=&\frac{1}{n}\sumi
\Delta_i\frac{\blam_2\{Z_i,{\bb^*}\trans\X_i,I(W_i\le Z_i),W_iI(W_i\le Z_i)\}}{\lambda\{Z_i,{\bb^*}\trans\X_i,I(W_i\le Z_i),W_iI(W_i\le Z_i)\}}\\
&&\quad\otimes\left[\X_{li}-
\frac{E\left\{\X_{li}
	Y_i(Z_i)\mid{\bb^*}\trans\X_i,I(W_i\le Z_i),W_iI(W_i\le Z_i)\right\}}
{E\left\{Y_i(Z_i)\mid{\bb^*}\trans\X_i,I(W_i\le Z_i),W_iI(W_i\le Z_i)\right\}}\right]+o_p(1)\\
&=&E\left(\Delta\frac{\blam_2\{Z,{\bb^*}\trans\X,I(W\le Z), WI(W\le Z)\}}{\lambda\{Z,{\bb^*}\trans\X,I(W\le Z), WI(W\le Z)\}}\right.\\
&&\left.\quad\otimes\left[\X_{l}-
\frac{E\left\{\X_{l}
	Y(Z)\mid{\bb^*}\trans\X,I(W\le Z),WI(W\le Z)\right\}}
{E\left\{Y(Z)\mid{\bb^*}\trans\X,I(W\le Z),WI(W\le Z)\right\}}\right]\right)+o_p(1)
\ese
under conditions \ref{assum:kernel}-\ref{assum:bandwidth} and
\ref{assum:bounded}. Note that
\bse
&&E\left(\Delta\frac{\blam_2\{Z,{\bb^*}\trans\X,I(W\le Z), WI(W\le Z)\}}{\lambda\{Z,{\bb^*}\trans\X,I(W\le Z), WI(W\le Z)\}}\right.\\
&&\quad\left.\otimes\left[\X_{l}-
\frac{E\left\{\X_{l}
	Y(Z)\mid{\bb^*}\trans\X,I(W\le Z),WI(W\le Z)\right\}}
{E\left\{Y(Z)\mid{\bb^*}\trans\X,I(W\le Z),WI(W\le Z)\right\}}\right]\right)
\ese
is a nonrandom quantity that does not depend on $n$, hence it is zero.
Thus the uniqueness requirement in Condition \ref{assum:unique}
ensures that
$\bb^*=\bb_0$.

We show by contradiction that the subsequence that converges includes all but a
finite number of $n$'s. If this were not true,  we could
obtain an infinite sequence of $\wh\bb_n$'s that did not converge
to
$\bb^*$. As this infinite sequence was in a compact set $\cal B$, we
could
obtain another subsequence that converged, say to
$\bb^{**}\ne\bb^*$. Using prior derivations would lead to
$\bb^{**}=\bb_0$, a contradiction to
$\bb^{**}\ne\bb^*$.
Thus we conclude
$
\wh\bb-\bb_0\to\0
$
in probability when $n\to\infty$ under Conditions
\ref{assum:kernel}-\ref{assum:bounded}.
\qed

\subsection{Proof of Theorem \ref{th:eff}}\label{app:asympofb}

We only provide the proof concerning (\ref{eq:eff});
the result concerning (\ref{eq:general}) 
follows by using a similar and simpler proof.

We first expand (\ref{eq:eff}) as
\be
\0
&=&n^{-1/2}\sumi
\Delta_i\frac{\wh\blam_2\{Z_i,\wh\bb\trans\X_i,I(W_i\le Z_i),W_iI(W_i\le Z_i)\}}{\wh\lambda\{Z_i,\wh\bb\trans\X_i,I(W_i\le Z_i),W_iI(W_i\le Z_i)\}}\\
&&\quad\otimes\left[\X_{li}-
\frac{\wh E\left\{\X_{li}
	Y_i(Z_i)\mid\wh\bb\trans\X_i,I(W_i\le Z_i),W_iI(W_i\le Z_i)\right\}}
{\wh
	E\left\{Y_i(Z_i)\mid\wh\bb\trans\X_i,I(W_i\le Z_i),W_iI(W_i\le Z_i)\right\}}\right]\nonumber\\
&=&n^{-1/2}\sumi
\Delta_i\frac{\wh\blam_2\{Z_i,\bb_0\trans\X_i,I(W_i\le Z_i),W_iI(W_i\le Z_i)\}}{\wh\lambda\{Z_i,\bb_0\trans\X_i,I(W_i\le Z_i),W_iI(W_i\le Z_i)\}}\\
&&\quad\otimes\left[\X_{li}-
\frac{\wh E\left\{\X_{li}
	Y_i(Z_i)\mid\bb_0\trans\X_i,I(W_i\le Z_i),W_iI(W_i\le Z_i)\right\}}
{\wh
	E\left\{Y_i(Z_i)\mid\bb_0\trans\X_i,I(W_i\le Z_i),W_iI(W_i\le Z_i)\right\}}\right]\n\\\label{eq:main}\\
&&+\frac{1}{n}\sumi
\left\{\frac{\partial}{\partial(\X_i\trans\bb)}
\left(\Delta_i\frac{\wh\blam_2\{Z_i,\bb\trans\X_i,I(W_i\le Z_i),W_iI(W_i\le Z_i)\}}{\wh\lambda\{Z_i,\bb\trans\X_i,I(W_i\le Z_i),W_iI(W_i\le Z_i)\}}\right.\right.\n\\
&&\quad\left.\left.\otimes\left[\X_{li}-
\frac{\wh E\left\{\X_{li}
	Y_i(Z_i)\mid\bb\trans\X_i,I(W_i\le Z_i),W_iI(W_i\le Z_i)\right\}}
{\wh
	E\left\{Y_i(Z_i)\mid\bb\trans\X_i,I(W_i\le Z_i),W_iI(W_i\le Z_i)\right\}}\right]\right)\otimes\X_{li}\trans\right\}\Bigg|_{\bb=\wt\bb}\label{eq:easy}\\
&&\times\sqrt{n}(\wh\bb-\bb_0),\nonumber
\ee
where $\wt\bb$ is on the line connecting $\bb_0$ and $\wh\bb$.

We first consider (\ref{eq:easy}). Because of Theorem
\ref{th:consistency} and Lemma \ref{lem:pre},
\be
&&\frac{1}{n}\sumi
\left\{\frac{\partial}{\partial(\X_i\trans\bb)}
\left(\Delta_i\frac{\wh\blam_2\{Z_i,\bb\trans\X_i,I(W_i\le Z_i),W_iI(W_i\le Z_i)\}}{\wh\lambda\{Z_i,\bb\trans\X_i,I(W_i\le Z_i),W_iI(W_i\le Z_i)\}}\right.\right.\n\\
&&\quad\left.\left.\otimes\left[\X_{li}-
\frac{\wh E\left\{\X_{li}
	Y_i(Z_i)\mid\bb\trans\X_i,I(W_i\le Z_i),W_iI(W_i\le Z_i)\right\}}
{\wh
	E\left\{Y_i(Z_i)\mid\bb\trans\X_i,I(W_i\le Z_i),W_iI(W_i\le Z_i)\right\}}\right]\right)\otimes\X_{li}\trans\right\}\Bigg|_{\bb=\wt\bb}\n\\
&=&\frac{1}{n}\sumi
\left\{\frac{\partial}{\partial(\X_i\trans\bb_0)}
\left(\Delta_i\frac{\wh\blam_2\{Z_i,\bb_0\trans\X_i,I(W_i\le Z_i),W_iI(W_i\le Z_i)\}}{\wh\lambda\{Z_i,\bb_0\trans\X_i,I(W_i\le Z_i),W_iI(W_i\le Z_i)\}}\right.\right.\n\\
&&\quad\left.\left.\otimes\left[\X_{li}-
\frac{\wh E\left\{\X_{li}
	Y_i(Z_i)\mid\bb_0\trans\X_i,I(W_i\le Z_i),W_iI(W_i\le Z_i)\right\}}
{\wh
	E\left\{Y_i(Z_i)\mid\bb_0\trans\X_i,I(W_i\le Z_i),W_iI(W_i\le Z_i)\right\}}\right]\right)\otimes\X_{li}\trans\right\}+o_p(1)\n\\
&=&-\frac{1}{n}\sumi \left(\Delta_i
\frac{\wh\blam_2^{\otimes2}\{Z_i,\bb_0\trans\X_i,I(W_i\le Z_i),W_iI(W_i\le Z_i)\}}{\wh\lambda^2\{Z_i,\bb_0\trans\X_i,I(W_i\le Z_i),W_iI(W_i\le Z_i)\}}\right.\n\\
&&\left.\quad\otimes\left[\X_{li}-
\frac{\wh E\left\{\X_{li}
	Y_i(Z_i)\mid\bb_0\trans\X_i,I(W_i\le Z_i),W_iI(W_i\le Z_i)\right\}}
{\wh
	E\left\{Y_i(Z_i)\mid\bb_0\trans\X_i,I(W_i\le Z_i),W_iI(W_i\le Z_i)\right\}}\right]\otimes\X_{li}\trans\right)\n\\\label{eq:easy1}\\
&&+\frac{1}{n}\sumi
\frac{\Delta_i}{\wh\lambda\{Z_i,\bb_0\trans\X_i,I(W_i\le Z_i),W_iI(W_i\le Z_i)\}}\n\\
&&\quad\frac{\partial}{\partial(\X_i\trans\bb_0)}
\left(\wh\blam_2\{Z_i,\bb_0\trans\X_i,I(W_i\le Z_i),W_iI(W_i\le Z_i)\}\right.\n\\
&&\quad\left.\otimes\left[\X_{li}-
\frac{\wh E\left\{\X_{li}
	Y_i(Z_i)\mid\bb_0\trans\X_i,I(W_i\le Z_i),W_iI(W_i\le Z_i)\right\}}
{\wh
	E\left\{Y_i(Z_i)\mid\bb_0\trans\X_i,I(W_i\le Z_i),W_iI(W_i\le Z_i)\right\}}\right]\right)\otimes\X_{li}\trans\n\\
\label{eq:easy2}\\
&&+o_p(1)\nonumber.
\ee
Because of Lemma \ref{lem:pre}, (\ref{eq:easy1}) converges
uniformly
in probability to
\bse
&&-E \left(\int
\frac{\blam_2^{\otimes2}\{s,\bb\trans\X,I(W\le s),WI(W\le s)\}}{\lambda^2\{s,\bb\trans\X,I(W\le s),WI(W\le s)\}}\right.\\
&&\left.\quad\otimes\left[\X_l-
\frac{ E\left\{\X_l
	Y(s)\mid\bb_0\trans\X,I(W\le s),WI(W\le s)\right\}}
{E\left\{Y(s)\mid\bb_0\trans\X,I(W\le s),WI(W\le s)\right\}}\right]\otimes\X_l\trans
dN(s)\right)\\
&=&-E \left(\int
\frac{\blam_2^{\otimes2}\{s,\bb\trans\X,I(W\le s),WI(W\le s)\}}{\lambda^2\{s,\bb\trans\X,I(W\le s),WI(W\le s)\}}\right.\\
&&\left.\otimes\left[\X_l-
\frac{ E\left\{\X_l
	Y(s)\mid\bb_0\trans\X,I(W\le s),WI(W\le s)\right\}}
{E\left\{Y(s)\mid\bb_0\trans\X,I(W\le s),WI(W\le s)\right\}}\right]\right.\\
&&\left.\otimes\X_l\trans
Y(s)\lambda\{s,\bb\trans\X,I(W\le s),WI(W\le s)\}ds\right)\\
&=&-E \left(\int
\frac{\blam_2^{\otimes2}\{s,\bb\trans\X,I(W\le s),WI(W\le s)\}}{\lambda\{s,\bb\trans\X,I(W\le s),WI(W\le s)\}}\right.\\
&&\left.\otimes\left[\X_l-
\frac{ E\left\{\X_l
	Y(s)\mid\bb_0\trans\X,I(W\le s),WI(W\le s)\right\}}
{E\left\{Y(s)\mid\bb_0\trans\X,I(W\le s),WI(W\le s)\right\}}\right]\right.\\
&&\left.\otimes
\left[\X_l
-\frac{ E\left\{\X_l
	Y(s)\mid\bb_0\trans\X,I(W\le s),WI(W\le s)\right\}}
{E\left\{Y(s)\mid\bb_0\trans\X,I(W\le s),WI(W\le s)\right\}}\right]
\trans
Y(s)ds\right)\\
&&-E \left(\int
\frac{\blam_2^{\otimes2}\{s,\bb\trans\X,I(W\le s),WI(W\le s)\}}{\lambda\{s,\bb\trans\X,I(W\le s),WI(W\le s)\}}\right.\\
&&\left.\otimes\left[\X_l-
\frac{ E\left\{\X_l
	Y(s)\mid\bb_0\trans\X,I(W\le s),WI(W\le s)\right\}}
{E\left\{Y(s)\mid\bb_0\trans\X,I(W\le s),WI(W\le s)\right\}}\right]\right.\\
&&\quad\left.\otimes
\frac{ E\left\{\X_l
	Y(s)\mid\bb_0\trans\X,I(W\le s),WI(W\le s)\right\}}
{E\left\{Y(s)\mid\bb_0\trans\X,I(W\le s),WI(W\le s)\right\}}\trans
Y(s)ds\right)\\
&=&-E\{\bS\eff(\Delta,Z,\X)^{\otimes2}\},
\ese
where the last equality is due to that the second term above is zero
by
first taking expectation conditional on $\bb_0\trans\X$.

Similarly, from Lemma \ref{lem:pre},  the term in
(\ref{eq:easy2})
converges uniformly in probability to the
limit of
\bse
&&E\left\{\frac{\Delta_i}{\lambda\{Z_i,\bb_0\trans\X_i,I(W_i\le Z_i),W_iI(W_i\le Z_i)\}}
\right.\\
&&\left.\times\frac{\partial}{\partial(\X_i\trans\bb_0)}
\left(\wh\blam_2\{Z_i,\bb_0\trans\X_i,I(W_i\le Z_i),W_iI(W_i\le Z_i)\}\right.\right.\\
&&\left.\left.\otimes \left[\X_{li}-
\frac{ E\left\{\X_{li} Y_i(Z_i)\mid\bb_0\trans\X_i,I(W_i\le Z_i),W_iI(W_i\le Z_i)\right\}}
{E\left\{Y_i(Z_i)\mid\bb_0\trans\X_i,I(W_i\le Z_i),W_iI(W_i\le Z_i)\right\}}
\right]
\right)\otimes\X_{li}\trans\right\}.
\ese
Now let $\wh\blam_{2,-i}(Z, \bb_0\trans\X)$ be the
leave-one-out version of $\wh\blam_2(Z, \bb_0\trans\X)$, i.e.
it
is constructed the same as $\wh\blam_2(Z, \bb_0\trans\X)$
except
that the $i$th observation is not used.
Obviously,
\bse
&&\frac{\Delta_i}{\lambda\{Z_i,\bb_0\trans\X_i,I(W_i\le Z_i),W_iI(W_i\le Z_i)\}}\\
&&\times\frac{\partial}{\partial(\X_i\trans\bb_0)}
\left(\wh\blam_2\{Z_i,\bb_0\trans\X_i,I(W_i\le Z_i),W_iI(W_i\le Z_i)\}\right.\\
&&\quad\left.\otimes \left[\X_{li}-
\frac{ E\left\{\X_{li}
	Y_i(Z_i)\mid\bb_0\trans\X_i,I(W_i\le Z_i),W_iI(W_i\le Z_i)\right\}}
{E\left\{Y_i(Z_i)\mid\bb_0\trans\X_i,I(W_i\le Z_i),W_iI(W_i\le Z_i)\right\}}\right]\right)\otimes\X_{li}\trans\\
&-&\frac{\Delta_i}{\lambda\{Z_i,\bb_0\trans\X_i,I(W_i\le Z_i),W_iI(W_i\le Z_i)\}}\\
&&\times\frac{\partial}{\partial(\X_i\trans\bb_0)}
\left(\wh\blam_{2,-i}\{Z_i,\bb_0\trans\X_i,I(W_i\le Z_i),W_iI(W_i\le Z_i)\}\right.\\
&&\quad\left.\otimes \left[\X_{li}-
\frac{ E\left\{\X_{li}
	Y_i(Z_i)\mid\bb_0\trans\X_i,I(W_i\le Z_i),W_iI(W_i\le Z_i)\right\}}
{E\left\{Y_i(Z_i)\mid\bb_0\trans\X_i,I(W_i\le Z_i),W_iI(W_i\le Z_i)\right\}}\right]\right)\otimes\X_{li}\trans\\
&=&o_p(1).
\ese
Let $E_i$ mean taking expectation with respect to the $i$th
observation conditional on all other observations, then
\bse
&&
E_i\left\{\frac{\Delta_i}{\lambda\{Z_i,\bb_0\trans\X_i,I(W_i\le Z_i),W_iI(W_i\le Z_i)\}}
\frac{\partial}{\partial(\X_i\trans\bb_0)}\right.\\
&&\left.\times\left(\wh\blam_{2,-i}\{Z_i,\bb_0\trans\X_i,I(W_i\le Z_i),W_iI(W_i\le Z_i)\}\right.\right.\\
&&\quad\left.\left.\otimes \left[\X_{li}-
\frac{ E\left\{\X_{li}
	Y_i(Z_i)\mid\bb_0\trans\X_i,I(W_i\le Z_i),W_iI(W_i\le Z_i)\right\}}
{E\left\{Y_i(Z_i)\mid\bb_0\trans\X_i,I(W_i\le Z_i),W_iI(W_i\le Z_i)\right\}}\right]\right)\otimes\X_{li}\trans\right\}\\
&=&E_i \left\{
\frac{\partial}{\partial\bb_0}\int
\wh\blam_{2,-i}\{s,\bb_0\trans\X_i,I(W_i\le s),W_iI(W_i\le s)\}\right.\\
&&\left.\quad\otimes \left[\X_{li}-
\frac{ E\left\{\X_{li}
	Y_i(s)\mid\bb_0\trans\X_i,I(W_i\le s),W_iI(W_i\le s)\right\}}
{E\left\{Y_i(s)\mid\bb_0\trans\X_i,I(W_i\le s),W_iI(W_i\le s)\right\}}\right]
E\{Y_i(s)\mid\X_i\}ds\right\}\\
&=&\frac{\partial}{\partial\bb_0} E_i \left\{
\int
\wh\blam_{2,-i}\{s,\bb_0\trans\X_i,I(W_i\le s),W_iI(W_i\le s)\}\right.\\
&&\quad\left.\otimes \left[\X_{li}-
\frac{ E\left\{\X_{li}
	Y_i(s)\mid\bb_0\trans\X_i,I(W_i\le s),W_iI(W_i\le s)\right\}}
{E\left\{Y_i(s)\mid\bb_0\trans\X_i,I(W_i\le s),W_iI(W_i\le s)\right\}}\right]
Y_i(s)ds\right\}\\
&=&\0.
\ese
The last equality  holds because the integrand has expectation zero
conditional on $\bb_0\trans\X_i$ and all other observations, and the third to last
equality follows because
the expectation is with respect to $\X_i$ and does not involve
$\bb_0$.
Therefore, the term in (\ref{eq:easy2}) converges in probability
uniformly to
\bse
&&E\left\{\frac{\Delta_i}{\lambda\{Z_i,\bb_0\trans\X_i,I(W_i\le Z_i),W_iI(W_i\le Z_i)\}}\right.\\
&&\quad\times\frac{\partial}{\partial(\X_i\trans\bb_0)}
\left(\wh\blam_{2,-i}\{Z_i,\bb_0\trans\X_i,I(W_i\le Z_i),W_iI(W_i\le Z_i)\}\right.\\
&&\quad\left.\left.\otimes \left[\X_{li}-
\frac{ E\left\{\X_{li}
	Y_i(Z_i)\mid\bb_0\trans\X_i,I(W_i\le Z_i),W_iI(W_i\le Z_i)\right\}}
{E\left\{Y_i(Z_i)\mid\bb_0\trans\X_i,I(W_i\le Z_i),W_iI(W_i\le Z_i)\right\}}\right]\right)\otimes\X_{li}\trans\right\}
=0
\ese
Combining the results concerning (\ref{eq:easy1}) and
(\ref{eq:easy2}), thus the expression in
(\ref{eq:easy})
is

\noindent$-E\{\bS\eff(\Delta,Z,\X)^{\otimes2}\}+o_p(1)$.

Next we decompose (\ref{eq:main}) into
\be\label{eq:Ts}
&&n^{-1/2}\sumi
\Delta_i\frac{\wh\blam_2\{Z_i,\bb_0\trans\X_i,I(W_i\le Z_i),W_iI(W_i\le Z_i)\}}{\wh\lambda\{Z_i,\bb_0\trans\X_i,I(W_i\le Z_i),W_iI(W_i\le Z_i)\}}\n\\
&&\quad\otimes\left[\X_{li}-
\frac{\wh E\left\{\X_{li}
	Y_i(Z_i)\mid\bb_0\trans\X_i,I(W_i\le Z_i),W_iI(W_i\le Z_i)\right\}}
{\wh E\left\{Y_i(Z_i)\mid\bb_0\trans\X_i,I(W_i\le Z_i),W_iI(W_i\le Z_i)\right\}}\right]\n\\
&=&\T_1+\T_2+\T_3+\T_4,
\ee
where
\bse
\T_1
&=&n^{-1/2}\sumi
\Delta_i\frac{\blam_2\{Z_i,\bb_0\trans\X_i,I(W_i\le Z_i),W_iI(W_i\le Z_i)\}}{\lambda\{Z_i,\bb_0\trans\X_i,I(W_i\le Z_i),W_iI(W_i\le Z_i)\}}\\
&&\otimes\left[\X_{li}-
\frac{E\left\{\X_{li}
	Y_i(Z_i)\mid\bb_0\trans\X_i,I(W_i\le Z_i),W_iI(W_i\le Z_i)\right\}}
{E\left\{Y_i(Z_i)\mid\bb_0\trans\X_i,I(W_i\le Z_i),W_iI(W_i\le Z_i)\right\}}\right],\\
\T_2&=&n^{-1/2}\sumi
\Delta_i\left[\frac{\wh\blam_2\{Z_i,\bb_0\trans\X_i,I(W_i\le Z_i),W_iI(W_i\le Z_i)\}}{\wh\lambda\{Z_i,\bb_0\trans\X_i,I(W_i\le Z_i),W_iI(W_i\le Z_i)\}}\right.\\
&&\left.-\frac{\blam_2\{Z_i,\bb_0\trans\X_i,I(W_i\le Z_i),W_iI(W_i\le Z_i)\}}{\lambda\{Z_i,\bb_0\trans\X_i,I(W_i\le Z_i),W_iI(W_i\le Z_i)\}}\right]\\
&&\otimes\left[\X_{li}-
\frac{E\left\{\X_{li}
	Y_i(Z_i)\mid\bb_0\trans\X_i,I(W_i\le Z_i),W_iI(W_i\le Z_i)\right\}}
{E\left\{Y_i(Z_i)\mid\bb_0\trans\X_i,I(W_i\le Z_i),W_iI(W_i\le Z_i)\right\}}\right],\\
\T_3&=&n^{-1/2}\sumi
\Delta_i\frac{\blam_2\{Z_i,\bb_0\trans\X_i,I(W_i\le Z_i),W_iI(W_i\le Z_i)\}}{\lambda\{Z_i,\bb_0\trans\X_i,I(W_i\le Z_i),W_iI(W_i\le Z_i)\}}\\
&&\otimes\left[\frac{E\left\{\X_{li}
	Y_i(Z_i)\mid\bb_0\trans\X_i,I(W_i\le Z_i),W_iI(W_i\le Z_i)\right\}}
{E\left\{Y_i(Z_i)\mid\bb_0\trans\X_i,I(W_i\le Z_i),W_iI(W_i\le Z_i)\right\}}\right.\\
&&\quad\left.-
\frac{\wh E\left\{\X_{li}
	Y_i(Z_i)\mid\bb_0\trans\X_i,I(W_i\le Z_i),W_iI(W_i\le Z_i)\right\}}
{\wh
	E\left\{Y_i(Z_i)\mid\bb_0\trans\X_i,I(W_i\le Z_i),W_iI(W_i\le Z_i)\right\}}\right],\\
\T_4&=&n^{-1/2}\sumi \Delta_i
\left[\frac{\wh\blam_2\{Z_i,\bb_0\trans\X_i,I(W_i\le Z_i),W_iI(W_i\le Z_i)\}}{\wh\lambda\{Z_i,\bb_0\trans\X_i,I(W_i\le Z_i),W_iI(W_i\le Z_i)\}}\right.\\
&&\left.-
\frac{\blam_2\{Z_i,\bb_0\trans\X_i,I(W_i\le Z_i),W_iI(W_i\le Z_i)\}}{\lambda\{Z_i,\bb_0\trans\X_i,I(W_i\le Z_i),W_iI(W_i\le Z_i)\}}\right]\\
&&\otimes\left[\frac{E\left\{\X_{li}
	Y_i(Z_i)\mid\bb_0\trans\X_i,I(W_i\le Z_i),W_iI(W_i\le Z_i)\right\}}
{E\left\{Y_i(Z_i)\mid\bb_0\trans\X_i,I(W_i\le Z_i),W_iI(W_i\le Z_i)\right\}}\right.\\
&&\quad\left.-
\frac{\wh E\left\{\X_{li}
	Y_i(Z_i)\mid\bb_0\trans\X_i,I(W_i\le Z_i),W_iI(W_i\le Z_i)\right\}}
{\wh
	E\left\{Y_i(Z_i)\mid\bb_0\trans\X_i,I(W_i\le Z_i),W_iI(W_i\le Z_i)\right\}}\right].
\ese

First,
\bse
\T_2&=&n^{-1/2}\sumi
\int\left[\frac{\wh\blam_2\{s,\bb_0\trans\X_i,I(W_i\le s),W_iI(W_i\le s)\}}{\wh\lambda\{s,\bb_0\trans\X_i,I(W_i\le s),W_iI(W_i\le s)\}}\right.\\
&&\quad\left.-\frac{\blam_2\{s,\bb_0\trans\X_i,I(W_i\le s),W_iI(W_i\le s)\}}{\lambda\{s,\bb_0\trans\X_i,I(W_i\le s),W_iI(W_i\le s)\}}\right]\\
&&\otimes\left[\X_{li}-
\frac{E\left\{\X_{li}
	Y_i(s)\mid\bb_0\trans\X_i,I(W_i\le s),W_iI(W_i\le s)\right\}}
{E\left\{Y_i(s)\mid\bb_0\trans\X_i,I(W_i\le s),W_iI(W_i\le s)\right\}}\right]dN_i(s)\\
&=&o_p\left(n^{-1/2}\sumi \int\left[\X_{li}-
\frac{E\left\{\X_{li}
	Y_i(s)\mid\bb_0\trans\X_i,I(W_i\le s),W_iI(W_i\le s)\right\}}
{E\left\{Y_i(s)\mid\bb_0\trans\X_i,I(W_i\le s),W_iI(W_i\le s)\right\}}\right]\right.\\
&&\quad\left.\times Y_i(s)\lambda(s,\bb_0\trans\X_{li})ds\right)\\
&=&o_p(1),
\ese
where the last equality holds  because the quantity inside the parentheses is
a mean zero normal random
quantity of order $O_p(1)$. Further,
\bse
&&\T_3\\
&=&n^{-1/2}\sumi
\Delta_i\frac{\blam_2\{Z_i,\bb_0\trans\X_i,I(W_i\le Z_i),W_iI(W_i\le Z_i)\}}{\lambda\{Z_i,\bb_0\trans\X_i,I(W_i\le Z_i),W_iI(W_i\le Z_i)\}}\\
&&\otimes\left(-
\frac{\wh E\left\{\X_{li}
	Y_i(Z_i)\mid\bb_0\trans\X_i,I(W_i\le Z_i),W_iI(W_i\le Z_i)\right\}}{E\left\{Y_i(Z_i)\mid\bb_0\trans\X_i,I(W_i\le Z_i),W_iI(W_i\le Z_i)\right\}}\right.\n\\
&&+\frac{\wh
	E\left\{Y_i(Z_i)\mid\bb_0\trans\X_i,I(W_i\le Z_i),W_iI(W_i\le Z_i)\right\}}
{[E\left\{Y_i(Z_i)\mid\bb_0\trans\X_i,I(W_i\le Z_i),W_iI(W_i\le Z_i)\right\}]}\\
&&\left. \times\frac{E\left\{\X_{li}Y_i(Z_i)\mid\bb_0\trans\X_i,I(W_i\le Z_i),W_iI(W_i\le Z_i)\right\}}
{[E\left\{Y_i(Z_i)\mid\bb_0\trans\X_i,I(W_i\le Z_i),W_iI(W_i\le Z_i)\right\}]}\right)
+o_p(1)\n\\
&=&
n^{-1/2}\sumi
\Delta_i\frac{\blam_2\{Z_i,\bb_0\trans\X_i,I(W_i\le Z_i),W_iI(W_i\le Z_i)\}}{\lambda\{Z_i,\bb_0\trans\X_i,I(W_i\le Z_i),W_iI(W_i\le Z_i)\}}\\
&&\otimes\left(-
\frac{
	n^{-1}\sumj
	K_h(\bb_0\trans\X_j-\bb_0\trans\X_i)\X_{lj}I(Z_j\ge
	Z_i)}{f_{\bb_0\trans\X}(\bb_0\trans\X_i)
	E\left\{Y_i(Z_i)\mid\bb_0\trans\X_i,I(W_i\le Z_i),W_iI(W_i\le Z_i)\right\}}\right.\n\\
&&\left.+\frac{E\left\{\X_{li}Y_i(Z_i)\mid\bb_0\trans\X_i,I(W_i\le Z_i),W_iI(W_i\le Z_i)\right\}}{f_{\bb_0\trans\X}(\bb_0\trans\X_i)
	E\left\{Y_i(Z_i)\mid\bb_0\trans\X_i,I(W_i\le Z_i),W_iI(W_i\le Z_i)\right\}}\right.\n\\
&&\times \left\{n^{-1}\sumj
K_h(\bb_0\trans\X_j-\bb_0\trans\X_i)-f_{\bb_0\trans\X}(\bb_0\trans\X_i)
\right\}\\
&&+\frac{E\left\{\X_{li}Y_i(Z_i)\mid\bb_0\trans\X_i,I(W_i\le Z_i),W_iI(W_i\le Z_i)\right\}}
{[E\left\{Y_i(Z_i)\mid\bb_0\trans\X_i,I(W_i\le Z_i),W_iI(W_i\le Z_i)\right\}]^2}
\\
&&\quad\otimes \left[\frac{
	n^{-1}\sumj K_h(\bb_0\trans\X_j-\bb_0\trans\X_i)I(Z_j\ge
	Z_i)}{f_{\bb_0\trans\X}(\bb_0\trans\X_i)}\right.\\
&&\left.-\frac{E\left\{Y_i(Z_i)\mid\bb_0\trans\X_i,I(W_i\le Z_i),W_iI(W_i\le Z_i)\right\}\}}{f_{\bb_0\trans\X}(\bb_0\trans\X_i) }\right.\n\\
&&\left.\left.\times\frac{\{n^{-1}\sumj
	K_h(\bb_0\trans\X_j-\bb_0\trans\X_i)-f_{\bb_0\trans\X}(\bb_0\trans\X_i)
	\}}{f_{\bb_0\trans\X}(\bb_0\trans\X_i) }\right]
\right) +o_p(1)\n\\
&=&n^{-3/2}\sumi\sumj
\Delta_i\frac{\blam_2\{Z_i,\bb_0\trans\X_i,I(W_i\le Z_i),W_iI(W_i\le Z_i)\}}{\lambda\{Z_i,\bb_0\trans\X_i,I(W_i\le Z_i),W_iI(W_i\le Z_i)\}}\\
&&\otimes\left[
-\frac{K_h(\bb_0\trans\X_j-\bb_0\trans\X_i)\X_{lj}I(Z_j\ge
	Z_i)}{f_{\bb_0\trans\X}(\bb_0\trans\X_i)
	E\left\{Y_i(Z_i)\mid\bb_0\trans\X_i,I(W_i\le Z_i),W_iI(W_i\le Z_i)\right\}}\right.\n\\
&&\left. +
\frac{E\left\{\X_{li}Y_i(Z_i)\mid\bb_0\trans\X_i,W_i\right\}
	K_h(\bb_0\trans\X_j-\bb_0\trans\X_i)I(Z_j\ge
	Z_i)}{f_{\bb_0\trans\X}(\bb_0\trans\X_i)
	[E\left\{Y_i(Z_i)\mid\bb_0\trans\X_i,I(W_i\le Z_i),W_iI(W_i\le Z_i)\right\}]^2
}\right] +o_p(1)\n\\
&=&\T_{31}+\T_{32}+\T_{33}+o_p(1),
\ese
where
\bse
\T_{31}&=&n^{-1/2}\sumi\Delta_i
\frac{\blam_2\{Z_i,\bb_0\trans\X_i,I(W_i\le Z_i),W_iI(W_i\le Z_i)\}}{\lambda\{Z_i,\bb_0\trans\X_i,I(W_i\le Z_i),W_iI(W_i\le Z_i)\}}\\
&&\otimes E\left[
-\frac{K_h(\bb_0\trans\X_j-\bb_0\trans\X_i)\X_{lj}I(Z_j\ge
	Z_i)}{f_{\bb_0\trans\X}(\bb_0\trans\X_i)
	E\left\{Y_i(Z_i)\mid\bb_0\trans\X_i,I(W_i\le Z_i),W_iI(W_i\le Z_i)\right\}}\right.\n\\
&&\quad\left. +
\frac{E\left\{\X_{li}Y_i(Z_i)\mid\bb_0\trans\X_i,W_i\right\}
	K_h(\bb_0\trans\X_j-\bb_0\trans\X_i)I(Z_j\ge
	Z_i)}{f_{\bb_0\trans\X}(\bb_0\trans\X_i)
	[E\left\{Y_i(Z_i)\mid\bb_0\trans\X_i,I(W_i\le Z_i),W_iI(W_i\le Z_i)\right\}]^2
}\right.\\
&&\left.\mid \Delta_i, Z_i, \X_i,I(W_i\le Z_i),W_iI(W_i\le Z_i)\right]\\
\T_{32}&=&n^{-1/2}\sumj E
\left(\Delta_i\frac{\blam_2\{Z_i,\bb_0\trans\X_i,I(W_i\le Z_i),W_iI(W_i\le Z_i)\}}{\lambda\{Z_i,\bb_0\trans\X_i,I(W_i\le Z_i),W_iI(W_i\le Z_i)\}}\right.\\
&&\left.\otimes\left[
-\frac{K_h(\bb_0\trans\X_j-\bb_0\trans\X_i)\X_{lj}I(Z_j\ge
	Z_i)}{f_{\bb_0\trans\X}(\bb_0\trans\X_i)
	E\left\{Y_i(Z_i)\mid\bb_0\trans\X_i,I(W_i\le Z_i),W_iI(W_i\le Z_i)\right\}}\right.\right.\n\\
&&\quad+
\frac{E\left\{\X_{li}Y_i(Z_i)\mid\bb_0\trans\X_i,I(W_i\le Z_i),W_iI(W_i\le Z_i)\right\}
}{f_{\bb_0\trans\X}(\bb_0\trans\X_i)
	[E\left\{Y_i(Z_i)\mid\bb_0\trans\X_i,I(W_i\le Z_i),W_iI(W_i\le Z_i)\right\}]^2
}\\
&&\left.\left.\quad\times K_h(\bb_0\trans\X_j-\bb_0\trans\X_i)I(Z_j\ge
Z_i)\right]\mid \Delta_j, Z_j,\X_j,I(W_i\le Z_i),W_iI(W_i\le Z_i)\right)\\
\T_{33}&=&-n^{1/2}E\left(\Delta_i
\frac{\blam_2\{Z_i,\bb_0\trans\X_i,I(W_i\le Z_i),W_iI(W_i\le Z_i)\}}{\lambda\{Z_i,\bb_0\trans\X_i,I(W_i\le Z_i),W_iI(W_i\le Z_i)\}}\right.\\
&&\left.\otimes E\left[
-\frac{K_h(\bb_0\trans\X_j-\bb_0\trans\X_i)\X_{lj}I(Z_j\ge
	Z_i)}{f_{\bb_0\trans\X}(\bb_0\trans\X_i)
	E\left\{Y_i(Z_i)\mid\bb_0\trans\X_i,I(W_i\le Z_i),W_iI(W_i\le Z_i)\right\}}\right.\right.\n\\
&&\quad\left.\left. +
\frac{E\left\{\X_{li}Y_i(Z_i)\mid\bb_0\trans\X_i,I(W_i\le Z_i),W_iI(W_i\le Z_i)\right\}
	K_h(\bb_0\trans\X_j-\bb_0\trans\X_i)I(Z_j\ge
	Z_i)}{f_{\bb_0\trans\X}(\bb_0\trans\X_i)
	[E\left\{Y_i(Z_i)\mid\bb_0\trans\X_i,I(W_i\le Z_i),W_iI(W_i\le Z_i)\right\}]^2
}\right]\right).
\ese
Here we used U-statistic property in the last equality.
Now when
$nh^4\to0$,
\bse
\T_{31}&=&n^{-1/2}\sumi\Delta_i
\frac{\blam_2\{Z_i,\bb_0\trans\X_i,I(W_i\le Z_i),W_iI(W_i\le Z_i)\}}{\lambda\{Z_i,\bb_0\trans\X_i,I(W_i\le Z_i),W_iI(W_i\le Z_i)\}}\\
&&\otimes \left[
-\frac{E\{\X_{li}Y_i(Z_i)\mid
	\bb_0\trans\X_i,I(W_i\le Z_i),W_iI(W_i\le Z_i)\}}{E\left\{Y_i(Z_i)\mid\bb_0\trans\X_i,I(W_i\le Z_i),W_iI(W_i\le Z_i)\right\}}\right.\\
&&+
\frac{E\left\{\X_{li}Y_i(Z_i)\mid\bb_0\trans\X_i,I(W_i\le Z_i),W_iI(W_i\le Z_i)\right\}}{
	[E\left\{Y_i(Z_i)\mid\bb_0\trans\X_i,I(W_i\le Z_i),W_iI(W_i\le Z_i)\right\}]^2
}\\
&&\left.\times E\left\{Y_i(Z_i)\mid\bb_0\trans\X_i,I(W_i\le Z_i),W_iI(W_i\le Z_i)\right\}\right]+O(n^{1/2}h^2)\\
&=&o_p(1).
\ese
Thus, $\T_{33}=o_p(1)$ as well.
Also,
\bse
&&\T_{32}\\
&=&n^{-1/2}\sumj E
\left(\Delta_i\frac{\blam_2\{Z_i,\bb_0\trans\X_i,I(W_i\le Z_i),W_iI(W_i\le Z_i)\}}{\lambda\{Z_i,\bb_0\trans\X_i,I(W_i\le Z_i),W_iI(W_i\le Z_i)\}}\right.\\
&&\left.\otimes\left[
-\frac{K_h(\bb_0\trans\X_j-\bb_0\trans\X_i)\X_{lj}I(Z_j\ge
	Z_i)}{f_{\bb_0\trans\X}(\bb_0\trans\X_i) E\left\{I(Z\ge
	Z_i)\mid\bb_0\trans\X=\bb_0\trans\X_i,Z_i,
	I(W_i\le Z_i),W_iI(W_i\le Z_i)\right\}}\right.\right.\n\\
&&\quad\left. \left.+
\frac{E\left\{\X_{l}I(Z\ge
	Z_i)\mid\bb_0\trans\X=\bb_0\trans\X_i, Z_i,I(W_i\le Z_i),W_iI(W_i\le Z_i)\right\}
	K_h(\bb_0\trans\X_j-\bb_0\trans\X_i)I(Z_j\ge
	Z_i)}{f_{\bb_0\trans\X}(\bb_0\trans\X_i) [E\left\{I(Z\ge
	Z_i)\mid\bb_0\trans\X=\bb_0\trans\X_i, Z_i,I(W_i\le Z_i),W_iI(W_i\le Z_i)\right\}]^2
}\right]\right.\\
&&\left.\mid \Delta_j, Z_j,\X_j,I(W_i\le Z_i),W_iI(W_i\le Z_i)\right)\\
&=&n^{-1/2}\sumj E\left\{E
\left(\Delta_i\frac{\blam_2\{Z_i,\bb_0\trans\X_i,I(W_i\le Z_i),W_iI(W_i\le Z_i)\}}{\lambda\{Z_i,\bb_0\trans\X_i,I(W_i\le Z_i),W_iI(W_i\le Z_i)\}}\right.\right.\\
&&\left.\left.\otimes\left[
-\frac{\x_{lj}I(z_j\ge
	Z_i)}{f_{\bb_0\trans\X}(\bb_0\trans\X_i) E\left\{I(Z\ge
	Z_i)\mid\bb_0\trans\X=\bb_0\trans\X_i,
	Z_i,I(W_i\le Z_i),W_iI(W_i\le Z_i)\right\}}\right.\right.\right.\n\\
&&\left.\left.+
\frac{E\left\{\X_{l}I(Z\ge
	Z_i)\mid\bb_0\trans\X=\bb_0\trans\X_i, Z_i,I(W_i\le Z_i),W_iI(W_i\le Z_i)\right\}
	I(z_j\ge
	Z_i)}{f_{\bb_0\trans\X}(\bb_0\trans\X_i) [E\left\{I(Z\ge
	Z_i)\mid\bb_0\trans\X=\bb_0\trans\X_i, Z_i,I(W_i\le Z_i),W_iI(W_i\le Z_i)\right\}]^2
}\right]\right.\\
&&\left.\left.\mid \bb_0\trans\X_i,I(W_i\le Z_i),W_iI(W_i\le Z_i)\right)K_h(\bb_0\trans\x_j-\bb_0\trans\X_i)\right\}\\
&=&n^{-1/2}\sumj
E\left(\int_0^{z_j}\frac{\blam_2\{s,\bb_0\trans\x_j,I(w_i\le s),w_iI(w_i\le s)\}}{E\left\{S_c(s,\X)\mid\bb_0\trans\X=\bb_0\trans\x_j\right\}}\right.\\
&&\left.\otimes\left[
\frac{E
	\left\{\X_lS_c(s,\X)\mid\bb_0\trans\X=\bb_0\trans\x_j\right\}
}{E\left\{S_c(s,\X)\mid\bb_0\trans\X=\bb_0\trans\x_j\right\}
}-\x_{lj}\right]
S_c(s,\X_i)ds
\mid\bb_0\trans\X_i=\bb_0\trans\x_j\right) +O_p(n^{1/2}h^2)\\
&=&n^{-1/2}\sumj \int
Y_j(s)\lambda\{s,\bb_0\trans\x_j,I(w_i\le s),w_iI(w_i\le s)\}\frac{\blam_2\{s,\bb_0\trans\x_j,I(w_i\le s),w_iI(w_i\le s)\}}{\lambda\{s,\bb_0\trans\x_j,I(w_i\le s),w_iI(w_i\le s)\}}\\
&&\otimes\left[
\frac{E\left\{\X_{lj}Y_j(s)\mid\bb_0\trans\x_j,I(w_i\le s),w_iI(w_i\le s)\right\}
}{E\left\{Y_j(s)\mid\bb_0\trans\x_j,I(w_i\le s),w_iI(w_i\le s)\right\}
}-\x_{lj}\right]ds +O_p(n^{1/2}h^2).
\ese
When $nh^4\to0$, plugging the results of $\T_1$ and $\T_{32}$ to
(\ref{eq:Ts}), the expression in (\ref{eq:main})
is
\bse
&&n^{-1/2}\sumi
\Delta_i\frac{\wh\blam_2\{Z_i,\bb_0\trans\X_i,I(W_i\le Z_i),W_iI(W_i\le Z_i)\}}{\wh\lambda\{Z_i,\bb_0\trans\X_i,I(W_i\le Z_i),W_iI(W_i\le Z_i)\}}\\
&&\quad\otimes\left[\X_{li}-
\frac{\wh E\left\{\X_{li}
	Y_i(Z_i)\mid\bb_0\trans\X_i,I(W_i\le Z_i),W_iI(W_i\le Z_i)\right\}}
{\wh E\left\{Y_i(Z_i)\mid\bb_0\trans\X_i,I(W_i\le Z_i),W_iI(W_i\le Z_i)\right\}}\right]\\
&=&n^{-1/2}\sumi
\int\frac{\blam_2\{t,\bb_0\trans\X_i,I(W_i\le t),W_iI(W_i\le t)\}}{\lambda\{t,\bb_0\trans\X_i,I(W_i\le t),W_iI(W_i\le t)\}}\\
&&\otimes\left[\X_{li}-
\frac{E\left\{\X_{li}
	Y_i(t)\mid\bb_0\trans\X_i,I(W_i\le Z_i),W_iI(W_i\le Z_i)\right\}}
{E\left\{Y_i(t)\mid\bb_0\trans\X_i,I(W_i\le Z_i),W_iI(W_i\le Z_i)\right\}}\right]\\
&&\times dM_i\{t,\bb_0\trans\X_i,I(W_i\le t),W_iI(W_i\le t)\}+o_p(1)\\
&=&n^{-1/2}\sumi\bS\eff\{\Delta_i,Z_i,\X_i,I(W_i\le t),W_iI(W_i\le Z_i)\}+o_p(1).
\ese

Finally,
\bse
\T_4&=&n^{-1/2}\sumi \Delta_i
\left[\frac{\wh\blam_2\{Z_i,\bb_0\trans\X_i,I(W_i\le Z_i),W_iI(W_i\le Z_i)\}}{\wh\lambda\{Z_i,\bb_0\trans\X_i,I(W_i\le Z_i),W_iI(W_i\le Z_i)\}}\right.\\
&&\quad\left.-
\frac{\blam_2\{Z_i,\bb_0\trans\X_i,I(W_i\le Z_i),W_iI(W_i\le Z_i)\}}{\lambda\{Z_i,\bb_0\trans\X_i,I(W_i\le Z_i),W_iI(W_i\le Z_i)\}}\right]\\
&&\times\left[\frac{E\left\{\X_{li}
	Y_i(Z_i)\mid\bb_0\trans\X_i,I(W_i\le Z_i),W_iI(W_i\le Z_i)\right\}}
{E\left\{Y_i(Z_i)\mid\bb_0\trans\X_i,I(W_i\le Z_i),W_iI(W_i\le Z_i)\right\}}\right.\\
&&\left.-
\frac{\wh E\left\{\X_{li}
	Y_i(Z_i)\mid\bb_0\trans\X_i,I(W_i\le Z_i),W_iI(W_i\le Z_i)\right\}}
{\wh
	E\left\{Y_i(Z_i)\mid\bb_0\trans\X_i,I(W_i\le Z_i),W_iI(W_i\le Z_i)\right\}}\right]\\
&=&o_p\left(
n^{-1/2}\sumi \Delta_i
\left[\frac{E\left\{\X_{li}
	Y_i(Z_i)\mid\bb_0\trans\X_i,I(W_i\le Z_i),W_iI(W_i\le Z_i)\right\}}
{E\left\{Y_i(Z_i)\mid\bb_0\trans\X_i,I(W_i\le Z_i),W_iI(W_i\le Z_i)\right\}}\right.\right.\\
&&\left.\left.-\frac{\wh E\left\{\X_{li}
	Y_i(Z_i)\mid\bb_0\trans\X_i,I(W_i\le Z_i),W_iI(W_i\le Z_i)\right\}}
{\wh E\left\{Y_i(Z_i)\mid\bb_0\trans\X_i,I(W_i\le Z_i),W_iI(W_i\le Z_i)\right\}}\right]\right)\\
&=&o_p\left(n^{-1/2}\sumi \int Y_i(s)\lambda\{s,\bb_0\trans\X_i,I(W_i\le s),W_iI(W_i\le s)\}\right.\\
&&\left.\times\left[
\frac{E\left\{\X_{li}Y_i(s)\mid\bb_0\trans\X_i,I(W_i\le Z_i),W_iI(W_i\le Z_i)\right\}
}{E\left\{Y_i(s)\mid\bb_0\trans\X_i,I(W_i\le Z_i),W_iI(W_i\le Z_i)\right\}
}-\x_{li}\right]ds\right)+o_p(n^{1/2}h^2)\\
&=&o_p(1),
\ese
where the last equality holds because the integrand has mean zero
conditional on $\bb_0\trans\X$, and the second to last equality follows
with the same derivation of $\T_3$.
Using these results in (\ref{eq:main}), combined with the
results on
(\ref{eq:easy}),  the theorem holds.
\qed

\subsection{Proof of Theorem \ref{th:m}}\label{app:m}
We first analyze $\wh{m}_N(t,\bb\trans\x)$. We only need to analyze
$\sqrt{nh}\left\{\wh{m}_N(t,\bb\trans\x)-{m}_N(t,\bb\trans\x)\right\}$
because	$\sqrt{nh}\left\{\wh{m}_N(t,\wh\bb\trans\x)-{\wh
	m}_N(t,\bb\trans\x)\right\}=O_p(\sqrt{h})=o_p(1)$ based on Theorems
\ref{th:consistency} and \ref{th:eff}.

We expand
$\sqrt{nh}\left\{\wh{m}_N(t,\bb\trans\x)-{m}_N(t,\bb\trans\x)\right\}$
as
\be
&&\sqrt{nh}\left\{\wh{m}_N(t,\bb\trans\x)-{m}_N(t,\bb\trans\x)\right\}\n\\
&=&\sqrt{nh}\left\{e^{{\wh\Lambda_N}(t,\bb\trans\x)}-
e^{{\Lambda_N}(t,\bb\trans\x)}\right\}\int_{t}^{\tau}
e^{-{\Lambda_N}(s,\bb\trans\x)}ds\label{eq:mnormal1}\\
&&+\sqrt{nh}e^{{\Lambda_N}(t,\bb\trans\x)}\int_{t}^{\tau}
\left\{e^{-{\wh\Lambda_N}(s,\bb\trans\x)}-e^{-{\Lambda_N}(s,\bb\trans\x)}\right\}ds.\label{eq:mnormal2}\\
&&+\sqrt{nh}\left\{e^{{\wh\Lambda_N}(t,\bb\trans\x)}-
e^{{\Lambda_N}(t,\bb\trans\x)}\right\}\int_{t}^{\tau}
\left\{e^{-{\wh\Lambda_N}(s,\bb\trans\x)}-e^{-{\Lambda_N}(s,\bb\trans\x)}\right\}ds\label{eq:mnormal3}.
\ee
It is easy to see that the term in (\ref{eq:mnormal3}) satisfies
\bse
&&\sqrt{nh}\left\{e^{{\wh\Lambda_N}(t,\bb\trans\x)}-
e^{{\Lambda_N}(t,\bb\trans\x)}\right\}\int_{t}^{\tau}
\left\{e^{-{\wh\Lambda_N}(s,\bb\trans\x)}-
e^{-{\Lambda_N}(s,\bb\trans\x)}\right\}ds\\
&=&\sqrt{nh}O_p\{\wh\Lambda_N(t,\bb\trans\x)-\Lambda_N(t,\bb\trans\x)\}
\int_{t}^{\tau}e^{-\Lambda_N(s,\bb\trans\x)}
O_p\{\wh\Lambda_N(s,\bb\trans\x)-\Lambda_N(s,\bb\trans\x)\}ds\\
&=&O_p(\sqrt{nh})O_p\{h^4+(nh)^{-1}\}\\
&=&o_p(1),
\ese
by Condition \ref{assum:bandwidth}.

We inspect the terms in (\ref{eq:mnormal1}) and
(\ref{eq:mnormal2}).
For (\ref{eq:mnormal1}), based on Lemma
\ref{lem:pre},
\bse
&&\sqrt{nh}\left\{e^{{\wh\Lambda_N}(t,\bb\trans\x)}-
e^{\Lambda_N(t,\bb\trans\x)}\right\}\int_{t}^{\tau}
e^{-{\Lambda_N}(s,\bb\trans\x)}ds\\
&=&\sqrt{nh}e^{\Lambda_N(t,\bb\trans\x)}
\left(\wh\Lambda_N(t,\bb\trans\x)-\Lambda_N(t,\bb\trans\x)+
O_p[\{\wh\Lambda_N(t,\bb\trans\x)-\Lambda_N(t,\bb\trans\x)\}^2]\right)\\
&&\times\int_{t}^{\tau}e^{-{\Lambda_N}(s,\bb\trans\x)}ds\\
&=&\sqrt{nh}e^{\Lambda_N(t,\bb\trans\x)}
\left\{\wh\Lambda_N(t,\bb\trans\x)-\Lambda_N(t,\bb\trans\x)\right\}
\int_{t}^{\tau}e^{-{\Lambda_N}(s,\bb\trans\x)}ds+o_p(1),
\ese
where the last step uses Condition
\ref{assum:bandwidth}.

For (\ref{eq:mnormal2}), using Condition \ref{assum:bandwidth}
as well, we get
\bse
&&\sqrt{nh}e^{{\Lambda_N}(t,\bb\trans\x)}\int_{t}^{\tau}
\left\{e^{-{\wh\Lambda_N}(s,\bb\trans\x)}-e^{-{\Lambda_N}(s,\bb\trans\x)}\right\}ds\\
&=&\sqrt{nh}e^{{\Lambda_N}(t,\bb\trans\x)}\int_{t}^{\tau}
\left[\wh\Lambda_N(s,\bb\trans\x)-\Lambda_N(s,\bb\trans\x)+O_p\{(nh)^{-1}+h^4\}\right]e^{-\Lambda_N(s,\bb\trans\x)}ds\\
&=&\sqrt{nh}e^{{\Lambda_N}(t,\bb\trans\x)}\int_{t}^{\tau}
\left\{\wh\Lambda_N(s,\bb\trans\x)-\Lambda_N(s,\bb\trans\x)\right\}
e^{-\Lambda_N(s,\bb\trans\x)}ds+o_p(1).
\ese

Now combine the leading terms in (\ref{eq:mnormal1}) and
(\ref{eq:mnormal2}) and use the expansion of
$\wh\Lambda_N(t,\bb\trans\x)-\Lambda_N(t,\bb\trans\x)$ in Lemma
\ref{th:Lambda},
\be
&&\sqrt{nh}e^{\Lambda_N(t,\bb\trans\x)}\left\{\wh\Lambda_N(t,\bb\trans\x)-\Lambda_N(t,\bb\trans\x)\right\}\int_{t}^{\tau}
e^{-{\Lambda_N}(s,\bb\trans\x)}ds\n\\
&&+\sqrt{nh}e^{{\Lambda_N}(t,\bb\trans\x)}\int_{t}^{\tau}
\left\{\wh\Lambda_N(s,\bb\trans\x)-\Lambda_N(s,\bb\trans\x)\right\}
e^{-\Lambda_N(s,\bb\trans\x)}ds\n\\
&=&e^{\Lambda_N(t,\bb\trans\x)}\sumi\int_{0}^{\tau}\sqrt{\frac{h}{n}}
\frac{I\left\{\phi_n(r,\bb\trans\x)>0\right\}K_h(\bb\trans\X_i-\bb\trans\x)}
{f_{\bb\trans\X}(\bb\trans\x)E\{I(Z\ge
	r)\mid\bb\trans\x\}}\n\\
&&\times\left\{I(r<t)\int_{t}^{\tau}
e^{-\Lambda_N(s,\bb\trans\x)}ds+\int_{\max(r,t)}^{\tau}
e^{-\Lambda_N(s,\bb\trans\x)}ds\right\}dM_i(r,\bb\trans\x)
+o_p(1).\label{eq:mnormallong}
\ee

Note that $I\left\{\phi_n(r,\bb\trans\x)>0\right\}=1$ almost
surely and
according to Lemma \ref{lem:pre}
\bse
\frac{1}{n}\sumi hK_h^2(\bb\trans\X_i-\bb\trans\x)Y_i(r)=
f_{\bb\trans\X}(\bb\trans\x)E\{I(Z\ge
r)\mid\bb\trans\x\}\int K^2(u)du +o_p(1).
\ese
The leading term in
(\ref{eq:mnormallong}) converges to
$N\{0,\sigma^2_N(t,\bb\trans\x)\}$ uniformly by martingale
central limit
theorem,
where
\bse
\sigma^2_N(t,\bb\trans\x)&=&e^{2\Lambda_N(t,\bb\trans\x)}\frac{\int
	K^2(u)du}{f_{\bb\trans\X}(\bb\trans\x)}
\int_{0}^{\tau}\frac{\lambda_N(r,\bb\trans\x)}
{E\{I(Z\ge r)\mid\bb\trans\x\}}\\
&&\times\left\{I(r<t)\int_{t}^{\tau}
e^{-\Lambda_N(s,\bb\trans\x)}ds+\int_{\max(r,t)}^{\tau}
e^{-\Lambda_N(s,\bb\trans\x)}ds\right\}^2dr.
\ese
Therefore
$\sqrt{nh}\left\{\wh{m}_N(t,\bb\trans\x)-{m}_N(t,\bb\trans\x)\right\}\to
N\{0,\sigma^2_N(t,\bb\trans\x)\}$ uniformly for all $t$ and
$\bb\trans\x$.

Similarly, $\sqrt{nh}\left\{\wh{m}_T(t-w,\bb\trans\x,w)-{m}_T(t-w,\bb\trans\x,w)\right\}\to
N\{0,\sigma^2_T(t-w,\bb\trans\x,w)\}$ uniformly for all $t$ and
$\bb\trans\x$ where
\bse
&&\sigma^2_T(t-w,\bb\trans\x,w)\\
&=&e^{2\Lambda_T(t-w,\bb\trans\x,w)}\frac{\int
	K^2(u)du}{f_{\bb\trans\X,W}(\bb\trans\x,w)}
\int_{0}^{\tau}\frac{\lambda_T(r,\bb\trans\x,w)}
{E\{I(Z\ge r)\mid\bb\trans\x,w\}}\\
&&\times\left\{I(r<t-w)\int_{t-w}^{\tau}
e^{-\Lambda_T(s-w,\bb\trans\x,w)}ds\right.\\
&&\left.+\int_{\max(r,t-w)}^{\tau}
e^{-\Lambda_T(s-w,\bb\trans\x,w)}ds\right\}^2dr.
\ese

Furthermore, $\wh{m}\{t,\bb\trans\x,I(w \le t),wI(w\le t)\} = \wh{m}_T(t-w,\bb\trans\x)I(w\le t)+\wh{m}_N(t,\bb\trans\x)\{1-I(w\le t)\}$ and $\sqrt{nh}\left\{\wh{m}\{t,\wh\bb\trans\x,I(w \le t),wI(w\le t)\}-{m}\{t,\wh\bb\trans\x,I(w \le t),wI(w\le t)\}\right\} \to 
N[0, \sigma^2_m\{t, \wh\bb\trans\x, I(w\le t), wI(w\le t)\}]$ uniformly for all $t$ and
$\bb\trans\x$ where
$\sigma^2_m\{t, \wh\bb\trans\x, I(w\le t), wI(w\le t)\} = \sigma^2_T(t-w,\bb\trans\x)I(w\le t)+\sigma^2_N(t,\bb\trans\x)\{1-I(w\le t)\}$.

\qed

\section{Relaxation of the Complete Follow-up Assumption}\label{sec:relax}

To weaken the complete follow-up assumption that the event time is supported on $[0, \tau]$, we relax the  compact
support assumption on the event time, while allow a sample size dependent end of
the study time $\tau_n$.  Because the estimation and inference of
$\bb, \Lambda\{t,\bb\trans\x,I(W\le t),WI(W\le t)\}$ does not rely on the complete follow-up
assumption, hence under the weakened assumptions,
the same analysis as in
the main text  leads to the same results for $\wh\bb-\bb$
as in Theorems \ref{th:consistency} and \ref{th:eff}.

We assume
\be\label{eq:cs}
\int_{\tau_n}^\infty
e^{-\Lambda\{s,\bb\trans\x,I(w\le s),wI(w\le s)\}}ds=o(n^{\frac{1-2\nu}{4\nu}})e^{-\Lambda\{\tau_n-\tau_0,\bb\trans\x,I(w\le \tau_n-\tau_0),wI(w\le \tau_n-\tau_0)\}},
\ee
where $\tau_0$ is a small positive constant.
(\ref{eq:cs}) is equivalent to
\bse
\lim_{n\to\infty}\frac{
\tau_n'e^{\Lambda\{\tau_n-\tau_0,\bb\trans\x,I(w\le \tau_n-\tau_0),wI(w\le \tau_n-\tau_0)\}-\Lambda\{\tau_n,\bb\trans\x,I(w\le \tau_n),wI(w\le \tau_n)\}}}
{\frac{2\nu-1}{4\nu} n^{\frac{1-6\nu}{4\nu}}+n^{\frac{1-2\nu}{4\nu}}
\lambda\{\tau_n-\tau_0,\bb\trans\x,I(w\le \tau_n-\tau_0),wI(w\le \tau_n-\tau_0)\}\tau_n'}=0,
\ese
where $\tau_n' \equiv d\tau_n/dn$.
Clearly, a sufficient condition for (\ref{eq:cs}) to hold is that the
cumulative hazard function $\Lambda(t,\bb\trans\x)\ge
C(\bb\trans\x)t^{1+\alpha}$ for some constant $\alpha>0$, and
$\tau_n=n^\omega$ for some constant $\omega>0$. We point out two facts
about this condition. First, $\omega$ can be very small as long as it
is fixed, for example, we can set $\omega=1/2$. This implies that
the end of the study time $\tau_n$, although required to increase with sample
size $n$, can increase very slowly. Second, the tail
condition on the distribution family of $T$ given $\bb\trans\x$ is very weak.
For example, all sub-Gaussian distributions  satisfy this
requirement, this naturally includes all the Cox proportional hazard
model with the baseline hazard $t^c$ for $c>0$ and the Weibull family
with increasing risk, which are often used as baseline to generate
various survival models in practice.

Let
\bse
&&m^*\{t,\bb\trans\x,I(W\le t),WI(W\le t)\}\\
&=&e^{\Lambda\{t,\bb\trans\x,I(W\le t),WI(W\le t)\}}\int_t^{\tau_n}
e^{-\Lambda\{s,\bb\trans\x,I(W\le s),WI(W\le s)\}}ds.
\ese
Then, under Condition \ref{assum:bandwidth},
\be\label{eq:small}
&&|m\{t,\bb\trans\x,I(W\le t),WI(W\le t)\}-m^*\{t,\bb\trans\x,I(W\le t),WI(W\le t)\}|\n\\
&= &o(n^{\frac{1-2\nu}{4\nu}})=o\{(nh)^{-1/2}\}.
\ee
Note that our estimator described in the main text, when viewed as an estimator
for $m^*\{t,\bb\trans\x,I(W\le t),WI(W\le t)\}$,
satisfies the same properties as described
in Theorem \ref{th:m}, i.e.
$\sqrt{nh}[\wh
m\{t,\bb\trans\x,I(W\le t),WI(W\le t)\}-m^*\{t,\bb\trans\x,I(W\le t),WI(W\le t)\}]/\sigma_{m^*}\{t,\bb\trans\x,I(W\le t),WI(W\le t)\}\to
N(0,1)$
in distribution when $n\to\infty$, where
\bse
&&\sigma^2_{m^*}\{t,\wh\bb\trans\x,I(W \le t),WI(W\le t)\}\\
&=&I(w\le t)e^{2\Lambda_T(t-w,\bb\trans\x,w)}\frac{\int
K^2(u)du}{f_{\bb\trans\X,W}(\bb\trans\x,w)}
\int_{0}^{\tau_n}\frac{\lambda_T(r-w,\bb\trans\x,w)}
{E\{I(Z\ge r)\mid\bb\trans\x,w\}}\\
&&\quad\times\left\{I(r<t-w)\int_{t-w}^{\tau_n}
e^{-\Lambda_T(s-w,\bb\trans\x,w)}ds+\int_{\max(r,t-w)}^{\tau_n}
e^{-\Lambda_T(s-w,\bb\trans\x,w)}ds\right\}dr\\
&&+\{1-I(w\le t)\}e^{2\Lambda_N(t,\bb\trans\x)}\frac{\int
K^2(u)du}{f_{\bb\trans\X}(\bb\trans\x)}
\int_{0}^{\tau_n}\frac{\lambda_N(r,\bb\trans\x)}
{E\{I(Z\ge r)\mid\bb\trans\x\}}\\
&&\quad\times\left\{I(r<t)\int_{t}^{\tau_n}
e^{-\Lambda_N(s,\bb\trans\x)}ds+\int_{\max(r,t)}^{\tau_n}
e^{-\Lambda_N(s,\bb\trans\x)}ds\right\}dr.
\ese
In fact, the proof of Theorem \ref{th:m} goes through
by substituting $\tau$ with $\tau_n$.
In combination with (\ref{eq:small}) and the fact that $\sigma_{m^*}(t,\bb\trans\x)$ converges to
\bse
&&\sigma^2_m\{t,\wh\bb\trans\x,I(W \le t),WI(W\le t)\}\\
&=&I(w\le t)e^{2\Lambda_T(t-w,\bb\trans\x,w)}\frac{\int
K^2(u)du}{f_{\bb\trans\X,W}(\bb\trans\x,w)}
\int_{0}^{\infty}\frac{\lambda_T(r-w,\bb\trans\x,w)}
{E\{I(Z\ge r)\mid\bb\trans\x,w\}}\\
&&\quad\times\left\{I(r<t-w)\int_{t-w}^{\infty}
e^{-\Lambda_T(s-w,\bb\trans\x,w)}ds+\int_{\max(r,t-w)}^{\infty}
e^{-\Lambda_T(s-w,\bb\trans\x,w)}ds\right\}dr\\
&&+\{1-I(w\le t)\}e^{2\Lambda_N(t,\bb\trans\x)}\frac{\int
K^2(u)du}{f_{\bb\trans\X}(\bb\trans\x)}
\int_{0}^{\infty}\frac{\lambda_N(r,\bb\trans\x)}
{E\{I(Z\ge r)\mid\bb\trans\x\}}\\
&&\quad\times\left\{I(r<t)\int_{t}^{\infty}
e^{-\Lambda_N(s,\bb\trans\x)}ds+\int_{\max(r,t)}^{\infty}
e^{-\Lambda_N(s,\bb\trans\x)}ds\right\}dr.
\ese
This further leads to
$\sqrt{nh}[\wh m\{t,\wh\bb\trans\x,I(W \le t),WI(W\le t)\}-m\{t,\wh\bb\trans\x,I(W \le t),WI(W\le t)\}]\to N[0,\sigma_m^2\{t,\wh\bb\trans\x,I(W \le t),WI(W\le t)\}]$ in distribution when $n\to\infty$.


\end{document}